\documentstyle[12pt]{article}
\newcommand{\svec}[1]{ \stackrel{\rightarrow}{#1} }
\newcommand{\dvec}[1]{ \stackrel{\leftrightarrow}{#1} }
\newcommand{\lvec}[1]{ \stackrel{\leftarrow}{#1} }
\newcommand{\swav}[1]{ \stackrel{\sim}{#1} }
\newcommand{\tdot}[1]{ \stackrel{\cdot}{#1} }
\newcommand{\uhat}{ \hat U }
\newcommand{\ehat}{ \hat U_{\epsilon} }
\newcommand{\mhat}[1]{ \hat U_{\epsilon_{#1}} }
\newcommand{\define}{ \stackrel{\triangle}{=} }

\begin{document}
\title{\bf Gauge Theory of Gravity }
\author{{Ning Wu}
\thanks{email address: wuning@heli.ihep.ac.cn}
\\
\\
{\small Institute of High Energy Physics, P.O.Box 918-1,
Beijing 100039, P.R.China}
\thanks{mailing address}}
\maketitle
\vskip 0.8in

~~

\vskip 0.8in

\begin{abstract}
The quantum gravity is formulated based on principle of local
gauge invariance.  The model discussed in this paper has local
gravitational gauge symmetry and gravitational field is
represented by gauge field. Path integral quantization of the
theory is discussed in the paper. A strict proof on the
renormalizability of the theory is also given. In leading order
approximation, the gravitational gauge field theory gives out
classical Newton's theory of gravity. In first order approximation
and for vacuum, the gravitational gauge field theory gives out
Einstein's general theory of relativity. Using this new quantum
gauge theory of gravity, we can explain some important puzzles of
Nature. This quantum gauge theory of gravity is a renormalizable
quantum theory.
\\

\end{abstract}

\newpage

\Roman{section}

\section{Introduction}

Gravity is an ancient topic in science. In ancient times, human has
known the existence of weight. Now we know that it is the
gravity between an object and earth. In 1686, Isaac Newton
published his epoch-making book {\it MATHEMATICAL
 PRINCIPLES OF NATURAL PHILOSOPHY}. In this
book, through studying the motion of planet in solar system, he
found that gravity obeys the inverse square law\cite{01}. The
Newton's classical theory of gravity is kept unchanged until 1916. At
that year, Einstein published his epoch-making paper on General
Relativity\cite{02,03}. In this great work, he founded a
relativistic theory on gravity, which is based on principle of
general relativity and equivalence principle. Newton's classical
theory for gravity appears as
a classical limit of  general relativity.\\

One of the biggest revolution in human kind in the last century is
the foundation of quantum theory. The quantum hypothesis was first
introduced into physics by Max Plank in 1900. Inspired by his
quantum hypothesis, Albert Einstein used it to explain the
photoelectric effect successfully and Niels Bohr used it to
explain the positions of spectral lines of Hydrogen. In 1923, Louis
de Broglie proposed the principle of wave-particle duality. In the
years 1925-1926, Werner Heisenberg, Max Born, Pascual Jordan and
Wolfgang Pauli develop matrix mechanics, and in 1926, Erwin
Schroedinger develop wave mechanics. Soon after, relativistic
quantum mechanics and quantum field theory are proposed by
P.A.M.Dirac, Oskar Klein, Walter Gordan and others. \\

In 1921, H.Weyl introduced the concept of gauge transformation
into physics\cite{d01,d05}, which is one of the most important
concepts in modern physics, though his original theory is not
successful. Later, V.Fock, H.Weyl and W.Pauli found that quantum
electrodynamics is a gauge invariant theory\cite{d02,d03,d04}. In
1954, Yang and Mills proposed non-Abel gauge field theory\cite{1}.
This theory was soon applied to elementary particle physics.
Unified electroweak theory\cite{2,3,4} and quantum chromodynamics
are all based on gauge field theory. The predictions of unified
electroweak theory have been confirmed in a large number of
experiments, and the intermediate gauge bosons $W^{\pm}$ and $Z^0$
which are predicted by unified electroweak model are also found in
experiments. The $U(1)$ part of the unified electroweak model,
quantum electrodynamics, now become one of the most accurate and
best-tested theories of modern physics. All these achievements of
gauge field theories suggest that gauge field theory is a
fundamental theory that describes fundamental interactions. Now,
it is generally  believed that four kinds of fundamental
interactions in Nature are all gauge interactions and they can be
described by gauge field theory. From theoretical point of view,
the principle of local gauge invariance plays a fundamental role
in particle's interaction theory.  \\

In 1916, Albert Einstein points out that quantum effects must
lead to modifications in the theory of general relativity\cite{5}.
Soon after the foundation of  quantum mechanics,
physicists try to found a theory that could describe the
quantum behavior of the full gravitational field. In the 70
years attempts, physicists have found two theories based
on quantum mechanics that attempt to unify general
relativity and quantum mechanics, one is canonical quantum
gravity and another is superstring theory. But for quantum
field theory, there are different kinds of mathematical
infinities that naturally occur in quantum descriptions of
fields. These infinities should be removed by the technique
of perturbative renormalization. However, the perturbative
renormalization does not work for the quantization of Einstein's
theory of gravity, especially in canonical quantum gravity. In
superstring theory, in order to make perturbative renormalization
to work, physicists have to introduce six extra dimensions. But
up to now, none of the extra dimensions have been observed.
To found a consistent theory that can unify general relativity and
quantum mechanics is a long dream for physicists. \\

The "relativity revolution" and the "quantum revolution" are among
the greatest successes of twentieth century physics, yet two
theories appears to be fundamentally incompatible. General
relativity remains a purely classical theory which describes the
geometry of space and time as smooth and continuous, on the
contrary, quantum mechanics divides everything into discrete
quanta. The underlying theoretical incompatibility between two
theories arises from the way  that they treat the geometry of
space and time. This situation makes some physicists still wonder
whether quantum theory is a truly fundamental theory of Nature, or
just a convenient description of some aspects of the microscopic
world. Some physicists even consider the twentieth century as the
century of the incomplete revolution. To set up a consistent
quantum theory of gravity is considered to be the last challenge
of quantum theory\cite{50,51}. In other words, combining general
relativity with quantum mechanics is considered to be the last
hurdle to be overcome in the "quantum revolution".
\\

Gauge treatment of gravity was suggested immediately after the
gauge theory birth itself\cite{b1,b2,b3}. In the traditional gauge
treatment of gravity, Lorentz group is localized, and the
gravitational field is not represented by gauge
potential\cite{b4,b5,b6}. It is represented by metric field. (In
this paper, we will find that metric field and gauge field are not
independent. We can use gauge field to determine metric field. )
The theory has beautiful mathematical forms, but up to now, its
renormalizability is not proved. In other words,
it is conventionally considered to be non-renormalizable. \\

I will not talk too much on the history of quantum gravity and
the incompatibilities between quantum mechanics and general
relativity here. Materials on these subject can be widely found
in literatures. Now we want to ask that, except for traditional
canonical quantum gravity and superstring theory, whether exists
another approach to set up a fundamental theory, in which general
relativity and quantum mechanics are compatible. \\

Recently, some new attempts were proposed to use Yang-Mills theory
to reformulate quantum gravity\cite{c1,c2,c3,c4}. In these new
approaches, the importance of gauge fields is emphasized. Some
physicists also try to use gauge potential to represent
gravitational field, some suggest that we should pay more
attention on translation group. In this paper, a completely new
attempt is proposed. In other words, we try to use completely new
notions and completely new methods to formulate quantum gravity.
Our goal is try to set up a renormalizable quantum gauge theory of
gravity. Its relation to the traditional quantum gravity is not
studied in this paper, which is an independent work that will be
done in the near future. Maybe this new formulation of quantum
gravity is equivalent to the traditional formulation, maybe they
are not equivalent. The main goal that we hope to formulate this
new quantum gauge theory of gravity is not to deny traditional
quantum gravity, but to prove the renormalizability of quantum
gravity, for the renormalizability of quantum gravity is easy to
be proved in this new formulation. It is known that the
mathematical formulation of traditional quantum gravity  is quite
different from that of the new quantum gauge theory of gravity
which is formulated in this paper, but it does not mean that they
are essentially  different in physics. In a meaning, maybe the
present situation is somewhat similar to that of the quantum
mechanics. It is well know that quantum mechanics can be
formulated in different representations, such as Schrodinger
representation, Heisenberg representation, $\cdots$.  Though
mathematical formulations of quantum mechanics are quite different
in different representations, they are essentially the same in
physics. Quantum gravity belongs to quantum mechanics, we can
imagine that it will of course have many different
representations. If some people can prove that the traditional
quantum gravity is equivalent to this new renormalizable quantum
gauge theory of gravity, then the renormalizability of the
traditional quantum gravity will automatically hold. If this is
true, then two kinds of quantum gravity can be regarded as two
different representations of quantum gravity. Collaborating with
Prof. Zhan Xu and Prof. Dahua Zhang, we have found the differential
geometrical formulation of the gravitational gauge theory of
gravity which is formulated in this paper\cite{c44}. The relation
between traditional quantum gravity and gravitational gauge theory
of gravity is under studying now.
\\

As we have mentioned above, gauge field theory provides a
fundamental tool to study fundamental interactions. In this paper,
we will use this tool to study quantum gravity. We will use a
completely new language to express the quantum theory of gravity.
In order to do this, we first need to introduce some
transcendental foundations of this new theory, which is the most
important thing to formulate the whole theory. Then we will
discuss a new kind of non-Abel gauge group, which will be the
fundamental symmetry of quantum gravity. For the  sake of
simplicity, we call this group gravitational gauge group.  After
that, we will construct a Lagrangian which has local gravitational
gauge symmetry. In this Lagrangian, gravitational field appears as
the gauge field of the gravitational gauge symmetry. Then we will
discuss the gravitational interactions between scalar field (or
Dirac field or vector field ) and gravitational field. Just as
what Albert Einstein had ever said in 1916 that quantum effects
must lead to modifications in the theory of general relativity,
there are indeed quantum modifications in this new quantum gauge
theory of gravity. In other words, the local gravitational gauge
symmetry requires some additional interaction terms other than
those given by general relativity. This new quantum  theory of
gravity can even give out an exact relationship between
gravitational fields and space-time metric in generally
relativity. The classical limit of this new quantum theory will
give out classical Newton's theory of gravity and general
relativity. In other words, the leading order approximation of the
new theory gives out classical Newton's theory of gravity, the
first order approximation of the new theory gives out Einstein's
general theory of relativity. This can be regarded as the first
test of the new theory. Then we will discuss quantization of
gravitational gauge field. Something most important is that this
new quantum theory of gravity is a renormalizable theory. A formal
strict proof on the renormalizability of this new quantum theory
of gravity is given in this paper. After that, we will discuss
some theoretical predictions of this new quantum theory of
gravity. In this chapter, we will find that some puzzles which can
not be explained in traditional theory can be explained by this
new quantum gauge theory of gravity. These explanations can be
regarded as the second test of the new theory. I hope that the
effort made in this paper will be beneficial to our understanding
on the quantum aspects of gravitational field. The relationship
between this new quantum theory of gravity and traditional
canonical quantum gravity or superstring theory is not study now,
and I hope that this work will be done in the near future. Because
the new quantum theory of gravity is logically independent of
traditional quantum gravity, we need not discuss traditional
quantum gravity first. Anyone who is familiar with traditional
non-Abel gauge field theory can understand the whole paper. In
other words, readers who never study anything on traditional
quantum gravity can understand this new quantum theory of gravity.
Now, let's begin our long journey to the realm of logos. \\

\section{The Transcendental Foundations}

It is known that action principle is one of the most important
fundamental principle in quantum field theory. Action principle
says that any quantum system is described by an action. The action
of the system contains all interaction information, contains all
information of the fundamental dynamics. The least of the action
gives out the classical equation of motion of a field. Action
principle is widely used in quantum field theory. We will accept
it as one of the most fundamental principles in this new quantum
theory of gravity. The rationality of action principle will not be
discussed here, but it is well know that the rationality of the
action principle has already been tested by a huge amount of
experiments. However, this principle is not a special principle
for quantum gravity, it is a ubiquitous principle in quantum field
theory. Quantum gravity discussed in this paper is a kind of
quantum field theory, it's naturally to accept action principle as
one of its fundamental principles. \\

We need a special fundamental principle to introduce quantum
gravitational field, which should be the foundation of all kinds
of fundamental interactions in Nature. This special principle is
gauge principle. In order to introduce this important principle,
let's first study some fundamental laws in some fundamental
interactions other than gravitational interactions. We know that,
except for gravitational interactions, there are strong
interactions, electromagnetic interactions and weak interactions,
which are described by quantum chromodynamics,  quantum
electrodynamics and unified electroweak theory respectively. Let's
study these three fundamental interactions one by one. \\

Quantum electrodynamics (QED) is one of the most successful theory
in physics which has been tested by most accurate experiments. Let's
study some logic in QED. It is know that QED theory has $U(1)$
gauge symmetry. According to Noether's theorem, there is a
conserved charge corresponding to the global $U(1)$ gauge
transformations. This conserved charge is just the ordinary
electric charge. On the other hand, in order to keep local $U(1)$
gauge symmetry of the system, we had to introduce a $U(1)$ gauge
field, which transmits electromagnetic interactions. This $U(1)$
gauge field is just the well-know electromagnetic field. The
electromagnetic interactions between charged particles and the
dynamics of electromagnetic field are completely determined by the
requirement of local $U(1)$ gauge symmetry. The source of this
electromagnetic field is just the conserved charge which is given
by Noether's theorem. After quantization of the field, this
conserved charge becomes the generator of the quantum $U(1)$ gauge
transformations. The quantum $U(1)$ gauge transformation has only
one generator, it has no generator other than the quantum electric
charge.
\\

Quantum chromodynamics (QCD) is a prospective fundamental theory
for strong interactions. QCD theory has $SU(3)$ gauge symmetry.
The global $SU(3)$ gauge symmetry of the system gives out
conserved charges of the theory, which are usually called color
charges. The local $SU(3)$ gauge symmetry of the system requires
introduction of a set of $SU(3)$ non-Abel gauge fields, and the
dynamics of non-Abel gauge fields and the strong interactions
between color charged particles and gauge fields are completely
determined by the requirement of local $SU(3)$ gauge symmetry of
the system. These $SU(3)$ non-Abel gauge fields are usually call
gluon fields. The sources of gluon fields are color charges. After
quantization, these color charges become generators of quantum
$SU(3)$ gauge transformation. Something which is different from
$U(1)$ Abel gauge symmetry is that gauge fields themselves carry
color charges. \\

Unified electroweak model is the fundamental theory for
electroweak interactions. Unified electroweak model is usually
called the standard model. It has $SU(2)_L \times U(1)_Y$
symmetry. The global $SU(2)_L \times U(1)_Y$ gauge symmetry of the
system also gives out conserved charges of the system, The
requirement of local $SU(2)_L \times U(1)_Y$ gauge symmetry needs
introducing a set of $SU(2)$ non-Abel gauge fields and one $U(1)$
Abel gauge field. These gauge fields transmit weak interactions
and electromagnetic interactions, which correspond to intermediate
gauge bosons $W^{\pm}$, $Z^0$ and photon. The sources of these
gauge fields are just the conserved Noether charges. After
quantization, these conserved charges become generators of
quantum $SU(2)_L \times U(1)_Y$ gauge transformation. \\

QED, QCD and the standard model are three fundamental theories
of three kinds of fundamental interactions. Now we want to
summarize some fundamental laws of Nature on interactions.
Let's first ruminate over above discussions. Then we will find
that our formulations on three different fundamental interaction
theories are almost completely the same, that is the global gauge
symmetry of the system gives out conserved Noether charges,
in order to keep local gauge symmetry of the system, we have to
introduce gauge field or a set of gauge fields, these gauge
fields transmit interactions, and the source of these gauge fields are
the conserved charges and these conserved Noether charges
become generators of quantum gauge transformations after
quantization. These will be the main content of gauge principle.
\\

Before we formulate gauge principle formally, we need to study
something more on symmetry. It is know that not all symmetries can
be localized, and not all symmetries can be regarded as gauge
symmetries and have corresponding gauge fields. For example, time
reversal symmetry, space reflection symmetry, $\cdots$ are those
kinds of symmetries. We can not find any gauge fields or
interactions which correspond to these symmetries. It suggests
that symmetries can be divided into two different classes in
nature. Gauge symmetry is a special kind of symmetry which has the
following properties: 1) it can be localized; 2) it has some
conserved charges related to it; 3) it has a kind of interactions
related to it; 4) it is usually a continuous symmetry. This
symmetry can completely determine the dynamical behavior of a kind
of fundamental interactions. For the sake of simplicity, we call
this kind of symmetry dynamical symmetry or gauge symmetry. Any
kind of fundamental interactions has a gauge  symmetry
corresponding to it. In QED, the $U(1)$ symmetry is a gauge
symmetry, in QCD, the color $SU(3)$ symmetry is a gauge symmetry
and in the standard model, the $SU(2)_L \times U(1)_Y$ symmetry is
also a gauge symmetry. The gravitational gauge symmetry which we
will discuss below is also a kind of gauge symmetry. The time
reversal symmetry and space reflection symmetry are not gauge
symmetries. Those global symmetries which can not be localized are
not gauge symmetries either. Gauge symmetry is a fundamental
concept for gauge principle. \\

Gauge principle can be formulated as follows: Any kind of
fundamental interactions has a gauge symmetry corresponding to it;
the gauge symmetry completely determines the forms of
interactions. In principle, the gauge principle has the following
four different contents:
\begin{enumerate}

\item {\bf Conservation Law:} the global gauge symmetry  gives out
conserved current and conserved charge;

\item {\bf Interactions:} the requirement of the local gauge
symmetry requires introduction of gauge field or a set of gauge
fields; the interactions between gauge fields and matter fields
are completely determined by the requirement of local gauge
symmetry; these gauge fields transmit the corresponding
interactions;

\item {\bf Source:} qualitative speaking,
the conserved charge given by global gauge  symmetry is the source
of gauge field; for non-Abel gauge field, gauge field is also the
source of itself;

\item {\bf Quantum Transformation:} the conserved charges given
by global gauge symmetry become generators of quantum gauge
transformation after quantization, and for this kind of of
interactions, the quantum transformation can not have generators
other than quantum conserved charges given by global gauge
symmetry.

\end{enumerate}
It is known that conservation law is the objective origin of gauge
symmetry, so gauge symmetry is the exterior exhibition of the
interior conservation law. The conservation law is the law that
exists in fundamental interactions, so fundamental interactions
are the logic precondition and foundation of the conservation law.
Gauge principle tells us how to study conservation law and
fundamental interactions through symmetry. Gauge principle is one
of the most important transcendental fundamental principles for
all kinds of fundamental interactions in Nature; it reveals the
common nature of all kinds of fundamental interactions in Nature.
It is also the transcendental foundation of the quantum gravity
which is formulated in this paper. It will help us to select the
gauge symmetry for quantum gravitational theory and help us to
determine the Lagrangian of the system. In a meaning, we can say
that without gauge principle, we can not set up this new
renormalizable quantum gauge theory of gravity.
\\

Another transcendental principle that widely used in quantum field
theory is the microscopic causality principle. The central idea of the
causality principle is that any changes in the objective world have
their causation. Quantum field theory is a relativistic theory. It is know
that, in the special theory of relativity, the limit
spread speed is the speed
of light. It means that, in a definite reference system, the limit
spread speed
of the causation of some changes is the speed of light. Therefore,
the special theory of relativity exclude the possibility of the existence
of any kinds of non-local interactions in a fundamental theory. Quantum
field theory inherits this basic idea and calls it the microscopic causality
principle. There are several expressions of the microscopic causality
principle in quantum field theory. One expression say that two events
which happen at the same time but in different space position are two
independent events. The mathematical formulation for microscopic
causality principle is that
$$
[ O_1(\svec{x},t)~~,~~O_2(\svec{y},t) ] =0,
\eqno{(2.1)}
$$
when $\svec{x} \not= \svec{y}$. In the above relation,
$ O_1(\svec{x},t)$ and $ O_2(\svec{y},t)$ are two different
arbitrary local bosonic operators. Another important expression
of the microscopic causality principle is that, in the Lagrangian
of a fundamental theory, all operators appear in the same point
of space-time. Gravitational interactions are a kind of physical
interactions, the fundamental theory of gravity should also obey
microscopic causality principle. This requirement is realized
in the construction of the Lagrangian for gravity. We will require
that all field operators in the Lagrangian should be at the same
point of space-time. \\

Because quantum field theory is a kind of relativistic theory, it
should obey some fundamental principles of the special theory of
relativity, such as principle of special relativity and principle
of invariance of light speed. These two principles conventionally
exhibit themselves through Lorentz invariance. So, in constructing
the Lagrangian of the quantum theory of gravity, we require that
it should have Lorentz invariance. This is also a transcendental
requirement for the quantum theory of gravity. But what we treat
here that is different from that of general relativity is that we
do not localize Lorentz transformation. Because gauge principle
forbids us to localize Lorentz transformation, asks us only to
localize gravitational gauge transformation. We will discuss this
topic in details later. However, it is important to remember that
global Lorentz invariance of the Lagrangian is a fundamental
requirement. The requirement of global Lorentz invariance can also
be treated as a transcendental principle of the quantum theory
of gravity.  \\

It is well-known that two transcendental principles of general
relativity are principle of general relativity and principle of
equivalence. It should be stated that, in the new gauge theory of
gravity, the principle of general relativity appears in  another
way, that is, it realized itself through local gravitational gauge
symmetry. From mathematical point of view, the local gravitational
gauge invariance is just the general covariance in general
relativity. In the new quantum theory of gravity, principle of
equivalence plays no role. In other words, we will not accept
principle of equivalence as a transcendental principle of the new
quantum theory of gravity, for gauge principle is enough for us to
construct quantum theory of gravity. We will discuss something
more about principle of equivalence later. \\

\section{Gravitational Gauge Group}

Before we start our mathematical formulation of gravitational
gauge theory, we have to determine  which group is the
gravitational gauge group, which is the starting point of the
whole theory. It is know that, in the traditional quantum gauge
theory of gravity, Lorentz group is localized. We will not follow
this way, for it contradicts with gauge principle. Now, we use
gauge principle to determine which group is the exact group for
gravitational gauge theory. \\

Some of the most important properties of gravity can be seen
from Newton's classical gravity. In this classical theory of
gravity, gravitational force between two point objects is given by:
$$
f = G \frac{m_1 m_2}{r^2}
\eqno{(3.1)}
$$
with $m_1$ and $m_2$ masses of two objects, $r$ the distance
between two objects. So, gravity is proportional to the masses of
both objects, in other words, mass is the source of gravitational
field. In general relativity, Einstein's gravitational equation is
the equation which gives out the relation between energy-momentum
tensor and space-time curvature, which is essentially the relation
between energy-momentum tensor and gravitational field. In the
Einstein's gravitational equation, energy-momentum is treated as
the source of gravity. This opinion is inherited in the new quantum
theory gauge of gravity. In other words, the starting point of
the new quantum gauge theory of gravity is that the energy-momentum is
the source of gravitational field. According to rule 3 and rule 1
of gauge principle, we know that, energy-momentum is the conserved
charges of the corresponding global symmetry, which is just the
symmetry for gravity. According to quantum field theory,
energy-momentum is the conserved charge of global space-time
translation, the corresponding conserved current is
energy-momentum tensor. Therefore, the global space-time
translation is the global gravitational gauge transformation.
According to rule 4, we know that, after quantization, the
energy-momentum operator becomes the generator of gravitational
gauge transformation. It also states that, except for energy-momentum
operator, there is no other generator for gravitational
gauge transformation, such as, angular momentum operator $M_{\mu
\nu}$ can not be the generator of gravitational gauge
transformation. This is the reason why we do not localize Lorentz
transformation in this new quantum gauge theory of gravity, for
the generator of Lorentz transformation is not energy-momentum
operator. According to rule 2 of gauge principle, the
gravitational interactions will be completely determined by the
requirement of the local gravitational gauge symmetry. These are
the basic ideas of the new quantum gauge theory of gravity, and
they are completely deductions of gauge principle. \\

We know that the generator of Lorentz group is angular momentum
operator $M_{\mu \nu}$. If we localize Lorentz group, according to
gauge principle, angular momentum will become source of a new
filed, which transmits direct spin interactions. This kind of
interactions does not belong to traditional Newton-Einstein
gravity. It is a new kind of interactions. Up to now, we do not
know that whether this kind of interactions exists in Nature or
not. Besides, spin-spin interaction is a kind of
non-renormalizable interaction. In other words, a quantum theory
which contains spin-spin interaction is a non-renormalizable
quantum theory. For these reasons, we will not localize Lorentz
group in this paper. We only localize translation group in this
paper. We will find that go along this way, we can set up a
consistent quantum gauge theory of gravity which is
renormalizable. In other words, only localizing space-time
translation group is enough for us to set up a
consistent quantum gravity. \\

From above discussions, we know that, from mathematical point of
view,  gravitational gauge transformation is the inverse
transformation of  space-time translation, and gravitational gauge
group is space-time translation group. Suppose that there is an
arbitrary function $\phi(x)$ of space-time coordinates $x^{\mu}$.
The global space-time translation is:
$$
x^{\mu} \to x'^{\mu} = x^{\mu} + \epsilon^{\mu}.
\eqno{(3.2)}
$$
The corresponding transformation for function $\phi(x)$ is
$$
\phi(x) \to \phi'(x')=\phi(x) = \phi(x' - \epsilon).
\eqno{(3.3)}
$$
According to Taylor series expansion, we have:
$$
\phi(x - \epsilon) = \left(1 + \sum_{n=1}^{\infty} \frac{(-1)^n}{n!}
\epsilon ^{\mu_1} \cdots \epsilon^{\mu^n}
\partial_{\mu_1} \cdots \partial_{\mu_n}  \right) \phi(x),
\eqno{(3.4)}
$$
where
$$
\partial_{\mu_i}=\frac{\partial}{\partial x^{\mu_i}}.
\eqno{(3.5)}
$$
\\

Let's define a special exponential operation here. Define
$$
E^{a^{\mu} \cdot b_{\mu}} \define 1 +
\sum_{n=1}^{\infty} \frac{1}{n!}
a^{\mu_1} \cdots a^{\mu_n} \cdot
b_{\mu_1} \cdots b_{\mu_n}.
\eqno{(3.6)}
$$
This definition is quite different from that of ordinary
exponential function. In general cases, operators $a^{\mu}$ and
$b_{\mu}$ do not commutate each other, so
$$
E^{a^{\mu} \cdot b_{\mu}} \not=
E^{b_{\mu} \cdot a^{\mu}},
\eqno{(3.7)}
$$
$$
E^{a^{\mu} \cdot b_{\mu}} \not=
e^{a^{\mu} \cdot b_{\mu}},
\eqno{(3.8)}
$$
where $ e^{a^{\mu} \cdot b_{\mu}}$ is the ordinary exponential
function whose definition is
$$
e^{a^{\mu} \cdot b_{\mu}} \equiv 1 +
\sum_{n=1}^{\infty} \frac{1}{n!}
(a^{\mu_1} \cdot b_{\mu_1}) \cdots
(a_{\mu_n} \cdot b_{\mu_n}).
\eqno{(3.9)}
$$
If operators $a^{\mu}$ and $b_{\mu}$ commutate each other, we will
have
$$
E^{a^{\mu} \cdot b_{\mu}}  =
E^{b_{\mu} \cdot a^{\mu}},
\eqno{(3.10)}
$$
$$
E^{a^{\mu} \cdot b_{\mu}} =
e^{a^{\mu} \cdot b_{\mu}}.
\eqno{(3.11)}
$$
The translation operator $\ehat$ can be defined by
$$
\ehat \equiv 1 + \sum_{n=1}^{\infty} \frac{(-1)^n}{n!}
\epsilon^{\mu_1} \cdots \epsilon^{\mu_n}
\partial_{\mu_1} \cdots \partial_{\mu_n}.
\eqno{(3.12)}
$$
Then we have
$$
\phi(x - \epsilon) = ( \uhat_{ \epsilon} \phi(x)).
\eqno{(3.13)}
$$
In order to have a good form which is similar to ordinary gauge
transformation operators, the form of $\ehat$ can also be
written as
$$
\ehat =  E^{- i \epsilon^{\mu} \cdot \hat{P}_{\mu}},
\eqno{(3.14)}
$$
where
$$
\hat{P}_{\mu} = -i \frac{\partial}{\partial x^{\mu}}.
\eqno{(3.15)}
$$
$\hat{P}_{\mu}$ is just the energy-momentum operator in
space-time coordinate space. In the definition of $\ehat$ of
eq.(3.14), $\epsilon^{\mu}$ can be independent of of space-time
coordinate or a function of space-time coordinate, in a ward, it
can be any functions of space time coordinate $x$.  \\

Some operation properties of translation operator $\ehat$ are
summarized below.
\begin{enumerate}

\item Operator $\ehat$ translate the space-time point of a field
from $x$ to $x - \epsilon$,
$$
\phi(x- \epsilon) = (\ehat \phi(x)),
\eqno{(3.16)}
$$
where $\epsilon^{\mu}$ can be any function of space-time
coordinate. This relation can also be regarded as the definition
of the translation operator $\ehat$.

\item If $\epsilon$ is a function of space-time coordinate, that is
$\partial_{\mu} \epsilon^{\nu} \not= 0$, then
$$
\ehat =  E^{- i \epsilon^{\mu} \cdot \hat{P}_{\mu}}
\not= E^{- i \hat{P}_{\mu} \cdot \epsilon^{\mu}  },
\eqno{(3.17)}
$$
and
$$
\ehat =  E^{- i \epsilon^{\mu} \cdot \hat{P}_{\mu}}
\not= e^{- i \epsilon^{\mu} \cdot \hat{P}_{\mu}}.
\eqno{(3.18)}
$$
If $\epsilon$ is a constant, that is
$\partial_{\mu} \epsilon^{\nu} = 0$, then
$$
\ehat =  E^{- i \epsilon^{\mu} \cdot \hat{P}_{\mu}}
= E^{- i \hat{P}_{\mu} \cdot \epsilon^{\mu}  },
\eqno{(3.19)}
$$
and
$$
\ehat =  E^{- i \epsilon^{\mu} \cdot \hat{P}_{\mu}}
= e^{- i \epsilon^{\mu} \cdot \hat{P}_{\mu}}.
\eqno{(3.20)}
$$

\item Suppose that $\phi_1(x)$ and $\phi_2(x)$ are two arbitrary
functions of space-time coordinate, then we have
$$
\left( \ehat (\phi_1(x) \cdot \phi_2(x))\right)=
(\ehat \phi_1(x)) \cdot (\ehat \phi_2(x))
\eqno{(3.21)}
$$

\item Suppose that $A^{\mu}$ and $B_{\mu}$ are two arbitrary
operators in Hilbert space, $\lambda$ is an arbitrary ordinary
c-number which is commutate with operators $A^{\mu}$ and
$B_{\mu}$, then we have
$$
\frac{\rm d}{{\rm d} \lambda}
E^{\lambda A^{\mu} \cdot B_{\mu}} =
A^{\mu} \cdot E^{\lambda A^{\mu} \cdot B_{\mu}} \cdot B_{\mu}.
\eqno{(3.22)}
$$

\item Suppose that $\epsilon$ is an arbitrary function of space-time
coordinate, then
$$
(\partial_{\mu} \ehat) =
-i (\partial_{\mu} \epsilon^{\nu} ) \ehat \hat{P}_{\nu}.
\eqno{(3.23)}
$$

\item Suppose that $A^{\mu}$ and $B_{\mu}$ are two arbitrary
operators in Hilbert space, then
$$
tr( E^{ A^{\mu} \cdot B_{\mu}} E^{ - B_{\mu} \cdot A^{\mu}} )=
tr {\bf I},
\eqno{(3.24)}
$$
where $tr$ is the trace operation and ${\bf I}$ is the unit operator
in the Hilbert space.

\item Suppose that $A^{\mu}$, $B_{\mu}$ and $C^{\mu}$ are three
operators in Hilbert space, but operators $A^{\mu}$ and $C^{\nu}$
commutate each other:
$$
[ A^{\mu}~~,~~C^{\nu} ] = 0,
\eqno{(3.25)}
$$
then
$$
tr( E^{ A^{\mu} \cdot B_{\mu}} E^{ B_{\nu} \cdot C^{\nu}} )=
tr( E^{ (A^{\mu}+ C^{\mu}) \cdot B_{\mu}}).
\eqno{(3.26)}
$$

\item Suppose that $A^{\mu}$, $B_{\mu}$ and $C^{\mu}$ are three
operators in Hilbert space, they satisfy
$$
\begin{array}{rcl}
\lbrack A^{\mu}~~,~~C^{\nu} \rbrack & = &  0 , \\
\lbrack B_{\mu}~~,~~C^{\nu} \rbrack & = & 0,
\end{array}
\eqno{(3.27)}
$$
then
$$
E^{ A^{\mu} \cdot B_{\mu}} E^{ C^{\nu} \cdot B_{\nu}}  =
E^{ (A^{\mu}+ C^{\mu}) \cdot B_{\mu}}.
\eqno{(3.28)}
$$

\item Suppose that $A^{\mu}$, $B_{\mu}$ and $C^{\mu}$ are three
operators in Hilbert space, they satisfy
$$
\begin{array}{rcl}
\lbrack A^{\mu}~~,~~C^{\nu} \rbrack & = & 0, \\
\lbrack  \lbrack B_{\mu}~~,~~C^{\nu}
\rbrack ~~,~~ A^{\rho} \rbrack &=& 0,  \\
\lbrack  \lbrack B_{\mu}~~,~~C^{\nu}
\rbrack ~~,~~ C^{\rho} \rbrack  &=& 0,
\end{array}
\eqno{(3.29)}
$$
then,
$$
E^{ A^{\mu} \cdot B_{\mu}} E^{ C^{\nu} \cdot B_{\nu}}  =
E^{ (A^{\mu}+ C^{\mu}) \cdot B_{\mu}}
+ [E^{ A^{\mu} \cdot B_{\mu}} ~~,~~C^{\sigma}]
E^{ C^{\nu} \cdot B_{\nu}} B_{\sigma}.
\eqno{(3.30)}
$$

\item Suppose that $\mhat{1}$ and $\mhat{2}$ are two arbitrary
translation operators, define
$$
\mhat{3}=\mhat{2} \cdot \mhat{1},
\eqno{(3.31)}
$$
then,
$$
\epsilon_3^{\mu}(x) = \epsilon_2^{\mu}(x) +
\epsilon_1^{\mu}(x- \epsilon_2(x) ).
\eqno{(3.32)}
$$
This property means that the product to two translation operator
satisfy closure property, which is one of the conditions that any group
must satisfy.

\item Suppose that $\ehat$ is a non-singular translation operator,
then
$$
\ehat^{-1} = E^{ i \epsilon^{\mu}(f(x)) \cdot \hat{P}_{\mu}},
\eqno{(3.33)}
$$
where $f(x)$ is defined by the following relations:
$$
f(x- \epsilon (x)) = x.
\eqno{(3.34)}
$$
$\ehat^{-1}$ is the inverse operator of $\ehat$, so
$$
\ehat^{-1} \ehat = \ehat \ehat^{-1} = {\rm\bf 1},
\eqno{(3.35)}
$$
where {\bf 1} is the unit element of the gravitational gauge group.

\item The product operation of translation also satisfies associative
law. Suppose that $\mhat{1}$ , $\mhat{2}$ and $\mhat{3}$ are three
arbitrary translation operators, then
$$
\mhat{3} \cdot ( \mhat{2} \cdot \mhat{1} )
 = ( \mhat{3} \cdot \mhat{2} ) \cdot \mhat{1} .
\eqno{(3.36)}
$$

\item Suppose that $\ehat$ is an arbitrary translation operator and
$\phi(x)$ is an arbitrary function of space-time coordinate, then
$$
\ehat \phi(x) \ehat^{-1} = f(x- \epsilon(x)).
\eqno{(3.37)}
$$
This relation is quite useful in following discussions.

\item Suppose that $\ehat$ is an arbitrary translation operator.
Define
$$
\Lambda^{\alpha}_{~~\beta} =
\frac{\partial x^{\alpha}}{\partial ( x - \epsilon (x) )^{\beta}},
\eqno{(3.38)}
$$
$$
\Lambda_{\alpha}^{~~\beta} =
\frac{\partial ( x - \epsilon (x))^{\beta}}{\partial x^{\alpha}}.
\eqno{(3.39)}
$$
They satisfy
$$
\Lambda_{\alpha}^{~~ \mu} \Lambda^{\alpha}_{~~ \nu}
= \delta^{\mu}_{\nu},
\eqno{(3.40)}
$$
$$
\Lambda_{\mu}^{ ~~\alpha} \Lambda^{\nu}_{~~ \alpha}
= \delta_{\mu}^{\nu}.
\eqno{(3.41)}
$$
Then we have following relations:
$$
\ehat \hat{P}_{\alpha} \ehat ^{-1}
= \Lambda^{\beta}_{~~\alpha}  \hat{P}_{\beta},
\eqno{(3.42)}
$$
$$
\ehat {\rm d}x^{\alpha} \ehat ^{-1}
= \Lambda_{\beta}^{~~\alpha}  {\rm d}x^{\beta}.
\eqno{(3.43)}
$$
These give out the the transformation laws of  $\hat{P}_{\alpha}$
and d$x^{\alpha}$ under local gravitational gauge transformations.

\end{enumerate}

Gravitational gauge group (GGG) is a transformation group which
consists of all non-singular translation operators $\ehat$. We can
easily see that gravitational gauge group is indeed a  group, for
\begin{enumerate}

\item the product of two arbitrary non-singular translation operators
is also a non-singular translation operator, which is also an element
of the gravitational gauge group. So, the product of the group satisfies
closure property which is expressed in eq(3.31);

\item the product of the gravitational gauge group also satisfies the
associative law which is expressed in eq(3.36);

\item the gravitational gauge group has
its unit element {\bf 1}, it satisfies
$$
{\rm\bf 1} \cdot \ehat = \ehat \cdot {\rm\bf 1} = \ehat;
\eqno{(3.44)}
$$

\item every non-singular element $\ehat$ has its inverse element
which is given by eqs(3.33) and (3.35).

\end{enumerate}
According to gauge principle, the gravitational gauge group is the symmetry
of gravitational interactions. The global invariance of gravitational gauge
transformation will give out conserved charges which is just the ordinary
energy-momentum; the requirement of local gravitational gauge invariance
needs introducing gravitational gauge field, and gravitational
interactions are completely determined by the local gravitational gauge
invariance. \\

The generators of gravitational gauge group is just the energy-momentum
operators $\hat{P}_{\alpha}$.  This is required by gauge principle. It can
also be seen from the form of infinitesimal transformations. Suppose that
$\epsilon$ is an infinitesimal quantity, then we have
$$
\ehat \simeq 1 - i \epsilon^{\alpha} \hat{P}_{\alpha}.
\eqno{(3.45)}
$$
Therefore,
$$
i \frac{\partial \ehat}{\partial \epsilon^{\alpha}} \mid _{\epsilon = 0}
\eqno{(3.46)}
$$
gives out  generators $\hat{P}_{\alpha}$ of gravitational gauge group.
It is known  that generators of gravitational gauge
group are commute each other
$$
\lbrack \hat{P}_{\alpha} ~~,~~ \hat{P}_{\beta} \rbrack = 0.
\eqno{(3.47)}
$$
However, the commutation property of generators does not mean that
gravitational gauge group is an Abel group, because two general elements
of gravitational gauge group do not commute:
$$
\lbrack \mhat{1} ~~,~~ \mhat{2} \rbrack \not= 0.
\eqno{(3.48)}
$$
Gravitational gauge group is a kind of non-Abel gauge group. The
non-Able nature of  gravitational gauge group will cause
self-interactions of gravitational gauge field. \\

In order to avoid confusion, we need to pay some attention to some
differences between two concepts: space-time translation group and
gravitational gauge group. Generally speaking, space-time
translation is a kind of coordinates transformation, that is, the
objects or fields in space-time are kept fixed while the
space-time coordinates that describe the motion of objective
matter undergo transformation. But gravitational gauge
transformation is a kind of system transformation rather than a
kind of coordinates transformation. In system transformation, the
space-time coordinate system is kept unchanged while objects or
fields undergo transformation. From mathematical point of view,
space-time translation and gravitational gauge transformation are
essentially the same, and the space-time translation is the
inverse transformation of the gravitational gauge transformation;
but from physical point of view, space-time translation and
gravitational gauge transformation are quite different, especially
when we discuss gravitational gauge transformation of
gravitational gauge field. For gravitational gauge field, its
gravitational gauge transformation is not the inverse
transformation of its space-time translation. In a meaning,
space-time translation is a kind of mathematical transformation,
which contains little dynamical information of interactions; while
gravitational gauge transformation is a kind of physical
transformation, which contains all dynamical information of
interactions and is convenient for us to study physical
interactions. Through gravitational gauge symmetry, we can
determine the whole gravitational interactions among various kinds
of fields. This is the reason why we do not call gravitational
gauge transformation space-time translation. This is important for
all of our discussions on gravitational
gauge transformations of various kinds of fields. \\

Suppose that $\phi(x)$ is an arbitrary scalar field. Its gravitational
gauge transformation is
$$
\phi(x) \to \phi'(x) = ( \ehat \phi(x)).
\eqno{(3.49)}
$$
Similar to ordinary $SU(N)$ non-able gauge field theory, there are
two kinds of scalars. For example, in chiral perturbative theory, the
ordinary $\pi$ mesons are scalar fields, but they are vector fields in
isospin space. Similar case exists in gravitational gauge field theory.
A Lorentz scalar can be a scalar or a vector or a tensor in the space
of gravitational gauge group. If $\phi (x)$ is a scalar in the space of
gravitational gauge group, we just simply denote it as $\phi (x)$ in
gauge group space. If it is a vector in the space of gravitational
gauge group, it can be expanded in the gravitational gauge group
space in the following way:
$$
\phi(x)  =  \phi^{\alpha}(x) \cdot \hat{P}_{\alpha}.
\eqno{(3.50)}
$$
The transformation of component field is
$$
\phi^{\alpha}(x)  \to \phi^{\prime  \alpha}(x) =
\Lambda^{\alpha}_{~~\beta} \ehat \phi^{\beta}(x) \ehat^{-1} .
\eqno{(3.51)}
$$
The important thing that we must remember is that, the $\alpha$
index is not a Lorentz index, it is just a group index. For
gravitation gauge group, it is quite special that a group index looks
like a Lorentz index. We must be carefully on this important thing.
This will cause some fundamental changes on quantum gravity.
Lorentz scalar $\phi(x)$ can also be a tensor in gauge group
space. suppose that it is a $n$th order tensor in gauge group
space, then it can be expanded as
$$
\phi(x)  =  \phi^{\alpha_1 \cdots \alpha_n}(x)
\cdot \hat{P}_{\alpha_1}  \cdots \hat{P}_{\alpha_n}.
\eqno{(3.52)}
$$
The transformation of component field is
$$
\phi^{\alpha_1 \cdots \alpha_n}(x)  \to
\phi ^{\prime \alpha_1 \cdots \alpha_n}(x) =
\Lambda^{\alpha_1}_{~~\beta_1} \cdots \Lambda^{\alpha_n}_{~~\beta_n}
\ehat \phi^{\beta_1 \cdots \beta_n}(x) \ehat^{-1} .
\eqno{(3.53)}
$$
\\

If $\phi (x)$ is a spinor field, the above discussion is also valid.
That is, a spinor can also be a scalar or a vector or a tensor in
the space of gravitational gauge group. The gravitational
gauge transformations of the component fields are also given by
eqs.(3.49-53). There is no transformations in spinor space, which
is different from that of the Lorentz transformation of a spinor.  \\

Suppose that $A_{\mu} (x)$ is an arbitrary vector field. Here,
the index $\mu$ is a Lorentz index. Its gravitational gauge
transformation is:
$$
A_{\mu}(x) \to A'_{\mu}(x) = ( \ehat A_{\mu}(x)).
\eqno{(3.54)}
$$
Please remember that there is no rotation in the space of Lorentz
index $\mu$, while in the general coordinates transformations of
general relativity, there is rotation in the space of Lorentz index
$\mu$. The reason is that gravitational gauge transformation is a
kind of system transformation, while in general relativity, the
general coordinates transformation is a kind of coordinates
transformation. If $A_{\mu}(x)$ is a scalar in the space of
gravitational gauge group, eq(3.54) is all for its gauge
transformation. If $A_{\mu}(x)$ is a vector in the space of
gravitational gauge group, it can be expanded as:
$$
A_{\mu}(x)  =  A_{\mu}^{\alpha}(x) \cdot \hat{P}_{\alpha}.
\eqno{(3.55)}
$$
The transformation of component field  is
$$
A_{\mu}^{\alpha}(x)  \to A_{ \mu}^{\prime\alpha}(x) =
\Lambda^{\alpha}_{~~\beta} \ehat A_{\mu}^{~~\beta}(x) \ehat^{-1} .
\eqno{(3.56)}
$$
If $A_{\mu}(x)$ is a $n$th order tensor in the space of
gravitational gauge group, then
$$
A_{\mu}(x)  =  A_{\mu}^{\alpha_1 \cdots \alpha_n}(x)
\cdot \hat{P}_{\alpha_1}  \cdots \hat{P}_{\alpha_n}.
\eqno{(3.57)}
$$
The transformation of component fields is
$$
A_{\mu}^{\alpha_1 \cdots \alpha_n}(x)  \to
A_{\mu}^{\prime\alpha_1 \cdots \alpha_n}(x) =
\Lambda^{\alpha_1}_{~~\beta_1} \cdots \Lambda^{\alpha_n}_{~~\beta_n}
\ehat A_{\mu}^{\beta_1 \cdots \beta_n}(x) \ehat^{-1} .
\eqno{(3.58)}
$$
Therefore, under gravitational gauge transformations, the behavior
of a group index is quite different from that of a Lorentz index. However,
they have the same behavior in global Lorentz transformations. \\

Generally speaking, suppose that $T^{\mu_1 \cdots \mu_n}
_{\nu_1 \cdots \nu_m} (x)$ is an arbitrary tensor, its
gravitational gauge transformations are:
$$
T^{\mu_1 \cdots \mu_n}_{\nu_1 \cdots \nu_m} (x)
\to T'^{\mu_1 \cdots \mu_n}_{\nu_1 \cdots \nu_m} (x)
 = ( \ehat T^{\mu_1 \cdots \mu_n}_{\nu_1 \cdots \nu_m} (x)).
\eqno{(3.59)}
$$
If it is a $p$th order tensor in group space, then
$$
T^{\mu_1 \cdots \mu_n}_{\nu_1 \cdots \nu_m} (x)
= T^{\mu_1 \cdots \mu_n ; \alpha_1 \cdots \alpha_p}
_{\nu_1 \cdots \nu_m} (x)
\cdot \hat{P}_{\alpha_1}  \cdots \hat{P}_{\alpha_p}.
\eqno{(3.60)}
$$
The transformation of component fields is
$$
T^{\mu_1 \cdots \mu_n ; \alpha_1 \cdots \alpha_p}
_{\nu_1 \cdots \nu_m} (x)  \to
T'^{\mu_1 \cdots \mu_n ; \alpha_1 \cdots \alpha_p}
_{\nu_1 \cdots \nu_m} (x) =
\Lambda^{\alpha_1}_{~~\beta_1} \cdots \Lambda^{\alpha_p}_{~~\beta_p}
\ehat T^{\mu_1 \cdots \mu_n ; \beta_1 \cdots \beta_p}
_{\nu_1 \cdots \nu_m} (x)\ehat^{-1}.
\eqno{(3.61)}
$$
\\

Finally, we give definitions of three useful tensors. $\eta_2$ is
a Lorentz scalar, but it is a second order tensor in group space.
That is,
$$
\eta_2  =  \eta_2 ^{\alpha \beta}
\cdot \hat{P}_{\alpha}  \hat{P}_{\beta}.
\eqno{(3.62)}
$$
The gravitational gauge transformation of $\eta_2$ is
$$
\eta_2  \to \eta '_2 = (\ehat \eta_2).
\eqno{(3.63)}
$$
The transformation  of component field  is
$$
\eta_2^{\alpha  \beta}  \to
\eta _2^{\prime \alpha  \beta}  =
\Lambda^{\alpha}_{~~\alpha_1}  \Lambda^{\beta}_{~~\beta_1}
\ehat \eta_2 ^{\alpha_1  \beta_1} \ehat^{-1} .
\eqno{(3.64)}
$$
As a symbolic operation, we use $\eta_2^{\alpha  \beta}$
to raise a group index or to descend a group index.
Covariant group metric tensor $\eta_{2 \alpha  \beta}$ is defined by
$$
\eta_{2 \alpha  \beta}   \eta_2^{\beta \gamma}
= \eta_2^{\gamma \beta} \eta_{2 \beta \alpha}
= \delta_{\alpha}^{\gamma}
\eqno{(3.65)}
$$
In a special representation of gravitational gauge group,
$\eta_2^{\alpha  \beta}$ is selected to be diagonal, that is:
$$
\begin{array}{rcl}
\eta_2^{0~0} &=& -1,  \\
\eta_2^{1~1} &=&  1,  \\
\eta_2^{2~2} &=&  1,  \\
\eta_2^{3~3} &=&  1,
\end{array}
\eqno{(3.66)}
$$
and other components of $\eta_2^{\alpha  \beta}$ vanish.
Group index of $\hat{P}_{\alpha}$ is raised by $\eta_2^{\alpha  \beta}$.
that is
$$
\hat{P}^{\alpha} = \eta_2^{\alpha \beta} \hat{P}_{\beta}.
\eqno{(3.67)}
$$
Eq.(3.66) is the simplest choice for $\eta_{2 \alpha \beta}$. In
this choice, the equation of motion of gravitational gauge field
has the simplest expressions. Another choice for $\eta_2^{\alpha
\beta}$ is given by the following relation\cite{c5},
$$
\eta_2^{\alpha \beta} = g^{\alpha \beta} = \eta^{\mu \nu} (
\delta^{\alpha}_{\mu} - g C^{\alpha}_{\mu} ) (
\delta^{\beta}_{\nu} - g C^{\beta}_{\nu} ),
 \eqno{(3.68)}
$$
where $\eta^{\mu \nu}$ is the Minkowski metric, and
$C_{\mu}^{\alpha}$ is the gravitational gauge field which will be
introduced in the next chapter. If we use eq.(3.68) to construct
the lagrangian of gravitational system, the equation of motion of
gravitational field will be much more complicated than that of
choice eq.(3.66)\cite{c5}. Eq.(3.68) is a possible choice. Because
of the traditional belief that fundamental theory of fundamental
interactions should be simple, we will choose the simplest choice
which has the most beautiful form in this paper. In other words,
in this paper, we select eq.(3.66) to
be the definition of $\eta_2^{\alpha \beta}$. $\eta_2^{\alpha
\beta}$ is not the metric in gravitational gauge group space when
we use eq.(3.66) as definition, it is only a mathematical
notation.
\\

$\eta_1^{\mu}$ is a Lorentz vector. It is also a vector in the space of
gravitational gauge group.
$$
\eta_1^{ \mu}= \eta_{ 1 \alpha}^{\mu} \cdot \hat{P}^{\alpha}
\eqno{(3.69)}
$$
For $\eta_{1 \alpha}^{\mu}$, the index $\mu$ is a Lorentz index and
the index $\alpha$ is a group index.
The gravitational gauge transformation of $\eta_1^{\mu}$ is
$$
\eta_1^{ \mu} \to \eta_1^{\prime \mu} = ( \ehat \eta_1^{\mu} ).
\eqno{(3.70)}
$$
The transformation of its component field  is
$$
\eta_{1 \alpha}^{\mu} \to \eta_{1 \alpha}^{\prime \mu} =
\Lambda_{\alpha}^{~~\beta} \ehat \eta_{1 \beta}^{\mu} \ehat^{-1}.
\eqno{(3.71)}
$$
In a special coordinate system and a special representation of gravitational
gauge group, $\eta_{1 \alpha}^{\mu}$ is selected to be $\delta$-function,
that is
$$
\eta_{1 \alpha}^{\mu} = \delta^{\mu}_{\alpha}.
\eqno{(3.72)}
$$
\\

$\eta^{\mu \nu}$ is a second order Lorentz tensor, but it is a scalar in
group space. It is the metric of the coordinate space. A Lorentz
index can be raised or descended by this metric tensor.
In a special coordinate system, it is selected to be:
$$
\begin{array}{rcl}
\eta^{0~0} &=& -1,  \\
\eta^{1~1} &=&  1,  \\
\eta^{2~2} &=&  1,  \\
\eta^{3~3} &=&  1,
\end{array}
\eqno{(3.73)}
$$
and other components of $\eta^{\mu \nu}$ vanish. $\eta^{\mu \nu}$
is the traditional Minkowski metric.
\\

\section{Pure Gravitational Gauge Fields}

Before we study gravitational field, we must determine which field
represents gravitational field. In the traditional gravitational
gauge theory, gravitational field is represented by space-time
metric tensor. If there is gravitational field in space-time, the
space-time metric will not be equivalent to Minkowski metric, and
space-time will become curved. In other words, in the traditional
gravitational gauge theory, quantum gravity is formulated in
curved space-time. In this paper, we will not follow this way. The
underlying point of view of this new quantum gauge theory of
gravity is that gravitational field is represented by gauge
potential, and curved space-time is considered to be a classical
effect of macroscopic gravitational field. When we study quantum
gravity, which mainly concerns microscopic gravitational
interactions, we will not inherit the conception of curved
space-time. In other words, if we put gravity into the structure
of space-time, the space-time will become curved and there will be
no physical gravity in space-time, because all gravitational
effects are put into space-time metric and gravity is geometrized.
But if we study physical gravitational interactions, it is better
to rescue gravity from space-time metric and treat gravity as a
physical field. In this case, space-time is flat and there is
physical gravity in Minkowski space-time. For this reason, we will
not introduce the concept of curved space-time to study quantum
gravity in this paper. The space-time is always flat,
gravitational field is represented by gauge potential and
gravitational interactions are always treated as physical
interactions. This point of view will be discussed again later.
Besides, according to gauge principle, it is required that
gravitational field is represented by gauge potential.
\\

Now, let's begin to construct the Lagrangian of gravitational
gauge theory. For the sake of simplicity, let's suppose that $\phi
(x)$ is a Lorentz scalar and gauge group scalar. According to
above discussions, Its gravitational gauge transformation is:
$$
\phi (x) \to \phi '(x) = (\ehat \phi (x)).
\eqno{(4.1)}
$$
Because
$$
(\partial_{\mu} \ehat ) \not= 0,
\eqno{(4.2)}
$$
partial differential of $\phi (x)$ does not transform covariantly
under gravitational gauge transformation:
$$
\partial_{\mu} \phi (x) \to \partial_{\mu} \phi '(x)
\not= ( \ehat \partial_{\mu} \phi (x) ).
\eqno{(4.3)}
$$
In order to construct an action which is invariant under local
gravitational gauge transformation, gravitational gauge
covariant derivative is highly necessary. The gravitational gauge
covariant derivative is defined by
$$
D_{\mu} = \partial_{\mu} - i g C_{\mu} (x),
\eqno{(4.4)}
$$
where $C_{\mu} (x)$ is the gravitational gauge field. It is a Lorentz
vector. Under gravitational gauge transformations, it transforms as
$$
C_{\mu}(x) \to  C'_{\mu}(x) =
\ehat (x) C_{\mu} (x) \ehat^{-1} (x)
+ \frac{i}{g} \ehat (x) (\partial_{\mu} \ehat^{-1} (x)).
\eqno{(4.5)}
$$
Using the original definition of $\ehat$, we can strictly proved
that
$$
\lbrack \partial_{\mu} ~~,~~ \ehat \rbrack
= (\partial_{\mu} \ehat).
\eqno{(4.6)}
$$
Therefor, we have
$$
\ehat \partial_{\mu} \ehat^{-1} =
\partial_{\mu}  + \ehat (\partial_{\mu} \ehat^{-1}),
\eqno{(4.7)}
$$
$$
\ehat D_{\mu} \ehat^{-1} =
\partial_{\mu} - i g  C'_{\mu} (x).
\eqno{(4.8)}
$$
So, under local gravitational gauge transformations,
$$
D_{\mu} \phi (x) \to
D'_{\mu} \phi '(x) =
(\ehat D_{\mu} \phi (x)) ,
\eqno{(4.9)}
$$
$$
D_{\mu} (x) \to D'_{\mu} (x)
= \ehat D_{\mu} (x) \ehat^{-1}.
\eqno{(4.10)}
$$
\\

Gravitational gauge field $C_{\mu} (x)$ is vector field, it is a
Lorentz vector. It is also a vector in gauge group space, so it can be
expanded as linear combinations of generators of gravitational gauge
group:
$$
C_{\mu} (x) = C_{\mu}^{\alpha}(x) \cdot \hat{P}_{\alpha}.
\eqno{(4.11)}
$$
$C_{\mu}^{\alpha}$ are component fields of gravitational gauge
field. They looks like a second rank tensor. But according to our
previous discussion, they are not tensor fields, they are vector
fields. The index $\alpha$ is not a Lorentz index, it is just a gauge
group index. Gravitational gauge field $C_{\mu}^{\alpha}$ has only
one Lorentz index, so it is a kind of vector field. This is an inevitable
results of gauge principle. The gravitational gauge transformation  of
component fields is:
$$
C_{\mu}^{\alpha}(x) \to C_{\mu}^{\prime \alpha}(x)=
\Lambda ^{\alpha}_{~~\beta} (\ehat C_{\mu}^{\beta}(x))
- \frac{1}{g} (\ehat  \partial_{\mu} \epsilon^{\alpha}(y)),
\eqno{(4.12)}
$$
where $y$ is a function of space-time coordinates which satisfy:
$$
( \ehat y(x) )=x.
\eqno{(4.13)}
$$
\\

The strength of gravitational gauge field is defined by
$$
F_{\mu \nu} = \frac{1}{-i g}
\lbrack D_{\mu} ~~,~~ D_{\nu} \rbrack,
\eqno{(4.14)}
$$
or
$$
F_{\mu \nu} = \partial_{\mu} C_{\nu}(x)
- \partial_{\nu} C_{\mu}(x)
- i g C_{\mu}(x) C_{\nu}(x)
+ i g  C_{\nu}(x) C_{\mu}(x).
\eqno{(4.15)}
$$
$F_{\mu \nu}$ is a second order Lorentz tensor. It is a vector is
group space, so it can be expanded in group space,
$$
F_{\mu \nu} (x) = F_{\mu \nu}^{\alpha} (x) \cdot \hat{P}_{\alpha}.
\eqno{(4.16)}
$$
The explicit form of component field strengths is
$$
F_{\mu \nu}^{\alpha} = \partial_{\mu} C_{\nu}^{\alpha}
- \partial_{\nu} C_{\mu}^{\alpha}
- g C_{\mu}^{\beta} \partial_{\beta} C_{\nu}^{\alpha}
+ g C_{\nu}^{\beta} \partial_{\beta} C_{\mu}^{\alpha}
\eqno{(4.17)}
$$
The strength of gravitational gauge fields transforms covariantly
under gravitational gauge transformation:
$$
F_{\mu \nu} \to F'_{\mu \nu} =
\ehat F_{\mu \nu} \ehat^{-1}.
\eqno{(4.18)}
$$
The gravitational gauge transformation of the field strength
of component field is
$$
F_{\mu \nu}^{\alpha} \to F_{\mu \nu}^{\prime \alpha} =
\Lambda^{\alpha}_{~~\beta} (\ehat F_{\mu \nu}^{\beta}).
\eqno{(4.19)}
$$
\\

Similar to traditional gauge field theory, the kinematical term for
gravitational gauge field can be selected as
$$
{\cal L}_0 = - \frac{1}{4} \eta^{\mu \rho} \eta^{\nu \sigma}
\eta_{2 ~ \alpha \beta}
F_{\mu \nu}^{\alpha} F_{\rho \sigma}^{\beta}.
\eqno{(4.20)}
$$
We can easily prove that this Lagrangian does not invariant under
gravitational gauge transformation, it transforms covariantly
$$
{\cal L}_0 \to {\cal L}'_0 = ( \ehat {\cal L}_0).
\eqno{(4.21)}
$$
\\

In order to resume the gravitational gauge symmetry of the action,
we introduce an extremely important factor, whose form is
$$
J(C) = e^{I(C)} = e^{g \eta_{1 \alpha}^{\mu} C_{\mu}^{\alpha}},
\eqno{(4.22)}
$$
where
$$
I(C) = g \eta_{1 \alpha}^{\mu} C_{\mu}^{\alpha}.
\eqno{(4.23)}
$$
The choice of $J(C)$ is not unique\cite{c5}. another choice of
$J(C)$ is
$$
J(C) = \sqrt{- {\rm det} g_{\alpha \beta} },
\eqno{(4.22a)}
$$
where $g_{\alpha \beta}$ is given by eq.(9.30).
Under gravitational gauge transformations, $g_{\alpha\beta}$
transforms as
$$
g_{\alpha \beta} \to g'_{\alpha \beta}
= \Lambda_{\alpha}^{~\alpha_1} \Lambda_{\beta}^{~\beta_1}
g_{\alpha_1 \beta_1}.
$$
Then $J(C)$ transforms as
$$
J(C) \to J'(C') = J \cdot (\ehat J(C)),
$$
where $J$ is the Jacobian of the transformation. Eq.(4.22) gives
out the simplest and most natural choice for $J(C)$, and in this
case, the equation of motion of gravitational gauge field has
simplest and most explicit expressions. Because
of the traditional belief that fundamental theory of fundamental
interactions should be simple, we will choose the simplest choice
which has the most beautiful form in this paper. On the other hand,
the choice given by eq.(4.22) can make the whole theory
renormalizable, therefore we use eq.(4.22) as the definition
of $J(C)$ in this paper. In fact,
$J(C)$ times any gravitational gauge
covariant functions of pure gravitational gauge field
$C_{\mu}^{\alpha}$ can be regarded as a possible choice of $J(C)$.
Some detailed discussions on the second choice can be found in
literature \cite{c5}.  When we use
eq.(4.22) as the definition of $J(C)$,  then, the Lagrangian for
gravitational gauge field is selected as
$$
{\cal L} = J(C) {\cal L}_0 = e^{I(C)} {\cal L}_0, \eqno{(4.24)}
$$
and the action for gravitational gauge field is
$$
S = \int {\rm d}^4 x {\cal L}.
\eqno{(4.25)}
$$
It can be proved that this action has gravitational gauge symmetry.
In other words, it is invariant under gravitational gauge
transformation,
$$
S \to S' = S.
\eqno{(4.26)}
$$
In order to prove the gravitational gauge symmetry of the action,
the following relation is important,
$$
\int {\rm d}^4 x J(\ehat f(x)) = \int {\rm d}^4 x f(x),
\eqno{(4.27)}
$$
where $f(x)$ is an arbitrary function of space-time coordinate and $J$
is the Jacobian of the corresponding transformation,
$$
J = det (\frac{\partial (x - \epsilon)^{\mu}}{\partial x^{\nu}}).
\eqno{(4.28)}
$$
\\

According to gauge principle, the global gauge symmetry will give out
conserved charges. Now, let's discuss the conserved charges of
global gravitational gauge transformation. Suppose that
$\epsilon^{\alpha}$ is an infinitesimal constant 4-vector. Then,
in the first order approximation, we have
$$
\ehat = 1 - \epsilon^{\alpha} \partial_{\alpha} + o(\epsilon^2).
\eqno{(4.29)}
$$
The first order variation of the gravitational gauge field is
$$
\delta C_{\mu}^{\alpha} (x) =
- \epsilon^{\nu} \partial_{\nu} C_{\mu}^{\alpha},
\eqno{(4.30)}
$$
and the first order variation of action is:
$$
\delta S = \int {\rm d}^4 x ~ \epsilon^{\alpha} \partial_{\mu}
T_{i \alpha}^{\mu},
\eqno{(4.31)}
$$
where $T_{i \alpha}^{\mu}$ is the inertial energy-momentum
tensor, whose definition is
$$
T_{i \alpha}^{\mu} \equiv e^{I(C)}
(- \frac{\partial {\cal L}_0}{\partial \partial_{\mu} C_{\nu}^{\beta}}
\partial_{\alpha} C_{\nu}^{\beta} + \delta^{\mu}_{\alpha} {\cal L}_0 ).
\eqno{(4.32)}
$$
It is a conserved current,
$$
\partial_{\mu} T_{i \alpha}^{\mu} = 0.
\eqno{(4.33)}
$$
Except for the factor $e^{I(C)}$, the form of the inertial
energy-momentum tensor is almost completely the same as that
in the traditional quantum field theory. It means that gravitational
interactions will change energy-momentum of matter fields, which
is what we expected in general relativity.\\

The Euler-Lagrange equation for gravitational gauge field is
$$
\partial_{\mu} \frac{\partial \cal L}{\partial \partial_{\mu} C_{\nu}^{\alpha}}
= \frac{\partial \cal L}{\partial C_{\nu}^{\alpha}}.
\eqno{(4.34)}
$$
This form is completely that same as what we have ever seen in
quantum field theory. But if we insert eq.(4.24) into it, we will get
$$
\partial_{\mu} \frac{\partial {\cal L }_0}{\partial \partial_{\mu} C_{\nu}^{\alpha}}
= \frac{\partial {\cal L}_0}{\partial C_{\nu}^{\alpha}}
+g \eta_{1 \alpha}^{\nu} {\cal L}_0 - g \partial_{\mu}
(\eta_{1 \gamma}^{\rho} C_{\rho}^{\gamma})
\frac{\partial {\cal L}_0}{\partial \partial_{\mu} C_{\nu}^{\alpha}}.
\eqno{(4.35)}
$$
Eq.(4.17) can be changed into
$$
F_{\mu \nu}^{\alpha} =
(D_{\mu} C_{\nu}^{\alpha})
- (D_{\nu} C_{\mu}^{\alpha}) ,
\eqno{(4.36)}
$$
so the Lagrangian ${\cal L}_0$ depends on gravitational gauge fields
completely through its covariant derivative. Therefore,
$$
\frac{\partial {\cal L}_0}{\partial C_{\nu}^{\alpha}}
= - g \frac{\partial {\cal L}_0}{\partial D_{\nu} C_{\mu}^{\beta}}
\partial_{\alpha} C_{\mu}^{\beta}.
\eqno{(4.37)}
$$
Using the above relations and
$$
\frac{\partial {\cal L}_0}{\partial \partial_{\mu} C_{\nu}^{\alpha}}
= - \eta^{\mu \lambda} \eta^{\nu \tau} \eta_{2 \alpha \beta}
F_{\lambda \tau}^{\beta}
- g \eta^{\nu \lambda} \eta^{\sigma \tau} \eta_{2 \alpha \beta}
F_{\lambda \tau}^{\beta} C_{\sigma}^{\mu},
\eqno{(4.38)}
$$
the above equation of motion of gravitational gauge fields
are changed into:
$$
\begin{array}{rcl}
\partial_{\mu} (\eta^{\mu \lambda} \eta^{\nu \tau} \eta_{2 \alpha \beta}
F_{\lambda \tau}^{\beta} ) &=&
-g ( - \frac{\partial {\cal L}_0}{\partial D_{\nu} C_{\mu}^{\beta}}
\partial_{\alpha} C_{\mu}^{\beta}
+ \eta_{1 \alpha}^{\nu} {\cal L}_0 ) \\
&& + g \partial_{\mu} (\eta_{1 \gamma}^{\rho} C_{\rho}^{\gamma})
\frac{\partial {\cal L}_0}{\partial \partial_{\mu} C_{\nu}^{\alpha}} \\
&& - g \partial_{\mu} (\eta^{\nu \lambda} \eta^{\sigma \tau} \eta_{2 \alpha \beta}
F_{\lambda \tau}^{\beta} C_{\sigma}^{\mu}).
\end{array}
\eqno{(4.39)}
$$
If we define
$$
\begin{array}{rcl}
T_{g \alpha}^{\nu}&=&
 - \frac{\partial {\cal L}_0}{\partial D_{\nu} C_{\mu}^{\beta}}
\partial_{\alpha} C_{\mu}^{\beta}
+ \eta_{1 \alpha}^{\nu} {\cal L}_0  \\
&& -  \partial_{\mu} (\eta_{1 \gamma}^{\rho} C_{\rho}^{\gamma})
\frac{\partial {\cal L}_0}{\partial \partial_{\mu} C_{\nu}^{\alpha}} \\
&& + \partial_{\mu} (\eta^{\nu \lambda} \eta^{\sigma \tau} \eta_{2 \alpha \beta}
F_{\lambda \tau}^{\beta} C_{\sigma}^{\mu}),
\end{array}
\eqno{(4.40)}
$$
Then we can get a simpler form of equation of motion,
$$
\partial_{\mu} (\eta^{\mu \lambda} \eta^{\nu \tau} \eta_{2 \alpha \beta}
F_{\lambda \tau}^{\beta} )  = - g T_{g \alpha}^{\nu}.
\eqno{(4.41)}
$$
$ T_{g \alpha}^{\nu}$ is also a conserved current, that is
$$
\partial_{\nu} T_{g \alpha}^{\nu} = 0,
\eqno{(4.42)}
$$
because
$$
\partial_{\nu} \partial_{\mu} (\eta^{\mu \lambda} \eta^{\nu \tau}
\eta_{2 \alpha \beta} F_{\lambda \tau}^{\beta} )  = 0.
\eqno{(4.43)}
$$
$ T_{g \alpha}^{\nu}$ is called gravitational energy-momentum
tensor, which is the source of gravitational gauge field. Now we get
two different energy-momentum tensors, one is the inertial energy-momentum
tensor $ T_{i \alpha}^{\nu}$ and another is the gravitational
energy-momentum tensor $ T_{g \alpha}^{\nu}$. They are similar,
but they are different.The inertial energy-momentum tensor
$ T_{i \alpha}^{\nu}$ is given by conservation law which is associate
with global gravitational gauge symmetry, it gives out an
energy-momentum 4-vector:
$$
P_{i \alpha} = \int {\rm d}^3 \svec{x} T_{i  \alpha}^{0}.
\eqno{(4.44)}
$$
It is a conserved charges,
$$
\frac{\rm d}{{\rm d} t}P_{i \alpha} =  0.
\eqno{(4.45)}
$$
The time component of $P_{i \alpha}$, that is
$P_{i 0}$,  gives out the Hamiltonian $H$ of the system,
$$
H = - P_{i~0} =  \int {\rm d}^3 \svec{x} e^{I(C)}
(\pi_{\alpha}^{\mu} \tdot{C} _{\mu}^{\alpha} - {\cal L}_0).
\eqno{(4.46)}
$$
According to our conventional belief, $H$ should be the inertial
energy of the system, therefore $P_{i \alpha}$ is the inertial
energy-momentum of the system. The gravitational energy-momentum
is given by equation of motion of gravitational gauge field, it is also
a conserved current. The space integration of the time component
of it gives out a conserved energy-momentum 4-vector,
$$
P_{g \alpha} = \int {\rm d}^3 \svec{x} T_{g  \alpha}^{0}.
\eqno{(4.47)}
$$
It is also a conserved charge,
$$
\frac{\rm d}{{\rm d} t}P_{g \alpha} =  0.
\eqno{(4.48)}
$$
The time component of it just gives out the gravitational energy
of the system. This can be easily seen. Set $\nu$  and $\alpha$
in eq.(4.41) to $0$, we get
$$
\partial^i F_{i 0}^0   = - g T_{g 0}^{0}.
\eqno{(4.49)}
$$
The field strength of gravitational field is defined by
$$
E^i = - F_{i 0}^0 .
\eqno{(4.50)}
$$
The space integration of eq.(4.49) gives out
$$
\oint {\rm d}\svec{\sigma} \cdot \svec{E}
=   g \int {\rm d}^3 \svec{x} T_{g  0}^0.
\eqno{(4.51)}
$$
According to Newton's classical theory of gravity,
$\int {\rm d}^3 \svec{x} T_{g  0}^0$ in the right hand term is
just the gravitational mass of the system. Denote the gravitational
mass of the system as $M_g$, that is
$$
M_g
= - \int {\rm d}^3 \svec{x} T_{g  0}^0.
\eqno{(4.52)}
$$
Then eq(4.51) is changed into
$$
\oint {\rm d}\svec{\sigma} \cdot \svec{E}
= - g M_g.
\eqno{(4.53)}
$$
This is just the classical Newton's law of universal gravitation.
It can be strictly proved that gravitational mass is different from
inertial mass. They are not equivalent. But their difference is at least
second order infinitesimal quantity if the gravitational field
$C_{\mu}^{\alpha}$ and the gravitational coupling constant $g$
are all first order infinitesimal quantities, for this difference is
proportional to $g C_{\mu}^{\alpha}$. So, this difference is
too small to be detected in experiments. But in the environment
of strong gravitational field, the difference will become relatively
larger and will be easier to be detected. Much more highly precise
measurement of this difference is strongly needed to test this
prediction and to test the validity of the equivalence principle.
In the chapter of classical limit of quantum gauge theory of
gravity, we will return to discuss this problem again. \\

Now, let's discuss self-coupling of gravitational
field. The Lagrangian of gravitational gauge field is given by
eq(4.24). Because
$$
e^{g \eta_{1 \alpha}^{\mu} C_{\mu}^{\alpha}}
= 1 + \sum_{n=1}^{\infty} \frac{1}{n!}
(g  \eta_{1 \alpha}^{\mu} C_{\mu}^{\alpha})^n,
\eqno{(4.54)}
$$
there are vertexes of  $n$ gravitational gauge fields in tree diagram
with $n$ can be arbitrary integer number that is greater than 3. This
property is important for renormalization of the theory. Because
the coupling constant of the gravitational gauge interactions has
negative mass dimension, any kind of regular vertex exists divergence.
In order to cancel these divergences, we need to introduce the
corresponding counterterms. Because of the existence of
the vertex of $n$ gravitational gauge fields in tree diagram in the
non-renormalized Lagrangian, we need not introduce any new
counterterm which does not exist in the non-renormalized
Lagrangian, what we need to do is to redefine gravitational coupling
constant $g$ and gravitational gauge field $C_{\mu}^{\alpha}$
in renormalization.
If there is no  $e^{I(C)}$ term in the original Lagrangian, then
we will have to introduce infinite counterterms in
renormalization, and therefore the theory is non-renormalizable.
Because of the existence of the factor $e^{I(C)}$, though
quantum gauge theory of gravity looks like a non-renormalizable
theory according to power counting law, it is indeed renormalizable.
In a word, the factor $e^{I(C)}$ is highly important for the
quantum gauge theory of gravity. \\

\section{Gravitational Interactions of Scalar Fields}

Now, let's start to discuss gravitational interactions of matter
fields. First, we discuss gravitational interactions of scalar fields.
For the sake of simplicity, we first discuss real scalar field. Suppose
that $\phi(x)$ is a real scalar field. The traditional Lagrangian for the
real scalar field is
$$
- \frac{1}{2} \eta^{\mu \nu} \partial_{\mu} \phi(x)
\partial_{\nu} \phi(x) - \frac{m^2}{2} \phi^2 (x),
\eqno{(5.1)}
$$
where $m$ is the mass of scalar field.
This is the Lagrangian for a free real scalar field. The Euler-Lagrangian
equation of motion of it is
$$
( \eta^{\mu \nu} \partial_{\mu} \partial_{\nu} - m^2 ) \phi(x) =0,
\eqno{(5.2)}
$$
which is the famous Klein-Gordan equation. \\

Now, replace the ordinary partial derivative $\partial_{\mu}$
with gauge covariant derivative $D_{\mu}$, and add the Lagrangian of
pure gravitational gauge field, we get
$$
{\cal L}_0 = -\frac{1}{2} \eta^{\mu \nu} (D_{\mu} \phi)( D_{\nu} \phi)
-\frac{m^2}{2} \phi^2
- \frac{1}{4} \eta^{\mu \rho} \eta^{\nu \sigma} \eta_{2 \alpha \beta }
F_{\mu \nu}^{\alpha} F_{\rho \sigma}^{\beta}.
\eqno{(5.3)}
$$
The full Lagrangian is selected to be
$$
{\cal L} = e^{I(C)} {\cal L}_0,
\eqno{(5.4)}
$$
and the action $S$ is defined by
$$
S = \int {\rm d}^4 x ~ {\cal L}.
\eqno{(5.5)}
$$
\\

Using our previous definitions of gauge covariant derivative $D_{\mu}$
and strength of gravitational gauge field $F_{\mu \nu}^{\alpha}$, we can
obtain an explicit form of Lagrangian ${\cal L}$,
$$
{\cal L} = {\cal L}_F  +  {\cal  L}_I,
\eqno{(5.6)}
$$
with ${\cal L}_F$ the free Lagrangian and ${\cal L}_I$  the
interaction Lagrangian. Their explicit expressions are
$$
{\cal L}_F =
- \frac{1}{2} \eta^{\mu \nu} \partial_{\mu} \phi(x)
\partial_{\nu} \phi(x) - \frac{m^2}{2} \phi^2 (x)
- \frac{1}{4} \eta^{\mu \rho} \eta^{\nu \sigma} \eta_{2 \alpha \beta }
F_{0 \mu \nu}^{\alpha} F_{0 \rho \sigma}^{\beta},
\eqno{(5.7)}
$$
$$
\begin{array}{rcl}
{\cal L}_I &=&
{\cal L}_F \cdot (\sum_{n=1}^{\infty} \frac{1}{n!}
( g \eta_{1 \alpha_1}^{\mu_1} C_{\mu_1}^{\alpha_1} )^n )  \\
&&\\
&& + g e^{I(C)} \eta^{\mu \nu} C_{\mu}^{\alpha}
(\partial_{\alpha} \phi)(\partial_{\nu} \phi)
- \frac{g^2}{2} e^{I(C)} \eta^{\mu \nu} C_{\mu}^{\alpha}
C_{\nu}^{\beta} (\partial_{\alpha} \phi)(\partial_{\beta} \phi)  \\
&&\\
&&  + g e^{I(C)} \eta^{\mu \rho} \eta^{\nu \sigma}
\eta_{2 \alpha \beta} (\partial_{\mu} C_{\nu }^{\alpha}
- \partial_{\nu} C_{\mu}^{\alpha})
C_{\rho}^{\delta} \partial_{\delta} C_{\sigma}^{\beta} \\
&&\\
&&  - \frac{1}{2} g^2 e^{I(C)} \eta^{\mu \rho} \eta^{\nu \sigma}
\eta_{2 \alpha \beta}
(C_{\mu}^{\delta} \partial_{\delta} C_{\nu}^{\alpha}
- C_{\nu}^{\delta} \partial_{\delta} C_{\mu}^{\alpha} )
C_{\rho}^{\epsilon} \partial_{\epsilon} C_{\sigma}^{\beta},
\end{array}
\eqno{(5.8)}
$$
where,
$$
F_{0 \mu \nu}^{\alpha} = \partial_{\mu} C_{\nu}^{\alpha}
- \partial_{\nu}  C_{\mu}^{\alpha}.
\eqno{(5.9)}
$$
From eq.(5.8), we can see that scalar field can directly couples
to any number of gravitational gauge fields. This is one of the most
important interaction properties of gravity. Other kinds of interactions,
such as strong interactions, weak interactions and electromagnetic
interactions do not have this kind of interaction properties. Because
the gravitational coupling constant has negative mass dimension,
renormalization of theory needs this kind of interaction properties.
In other words, if matter field can not directly couple to any
number of gravitational gauge fields, the theory will be
non-renormalizable. \\

The symmetries of the theory can be easily seen from eq.(5.3).
First, let's discuss Lorentz symmetry. In eq.(5.3), some indexes
are Lorentz indexes and some are group indexes. Lorentz indexes
and group indexes have different transformation law under
gravitational gauge transformation, but they have the same
transformation law under Lorentz transformation. Therefor,
it can be easily seen that both ${\cal L}_0$ and $I(C)$ are
Lorentz scalars, the Lagrangian ${\cal  L}$ and action $S$
are invariant under global Lorentz transformation. \\

Under gravitational gauge transformations, real scalar field
$\phi (x)$ transforms as
$$
\phi (x) \to \phi '(x) = ( \ehat \phi (x) ),
\eqno{(5.10)}
$$
therefore,
$$
D_{\mu} \phi (x) \to D'_{\mu} \phi '(x) = ( \ehat D_{\mu} \phi (x) ).
\eqno{(5.11)}
$$
It can be easily proved that ${\cal L}_0$ transforms covariantly
$$
{\cal L}_0 \to {\cal L}'_0  = (\ehat {\cal L}_0),
\eqno{(5.12)}
$$
and the action eq.(5.5) of the system is invariant,
$$
S \to S' =  S.
\eqno{(5.13)}
$$
Please remember that eq.(4.27) is an important relation to be used
in the proof of the gravitational gauge symmetry of the action. \\

Global gravitational gauge symmetry gives out conserved charges.
Suppose that $\ehat$ is an infinitesimal gravitational gauge
transformation, it will have the form of eq.(4.29). The first order
variations of fields are
$$
\delta C_{\mu}^{\alpha} (x) =
- \epsilon^{\nu} (\partial_{\nu} C_{\mu}^{\alpha} (x)),
\eqno{(5.14)}
$$
$$
\delta \phi (x) =
- \epsilon^{\nu} (\partial_{\nu} \phi (x)),
\eqno{(5.15)}
$$
Using Euler-Lagrange equation of motions for scalar fields and
gravitational gauge fields, we can obtain that
$$
\delta S = \int {\rm d}^4 x
\epsilon^{\alpha} \partial_{\mu} T_{i \alpha}^{\mu},
\eqno{(5.16)}
$$
where
$$
T_{i \alpha}^{\mu} \equiv e^{I(C)}
( - \frac{\partial {\cal L}_0}{\partial \partial_{\mu} \phi} \partial_{\alpha} \phi
- \frac{\partial {\cal L}_0}{\partial \partial_{\mu} C_{\nu}^{\beta}}
\partial_{\alpha} C_{\nu}^{\beta} + \delta^{\mu}_{\alpha} {\cal L}_0 ).
\eqno{(5.17)}
$$
Because action is invariant under global gravitational gauge
transformation,
$$
\delta S = 0,
\eqno{(5.18)}
$$
and $\epsilon^{\alpha}$ is an  arbitrary  infinitesimal
constant 4-vector,   we obtain,
$$
\partial_{\mu} T_{i \alpha}^{\mu} = 0.
\eqno{(5.19)}
$$
This is the conservation equation for inertial energy-momentum
tensor. $T_{i \alpha}^{\mu}$ is the conserved current which
corresponds to the global gravitational gauge symmetry.
The space integration of the time component of inertial
energy-momentum tensor gives out the conserved charge,
which is just the inertial energy-momentum of the system.
The time component of the conserved charge is the Hamilton of
the system, which is
$$
H = - P_{i~0} =  \int {\rm d}^3 \svec{x} e^{I(C)}
( \pi_{\phi} \tdot{\phi} +
\pi_{\alpha}^{\mu} \tdot{C} _{\mu}^{\alpha} - {\cal L}_0),
\eqno{(5.20)}
$$
where $\pi_{\phi}$ is the canonical conjugate momentum of
the real scalar field. The inertial space momentum of the system
is given by
$$
P^i  = P_{i ~i} =  \int {\rm d}^3 \svec{x} e^{I(C)}
( - \pi_{\phi} \partial_i {\phi} -
\pi_{\alpha}^{\mu} \partial_i {C} _{\mu}^{\alpha} ).
\eqno{(5.21)}
$$
According to gauge principle, after quantization, they will become
generators of quantum gravitational gauge transformation. \\

There is an important and interesting relation that can be easily obtained
from the Lagrangian which is given by eq.(5.3). Define
$$
G_{\mu}^{\alpha}
=\delta_{\mu}^{\alpha} - g C_{\mu}^{\alpha}.
\eqno{(5.22)}
$$
Then Lagrangian given by eq(5.3) can be changed into
$$
{\cal L}_0 = -\frac{1}{2} g^{\alpha \beta} \partial_{\alpha} \phi
\partial_{\beta} \phi  -  \frac{m^2}{2} \phi^2
- \frac{1}{4} \eta^{\mu \rho} \eta^{\nu \sigma} \eta_{2 \alpha \beta }
F_{\mu \nu}^{\alpha} F_{\rho \sigma}^{\beta},
\eqno{(5.23)}
$$
where
$$
g^{\alpha \beta} \equiv \eta^{\mu \nu}
G_{\mu}^{\alpha} G_{\nu}^{\beta}.
\eqno{(5.24)}
$$
$g^{\alpha \beta}$ is the metric tensor of curved space-time in the
classical limit of the quantum gauge theory of gravity. We can easily
see that, when there is no gravitational field in space-time, that is,
$$
C_{\mu}^{\alpha} = 0,
\eqno{(5.25)}
$$
the classical space-time will be flat
$$
g^{ \alpha \beta} = \eta^{\alpha \beta}.
\eqno{(5.26)}
$$
This is what we expected in general relativity. We do not talk to much
on this problem here, for we will discuss this problem again
in details in the chapter on the classical limit of the quantum
gauge theory of gravity.  \\

Euler-Lagrange equations of motion can be easily deduced from
action principle. Keep gravitational gauge field $C_{\mu}^{\alpha}$
fixed and let real scalar field vary infinitesimally, then the first order
infinitesimal variation of action is
$$
\delta S =
\int {\rm d}^4 x  e^{I(C)}
\left( \frac{\partial {\cal L}_0}{\partial \phi}
- \partial_{\mu} \frac{\partial {\cal L}_0}{\partial \partial_{\mu} \phi}
- g \partial_{\mu} (\eta_{1 \alpha}^{\nu} C_{\nu}^{\alpha})
\frac{\partial {\cal L}_0}{\partial \partial_{\mu} \phi} \right) \delta \phi .
\eqno{(5.27)}
$$
Because $\delta \phi$ is an arbitrary variation of scalar field, according
to action principle, we get
$$
\frac{\partial {\cal L}_0}{\partial \phi}
- \partial_{\mu} \frac{\partial {\cal L}_0}{\partial \partial_{\mu} \phi}
- g \partial_{\mu} (\eta_{1 \alpha}^{\nu} C_{\nu}^{\alpha})
\frac{\partial {\cal L}_0}{\partial \partial_{\mu} \phi} = 0.
\eqno{(5.28)}
$$
 Because of the existence of the factor $e^{I(C)}$,
the equation of motion for scalar field is quite different from the
traditional form in quantum field theory. But the difference is
a second order infinitesimal quantity if we suppose that both gravitational
coupling constant and gravitational gauge field are first order
infinitesimal quantities. Because
$$
\frac{\partial {\cal L}_0}{\partial \partial_{\alpha} \phi}
= - g^{\alpha \beta} \partial_{\beta} \phi,
\eqno{(5.29)}
$$
$$
\frac{\partial {\cal L}_0}{\partial \phi}
= - m^2 \phi,
\eqno{(5.30)}
$$
the explicit form of the equation of motion of scalar field is
$$
g^{\alpha \beta} \partial_{\alpha} \partial_{\beta} \phi
- m^2 \phi + (\partial_{\alpha} g^{\alpha \beta})
\partial_{\beta} \phi
+ g g^{\alpha \beta} (\partial_{\beta} \phi )
\partial_{\alpha} (\eta_{1 \gamma }^{\nu} C_{\nu}^{\gamma}) = 0.
\eqno{(5.31)}
$$
The equation of motion for gravitational gauge field is:
$$
\partial_{\mu} (\eta^{\mu \lambda} \eta^{\nu \tau} \eta_{2 \alpha \beta}
F_{\lambda \tau}^{\beta} )  = - g T_{g \alpha}^{\nu},
\eqno{(5.32)}
$$
where $ T_{g \alpha}^{\nu}$ is the gravitational energy-momentum
tensor, whose definition is:
$$
\begin{array}{rcl}
T_{g \alpha}^{\nu}&=&
 - \frac{\partial {\cal L}_0}{\partial D_{\nu} C_{\mu}^{\beta}}
\partial_{\alpha} C_{\mu}^{\beta}
- \frac{\partial {\cal L}_0}{\partial D_{\nu} \phi}
\partial_{\alpha} \phi
+ \eta_{1 \alpha}^{\nu} {\cal L}_0  \\
&&\\
&& -  \partial_{\mu} (\eta_{1 \gamma}^{\rho} C_{\rho}^{\gamma})
\frac{\partial {\cal L}_0}{\partial \partial_{\mu} C_{\nu}^{\alpha}}
 + \partial_{\mu} (\eta^{\nu \lambda}
\eta^{\sigma \tau} \eta_{2 \alpha \beta}
F_{\lambda \tau}^{\beta} C_{\sigma}^{\mu}),
\end{array}
\eqno{(5.33)}
$$
We can see again that, for matter field, its inertial energy-momentum
tensor is also different from the gravitational energy-momentum
tensor, this difference completely originate from the influences
of gravitational gauge field. Compare eq.(5.33) with eq.(5.17),
and set gravitational gauge field to zero, that is
$$
D_{\mu} \phi = \partial_{\mu} \phi ,
\eqno{(5.34)}
$$
$$
I(C) = 0,
\eqno{(5.35)}
$$
then we find that two energy-momentum tensors are completely
the same:
$$
T_{i \alpha}^{\mu } = T_{g \alpha}^{\mu }.
\eqno{(5.36)}
$$
It means that the equivalence principle only strictly hold in a space-time
where there is no gravitational field. In the environment of strong
gravitational field, such as in black hole, the equivalence principle
will be strongly violated. \\

Define
$$
{\bf L} = \int {\rm d}^3 \svec{x} {\cal L}
= \int {\rm d}^3 \svec{x} e^{I(C)} {\cal L}_0.
\eqno{(5.37)}
$$
Then, we can easily prove that
$$
\frac{\delta {\bf L} }{\delta \phi} =
e^{I(C)} \left( \frac{\partial {\cal L}_0 }{\partial \phi}
- \partial_i \frac{\partial {\cal L}_0}{\partial \partial_i \phi}
-g \partial_i (\eta_{1 \alpha}^{\mu} C_{\mu}^{\alpha})
\frac{\partial {\cal L}_0}{\partial \partial_i \phi} \right),
\eqno{(5.38)}
$$
$$
\frac{\delta {\bf L} }{\delta \tdot{\phi} } =
e^{I(C)} \frac{\partial {\cal L}_0 }{\partial \tdot{\phi} } ,
\eqno{(5.39)}
$$
$$
\frac{\delta {\bf L} }{\delta C_{\nu}^{\alpha}} =
e^{I(C)} \left( \frac{\partial {\cal L}_0 }{\partial C_{\nu}^{\alpha}}
- \partial_i \frac{\partial
{\cal L}_0}{\partial \partial_i C_{\nu}^{\alpha}}
+ g \eta_{1 \alpha}^{\nu} {\cal L}_0
-g \partial_i (\eta_{1 \beta}^{\mu} C_{\mu}^{\beta})
\frac{\partial {\cal L}_0}{\partial \partial_i C_{\nu}^{\alpha} } \right),
\eqno{(5.40)}
$$
$$
\frac{\delta {\bf L} }{\delta \tdot{ C_{\nu}^{\alpha}} } =
e^{I(C)} \frac{\partial {\cal L}_0 }{\partial \tdot{ C_{\nu}^{\alpha}} } .
\eqno{(5.41)}
$$
Then, Hamilton's action principle gives out the following equations
of motion:
 $$
\frac{\delta {\bf L} }{\delta \phi}
- \frac{\rm d}{{\rm d}t}
\frac{\delta {\bf L} }{\delta \tdot{\phi}} = 0 ,
\eqno{(5.42)}
$$
$$
\frac{\delta {\bf L} }{\delta C_{\nu}^{\alpha} }
- \frac{\rm d}{{\rm d}t}
\frac{\delta {\bf L} }{\delta \tdot{ C_{\nu}^{\alpha} }} = 0 .
\eqno{(5.43)}
$$
These two equations of motion are essentially the same as the
Euler-Lagrange equations of motion which we have obtained before.
But these two equations have more beautiful forms. \\

The Hamiltonian of the system is given by a Legendre transformation,
$$
\begin{array}{rcl}
H  &= & \int {\rm d}^3 \svec{x}
(\frac{\delta {\bf L} }{\delta \tdot{ \phi} }  \tdot{\phi}
+ \frac{\delta {\bf L} }{\delta \tdot{ C_{\mu}^{\alpha}} }
\tdot{ C_{\mu}^{\alpha}} ) - {\bf L}  \\
&=&  \int {\rm d}^3 \svec{x} e^{I(C)}
( \pi_{\phi} \tdot{\phi} +
\pi_{\alpha}^{\mu} \tdot{C} _{\mu}^{\alpha} - {\cal L}_0),
\end{array}
\eqno{(5.44)}
$$
where,
$$
\pi_{\phi} =  \frac{\partial {\cal L}_0 }{\partial \tdot{\phi} } ,
\eqno{(5.45)}
$$
$$
\pi_{\alpha}^{\mu} =
\frac{\partial {\cal L}_0 }{\partial \tdot{C_{\mu}^{\alpha}} } .
\eqno{(5.46)}
$$
It can be easily seen that the Hamiltonian given by Legendre
transformation is completely the same as that given by
inertial energy-momentum tensor. After Legendre transformation,
$\phi$, $C_{\mu}^{\alpha}$, $e^{I(C)} \pi_{\phi}$ and
$e^{I(C)} \pi_{\alpha}^{\mu}$ are canonical independent
variables. Let these variables vary infinitesimally, we can get
$$
\frac{\delta H }{\delta \phi}
= - \frac{\delta {\bf L} }{\delta \phi} ,
\eqno{(5.47)}
$$
$$
\frac{\delta H }{\delta ( e^{I(C)} \pi_{\phi})}
= \tdot{\phi} ,
\eqno{(5.48)}
$$
$$
\frac{\delta H }{\delta C_{\nu}^{\alpha}}
= - \frac{\delta {\bf L} }{\delta C_{\nu}^{\alpha}} ,
\eqno{(5.49)}
$$
$$
\frac{\delta H }{\delta ( e^{I(C)} \pi_{\alpha}^{\nu})}
= \tdot{ C_{\nu}^{\alpha}}.
\eqno{(5.50)}
$$
Then, Hamilton's equations of motion read:
$$
\frac{\rm d}{{\rm d}t} \phi =
\frac{\delta H }{\delta ( e^{I(C)} \pi_{\phi})} ,
\eqno{(5.51)}
$$
$$
\frac{\rm d}{{\rm d}t} (e^{I(C)} \pi_{\phi} ) =
- \frac{\delta H }{\delta \phi} ,
\eqno{(5.52)}
$$
$$
\frac{\rm d}{{\rm d}t} C_{\nu}^{\alpha} =
\frac{\delta H }{\delta ( e^{I(C)} \pi_{\alpha}^{\nu})} ,
\eqno{(5.53)}
$$
$$
\frac{\rm d}{{\rm d}t} (e^{I(C)} \pi_{\alpha}^{\nu} ) =
- \frac{\delta H }{\delta C_{\nu}^{\alpha}} .
\eqno{(5.54)}
$$
The  forms of the Hamilton's equations of motion are completely
the same as those appears in usual quantum field theory and usual
classical analytical mechanics. Therefor, the introduction of the
factor $e^{I(C)}$ does not affect the forms of Lagrange equations
of motion and Hamilton's equations of motion. \\

The Poisson brackets of two general functional of canonical arguments
can be defined by
$$
\begin{array}{rcl}
\lbrace  A ~~,~~ B \rbrace  &=& \int {\rm d}^3 \svec{x}
( \frac{\delta A}{\delta \phi} \frac{\delta B}{\delta (e^{I(C)} \pi_{\phi})}
- \frac{\delta A}{\delta (e^{I(C)} \pi_{\phi})}
\frac{\delta B}{\delta \phi}  \\
&&\\
&&+\frac{\delta A}{\delta C_{\nu}^{\alpha}}
\frac{\delta B}{\delta (e^{I(C)} \pi_{\alpha}^{\nu})}
- \frac{\delta A}{\delta (e^{I(C)} \pi_{\alpha}^{\nu})}
\frac{\delta B}{\delta C_{\nu}^{\alpha}} ).
\end{array}
\eqno{(5.55)}
$$
According to this definition, we have
$$
\lbrace  \phi(\svec{x},t) ~~,
~~ (e^{I(C)} \pi_{\phi})(\svec{y},t) \rbrace
= \delta^3(\svec{x} - \svec{y}),
\eqno{(5.56)}
$$
$$
\lbrace  C_{\nu}^{\alpha}(\svec{x},t) ~~,
~~ (e^{I(C)} \pi_{\beta}^{\mu})(\svec{y},t) \rbrace
= \delta_{\nu}^{\mu} \delta^{\alpha}_{\beta}
\delta^3(\svec{x} - \svec{y}).
\eqno{(5.57)}
$$
These two relations can be used as the starting point of
canonical quantization of quantum gravity.\\

Using Poisson brackets, the Hamilton's equations of motion can
be expressed in another forms,
$$
\frac{\rm d}{{\rm d}t} \phi (\svec{x},t) =
\lbrace  \phi(\svec{x},t) ~~,~~ H \rbrace  ,
\eqno{(5.58)}
$$
$$
\frac{\rm d}{{\rm d}t} (e^{I(C)} \pi_{\phi}) (\svec{x},t) =
\lbrace  (e^{I(C)} \pi_{\phi}) (\svec{x},t) ~~,~~ H \rbrace  ,
\eqno{(5.59)}
$$
$$
\frac{\rm d}{{\rm d}t} C_{\nu}^{\alpha} (\svec{x},t) =
\lbrace C_{\nu}^{\alpha}(\svec{x},t) ~~,~~ H \rbrace  ,
\eqno{(5.60)}
$$
$$
\frac{\rm d}{{\rm d}t} (e^{I(C)} \pi_{\alpha}^{\nu}) (\svec{x},t) =
\lbrace  (e^{I(C)} \pi_{\alpha}^{\nu} ) (\svec{x},t) ~~,~~ H \rbrace  .
\eqno{(5.61)}
$$
Therefore, if $A$ is an arbitrary functional of the canonical arguments
$\phi$, $C_{\mu}^{\alpha}$, $e^{I(C)} \pi_{\phi}$ and
$e^{I(C)} \pi_{\alpha}^{\mu}$, then we have
$$
\tdot{A} =
\lbrace  A ~~,~~ H \rbrace  .
\eqno{(5.62)}
$$
After quantization, this equation will become the Heisenberg equation.
\\

If $\phi(x)$ is a complex scalar field, its traditional Lagrangian is
$$
-  \eta^{\mu \nu} \partial_{\mu} \phi(x)
\partial_{\nu} \phi^{*}(x) - m^2 \phi(x) \phi^{*}(x).
\eqno{(5.63)}
$$
Replace ordinary partial derivative with gauge covariant derivative,
and add into the Lagrangian for pure gravitational gauge field,
we get,
$$
{\cal L}_0 = - \eta^{\mu \nu} D_{\mu} \phi  (D_{\nu} \phi)^*
- m^2\phi \phi^{*}
- \frac{1}{4} \eta^{\mu \rho} \eta^{\nu \sigma} \eta_{2 \alpha \beta }
F_{\mu \nu}^{\alpha} F_{\rho \sigma}^{\beta}.
\eqno{(5.64)}
$$
Repeating all above discussions, we can get the whole theory
for gravitational interactions of complex scalar fields. We will not
repeat this discussion here.\\

\section{Gravitational Interactions of Dirac Field}

In the usual quantum field theory, the Lagrangian for Dirac field is
$$
- \bar{\psi} (\gamma^{\mu} \partial_{\mu} + m) \psi.
\eqno{(6.1)}
$$
Replace ordinary partial derivative with gauge covariant derivative,
and add into the Lagrangian of pure gravitational gauge field,
we get,
$$
{\cal L}_0 =
- \bar{\psi} (\gamma^{\mu} D_{\mu} + m) \psi
- \frac{1}{4} \eta^{\mu \rho} \eta^{\nu \sigma} \eta_{2 \alpha \beta }
F_{\mu \nu}^{\alpha} F_{\rho \sigma}^{\beta}.
\eqno{(6.2)}
$$
The full Lagrangian of the system is
$$
{\cal L} = e^{I(C)} {\cal L}_0 ,
\eqno{(6.3)}
$$
and the corresponding action is
$$
S =  \int {\rm d}^4 x {\cal L}
=\int {\rm d}^4 x  ~~ e^{I(C)} {\cal L}_0 .
\eqno{(6.4)}
$$
This Lagrangian can be separated into two parts,
$$
{\cal L} = {\cal L}_F + {\cal L}_I ,
\eqno{(6.5)}
$$
with ${\cal L}_F$ the free Lagrangian and ${\cal L}_I$ the interaction
Lagrangian. Their explicit forms are
$$
{\cal L}_F =
- \bar{\psi} (\gamma^{\mu} \partial_{\mu} + m) \psi
- \frac{1}{4} \eta^{\mu \rho} \eta^{\nu \sigma} \eta_{2 \alpha \beta }
F_{0 \mu \nu}^{\alpha} F_{0 \rho \sigma}^{\beta},
\eqno{(6.6)}
$$
$$
\begin{array}{rcl}
{\cal L}_I &=&
{\cal L}_F \cdot (\sum_{n=1}^{\infty} \frac{1}{n!}
( g \eta_{1 \alpha_1}^{\mu_1} C_{\mu_1}^{\alpha_1} )^n )
 + g e^{I(C)} \bar{\psi} \gamma^{\mu} (\partial_{\alpha}
\psi) C_{\mu}^{\alpha}  \\
&&\\
&&  + g e^{I(C)} \eta^{\mu \rho} \eta^{\nu \sigma}
\eta_{2 \alpha \beta} (\partial_{\mu} C_{\nu }^{\alpha}
- \partial_{\nu} C_{\mu}^{\alpha})
C_{\rho}^{\delta} \partial_{\delta} C_{\sigma}^{\beta} \\
&&\\
&&  - \frac{1}{2} g^2 e^{I(C)} \eta^{\mu \rho} \eta^{\nu \sigma}
\eta_{2 \alpha \beta}
(C_{\mu}^{\delta} \partial_{\delta} C_{\nu}^{\alpha}
- C_{\nu}^{\delta} \partial_{\delta} C_{\mu}^{\alpha} )
C_{\rho}^{\epsilon} \partial_{\epsilon} C_{\sigma}^{\beta}.
\end{array}
\eqno{(6.7)}
$$
From ${\cal L}_I$, we can see that Dirac field can directly
couple to any number of gravitational gauge fields, the mass
term of Dirac field also take part in gravitational interactions. All these
interactions are completely determined by the requirement of
gravitational gauge symmetry. The Lagrangian function before
renormalization almost contains all kind of divergent vertex,
which is important in the renormalization of the theory.
Besides, from eq.(6.7), we can directly write out Feynman
rules of the corresponding interaction vertexes.
\\

Because the traditional Lagrangian function eq.(6.1) is invariant
under global Lorentz transformation, which is already proved in the
traditional quantum field theory, and the covariant derivative
has the same behavior as partial derivative under global Lorentz
transformation, the first two terms of Lagrangian ${\cal L}$ are
global Lorentz invariant. We have already prove that the Lagrangian
function for pure gravitational gauge field is invariant under
global Lorentz transformation. Therefor, ${\cal L}$ has global
Lorentz symmetry.  \\

The gravitational gauge transformation of Dirac field is
$$
\psi(x) \to \psi ' (x) = (\ehat \psi (x) ).
\eqno{(6.8)}
$$
$\bar{\psi}$ transforms similarly,
$$
\bar{\psi}(x) \to \bar{\psi} ' (x) = (\ehat \bar{\psi} (x) ).
\eqno{(6.9)}
$$
Dirac $\gamma$-matrices is not a physical field, so it keeps unchanged
under gravitational gauge transformation,
$$
\gamma^{\mu} \to \gamma^{\mu}.
\eqno{(6.10)}
$$
It can be proved that, under gravitational gauge transformation,
${\cal L}_0$ transforms as
$$
{\cal L}_0 \to {\cal L}'_0 = (\ehat {\cal L}_0 ).
\eqno{(6.11)}
$$
So,
$$
{\cal L} \to {\cal L}'  =  J (\ehat {\cal L}_0 ),
\eqno{(6.12)}
$$
where $J$ is the Jacobi of the corresponding space-time
translation. Then using eq.(4.27), we can prove that the action $S$
has gravitational gauge symmetry. \\

Suppose that $\ehat$ is an infinitesimal global transformation, then
the first order infinitesimal variations of Dirac field are
$$
\delta \psi = - \epsilon^{\nu} \partial_{\nu} \psi,
\eqno{(6.13)}
$$
$$
\delta \bar{\psi}  = - \epsilon^{\nu} \partial_{\nu} \bar{\psi}.
\eqno{(6.14)}
$$
The first order variation of action is
$$
\delta S = \int {\rm d}^4 x
\epsilon^{\alpha} \partial_{\mu} T_{i \alpha}^{\mu},
\eqno{(6.15)}
$$
where $ T_{i \alpha}^{\mu}$ is the inertial energy-momentum
tensor whose definition is,
$$
T_{i \alpha}^{\mu} \equiv e^{I(C)}
\left( - \frac{\partial {\cal L}_0}{\partial \partial_{\mu} \psi}
\partial_{\alpha} \psi
- \frac{\partial {\cal L}_0}{\partial \partial_{\mu} C_{\nu}^{\beta}}
\partial_{\alpha} C_{\nu}^{\beta}
+ \delta^{\mu}_{\alpha} {\cal L}_0 \right).
\eqno{(6.16)}
$$
The global gravitational gauge symmetry of action gives out conservation
equation of the inertial energy-momentum tensor,
$$
\partial_{\mu} T_{i \alpha}^{\mu} = 0.
\eqno{(6.17)}
$$
The inertial energy-momentum tensor is the conserved current
which expected by gauge principle. The space integration of its
time component gives out the conserved energy-momentum of the
system,
$$
H = - P_{i~0} =  \int {\rm d}^3 \svec{x} e^{I(C)}
( \pi_{\psi} \tdot{\psi} +
\pi_{\alpha}^{\mu} \tdot{C} _{\mu}^{\alpha} - {\cal L}_0),
\eqno{(6.18)}
$$
$$
P^i  = P_{i~i} =  \int {\rm d}^3 \svec{x} e^{I(C)}
( - \pi_{\psi} \partial_i {\psi} -
\pi_{\alpha}^{\mu} \partial_i {C} _{\mu}^{\alpha} ),
\eqno{(6.19)}
$$
where
$$
\pi_{\psi} = \frac{\partial {\cal L}_0}{\partial \tdot{\psi}}.
\eqno{(6.20)}
$$

The equation of motion for Dirac field is
$$
(\gamma^{\mu} D_{\mu} + m) \psi = 0.
\eqno{(6.21)}
$$
From this expression, we can see that the factor $e^{I(C)}$ does not
affect the equation of motion of Dirac field. This is caused by the
asymmetric form of the Lagrangian. If we use a symmetric form of
Lagrangian, the factor $e^{I(C)}$ will also affect the equation
of motion of Dirac field, which will be discussed later. \\

The equation of motion for gravitational fields can be easily
deduced,
$$
\partial_{\mu} (\eta^{\mu \lambda} \eta^{\nu \tau} \eta_{2 \alpha \beta}
F_{\lambda \tau}^{\beta} )  = - g T_{g \alpha}^{\nu},
\eqno{(6.22)}
$$
where $ T_{g \alpha}^{\nu}$ is the gravitational energy-momentum
tensor, whose definition is:
$$
\begin{array}{rcl}
T_{g \alpha}^{\nu}&=&
 - \frac{\partial {\cal L}_0}{\partial D_{\nu} C_{\mu}^{\beta}}
\partial_{\alpha} C_{\mu}^{\beta}
- \frac{\partial {\cal L}_0}{\partial D_{\nu} \psi}
\partial_{\alpha} \psi
+ \eta_{1 \alpha}^{\nu} {\cal L}_0  \\
&&\\
&& -  \partial_{\mu} (\eta_{1 \gamma}^{\rho} C_{\rho}^{\gamma})
\frac{\partial {\cal L}_0}{\partial \partial_{\mu} C_{\nu}^{\alpha}}
 + \partial_{\mu} (\eta^{\nu \lambda}
\eta^{\sigma \tau} \eta_{2 \alpha \beta}
F_{\lambda \tau}^{\beta} C_{\sigma}^{\mu}).
\end{array}
\eqno{(6.23)}
$$
We see again that the gravitational energy-momentum tensor is
different from the  inertial energy-momentum tensor. \\

In usual quantum field theory, the Lagrangian for Dirac field
has a more symmetric form, which is
$$
- \bar{\psi} (\gamma^{\mu} \dvec{\partial}_{\mu} + m) \psi,
\eqno{(6.24)}
$$
where
$$
 \dvec{\partial}_{\mu} =
\frac{\partial_{\mu} - \lvec{\partial_{\mu}}}{2}.
\eqno{(6.25)}
$$
The Euler-Lagrange equation of motion of eq.(6.24) also
gives out the conventional Dirac equation. \\

Now replace ordinary space-time partial derivative with
covariant derivative, and add into the Lagrangian of pure
gravitational gauge field, we get,
$$
{\cal L}_0 =
- \bar{\psi} (\gamma^{\mu} \dvec{D}_{\mu} + m) \psi
- \frac{1}{4} \eta^{\mu \rho} \eta^{\nu \sigma} \eta_{2 \alpha \beta }
F_{\mu \nu}^{\alpha} F_{\rho \sigma}^{\beta},
\eqno{(6.26)}
$$
where $\dvec{D}_{\mu}$ is defined by
$$
 \dvec{D}_{\mu} =
\frac{D_{\mu} - \lvec{D}_{\mu}}{2}.
\eqno{(6.27)}
$$
Operator $\lvec{D}_{\mu}$ is understood in the following way
$$
f(x) \lvec{D}_{\mu} g(x) =
(D_{\mu} f(x)) g(x),
\eqno{(6.28)}
$$
with $f(x)$ and $g(x)$ two arbitrary functions. The Lagrangian density
${\cal L}$ and action $S$ are also defined by eqs.(6.3-4). In this case,
the free Lagrangian ${\cal L}_F$ and interaction Lagrangian ${\cal L}_I$
are given by
$$
{\cal L}_F =
- \bar{\psi} (\gamma^{\mu} \dvec{\partial}_{\mu} + m) \psi
- \frac{1}{4} \eta^{\mu \rho} \eta^{\nu \sigma} \eta_{2 \alpha \beta }
F_{0 \mu \nu}^{\alpha} F_{0 \rho \sigma}^{\beta},
\eqno{(6.29)}
$$
$$
\begin{array}{rcl}
{\cal L}_I &=&
{\cal L}_F \cdot (\sum_{n=1}^{\infty} \frac{1}{n!}
( g \eta_{1 \alpha_1}^{\mu_1} C_{\mu_1}^{\alpha_1} )^n )
 + g e^{I(C)}( \bar{\psi} \gamma^{\mu} \dvec{\partial}_{\alpha}
\psi) C_{\mu}^{\alpha}  \\
&&\\
&&  + g e^{I(C)} \eta^{\mu \rho} \eta^{\nu \sigma}
\eta_{2 \alpha \beta} (\partial_{\mu} C_{\nu }^{\alpha}
- \partial_{\nu} C_{\mu}^{\alpha})
C_{\rho}^{\delta} \partial_{\delta} C_{\sigma}^{\beta} \\
&&\\
&&  - \frac{1}{2} g^2 e^{I(C)} \eta^{\mu \rho} \eta^{\nu \sigma}
\eta_{2 \alpha \beta}
(C_{\mu}^{\delta} \partial_{\delta} C_{\nu}^{\alpha}
- C_{\nu}^{\delta} \partial_{\delta} C_{\mu}^{\alpha} )
C_{\rho}^{\epsilon} \partial_{\epsilon} C_{\sigma}^{\beta}.
\end{array}
\eqno{(6.30)}
$$
\\

The Euler-Lagrange equation of motion for Dirac field is
$$
\frac{\partial {\cal L}_0}{\partial \bar{\psi}}
- \partial_{\mu} \frac{\partial {\cal L}_0}{\partial
\partial_{\mu} \bar{\psi}}
- g \partial_{\mu} (\eta_{1 \alpha}^{\nu} C_{\nu}^{\alpha})
\frac{\partial {\cal L}_0}{\partial
\partial_{\mu} \bar{\phi}} = 0.
\eqno{(6.31)}
$$
Because
$$
\frac{\partial {\cal L}_0}{\partial \bar{\psi}}
= e^{I(C)} ( -\frac{1}{2} \gamma^{\mu} D_{\mu} \psi
- m \psi ),
\eqno{(6.32)}
$$
$$
\frac{\partial {\cal L}_0}{\partial \partial_{\mu} \bar{\psi}}
= e^{I(C)} ( \frac{1}{2} \gamma^{\alpha} G_{\alpha}^{\mu} \psi  ),
\eqno{(6.33)}
$$
eq.(6.31) will be changed into
$$
(\gamma^{\mu} D_{\mu} + m) \psi =
- \frac{1}{2}\gamma^{\mu}
(\partial_{\alpha} G_{\mu}^{\alpha}) \psi
-  g \gamma^{\mu} \psi
D_{\mu} (\eta_{1 \beta}^{\nu} C_{\nu}^{\beta} ).
\eqno{(6.34)}
$$
If gravitational gauge field vanishes, this equation of motion
will return to the traditional Dirac equation. \\

The inertial energy-momentum tensor now becomes
$$
T_{i \alpha}^{\mu} = e^{I(C)}
( - \frac{\partial {\cal L}_0}{\partial \partial_{\mu} \psi}
\partial_{\alpha} \psi
- (\partial_{\alpha} \bar{\psi})
\frac{\partial {\cal L}_0}{\partial \partial_{\mu} \bar{\psi}}
- \frac{\partial {\cal L}_0}{\partial \partial_{\mu} C_{\nu}^{\beta}}
\partial_{\alpha} C_{\nu}^{\beta} + \delta^{\mu}_{\alpha} {\cal L}_0 ),
\eqno{(6.35)}
$$
and the gravitational energy-momentum tensor becomes
$$
\begin{array}{rcl}
T_{g \alpha}^{\nu}&=&
 - \frac{\partial {\cal L}_0}{\partial D_{\nu} C_{\mu}^{\beta}}
\partial_{\alpha} C_{\mu}^{\beta}
- \frac{\partial {\cal L}_0}{\partial D_{\nu} \psi}
\partial_{\alpha} \psi
- (\partial_{\alpha} \bar{\psi})
\frac{\partial {\cal L}_0}{\partial D_{\nu} \bar{\psi}}
+ \eta_{1 \alpha}^{\nu} {\cal L}_0  \\
&&\\
&& -  \partial_{\mu} (\eta_{1 \gamma}^{\rho} C_{\rho}^{\gamma})
\frac{\partial {\cal L}_0}{\partial \partial_{\mu} C_{\nu}^{\alpha}}
 + \partial_{\mu} (\eta^{\nu \lambda}
\eta^{\sigma \tau} \eta_{2 \alpha \beta}
F_{\lambda \tau}^{\beta} C_{\sigma}^{\mu}).
\end{array}
\eqno{(6.36)}
$$
Both of them are conserved energy-momentum tensor.  But they
are not equivalent.
\\

\section{Gravitational Interactions of Vector Field}

The traditional Lagrangian for vector field is
$$
- \frac{1}{4} \eta^{\mu \rho} \eta^{\nu \sigma}
A_{\mu \nu} A_{\rho \sigma}
- \frac{m^2}{2}  \eta^{\mu \nu} A_{\mu} A_{\nu},
\eqno{(7.1)}
$$
where $A_{\mu \nu}$ is the strength of vector field which is
given by
$$
 \partial_{\mu} A_{\nu}
- \partial_{\nu} A_{\mu}.
\eqno{(7.2)}
$$
The Lagrangian ${\cal L}_0$ that describes gravitational interactions
between vector field and gravitational fields is
$$
{\cal L}_0 =
- \frac{1}{4} \eta^{\mu \rho} \eta^{\nu \sigma}
A_{\mu \nu} A_{\rho \sigma}
- \frac{m^2}{2}  \eta^{\mu \nu} A_{\mu} A_{\nu}
- \frac{1}{4} \eta^{\mu \rho} \eta^{\nu \sigma} \eta_{2 \alpha \beta }
F_{\mu \nu}^{\alpha} F_{\rho \sigma}^{\beta}.
\eqno{(7.3)}
$$
In eq.(7.3), the definition of strength $A_{\mu \nu}$ is not
given by eq.(7.2), it is given by
$$
\begin{array}{rcl}
A_{\mu \nu} &=& D_{\mu} A_{\nu}
- D_{\nu} A_{\mu}  \\
&=&  \partial_{\mu} A_{\nu} - \partial_{\nu} A_{\mu}
- g C_{\mu}^{\alpha} \partial_{\alpha} A_{\nu}
+ g C_{\nu}^{\alpha} \partial_{\alpha} A_{\mu},
\end{array}
\eqno{(7.4)}
$$
where $D_{\mu}$ is the gravitational gauge
 covariant derivative, whose definition
is given by eq.(4.4). The full Lagrangian ${\cal L}$ is given by,
$$
{\cal L} = e^{I(C)} {\cal L}_0.
\eqno{(7.5)}
$$
The action $S$ is defined by
$$
S = \int {\rm d}^4 x ~ {\cal L}.
\eqno{(7.6)}
$$
\\

The Lagrangian ${\cal L}$ can be separated into two parts:
the free Lagrangian ${\cal L}_F$ and interaction Lagrangian
${\cal L}_I$. The explicit forms of them are
$$
{\cal L}_F =
- \frac{1}{4} \eta^{\mu \rho} \eta^{\nu \sigma}
A_{0 \mu \nu} A_{0 \rho \sigma}
- \frac{m^2}{2}  \eta^{\mu \nu} A_{\mu} A_{\nu}
- \frac{1}{4} \eta^{\mu \rho} \eta^{\nu \sigma} \eta_{2 \alpha \beta }
F_{0 \mu \nu}^{\alpha} F_{0 \rho \sigma}^{\beta},
\eqno{(7.7)}
$$
$$
\begin{array}{rcl}
{\cal L}_I &=&
{\cal L}_F \cdot (\sum_{n=1}^{\infty} \frac{1}{n!}
( g \eta_{1 \alpha_1}^{\mu_1} C_{\mu_1}^{\alpha_1} )^n )
 + g e^{I(C)} \eta^{\mu \rho} \eta^{\nu \sigma}
A_{0 \mu \nu} C_{\rho}^{\alpha} \partial_{\alpha} A_{\sigma} \\
&&\\
&& - \frac{g^2}{2} e^{I(C)} \eta^{\mu \rho} \eta^{\nu \sigma}
( C_{\mu}^{\alpha} C_{\rho}^{\beta} (\partial_{\alpha} A_{\nu} )
(\partial_{\beta} A_{\sigma} )
- C_{\nu}^{\alpha} C_{\rho}^{\beta} (\partial_{\alpha} A_{\mu} )
(\partial_{\beta} A_{\sigma} )  )  \\
&&\\
&&  + g e^{I(C)} \eta^{\mu \rho} \eta^{\nu \sigma}
\eta_{2 \alpha \beta} (\partial_{\mu} C_{\nu }^{\alpha}
- \partial_{\nu} C_{\mu}^{\alpha})
C_{\rho}^{\delta} \partial_{\delta} C_{\sigma}^{\beta} \\
&&\\
&&  - \frac{1}{2} g^2 e^{I(C)} \eta^{\mu \rho} \eta^{\nu \sigma}
\eta_{2 \alpha \beta}
(C_{\mu}^{\delta} \partial_{\delta} C_{\nu}^{\alpha}
- C_{\nu}^{\delta} \partial_{\delta} C_{\mu}^{\alpha} )
C_{\rho}^{\epsilon} \partial_{\epsilon} C_{\sigma}^{\beta},
\end{array}
\eqno{(7.8)}
$$
where
$A_{0 \mu \nu} = \partial_{\mu} A_{\nu} - \partial_{\nu} A_{\mu}$
The first two lines of ${\cal L}_I$ contain interactions between
vector field and gravitational gauge fields. It can be seen that
the vector field can also directly couple to arbitrary number of
gravitational gauge fields, which is one of the most important
properties of gravitational gauge interactions. This interaction
properties are required and determined by local gravitational
gauge symmetry.  \\

Under Lorentz transformations, group index and Lorentz index have
the same behavior. Therefor every term in the Lagrangian ${\cal L}$
are Lorentz scalar, and the whole Lagrangian ${\cal L}$ and
action $S$ have Lorentz symmetry. \\

Under gravitational gauge transformations, vector field $A_{\mu}$
transforms as
$$
A_{\mu} (x) \to A'_{\mu}(x) = (\ehat A_{\mu} (x)).
\eqno{(7.9)}
$$
$D_{\mu} A_{\nu}$ and $A_{\mu \nu}$ transform covariantly,
$$
D_{\mu} A_{\nu} \to D'_{\mu} A'_{\nu} =
(\ehat D_{\mu} A_{\nu}) ,
\eqno{(7.10)}
$$
$$
A_{\mu \nu} \to A'_{\mu \nu} =
(\ehat A_{\mu \nu}).
\eqno{(7.11)}
$$
So, the gravitational gauge transformations of ${\cal L}_0$
and ${\cal L}$ respectively are
$$
{\cal L}_0 \to {\cal L}'_0 = (\ehat {\cal L}_0 ),
\eqno{(7.12)}
$$
$$
{\cal L} \to {\cal L}'  =  J (\ehat {\cal L}_0 ).
\eqno{(7.13)}
$$
The action of the system is gravitational gauge invariant.
\\

The global gravitational gauge transformation gives out conserved
current of gravitational gauge  symmetry.
Under infinitesimal global gravitational gauge
transformation, the vector field $A_{\mu}$ transforms as
$$
\delta A_{\mu} = - \epsilon^{\alpha} \partial_{\alpha} A_{\mu}.
\eqno{(7.14)}
$$
The first order variation of action is
$$
\delta S = \int {\rm d}^4 x
\epsilon^{\alpha} \partial_{\mu} T_{i \alpha}^{\mu},
\eqno{(7.15)}
$$
where $ T_{i \alpha}^{\mu}$ is the inertial energy-momentum
tensor whose definition is,
$$
T_{i \alpha}^{\mu} = e^{I(C)}
\left( - \frac{\partial {\cal L}_0}{\partial \partial_{\mu} A_{\nu}}
\partial_{\alpha} A_{\nu}
- \frac{\partial {\cal L}_0}{\partial \partial_{\mu} C_{\nu}^{\beta}}
\partial_{\alpha} C_{\nu}^{\beta} +
\delta^{\mu}_{\alpha} {\cal L}_0 \right).
\eqno{(7.16)}
$$
$T_{i \alpha}^{\mu}$ is a conserved current. The space integration
of its time component gives out inertial energy-momentum of the
system,
$$
H = - P_{i~0} =  \int {\rm d}^3 \svec{x} e^{I(C)}
( \pi^{\mu} \tdot{A_{\mu}} +
\pi_{\alpha}^{\mu} \tdot{C} _{\mu}^{\alpha} - {\cal L}_0),
\eqno{(7.17)}
$$
$$
P^i  = P_{i~i} =  \int {\rm d}^3 \svec{x} e^{I(C)}
( - \pi^{\mu} \partial_i A_{\mu} -
\pi_{\alpha}^{\mu} \partial_i {C} _{\mu}^{\alpha} ),
\eqno{(7.18)}
$$
where
$$
\pi^{\mu} = \frac{\partial {\cal L}_0}{\partial \tdot{A_{\mu}}}.
\eqno{(7.19)}
$$
\\

The equation of motion for vector field is
$$
\frac{\partial {\cal L}_0}{\partial A_{\nu}}
- \partial_{\mu}
\frac{\partial {\cal L}_0}{\partial \partial_{\mu} A_{\nu}}
- g \partial_{\mu}
(\eta_{1 \alpha}^{\lambda} C_{\lambda}^{\alpha})
\frac{\partial {\cal L}_0}{\partial \partial_{\mu} A_{\nu}} = 0.
\eqno{(7.20)}
$$
From eq.(7.3), we can obtain
$$
\frac{\partial {\cal L}_0}{\partial \partial_{\mu} A_{\nu}}
= - \eta^{\lambda \rho} \eta^{\nu \sigma}
G_{\lambda}^{\mu} A_{\rho \sigma},
\eqno{(7.21)}
$$
$$
\frac{\partial {\cal L}_0}{\partial A_{\nu}}
= - m^2 \eta^{\lambda \nu} A_{\lambda}.
\eqno{(7.22)}
$$
Then, eq.(7.20) is changed into
$$
\eta^{\mu \rho} \eta^{\nu \sigma} D_{\mu} A_{\rho \sigma}
- m^2 \eta^{\mu \nu} A_{\mu}
= - \eta^{\lambda \rho} \eta^{\nu \sigma}
( \partial_{\mu} G_{\lambda}^{\mu} ) A_{\rho \sigma}
- g \eta^{\mu \rho} \eta^{\nu \sigma}
 A_{\rho \sigma} D_{\mu}
(\eta_{1 \alpha}^{\mu} C_{\mu}^{\alpha}).
\eqno{(7.23)}
$$
The equation of motion of gravitational gauge field is
$$
\partial_{\mu} (\eta^{\mu \lambda} \eta^{\nu \tau} \eta_{2 \alpha \beta}
F_{\lambda \tau}^{\beta} )  = - g T_{g \alpha}^{\nu},
\eqno{(7.24)}
$$
where $ T_{g \alpha}^{\nu}$ is the gravitational energy-momentum
tensor,
$$
\begin{array}{rcl}
T_{g \alpha}^{\nu}&=&
 - \frac{\partial {\cal L}_0}{\partial D_{\nu} C_{\mu}^{\beta}}
\partial_{\alpha} C_{\mu}^{\beta}
- \frac{\partial {\cal L}_0}{\partial D_{\nu} A_{\mu}}
\partial_{\alpha} A_{\mu}
+ \eta_{1 \alpha}^{\nu} {\cal L}_0  \\
&&\\
&& -  \partial_{\mu} (\eta_{1 \gamma}^{\rho} C_{\rho}^{\gamma})
\frac{\partial {\cal L}_0}{\partial \partial_{\mu} C_{\nu}^{\alpha}}
 + \partial_{\mu} (\eta^{\nu \lambda}
\eta^{\sigma \tau} \eta_{2 \alpha \beta}
F_{\lambda \tau}^{\beta} C_{\sigma}^{\mu}).
\end{array}
\eqno{(7.25)}
$$
$T_{g \alpha}^{\nu}$ is also a conserved current. The space integration
of its time component gives out the gravitational energy-momentum
which is the source of gravitational interactions.
It can be also seen that inertial energy-momentum tensor and
gravitational energy-momentum tensor are not equivalent.   \\

\section{Gravitational Interactions of Gauge Fields}

It is know that QED, QCD and unified electroweak theory are
all gauge theories. In this chapter, we will
discuss how to unify these gauge theories with gravitational
gauge theory, and how to unify four different kinds of
fundamental interactions formally. \\

First, let's discuss QED theory. As an example, let's discuss
electromagnetic interactions of Dirac field. The traditional
electromagnetic interactions between Dirac field $\psi$ and
electromagnetic field $A_{\mu}$ is
$$
- \frac{1}{4} \eta^{\mu \rho} \eta^{\nu \sigma}
A_{\mu \nu} A_{\rho \sigma}
- \bar{\psi}
( \gamma^{\mu} ( \partial_{\mu} - i e A_{\mu}  ) + m ) \psi.
\eqno{(8.1)}
$$
The Lagrangian that describes gravitational gauge interactions
between gravitational gauge field and Dirac field or electromagnetic
field and describes electromagnetic interactions between Dirac field
and electromagnetic field is
$$
{\cal L}_0 =
- \bar{\psi} (\gamma^{\mu} (D_{\mu} - i e A_{\mu}  ) + m ) \psi
- \frac{1}{4} \eta^{\mu \rho} \eta^{\nu \sigma}
{\mathbf A}_{\mu \nu} {\mathbf A}_{\rho \sigma}
- \frac{1}{4} \eta^{\mu \rho} \eta^{\nu \sigma}
\eta_{2 \alpha \beta } F_{\mu\nu}^{\alpha} F_{\rho \sigma}^{\beta},
\eqno{(8.2)}
$$
where $D_{\mu}$ is the gravitational gauge covariant derivative
which is given by eq.(4.4) and the strength of electromagnetic
field $A_{\mu}$ is
$$
{\mathbf A}_{\mu \nu} = A_{\mu \nu} + g G^{-1 \lambda}_{\alpha}
A_{\lambda} F_{\mu \nu}^{\alpha},
\eqno{(8.3)}
$$
where $A_{\mu \nu}$ is given by eq.(7.4) and $G^{-1}$ is given by
eq.(9.24). The full Lagrangian density and the action of the
system are respectively given by,
$$
{\cal L} = e^{I(C)} {\cal L}_0,
\eqno{(8.4)}
$$
$$
S = \int {\rm d}^4 x ~ {\cal L}.
\eqno{(8.5)}
$$
\\

The system given by above Lagrangian has both $U(1)$ gauge
symmetry and gravitational gauge symmetry. Under $U(1)$ gauge
transformations,
$$
\psi(x) \to \psi'(x) = e^{-i \alpha(x)} \psi(x), \eqno{(8.6)}
$$
$$
A_{\mu}(x) \to A'_{\mu}(x) = A_{\mu}(x) - \frac{1}{e} D_{\mu}
\alpha(x), \eqno{(8.7)}
$$
$$
C_{\mu}^{\alpha}(x) \to C^{\prime \alpha}_{\mu}(x) =
C_{\mu}^{\alpha}(x). \eqno{(8.8)}
$$
It can be proved that the Lagrangian ${\cal L}$ is invariant
under $U(1)$ gauge transformation.
Under gravitational gauge transformations,
$$
\psi(x) \to \psi'(x) = (\ehat \psi(x)), \eqno{(8.9)}
$$
$$
A_{\mu}(x) \to A'_{\mu}(x) = (\ehat A_{\mu}(x)), \eqno{(8.10)}
$$
$$
C_{\mu}(x) \to  C'_{\mu}(x) = \ehat (x) C_{\mu} (x) \ehat^{-1} (x)
+ \frac{i}{g} \ehat (x) (\partial_{\mu} \ehat^{-1} (x)).
\eqno{(8.11)}
$$
The action $S$ given by eq.(8.4) is invariant under gravitational
gauge transformation. \\

Lagrangian ${\cal L}$ can be separated into free Lagrangian ${\cal L}_F$
and interaction Lagrangian ${\cal L}_I$,
$$
{\cal L} = {\cal L}_F + {\cal L}_I,
\eqno{(8.12)}
$$
where
$$ {\cal L}_F = - \frac{1}{4} \eta^{\mu \rho} \eta^{\nu
\sigma} A_{0 \mu \nu} A_{0 \rho \sigma} - \bar{\psi} (
\gamma^{\mu}
\partial_{\mu}  + m ) \psi - \frac{1}{4} \eta^{\mu \rho} \eta^{\nu
\sigma} \eta_{2 \alpha \beta } F_{0 \mu \nu}^{\alpha} F_{0 \rho
\sigma}^{\beta},
\eqno{(8.13)}
$$
$$
\begin{array}{rcl}
{\cal L}_I &=&
{\cal L}_F \cdot (\sum_{n=1}^{\infty} \frac{1}{n!}
( g \eta_{1 \alpha_1}^{\mu_1} C_{\mu_1}^{\alpha_1} )^n )
+ i e \cdot e^{I(C)} \bar{\psi} \gamma^{\mu} \psi A_{\mu} \\
&&\\
&& + g e^{I(C)} \bar\psi \gamma^{\mu} \partial_{\alpha} \psi
C_{\mu}^{\alpha}
 + g e^{I(C)} \eta^{\mu \rho} \eta^{\nu \sigma}
A_{0 \mu \nu} C_{\rho}^{\alpha} \partial_{\alpha} A_{\sigma} \\
&&\\
&&- \frac{g}{2} e^{I(C)} \eta^{\mu \rho} \eta^{\nu \sigma} A_{\mu
\nu} G^{-1 \lambda }_{\alpha} A_{\lambda} F_{\rho \sigma}^{\alpha}\\
&&\\
&& -\frac{g^2}{4} e^{I(C)} \eta^{\mu \rho} \eta^{\nu \sigma} G^{-1
\kappa}_{\alpha} G^{-1 \lambda }_{\beta} A_{\kappa} A_{\lambda}
F_{\mu \nu}^{\alpha} F_{\rho \sigma}^{\beta} \\
&&\\
&& - \frac{g^2}{2} e^{I(C)} \eta^{\mu \rho} \eta^{\nu \sigma}
( C_{\mu}^{\alpha} C_{\rho}^{\beta} (\partial_{\alpha} A_{\nu} )
(\partial_{\beta} A_{\sigma} )
- C_{\nu}^{\alpha} C_{\rho}^{\beta} (\partial_{\alpha} A_{\mu} )
(\partial_{\beta} A_{\sigma} )  )  \\
&&\\
&&  + g e^{I(C)} \eta^{\mu \rho} \eta^{\nu \sigma}
\eta_{2 \alpha \beta} (\partial_{\mu} C_{\nu }^{\alpha}
- \partial_{\nu} C_{\mu}^{\alpha})
C_{\rho}^{\delta} \partial_{\delta} C_{\sigma}^{\beta} \\
&&\\
&&  - \frac{1}{2} g^2 e^{I(C)} \eta^{\mu \rho} \eta^{\nu \sigma}
\eta_{2 \alpha \beta}
(C_{\mu}^{\delta} \partial_{\delta} C_{\nu}^{\alpha}
- C_{\nu}^{\delta} \partial_{\delta} C_{\mu}^{\alpha} )
C_{\rho}^{\epsilon} \partial_{\epsilon} C_{\sigma}^{\beta}.
\end{array}
\eqno{(8.14)}
$$
\\

The traditional Lagrangian for QCD is
$$
- \sum_n \bar{\psi}_n \lbrack \gamma^{\mu} (\partial_{\mu} -i g_c
A^i_{\mu } \frac{\lambda_i}{2} ) + m_n \rbrack \psi_n -\frac{1}{4}
\eta^{\mu \rho} \eta^{\nu \sigma} A^i_{\mu \nu } A^i_{\rho \sigma
},
\eqno{(8.15)}
$$
where $\psi_n$ is the quark color triplet of the $n$th
flavor, $A_{\mu \alpha}$ is the color gauge vector
potential, $A_{\alpha \mu \nu }$ is the color gauge
covariant field strength tensor, $g_c$ is the strong
coupling constant, $\lambda_{\alpha}$ is the Gell-Mann
matrix and $m_n$ is the quark mass of the $n$th flavor.
In gravitational gauge theory, this Lagrangian should be
changed into
$$
\begin{array}{rcl}
{\cal L}_0 &=&
- \sum_n \bar{\psi}_n
\lbrack \gamma^{\mu} (D_{\mu} -i g_c A^i_{\mu}
\frac{\lambda^i}{2} ) + m_n
\rbrack \psi_n
-\frac{1}{4} \eta^{\mu \rho} \eta^{\nu \sigma}
{\mathbf A}^i_{\mu \nu } {\mathbf A}^i_{ \rho \sigma } \\
&&\\
&&- \frac{1}{4} \eta^{\mu \rho} \eta^{\nu \sigma} \eta_{2 \alpha \beta }
F_{\mu \nu}^{\alpha} F_{\rho \sigma}^{\beta},
\end{array}
\eqno{(8.16)}
$$
where
$$
{\mathbf A}^i_{\mu \nu } = A^i_{\mu \nu} + g G^{-1
\lambda}_{\sigma} A^i_{\lambda} F^{\sigma}_{\mu\nu},
\eqno{(8.17)}
$$
$$
A^i_{\mu \nu } =
D_{\mu} A^i_{\nu } - D_{\nu} A^i_{ \mu }
+ g_c f_{i j k}
A^j_{ \mu } A^k_{\nu }.
\eqno{(8.18)}
$$
It can be proved that this system has both $SU(3)_c$
gauge symmetry and gravitational gauge symmetry.
The unified electroweak model can be discussed in
similar way. \\

Now, let's try to construct a theory which can describe
all kinds of fundamental interactions in Nature. First we
know that the fundamental particles that we know are
fundamental fermions(such as leptons and quarks),
gauge bosons(such as photon, gluons, gravitons and
intermediate gauge bosons $W^{\pm}$ and $Z^0$), and
possible Higgs bosons. According to the Standard Model,
leptons form left-hand doublets and right-hand singlets.
Let's denote
$$
\psi^{(1)}_L =\left (
\begin{array}{c}
\nu_e  \\
e
\end{array}
\right )_L
~~~,~~~
\psi^{(2)}_L =\left (
\begin{array}{c}
\nu_{\mu}  \\
\mu
\end{array}
\right )_L
~~~,~~~
\psi^{(3)}_L =\left (
\begin{array}{c}
\nu_{\tau}  \\
\tau
\end{array}
\right )_L ,
\eqno{(8.19)}
$$
$$
\psi^{(1)}_R=e_R
~~~,~~~
\psi^{(2)}_R=\mu_R
~~~,~~~
\psi^{(3)}_R=\tau_R.
\eqno{(8.20)}
$$
Neutrinos have no right-hand singlets. The weak hypercharge
for left-hand doublets $\psi^{(i)}_L$ is $-1$ and for right-hand
singlet $\psi^{(i)}_R$ is $-2$. All leptons carry no color
charge. In order to define the wave
function for quarks, we have to introduce
Kabayashi-Maskawa mixing matrix first, whose general form is,
$$
K =
\left (
\begin{array}{ccc}
c_1 & s_1 c_3 & s_1 s_3 \\
-s_1 c_2 & c_1 c_2 c_3 - s_2 s_3 e^{i \delta}
& c_1 c_2 s_3 + s_2 c_3 e^{i \delta}   \\
s_1 s_2 & -c_1 s_2 c_3 -c_2 s_3 e^{i \delta}
& -c_1 s_2 s_3 +c_2 c_3 e^{i \delta}
\end{array}
\right )
\eqno{(8.21)}
$$
where
$$
c_i = {\rm cos} \theta_i ~~,~~~~ s_i = {\rm sin} \theta_i ~~~(i=1,2,3)
\eqno{(8.22)}
$$
and $\theta_i$ are generalized Cabibbo angles. The mixing between
three different quarks $d,s$ and $b$ is given by
$$
\left (
\begin{array}{c}
d_{\theta} \\
s_{\theta}  \\
b_{\theta}
\end{array}
\right )
= K
\left (
\begin{array}{c}
d\\
s\\
b
\end{array}
\right ).
\eqno{(8.23)}
$$
Quarks also form left-hand doublets and right-hand singlets,
$$
q_L^{(1)a} =
\left (
\begin{array}{c}
u_L^a \\
d_{\theta L}^a
\end{array}
\right ) ,~~
q_L^{(2)a} =
\left (
\begin{array}{c}
c_L^a \\
s_{\theta L}^a
\end{array}
\right ),~~
q_L^{(3)a} =
\left (
\begin{array}{c}
t_L^a \\
b_{\theta L}^a
\end{array}
\right ) ,
\eqno{(8.24)}
$$
$$
\begin{array}{ccc}
q_u^{(1)a}= u_R^a
& q_u^{(2)a}= c_R^a
& q_u^{(3)a}= t_R^a \\
q_{ \theta d}^{(1)a}= d_{\theta R}^a
& q_{ \theta d}^{(2)a}= s_{\theta R}^a
& q_{ \theta d}^{(3)a}= b_{\theta R}^a,
\end{array}
\eqno{(8.25)}
$$
where index $a$ is color index.
It is known that left-hand doublets have weak isospin
$\frac{1}{2}$ and weak hypercharge  $\frac{1}{3}$,
right-hand singlets  have no weak isospin, $q_u^{(j)a}$s
have weak hypercharge $\frac{4}{3}$ and $q_{\theta d}^{(j)a}$s
have weak hypercharge $ - \frac{2}{3}$. \\

For gauge bosons, gravitational gauge field is also denoted
by $C_{\mu}^{\alpha}$. The gluon field is denoted $A_{\mu}$,
$$
A_{\mu} =
A_{\mu}^i \frac{\lambda^i}{2}.
\eqno{(8.26)}
$$
The color gauge covariant field strength tensor is also
given by eq.(8.18). The $U(1)_Y$ gauge field is denoted
by $B_{\mu}$ and $SU(2)$ gauge field is denoted by
$F_{\mu}$
$$
F_{\mu} =
F^n_{\mu} \frac{\sigma_n}{2},
\eqno{(8.27)}
$$
where $\sigma_n$ is the Pauli matrix. The $U(1)_Y$ gauge field strength
tensor is given by
$$
{\mathbf B}_{\mu\nu} = B_{\mu\nu} + g G^{-1 \lambda}_{\alpha}
B_{\lambda} F^{\alpha}_{\mu\nu},
\eqno{(8.28a)}
$$
where
$$ B_{ \mu \nu} = D_{\mu} B_{ \nu} - D_{\nu} B_{\mu},
\eqno{(8.28b)}
$$
and the $SU(2)$ gauge field strength tensor is given by
$$
{\mathbf F}^n_{\mu\nu} = F^n_{\mu\nu} + g G^{-1 \lambda}_{\alpha}
F^n_{\lambda} F^{\alpha}_{\mu\nu}, \eqno{(8.29a)}
$$
$$
F_{\mu \nu}^n = D_{\mu} F_{\nu}^n - D_{\nu} F_{\mu}^n + g_w
\epsilon _{lmn} F_{\mu}^l    F_{\nu}^m,
\eqno{(8.29b)}
$$
where $g_w$ is the coupling constant for $SU(2)$ gauge interactions
and the coupling constant for $U(1)_Y$ gauge interactions is $g'_w$.
\\

If there exist Higgs particles in Nature, the Higgs fields is
represented by a complex scalar $SU(2)$ doublet,
$$
\phi =
\left (
\begin{array}{c}
\phi^{\dagger} \\
\phi^0
\end{array}
\right ).
\eqno{(8.30)}
$$
The hypercharge of Higgs field $\phi$ is $1$. \\

The Lagrangian ${\cal L}_0$ that describes four kinds of fundamental
interactions is given by
$$
\begin{array}{rcl}
{\cal L}_0 &=&
-\sum_{j=1}^{3} \overline{\psi}_L^{(j)} \gamma ^{\mu}
(D_{\mu}+ \frac{i}{2} g'_w  B_{\mu} -ig_w F_{\mu} ) \psi_L^{(j)} \\
&&\\
&&- \sum_{j=1}^{3}\overline{e}_R^{(j)} \gamma ^{\mu}
(D_{\mu}+ ig'_w  B_{\mu} ) e_R^{(j)}  \\
&&\\
&&-\sum_{j=1}^{3} \overline{q}_L^{(j)a} \gamma^{\mu}
\left( (D_{\mu}-ig_w F_{\mu}- \frac{i}{6}g'_w B_{\mu} )\delta_{ab}
-i g_c A^k_{\mu} (\frac{\lambda^k}{2})_{ab} \right) q_L^{(j)b} \\
&&\\
&&-\sum_{j=1}^{3} \overline{q}_u^{(j)a} \gamma^{\mu}
\left( (D_{\mu}-i \frac{2}{3} g'_w B_{\mu} )\delta_{ab}
-i g_c A^k_{\mu} (\frac{\lambda^k}{2})_{ab} \right) q_u^{(j)b}  \\
&&\\
&& -\sum_{j=1}^{3} \overline{q}_{\theta d}^{(j)a} \gamma^{\mu}
\left( (D_{\mu} + i \frac{1}{3} g'_w B_{\mu} )\delta_{ab}
-i g_c A^k_{\mu} (\frac{\lambda^k}{2})_{ab} \right) q_{\theta d}^{(j)b} \\
&& \\
&&-\frac{1}{4}  \eta^{\mu \rho} \eta^{\nu \sigma}
{\mathbf F}^{n}_{ \mu \nu} {\mathbf F}^n_{\rho \sigma}
-\frac{1}{4} \eta^{\mu \rho} \eta^{\nu \sigma}
{\mathbf B}_{\mu \nu} {\mathbf B}_{\rho \sigma} \\
&& \\
&& -\frac{1}{4} \eta^{\mu \rho} \eta^{\nu \sigma}
{\mathbf A}^i_{\mu \nu } {\mathbf A}^i_{ \rho \sigma }
- \frac{1}{4} \eta^{\mu \rho} \eta^{\nu \sigma} \eta_{2 \alpha \beta }
F_{\mu \nu}^{\alpha} F_{\rho \sigma}^{\beta} \\
&&\\
&& -\left\lbrack (D_{\mu}- \frac{i}{2}
g'_w  B_{\mu} -ig_w F_{\mu}) \phi \right\rbrack ^{\dagger}
\cdot \left\lbrack (D_{\mu}- \frac{i}{2}
g'_w  B_{\mu} -ig_w F_{\mu}) \phi \right\rbrack \\
&&\\
&&   - \mu^2 \phi^{\dagger} \phi
+ \lambda (\phi^{\dagger} \phi)^2  \\
&&\\
&& - \sum_{j=1}^{3} f^{(j)}
\left(\overline{e}_R^{(j)} \phi^{\dag} \psi_L^{(j)}
+\overline{\psi}_L^{(j)} \phi  e_R^{(j)}\right)  \\
&&\\
&& -\sum_{j=1}^{3} \left( f_u^{(j)} \overline{q}_L^{(j)a}
\overline{\phi}
q_u^{(j)a} + f_u^{(j) \ast} \overline{q}_u^{(j)a}
\overline{\phi}^{\dag} q_L^{(j)a} \right)   \\
&&\\
&&-\sum_{j,k=1}^{3} \left( f_d^{(jk)} \overline{q}_L^{(j)a} \phi
q_{\theta d}^{(k)a}
+ f_d^{(jk) \ast} \overline{q}_{\theta d}^{(k)a}
\phi^{\dag} q_L^{(j)a} \right),
\end{array}
\eqno{(8.31)}
$$
where
$$
\overline{\phi} = i \sigma_{2} \phi^{\ast} =
\left (
\begin{array}{c}
\phi^{0 \dag} \\
- \phi
\end{array}
\right ).
\eqno{(8.32)}
$$
The full Lagrangian is given by
$$
{\cal L} = e^{I(C)} {\cal L}_0.
\eqno{(8.33)}
$$
This Lagrangian describes four kinds of  fundamental interactions
in Nature. It has ($SU(3) \times SU(2) \times U(1)) \otimes_s
Gravitational~ Gauge ~Group$ symmetry\cite{7}. Four kinds of fundamental
interactions are formally unified in this Lagrangian. However,
this unification is not a genuine unification. Finally, an
important and fundamental problem is that, can we genuine unify
four kinds of fundamental interactions in a single group, in which
there is only one coupling constant for all kinds
of fundamental interactions? This theory may exist. \\

\section{Classical Limit of Quantum Gravity}

It is known that both Newton's theory of gravity and
Einstein's general relativity
obtain immense achievements in astrophysics and cosmology. Any
correct quantum theory of gravity should return to these two theories
in classical limit. In this chapter, we will discuss the classical limit
of the quantum gauge theory of gravity. \\

First, we discuss an important problem qualitatively. It is know that, in
usual gauge theory, such as QED, the coulomb force between two objects
which carry like electric charges is always mutual repulsive. Gravitational
gauge theory is also a kind of gauge theory, is the force between two static
massive objects attractive or repulsive? For the sake of simplicity, we use
Dirac field as an example to discuss this problem. The discussions for
other kinds of fields can be proceeded similarly.  \\

Suppose that the gravitational field is very weak, so both the gravitational
field and the gravitational coupling constant are first order infinitesimal
quantities. Then in leading order approximation, both inertial
energy-momentum tensor and gravitational energy-momentum tensor
give the same results, which we denoted as
$$
T^{\mu}_{\alpha} =
\bar{\psi} \gamma^{\mu} \partial_{\alpha} \psi.
\eqno{(9.1)}
$$
The time component of the current is
$$
T^{0}_{\alpha} =
-i \psi^{\dagger} \partial_{\alpha} \psi = \psi^{\dagger} \hat{P}_{\alpha} \psi.
\eqno{(9.2)}
$$
Its space integration  gives out the energy-momentum of
the system. The interaction Lagrangian between Dirac field and
gravitational field is given by eq.(6.7). After considering the
equation of motion of Dirac field, the coupling
between Dirac field and Gravitational gauge field in the leading order is:
$$
{\cal L}_I \approx g T_{\alpha}^{\mu} C_{\mu}^{\alpha}.
\eqno{(9.3)}
$$
The the leading order interaction Hamiltonian density is given by
$$
{\cal H}_I  \approx - {\cal L}_I \approx
 - g T_{\alpha}^{\mu} C_{\mu}^{\alpha}.
\eqno{(9.4)}
$$
The equation of motion of gravitational gauge field in the leading order is:
$$
\partial_{\lambda} \partial^{\lambda}
( \eta^{\nu \tau} \eta_{2 \alpha \beta}  C_{\tau}^{\beta})
- \partial^{\lambda} \partial^{\nu}
(  \eta_{2 \alpha \beta}  C_{\lambda}^{\beta})
= - g T^{\nu}_{\alpha}.
\eqno{(9.5)}
$$
As a classical limit approximation,
let's consider static gravitational interactions
between two static objects. In this case,
the leading order component of energy-momentum tensor is $T^0_0$, other
components of energy-momentum tensor is a first order infinitesimal
quantity. So, we only need to consider the equation of motion of
$\nu = \alpha = 0$ of eq.(9.5), which now becomes
$$
\partial_{\lambda} \partial^{\lambda} C_0^0
- \partial^{\lambda} \partial^0 C_{\lambda}^{0}
= - g T^0_0.
\eqno{(9.6)}
$$
For static problems, all time derivatives vanish. Therefor,
the above equation is changed into
$$
\nabla ^2 C_0^0 = - g T^0_0.
\eqno{(9.7)}
$$
This is just the Newton's equation of gravitational field. Suppose
that there is only one point object at the origin of the coordinate
system. Because $T^0_0$ is the negative value of energy density,
we can let
$$
T^0_0 = - M \delta(\svec{x}) .
\eqno{(9.8)}
$$
Applying
$$
\nabla ^2  \frac{1}{r} = - 4 \pi \delta(\svec{x}),
\eqno{(9.9)}
$$
with $r =  |\svec{x} |$, we get
$$
C^0_0  = - \frac{gM}{4 \pi r}.
\eqno{(9.10)}
$$
This is just the gravitational potential which is expected in Newton's
theory of gravity. \\

Suppose that there is another point object at the position of
point $\svec{x}$ with mass $m$. The gravitational potential energy
between these two objects is that
$$
V(r) = \int {\rm d}^3 \svec{y} {\cal H}_I
=  -g \int {\rm d}^3 \svec{y} T^0_{2~0} (\svec{x}) C_0^0,
\eqno{(9.11)}
$$
with $C^0_0$ is the gravitational potential generated by the first
point object, and $T^0_{2~0}$ is the $(0,0)$ component of the
energy-momentum tensor of the second object,
$$
T^0_{2~0} (\svec{y})= - m  \delta(\svec{y} - \svec{x}) .
\eqno{(9.12)}
$$
The final result for gravitational potential energy between two
point objects is
$$
V(r) = - \frac{g^2 M m}{4 \pi r}.
\eqno{(9.13)}
$$
The gravitational potential energy between two point objects is always
negative, which is what expected by Newton's theory of gravity and is
the inevitable result of the attractive nature of gravitational
interactions.
\\

The gravitational force that the first point object acts on the second
point object is
$$
\svec{f} = - \nabla V(r) = - \frac{g^2 M m}{4 \pi r^2} \hat{r},
\eqno{(9.14)}
$$
where $\hat r = \svec{r} / r $. This is the famous formula
of Newton's gravitational force.  Therefore, in the classical limit,
the gravitational gauge theory can return to Newton's theory of
gravity. Besides, from eq.(9.14), we can clearly see that the
gravitational interaction force between two point objects is
attractive.
\\

Now, we want to ask a problem: why in QED, the force between two
like electric charges is always repulsive, while in gravitational gauge theory,
the force between two like gravitational charges is always attractive?
A simple answer to this fundamental problem is that the attractive nature
of the gravitational force is an inevitable result of
the global Lorentz symmetry of the system. Because of the requirement
of global Lorentz symmetry, the Lagrangian
function given by eq.(4.20) must use  $\eta_{2 \alpha \beta}$ ,
 can not use the ordinary $\delta$-function $\delta_{\alpha \beta}$.
It can be easily prove that, if we use $\delta_{\alpha \beta}$ instead of
$\eta_{2 \alpha \beta}$ in eq.(4.20), the Lagrangian of pure gravitational
gauge field is not invariant under global Lorentz transformation.
On the other hand,
if we use $\delta_{\alpha \beta}$ instead of  $\eta_{2 \alpha \beta}$
in eq.(4.20), the gravitational force will be repulsive which obviously
contradicts with experiment results. In QED, $\delta_{a b}$ is
used to construct the Lagrangian for electromagnetic fields, therefore,
the interaction force between two like electric charges is
always repulsive.
\\

One fundamental influence of using the metric $\eta_{2 \alpha \beta}$
in the Lagrangian of pure gravitational field is that the kinematic
energy term of gravitation field $C_{\mu}^0$ is always negative.
This result is novels, but it is not surprising, for gravitational interaction
energy is always negative. In a meaning, it is the reflection of the
negative nature of graviton's kinematic energy. Though the kinematic
energy term of gravitation field $C_{\mu}^0$ is always negative,
the kinematic energy term of gravitation field $C_{\mu}^i$ is always
positive. The negative energy problem of graviton does not cause any
trouble in quantum gauge theory of gravity. Contrarily, it will help us
to understand some puzzle phenomena of Nature. From theoretical
point of view, the negative nature of graviton's kinematic energy
is essentially an inevitable result of global Lorentz symmetry.
Global Lorentz symmetry of the system,
attractive nature of gravitational interaction
force and negative nature of graviton's kinematical energy are essentially
related to each other, and they have the same origin in nature.
We will return to discuss the negative energy problem again later.  \\

In general relativity, gravitational field obeys Einstein field equation, which
is usually written in the following form,
$$
R_{\mu \nu} -\frac{1}{2} g_{\mu \nu} R + \lambda g_{\mu \nu}
= - 8 \pi G T_{\mu \nu},
\eqno{(9.15)}
$$
where $R_{\mu \nu}$ is Ricci tensor, $R$ is curvature, $G$ is Newton
gravitational constant and $\lambda$ is cosmology constant. The classical
limit of Einstein field equation is
$$
\nabla^2 g_{00} =
  - 8 \pi G T_{0 0}.
\eqno{(9.16)}
$$
Compare this equation with eq.(9.7) and use eq.(5.24), we get
$$
g^2 = 4 \pi G.
\eqno{(9.17)}
$$
In order to get eq.(9.17), the following relations are used
$$
T_{0 0} = - T^0_0,~~~~
g_{00} \simeq -(1 + 2 g C_0^0).
\eqno{(9.18)}
$$
\\

In general relativity, Einstein field equation transforms covariantly
under general coordinates transformation, in other words, it is a general
covariant equation.  In gravitational gauge theory, the system has
local gravitational gauge symmetry. From mathematical point of view,
general coordinates transformation is equivalent to local gravitational
gauge transformation. Therefore, it seems that two theories have
the same symmetry. On the other hand, both theories have
global Lorentz symmetry.\\

In the previous discussions, we have given out the relations between
space-time metric and gravitational gauge fields, which is shown in
eq.(5.24). For scalar field, if the gravitational field is not so strong and
its variations with space-time is not so great,
we can select a local  inertial reference
system where gravity is completely shielded, which is shown in
eq.(5.23), where we can not directly see any coupling between
gravitational gauge field and scalar field. For a general macroscopic
object, its dynamical properties are more like real scalar field. Therefore,
in macroscopic world, we can always select a local  inertial reference
system where all macroscopic effects of gravitational interactions are
put into the structure space-time, which is the main point of view of
general relativity. Now, let's take this point of view to study structures
of space-time in the classical and macroscopic limit.\\

Let's omit Lagrangian of pure gravitational gauge fields, then eq.(5.23)
is changed into
$$
{\cal L} = -\frac{1}{2} g^{\alpha \beta} \partial_{\alpha} \phi
\partial_{\beta} \phi  -  \frac{m^2}{2} \phi^2.
\eqno{(9.19)}
$$
This is the Lagrangian for curved space-time. The local inertial reference
system is given by following coordinates transformation,
$$
{\rm d}x^{\mu}  \to {\rm d}x^{\prime \mu}
= \frac{\partial x^{\prime \mu}}{\partial x^{\nu}} {\rm d}x^{\nu},
\eqno{(9.20)}
$$
where $\frac{\partial x^{\prime \mu}}{\partial x^{\nu}}$ is given by,
$$
\frac{\partial x^{\prime \mu}}{\partial x^{\nu}}
= (G^{-1}) ^{\mu}_{\nu}.
\eqno{(9.21)}
$$
Define matrix $G$ as
$$
G = (G_{\mu}^{\alpha}) = ( \delta_{\mu}^{\alpha} - g C_{\mu}^{\alpha}).
\eqno{(9.22)}
$$
A simple form for matrix $G$ is
$$
G = I - gC,
\eqno{(9.23)}
$$
where $I$ is a unit matrix and $C= (C_{\mu}^{\alpha})$. Therefore,
$$
G^{-1} = \frac{1}{I-gC}.
\eqno{(9.24)}
$$
If gravitational gauge fields are weak enough, we have
$$
G^{-1} = \sum_{n=0}^{\infty} (g C)^n.
\eqno{(9.25)}
$$
$G^{-1}$ is the inverse matrix of $G$, it satisfies
$$
(G^{-1})^{\mu}_{\beta} G^{\alpha}_{\mu}
= \delta_{\beta}^{\alpha},
\eqno{(9.26)}
$$
$$
G^{\alpha}_{\mu} (G^{-1})^{\nu}_{\alpha}
= \delta^{\nu}_{\mu}.
\eqno{(9.27)}
$$
Using all these relations, we can prove that
$$
\begin{array}{rcl}
g^{\alpha \beta} \to g^{\prime \alpha \beta}
& = & \frac{\partial x^{\prime \alpha}}{\partial x^{\mu}}
\frac{\partial x^{\prime \beta}}{\partial x^{\nu}} g^{\mu \nu}  \\
& = & \eta^{\alpha \beta}.
\end{array}
\eqno{(9.28)}
$$
Therefore, under this coordinates transformation, the space-time
metric becomes flat, in other words, we go into local inertial
reference system. In this local inertial reference system, the Lagrangian
eq.(9.19) becomes
$$
{\cal L} = -\frac{1}{2} \eta^{\alpha \beta} \partial_{\alpha} \phi
\partial_{\beta} \phi  -  \frac{m^2}{2} \phi^2.
\eqno{(9.29)}
$$
Eq.(9.29) is just the Lagrangian for real scalar fields in flat Minkowski
space-time. \\

Define covariant metric tensor $g_{\alpha \beta}$ as
$$
g_{\alpha \beta} \define \eta_{\mu \nu}
(G^{-1})_{\alpha}^{\mu} (G^{-1})_{\beta}^{\nu}.
\eqno{(9.30)}
$$
It can be easily proved that
$$
g_{\alpha \beta} g^{\beta \gamma} = \delta_{\alpha}^{\gamma},
\eqno{(9.31)}
$$
$$
g^{\alpha \beta} g_{\beta \gamma} = \delta^{\alpha}_{\gamma}.
\eqno{(9.32)}
$$
\\

The affine connection $\Gamma^{\lambda}_{\mu \nu}$ is defined by
$$
\Gamma^{\lambda}_{\mu \nu}
= \frac{1}{2} g^{\lambda \sigma}
\left( \frac{\partial g_{\mu \sigma}}{\partial x^{\nu}}
+ \frac{\partial g_{\nu \sigma}}{\partial x^{\mu}}
-\frac{\partial g_{\mu \nu}}{\partial x^{\sigma}} \right).
\eqno{(9.33)}
$$
Using the following relation,
$$
g F^{\lambda}_{\rho \sigma}
= G^{\nu}_{\rho} G^{\mu}_{\sigma}
\lbrack ( G^{-1} \partial_{\mu} G)^{\lambda}_{\nu}
- ( G^{-1} \partial_{\nu} G)^{\lambda}_{\mu} \rbrack,
\eqno{(9.34)}
$$
where $ F^{\lambda}_{\rho \sigma}$ is the component field strength of
gravitational gauge field, we get
$$
\begin{array}{rcl}
\Gamma^{\lambda}_{\mu \nu}
&=& - \frac{1}{2}
\lbrack ( G^{-1} \partial_{\mu} G)^{\lambda}_{\nu}
+ ( G^{-1} \partial_{\nu} G)^{\lambda}_{\mu} \rbrack \\
&& + \frac{1}{2} g \eta^{\alpha_1 \beta_1} \eta_{\alpha \beta}
F^{\rho}_{\mu_1 \beta_1} G^{\lambda}_{\alpha_1} G^{-1 \alpha}_{\rho}
( G^{-1 \beta}_{\nu} G^{-1 \mu_1}_{\mu}
+ G^{-1 \beta}_{\mu} G^{-1 \mu_1}_{\nu} ).
\end{array}
\eqno{(9.35)}
$$
From this expression, we can see that, if there is no gravity in space-time,
that is
$$
C_{\mu}^{\alpha} = 0~~,~~
F^{\lambda}_{\mu  \nu} = 0,
\eqno{(9.36)}
$$
then the affine connection $\Gamma^{\lambda}_{\mu \nu}$ will vanish,
which is what we expect in general relativity.  \\

The curvature tensor $R^{\lambda}_{\mu \nu \kappa}$ is defined by
$$
R^{\lambda}_{\mu \nu \kappa}
\define \partial_{\kappa} \Gamma^{\lambda}_{\mu \nu}
-\partial_{\nu} \Gamma^{\lambda}_{\mu \kappa}
+\Gamma^{\eta}_{\mu \nu} \Gamma^{\lambda}_{\kappa \eta}
- \Gamma^{\eta}_{\mu \kappa} \Gamma^{\lambda}_{\nu \eta},
\eqno{(9.37)}
$$
the Ricci tensor $R_{\mu \kappa}$ is defined by
$$
R_{\mu \kappa} \define
R^{\lambda}_{\mu \lambda \kappa},
\eqno{(9.38)}
$$
and the curvature scalar $R$ is defined by
$$
R \define g^{\mu \kappa} R_{\mu \kappa}.
\eqno{(9.39)}
$$
The explicit expression for Ricci tensor $R_{\mu \kappa}$ is
$$
\begin{array}{rcl}
R_{\mu \kappa} &=&
-(\partial_{\kappa} \partial_{\mu} G\cdot G^{-1}) ^{\alpha}_{\alpha}
+ 2 ( \partial_{\kappa} G \cdot G^{-1}
\cdot \partial_{\mu} G \cdot G^{-1} )^{\alpha}_{\alpha}  \\
&& + \eta^{\rho \sigma} \eta_{\alpha \beta}
( \partial_{\mu} G \cdot G^{-1} )^{\alpha}_{\rho}
( \partial_{\kappa} G \cdot G^{-1} )^{\beta}_{\sigma} \\
&&+ \frac{1}{2} g^{\lambda \nu} \eta_{\alpha \beta}
( G^{-1} \cdot \partial_{\nu} G \cdot G^{-1} \cdot \partial_{\lambda} G
\cdot G^{-1} )^{\alpha}_{\mu} G^{-1 \beta}_{\kappa}  \\
&&  - \frac{1}{2} g^{\lambda \nu} \eta_{\alpha \beta}
( G^{-1} \cdot \partial_{\nu} \partial_{\lambda} G
\cdot G^{-1} )^{\alpha}_{\mu} G^{-1 \beta}_{\kappa} \\
&& + \frac{1}{2} g^{\lambda \nu} \eta_{\alpha \beta}
( G^{-1} \cdot \partial_{\lambda}G \cdot G^{-1} \cdot
\partial_{\nu} G \cdot G^{-1} ) ^{\alpha}_{\mu} G^{-1 \beta}_{\kappa} \\
&&+ \frac{1}{2} g^{\lambda \nu} \eta_{\alpha \beta}
G^{-1 \alpha }_{\mu}
( G^{-1} \cdot \partial_{\nu} G \cdot G^{-1} \cdot \partial_{\lambda} G
\cdot G^{-1} )^{\beta}_{\kappa} \\
&& -\frac{1}{2} g^{\lambda \nu} \eta_{\alpha \beta}
G^{-1 \alpha }_{\mu}
( G^{-1} \cdot \partial_{\nu} \partial_{\lambda}G
\cdot G^{-1} )^{\beta}_{\kappa} \\
&& + \frac{1}{2} g^{\lambda \nu} \eta_{\alpha \beta}
G^{-1 \alpha }_{\mu}
( G^{-1} \cdot \partial_{\lambda} G \cdot G^{-1} \cdot \partial_{\nu} G
\cdot G^{-1} )^{\beta}_{\kappa}  \\
&&+  g^{\lambda \nu} \eta_{\alpha \beta}
( G^{-1} \cdot \partial_{\nu} G \cdot G^{-1} ) ^{\alpha}_{\mu}
( G^{-1} \cdot \partial_{\lambda} G \cdot G^{-1} )^{\beta}_{\kappa} \\
&& -\frac{1}{2} ( G^{-1} \cdot \partial_{\kappa} G \cdot G^{-1}
\cdot \partial_{\lambda} G ) ^{\lambda}_{\mu}
-\frac{1}{2} ( G^{-1} \cdot \partial_{\lambda} G \cdot G^{-1}
\cdot \partial_{\kappa} G ) ^{\lambda}_{\mu}
+ \frac{1}{2} ( G^{-1} \cdot \partial_{\kappa}
\partial_{\lambda} G ) ^{\lambda}_{\mu} \\
&& - \eta^{\rho \sigma} \eta_{\alpha \beta} G^{\lambda}_{\rho}
( G^{-1} \cdot \partial_{\lambda} G \cdot G^{-1} )^{\alpha}_{\mu}
(\partial_{\kappa} G \cdot G^{-1})^{ \beta}_{\sigma} \\
&& - \frac{1}{2} \eta^{\rho \sigma} \eta_{\alpha \beta}
G^{\lambda}_{\rho} G^{-1 \alpha }_{\mu}
 ( \partial_{\kappa} G \cdot G^{-1} \cdot \partial_{\lambda} G
\cdot G^{-1} ) ^{\beta}_{\sigma}  \\
&& + \frac{1}{2}\eta^{\rho \sigma} \eta_{\alpha \beta} G^{\lambda}_{\rho}
G^{-1 \alpha }_{\mu}
( \partial_{\kappa} \partial_{\lambda} G \cdot G^{-1} )^{\beta}_{\sigma}  \\
&& - \frac{1}{2} \eta^{\rho \sigma} \eta_{\alpha \beta}
G^{\lambda}_{\rho} G^{-1 \alpha }_{\mu}
 ( \partial_{\lambda} G \cdot G^{-1} \cdot \partial_{\kappa} G
\cdot G^{-1} ) ^{\beta}_{\sigma}  \\
&&-\frac{1}{2} ( G^{-1} \cdot \partial_{\nu} G \cdot G^{-1}
\cdot \partial_{\mu} G ) ^{\nu}_{\kappa}
-\frac{1}{2} ( G^{-1} \cdot \partial_{\mu} G \cdot G^{-1}
\cdot \partial_{\nu} G ) ^{\nu}_{\kappa}
+ \frac{1}{2} ( G^{-1} \cdot \partial_{\nu}
\partial_{\mu} G ) ^{\nu}_{\kappa} \\
&&- \eta^{\rho \sigma} \eta_{\alpha \beta} G^{\nu}_{\sigma}
( G^{-1} \cdot \partial_{\mu} G \cdot G^{-1} )^{\alpha}_{\kappa}
(\partial_{\nu} G \cdot G^{-1})^{ \beta}_{\rho} \\
&&- \frac{1}{2} \eta^{\rho \sigma} \eta_{\alpha \beta}
G^{\nu}_{\sigma} G^{-1 \alpha }_{\kappa}
 ( \partial_{\nu} G \cdot G^{-1} \cdot \partial_{\mu} G
\cdot G^{-1} ) ^{\beta}_{\rho}  \\
&& + \frac{1}{2}\eta^{\rho \sigma} \eta_{\alpha \beta} G^{\nu}_{\sigma}
G^{-1 \alpha }_{\kappa}
( \partial_{\nu} \partial_{\mu} G \cdot G^{-1} )^{\beta}_{\rho}
- \frac{1}{2} \eta^{\rho \sigma} \eta_{\alpha \beta}
G^{\nu}_{\sigma} G^{-1 \alpha }_{\kappa}
 ( \partial_{\mu} G \cdot G^{-1} \cdot \partial_{\nu} G
\cdot G^{-1} ) ^{\beta}_{\rho}  \\
&& + \frac{1}{2} \eta_{\alpha \beta} \eta^{\alpha_1 \beta_1}
G^{\nu}_{\beta_1} ( \partial_{\nu} G \cdot G^{-1})^{\alpha}_{\alpha_1}
( G^{-1} \cdot \partial_{\mu} G \cdot G^{-1})^{\beta}_{\kappa} \\
&&+ \frac{1}{2} \eta_{\alpha \beta} \eta^{\alpha_1 \beta_1}
G^{\nu}_{\beta_1} ( \partial_{\nu} G \cdot G^{-1})^{\alpha}_{\alpha_1}
( G^{-1} \cdot \partial_{\kappa} G \cdot G^{-1})^{\beta}_{\mu}  \\
&& - \frac{1}{4} \eta_{\alpha \beta} \eta^{\alpha_1 \beta_1}
( \partial_{\kappa} G \cdot G^{-1})^{\alpha}_{\alpha_1}
( \partial_{\mu} G \cdot G^{-1})^{\beta}_{\beta_1} \\
&&- \frac{1}{4} \eta_{\alpha \beta} \eta^{\alpha_1 \beta_1} G^{\nu}_{\beta_1}
( \partial_{\kappa} G \cdot G^{-1})^{\alpha}_{\alpha_1}
( G^{-1} \cdot \partial_{\nu} G \cdot G^{-1})^{\beta}_{\mu}\\
&& - \frac{1}{4} \eta_{\alpha \beta} \eta^{\alpha_1 \beta_1} G^{\lambda}_{\alpha_1}
(G^{-1} \cdot \partial_{\lambda} G \cdot G^{-1})^{\alpha}_{\kappa}
( \partial_{\mu} G \cdot G^{-1})^{\beta}_{\beta_1} \\
&& - \frac{1}{4} \eta_{\alpha \beta} \eta^{\alpha_1 \beta_1}
G^{\lambda}_{\alpha_1}G^{\nu}_{\beta_1}
( G^{-1} \cdot \partial_{\lambda} G \cdot G^{-1})^{\alpha}_{\kappa}
( G^{-1} \cdot \partial_{\nu} G \cdot G^{-1})^{\beta}_{\mu}\\
&& - \frac{g}{2} \eta_{\alpha \beta} \eta^{\alpha_3 \beta_3}
F^{\rho}_{\mu_1 \beta_1} G^{\nu}_{\beta_3} G^{-1 \alpha}_{\rho}
G^{-1 \beta}_{\kappa} G^{-1 \mu_1}_{\mu}
( \partial_{\nu} G \cdot G^{-1} )^{\beta_1}_{\alpha_3}  \\
&& - \frac{g}{2} \eta_{\alpha \beta} \eta^{\alpha_3 \beta_3}
F^{\rho}_{\mu_1 \beta_1} G^{\nu}_{\beta_3} G^{-1 \alpha}_{\rho}
G^{-1 \beta}_{\mu} G^{-1 \mu_1}_{\kappa}
( \partial_{\nu} G \cdot G^{-1} )^{\beta_1}_{\alpha_3} \\
&& - \frac{g}{2} F^{\rho}_{\alpha \beta} G^{-1 \alpha}_{\rho}
( G^{-1} \cdot \partial_{\mu} G \cdot G^{-1})^{\beta}_{\kappa}
- \frac{g}{2} F^{\rho}_{\alpha \beta} G^{-1 \alpha}_{\rho}
( G^{-1} \cdot \partial_{\kappa} G \cdot G^{-1})^{\beta}_{\mu}\\
&& + \frac{g}{4} \eta^{\alpha_3 \mu_1} \eta_{\alpha \beta}
F^{\rho}_{\mu_1 \beta_1} G^{-1 \alpha}_{\rho} G^{-1 \beta}_{\mu}
(\partial_{\kappa}G \cdot G^{-1} )^{\beta_1}_{\alpha_3}
+ \frac{g}{4} F^{\rho}_{\alpha \beta}
(G^{-1} \cdot \partial_{\kappa} G \cdot G^{-1} )^{\beta}_{\rho}
G^{-1 \alpha}_{\mu}  \\
&&+ \frac{g}{4} \eta^{\alpha_3 \mu_1} \eta_{\alpha \beta}
F^{\rho}_{\mu_1 \beta_1} G^{-1 \alpha}_{\rho} G^{-1 \beta}_{\mu}
G^{\lambda}_{\alpha_3}
(G^{-1} \cdot \partial_{\lambda}G \cdot G^{-1} )^{\beta_1}_{\kappa} \\
&&+ \frac{g}{4} F^{\rho}_{\alpha \beta}
(G^{-1} \cdot \partial_{\rho} G \cdot G^{-1} )^{\beta}_{\kappa}
G^{-1 \alpha}_{\mu}
+ \frac{g}{4} F^{\rho}_{\alpha \beta}
(G^{-1} \cdot \partial_{\rho} G \cdot G^{-1} )^{\beta}_{\mu}
G^{-1 \alpha}_{\kappa} \\
&&+ \frac{g}{4} \eta^{\alpha_3 \mu_1} \eta_{\alpha \beta}
F^{\rho}_{\mu_1 \beta_1} G^{-1 \alpha}_{\rho} G^{-1 \beta}_{\kappa}
 (\partial_{\mu}G \cdot G^{-1} )^{\beta_1}_{\alpha_3}
+ \frac{g}{4} F^{\rho}_{\alpha \beta}
(G^{-1} \cdot \partial_{\mu} G \cdot G^{-1} )^{\beta}_{\rho}
G^{-1 \alpha}_{\kappa}  \\
&&+ \frac{g}{4} \eta^{\alpha_3 \mu_1} \eta_{\alpha \beta}
F^{\rho}_{\mu_1 \beta_1} G^{-1 \alpha}_{\rho} G^{-1 \beta}_{\kappa}
G^{\lambda}_{\alpha_3}
(G^{-1} \cdot \partial_{\lambda}G \cdot G^{-1} )^{\beta_1}_{\mu} \\
&& + \frac{g^2}{2} \eta^{\sigma \beta_1} \eta_{\alpha_2 \beta_2}
F^{\rho}_{\alpha \sigma} F^{\rho_2}_{\mu_2 \beta_1}
G^{-1 \alpha}_{\rho} G^{-1 \alpha_2}_{\rho_2}
( G^{-1 \beta_2}_{\mu} G^{-1 \mu_2}_{\kappa}
+ G^{-1 \beta_2}_{\kappa} G^{-1 \mu_2}_{\mu}  )  \\
&& - \frac{g^2}{4} \eta_{\alpha \beta} \eta_{\alpha_2 \beta_2}
\eta^{\alpha_1 \beta_1} \eta^{\mu_1 \sigma_2}
F^{\rho}_{\mu_1 \beta_1} F^{\rho_1}_{\sigma_2 \alpha_1}
G^{-1 \alpha }_{\rho}  G^{-1 \alpha_2}_{\rho_1}
G^{-1 \beta}_{\kappa}  G^{-1 \beta_2}_{\mu} \\
&&- \frac{g^2}{4} \eta_{\alpha \beta} \eta^{\alpha_1 \beta_1}
F^{\rho}_{\mu_1 \beta_1} F^{\rho_1}_{\mu_3 \alpha_1}
G^{-1 \alpha }_{\rho}  G^{-1 \mu_1}_{\rho_1}
G^{-1 \beta}_{\kappa}  G^{-1 \mu_3}_{\mu}\\
&& - \frac{g^2}{4}  \eta_{\alpha_2 \beta_2} \eta^{\alpha_1 \beta_1}
F^{\rho}_{\mu_1 \beta_1} F^{\rho_1}_{\alpha \alpha_1}
G^{-1 \alpha }_{\rho}  G^{-1 \alpha_2}_{\rho_1}
G^{-1 \mu_1}_{\kappa}  G^{-1 \beta_2}_{\mu} \\
&&- \frac{g^2}{4} \eta_{\alpha \beta} \eta^{\alpha_1 \beta_1}
F^{\rho}_{\mu_1 \beta_1} F^{\rho_1}_{\mu_3 \alpha_1}
G^{-1 \alpha }_{\rho}  G^{-1 \beta}_{\rho_1}
G^{-1 \mu_1}_{\kappa}  G^{-1 \mu_3}_{\mu}
\end{array}
\eqno{(9.40)}
$$
The explicit expression for scalar curvature $R$ is
$$
\begin{array}{rcl}
R & = &
4 g^{\mu \kappa} ( \partial_{\mu} G \cdot G^{-1}
\cdot \partial_{\kappa} G \cdot G^{-1} ) ^{\alpha}_{\alpha}
- 2 g^{\mu \kappa} ( \partial_{\mu} \partial_{\kappa}G
 \cdot G^{-1} ) ^{\alpha}_{\alpha} \\
&& + \frac{3}{2} \eta^{\rho \sigma } \eta_{\alpha \beta} g^{\mu \kappa}
( \partial_{\mu} G  \cdot G^{-1} ) ^{\alpha}_{\rho}
( \partial_{\kappa} G  \cdot G^{-1} ) ^{\beta}_{\sigma}
 - 2 \eta^{\alpha \beta} G^{\kappa}_{\beta}
( \partial_{\kappa} G \cdot G^{-1} \cdot
\partial_{\lambda} G )^{\lambda}_{\alpha}\\
&&- 2 \eta^{\alpha \beta} G^{\kappa}_{\beta}
( \partial_{\lambda} G \cdot G^{-1} \cdot
\partial_{\kappa} G )^{\lambda}_{\alpha}
 + 2 \eta^{\alpha \beta} G^{\kappa}_{\beta}
( \partial_{\kappa} \partial_{\lambda} G )^{\lambda}_{\alpha}\\
&& - \frac{5}{2} \eta_{\alpha \beta} \eta^{\rho \sigma} \eta^{\mu \nu}
G^{\kappa}_{\nu} G^{\lambda}_{\rho}
( \partial_{\lambda} G  \cdot G^{-1} ) ^{\alpha}_{\mu}
( \partial_{\kappa} G  \cdot G^{-1} ) ^{\beta}_{\sigma} \\
&& + \eta_{\alpha \beta} \eta^{\alpha_1 \beta_1} \eta^{\mu_1 \nu_1}
G^{\nu}_{\beta_1} G^{\mu}_{\mu_1}
( \partial_{\nu} G  \cdot G^{-1} ) ^{\alpha}_{\alpha_1}
( \partial_{\mu} G  \cdot G^{-1} ) ^{\beta}_{\nu_1 } \\
&&- 2 g \eta^{\alpha \beta} F^{\rho}_{\alpha_1 \beta_1}
G^{\nu}_{\beta} G^{-1 \alpha_1}_{\rho}
( \partial_{\nu} G  \cdot G^{-1} ) ^{\beta_1}_{\alpha} \\
&& + g \eta^{\alpha \mu} F^{\kappa}_{\mu \beta}
 ( \partial_{\kappa} G  \cdot G^{-1} ) ^{\beta}_{\alpha} \\
&& + g \eta^{\alpha \mu} F^{\rho}_{\alpha \beta}
G^{\kappa}_{\mu}  (G^{-1} \cdot
\partial_{\kappa} G  \cdot G^{-1} ) ^{\beta}_{\rho} \\
&& + g^2 \eta^{\beta \beta_1} F^{\rho}_{\alpha \beta}
F^{\rho_1}_{\alpha_1 \beta_1} G^{-1 \alpha}_{\rho}
G ^{-1 \alpha_1}_{\rho_1} \\
&& - \frac{g^2}{2}  \eta_{\alpha \beta} \eta^{\alpha_1 \beta_1}
\eta^{\mu \sigma}  F^{\rho}_{\mu \beta_1}
F^{\rho_1}_{\sigma \alpha_1 } G^{-1 \alpha}_{\rho}
G ^{-1 \beta}_{\rho_1} \\
&&- \frac{g^2}{2} \eta^{\alpha \beta } F^{\rho}_{\mu \beta}
F^{\rho_1}_{\alpha_1 \alpha} G^{-1 \alpha_1}_{\rho}
G ^{-1 \mu}_{\rho_1}
\end{array}
\eqno{(9.41)}
$$
From these expressions, we can see that, if there is no gravity,
that is $C_{\mu}^{\alpha} $ vanishes, then $R_{\mu \nu}$ and
$R$ all vanish. It means that, if there is gravity, the space-time
is flat, which is what we expected in general relativity.
\\

Now, let's discuss some transformation properties of these tensors
under general coordinates transformation. Make a special kind of
local coordinates translation,
$$
x^{\mu }  \to x^{\prime \mu}
= x^{\mu} + \epsilon^{\mu} (x').
\eqno{(9.42)}
$$
Under this transformation, the covariant derivative and gravitational
gauge fields transform as
$$
D_{\mu} (x) \to D'_{\mu} (x')
= \ehat (x') D_{\mu} (x') \ehat^{-1} (x') ,
\eqno{(9.43)}
$$
$$
C_{\mu} (x) \to C'_{\mu} (x')
= \ehat (x') C_{\mu} (x') \ehat^{-1} (x') ,
+ \frac{i}{g} \ehat (x')
(\frac{\partial}{\partial x^{\prime \mu}} \ehat^{-1} (x')).
\eqno{(9.44)}
$$
It can be proved that
$$
G^{\alpha}_{\mu} (x) \to G^{\prime \alpha}_{\mu} (x')
= \Lambda^{\alpha}_{~\beta} G^{\beta}_{\mu}(x),
\eqno{(9.45)}
$$
where
$$
\Lambda^{\alpha}_{~\beta} =
\frac{\partial x^{\prime \alpha}}{ \partial x^{\beta}}.
\eqno{(9.46)}
$$
We can see that Lorentz index $\mu$ does not take part in transformation.
In fact, all Lorentz indexes do not take part in this kind of transformation.
Therefore,
$$
\eta^{\mu \nu} \to \eta^{\prime \mu \nu} = \eta^{\mu \nu}.
\eqno{(9.47)}
$$
Using all these relations, we can prove that
$$
g^{\alpha \beta}(x)  \to g^{\prime \alpha \beta}(x')
= \Lambda^{\alpha}_{~\alpha_1} \Lambda^{\beta}_{~\beta_1}
g^{\alpha_1 \beta_1}(x),
\eqno{(9.48)}
$$
$$
R_{\alpha \beta \gamma \delta}(x)  \to
R_{\prime \alpha \beta \gamma \delta}(x')
= \Lambda_{\alpha}^{~\alpha_1} \Lambda_{\beta}^{~\beta_1}
\Lambda_{\gamma}^{~\gamma_1} \Lambda_{\delta}^{~\delta_1}
g_{\alpha_1 \beta_1\gamma_1 \delta_1}(x),
\eqno{(9.49)}
$$
$$
R_{\alpha \beta }(x)  \to
R'_{ \alpha \beta }(x')
= \Lambda_{\alpha}^{~\alpha_1} \Lambda_{\beta}^{~\beta_1}
g_{\alpha_1 \beta_1}(x),
\eqno{(9.50)}
$$
$$
R(x) \to R'(x') = R(x).
\eqno{(9.51)}
$$
These transformation properties are just what we
expected in general relativity.\\

In vacuum, there is no matter field and the energy momentum tensor
of matter fields vanishes. Suppose that the self-energy of
gravitational field is small
enough to be neglected, that is
$$
T_{\mu \nu} \approx 0,
\eqno{(9.52)}
$$
then Einstein field equation becomes
$$
G_{\mu \nu} - \frac{1}{2} g_{\mu \nu} R = 0.
\eqno{(9.53)}
$$
We will prove that, in the classical limit of gravitational gauge field theory,
the Einstein field equation of vacuum holds in the first order approximation
of $gC_{\mu}^{\alpha}$. Using relation eq.(9.22), we get
$$
\partial_{\mu} G = - g \partial_{\mu}C,
\eqno{(9.54)}
$$
which is a first order quantity of $g C_{\mu}^{\alpha}$. In first order
approximations, eq.(9.40) and eq.(9.41) are changed into
$$
\begin{array}{rcl}
R_{\mu \kappa} & = &
g \partial_{\kappa} \partial_{\mu} C_{\alpha}^{\alpha}
-\frac{g}{2} \partial_{\kappa} \partial_{\lambda} C_{\mu}^{\lambda}
-\frac{g}{2} \partial_{\mu} \partial_{\lambda} C_{\kappa}^{\lambda} \\
&& + \frac{g}{2} \eta^{\lambda \nu} \eta_{\alpha \kappa}
\partial_{\nu} \partial_{\lambda} C_{\mu}^{\alpha}
+ \frac{g}{2} \eta^{\lambda \nu} \eta_{\alpha \mu}
\partial_{\nu} \partial_{\lambda} C_{\kappa}^{\alpha} \\
&& - \frac{g}{2} \eta^{\alpha \sigma} \eta_{\mu \beta}
\partial_{\kappa} \partial_{\alpha} C_{\sigma}^{\beta}
- \frac{g}{2} \eta^{\alpha \sigma} \eta_{\kappa \beta}
\partial_{\mu} \partial_{\alpha} C_{\sigma}^{\beta} +o((gC)^2)  ,
\end{array}
\eqno{(9.55)}
$$
$$
R = 2 g \eta^{\alpha \beta}
\partial_{\alpha} \partial_{\beta} C_{\sigma}^{\sigma}
-  2 g \eta^{\alpha \beta}
\partial_{\beta} \partial_{\sigma} C_{\alpha}^{\sigma}+o((gC)^2) .
\eqno{(9.56)}
$$
The equation of motion eq.(9.5) of gravitational gauge fields
gives out the following constraint on gravitational gauge field
in vacuum:
$$
\partial^{\lambda} \partial_{\lambda} C^{\beta}_{\tau} =
\partial^{\lambda} \partial_{\tau} C^{\beta}_{\lambda}.
\eqno{(9.57)}
$$
Set $\beta = \tau$, eq.(9.57) gives out
$$
\partial^{\lambda} \partial_{\lambda} C^{\beta}_{\beta} =
\partial^{\lambda} \partial_{\beta} C^{\beta}_{\lambda}.
\eqno{(9.58)}
$$
Using these two constraints, we can prove that $R_{\mu \nu}$
and $R$ given by eqs.(9.55-56) vanish:
$$
R_{\mu \nu} = 0,
\eqno{(9.59)}
$$
$$
R = 0.
\eqno{(9.60)}
$$
Therefore, Einstein field equation of vacuum indeed holds in first
order approximation of $gC_{\mu}^{\alpha}$, which is indeed a second
order infinitesimal quantity.
 \\

Equivalence principle is one of the most important fundamental
principles of general relativity. But, as we have studied in previous
chapters, the inertial energy-momentum is not equivalent to the
gravitational energy-momentum in gravitational gauge theory.
This result is an inevitable result of gauge principle. But all these
differences are caused by gravitational gauge field. In leading term
approximation, the inertial energy-momentum tensor
$T_{i \alpha}^{\mu}$ is the same as the gravitational energy-momentum
tensor $T_{g \alpha}^{\mu}$. Because gravitational coupling
constant $g$ is extremely small and the strength of gravitational
field is also weak, it is hard to detect the difference between
inertial mass and gravitational mass. Using gravitational gauge
field theory, we can calculate the difference of inertial mass
and gravitational mass for different kinds of matter and help us
to test the validity of equivalence principle. This is a fundamental
problem which will help us to understand the nature of gravitational
interactions.  We will return to this problem in chapter 12.\\

Through above discussions, we can make following two important
conclusions: 1) the leading order approximation of gravitational gauge
theory gives out Newton's theory of gravity; 2) in the first order
approximation, equation of motion of gravitational gauge fields
in vacuum gives out Einstein field equation of vacuum. \\

\section{Path Integral Quantization of Gravitational\\ Gauge Fields}

For the sake of simplicity, in this chapter and the next chapter, we only
discuss pure gravitational gauge field. For pure gravitational gauge
field, its Lagrangian function is
$$
{\cal L} = - \frac{1}{4} \eta^{\mu \rho} \eta^{\nu \sigma}
\eta_{2 \alpha \beta} e^{I(C)}
F_{\mu \nu}^{\alpha} F_{\rho \sigma}^{\beta}.
\eqno{(10.1)}
$$
Its space-time integration gives out the action of the system
$$
S = \int {\rm d}^4 x  {\cal L}.
\eqno{(10.2)}
$$
This action has local gravitational gauge symmetry. Gravitational
gauge field $C_{\mu}^{\alpha}$ has $4 \times 4 = 16$ degrees of
freedom. But, if gravitons are massless, the system has only
$2 \times 4 = 8$ degrees of freedom. There are gauge degrees
of freedom in the theory. Because only physical degrees of
freedom can be quantized, in order to quantize the system, we
have to introduce gauge conditions to eliminate un-physical degrees
of freedom. For the sake of convenience, we take temporal gauge
conditions
$$
C_0^{\alpha} = 0, ~~~(\alpha = 0,1,2,3).
\eqno{(10.3)}
$$
\\

In temporal gauge, the generating functional $W\lbrack J \rbrack$ is
given by
$$
W\lbrack J \rbrack = N \int \lbrack {\cal D} C\rbrack
\left(\prod_{\alpha, x} \delta( C_0^{\alpha}(x))\right)
exp \left\lbrace i \int {\rm d}^4 x ( {\cal L}
+ J^{\mu}_{\alpha} C^{\alpha}_{\mu}),
\right\rbrace
\eqno{(10.4)}
$$
where $N$ is the normalization constant, $J^{\mu}_{\alpha}$
is a fixed external source and $ \lbrack {\cal D} C\rbrack $ is the
integration measure,
$$
\lbrack {\cal D} C\rbrack
= \prod_{\mu=0}^{3} \prod_{\alpha= 0}^{3} \prod_j
\left(\varepsilon {\rm d}C_{\mu}^{\alpha} (\tau_j)
/ \sqrt{2 \pi i \hbar} \right).
\eqno{(10.5)}
$$
We use this generation functional as our starting
point of the path integral quantization of gravitational gauge field. \\

Generally speaking, the action of the system has local gravitational gauge
symmetry, but the gauge condition has no local gravitational gauge
symmetry. If we make a local gravitational gauge transformations,
the action of the system is kept unchanged while gauge condition will
be changed. Therefore, through local gravitational gauge transformation,
we can change one gauge condition to another gauge condition. The most
general gauge condition is
$$
f^{\alpha} (C(x)) - \varphi^{\alpha} (x) = 0,
\eqno{(10.6)}
$$
where $\varphi^{\alpha}(x)$ is an arbitrary space-time function.
The Fadeev-Popov determinant $\Delta_f(C)$ \cite{f01} is defined by
$$
\Delta_f^{-1} (C) \equiv
\int \lbrack {\cal D} g \rbrack
\prod_{x, \alpha} \delta \left(f^{\alpha} ( ^gC(x))
- \varphi^{\alpha}(x) \right),
\eqno{(10.7)}
$$
where $g$ is an element of gravitational gauge group, $^gC$ is the gravitational
gauge field after gauge transformation $g$ and
$\lbrack {\cal D} g \rbrack $ is the integration measure on
gravitational gauge group
$$
\lbrack {\cal D} g \rbrack
= \prod_x {\rm d}^4 \epsilon(x),
\eqno{(10.8)}
$$
where $\epsilon (x)$ is the transformation parameter of $\ehat$.
Both $\lbrack {\cal D} g \rbrack $ and $\lbrack {\cal D} C \rbrack $
are not invariant under gravitational gauge transformation. Suppose
that,
$$
\lbrack {\cal D} (gg') \rbrack
= J_1(g') \lbrack {\cal D} g \rbrack,
\eqno{(10.9)}
$$
$$
\lbrack {\cal D} ~^gC \rbrack
= J_2(g) \lbrack {\cal D} C \rbrack.
\eqno{(10.10)}
$$
$J_1(g)$ and $J_2(g)$ satisfy the following relations
$$
J_1(g) \cdot J_1(g^{-1}) = 1,
\eqno{(10.11)}
$$
$$
J_2(g) \cdot J_2(g^{-1}) = 1.
\eqno{(10.12)}
$$
It can be proved that, under gravitational gauge transformations,
the Fadeev-Popov determinant transforms as
$$
\Delta_f^{-1} ( ^{g'}C ) = J_1^{-1}(g') \Delta_f^{-1} (C).
\eqno{(10.13)}
$$
\\

Insert eq.(10.7) into eq.(10.4), we get
$$
\begin{array}{rcl}
W\lbrack J \rbrack &=& N \int \lbrack {\cal D} g \rbrack
\int \lbrack {\cal D} C\rbrack~~
\left\lbrack \prod_{\alpha, y}
\delta( C_0^{\alpha}(y)) \right\rbrack \cdot
\Delta_f  (C) \\
&&\\
&&\cdot \left\lbrack \prod_{\beta, z}
\delta ( f^{\beta} ( ^gC(z))- \varphi^{\beta}(z) )\right\rbrack
 \cdot exp \left\lbrace i \int {\rm d}^4 x ( {\cal L}
+ J^{\mu}_{\alpha} C^{\alpha}_{\mu})
\right\rbrace .
\end{array}
\eqno{(10.14)}
$$
Make a gravitational gauge transformation,
$$
C(x)~~ \to ~~^{g^{-1}} C(x),
\eqno{(10.15)}
$$
then,
$$
^g C(x)~~ \to
~~^{gg^{-1}} C(x).
\eqno{(10.16)}
$$
After this transformation, the generating functional is changed
into
$$
\begin{array}{rcl}
W\lbrack J \rbrack &=& N \int \lbrack {\cal D} g \rbrack
\int \lbrack {\cal D} C\rbrack~~
J_1(g) J_2(g^{-1}) \cdot \left\lbrack \prod_{\alpha, y}
\delta( ^{g^{-1}}C_0^{\alpha}(y)) \right\rbrack \cdot
\Delta_f  (C) \\
&&\\
&&\cdot \left\lbrack \prod_{\beta, z}
\delta ( f^{\beta} ( C(z))- \varphi^{\beta}(z) )\right\rbrack
 \cdot exp \left\lbrace i \int {\rm d}^4 x ( {\cal L}
+ J^{\mu}_{\alpha} \cdot  ^{g^{-1}} \!\!\! C^{\alpha}_{\mu})
\right\rbrace.
\end{array}
\eqno{(10.17)}
$$
\\

Suppose that the gauge transformation $g_0(C)$ transforms general
gauge condition $f^{\beta}(C) - \varphi^{\beta} = 0$ to temporal
gauge condition $C_0^{\alpha} = 0$, and suppose that this transformation
$g_0(C)$ is unique. Then two $\delta$-functions in eq.(10.17) require
that the integration on gravitational gauge group must be in the
neighborhood of $g^{-1}_0(C)$. Therefore eq.(10.17) is changed into
$$
\begin{array}{rcl}
W\lbrack J \rbrack &=& N \int \lbrack {\cal D} C\rbrack~~
\Delta_f  (C)  \cdot \left\lbrack \prod_{\beta, z}
\delta ( f^{\beta} ( C(z))- \varphi^{\beta}(z) )\right\rbrack \\
&&\\
&& \cdot exp \left\lbrace i \int {\rm d}^4 x ( {\cal L}
+ J^{\mu}_{\alpha} \cdot  ^{g_0}\!C^{\alpha}_{\mu})
\right\rbrace \\
&&\\
&& \cdot J_1(g_0^{-1}) J_2(g_0) \cdot \int \lbrack {\cal D} g \rbrack
\left\lbrack \prod_{\alpha, y}
\delta( ^{g^{-1}}C_0^{\alpha}(y))\right\rbrack.
\end{array}
\eqno{(10.18)}
$$
The last line in eq.(10.18) will cause no trouble in renormalization,
and if we consider the contribution from ghost fields which will
be introduced below, it will becomes a quantity which is independent
of gravitational gauge field. So, we put it into normalization constant
$N$ and still denote the new normalization constant as $N$. We also
change $J^{\mu}_{\alpha} ~  ^{g_0}\!C^{\alpha}_{\mu}$
into $J^{\mu}_{\alpha} C^{\alpha}_{\mu}$, this will cause no
trouble in renormalization. Then we get
$$
\begin{array}{rcl}
W\lbrack J \rbrack &=& N \int \lbrack {\cal D} C\rbrack~~
\Delta_f  (C)  \cdot \lbrack \prod_{\beta, z}
\delta ( f^{\beta} ( C(z))- \varphi^{\beta}(z) )\rbrack \\
&& \cdot exp \lbrace i \int {\rm d}^4 x ( {\cal L}
+ J^{\mu}_{\alpha} C^{\alpha}_{\mu}) \rbrace.
\end{array}
\eqno{(10.19)}
$$
In fact, we can use this formula as our start-point of path integral
quantization of gravitational gauge field, so we need not worried
about the influences of the third lines of eq.(10.18).
\\

Use another functional
$$
exp \left\lbrace - \frac{i}{2 \alpha}
\int {\rm d}^4 x \eta_{\alpha \beta}
\varphi^{\alpha}(x) \varphi^{\beta}(x) \right\rbrace,
\eqno{(10.20)}
$$
times both sides of eq.(10.19) and then make functional
integration
$\int \lbrack {\cal D} \varphi  \rbrack$,
we get
$$
W\lbrack J \rbrack  = N \int \lbrack {\cal D} C\rbrack~~
\Delta_f (C) \cdot exp \left\lbrace i \int {\rm d}^4 x ( {\cal L}
- \frac{1}{2 \alpha} \eta_{\alpha \beta} f^{\alpha} f^{\beta}
+ J^{\mu}_{\alpha} C^{\alpha}_{\mu}) \right\rbrace.
\eqno{(10.21)}
$$
Now, let's discuss the contribution from $\Delta_f (C)$ which
is related to the ghost fields. Suppose that $g = \ehat$
is an infinitesimal gravitational gauge transformation. Then
eq.(4.12) gives out
$$
^gC_{\mu}^{\alpha} (x)
= C_{\mu}^{\alpha} (x)
- \frac{1}{g} {\mathbf D}_{\mu~\sigma}^{\alpha} \epsilon^{\sigma},
\eqno{(10.22)}
$$
where
$$
{\mathbf D}_{\mu~\sigma}^{\alpha}
=\delta^{\alpha}_{\sigma} \partial_{\mu}
- g \delta^{\alpha}_{\sigma} C_{\mu}^{\beta} \partial_{\beta}
+ g \partial_{\sigma} C_{\mu}^{\alpha}.
\eqno{(10.23)}
$$
In order to deduce eq.(10.22), the following relation is used
$$
\Lambda^{\alpha}_{~\beta}
= \delta^{\alpha}_{\beta}
+ \partial_{\beta} \epsilon^{\alpha}
+ o( \epsilon^2).
\eqno{(10.24)}
$$
${\mathbf D}_{\mu}$ can be regarded as the covariant derivate
in adjoint representation, for
$$
{\mathbf D}_{\mu} \epsilon
= \lbrack D_{\mu} ~~~,~~~ \epsilon \rbrack,
\eqno{(10.25)}
$$
$$
({\mathbf D}_{\mu} \epsilon)^{\alpha}
= {\mathbf D}_{\mu~\sigma}^{\alpha} \epsilon^{\sigma}.
\eqno{(10.26)}
$$
Using all these relations,  we have,
$$
f^{\alpha} (^gC(x)) = f^{\alpha} (C)
- \frac{1}{g} \int {\rm d}^4 y
\frac{\delta f^{\alpha}(C(x))}{\delta C_{\mu}^{\beta}(y)}
{\mathbf D}_{\mu~\sigma}^{\beta}(y) \epsilon^{\sigma}(y)
+ o(\epsilon^2).
\eqno{(10.27)}
$$
Therefore, according to eq.(10.7) and eq.(10.6), we get
$$
\Delta_f^{-1} (C) =
\int \lbrack {\cal D} \epsilon \rbrack
\prod_{x, \alpha}
\delta \left( - \frac{1}{g}     \int {\rm d}^4 y
\frac{\delta f^{\alpha}(C(x))}{\delta C_{\mu}^{\beta}(y)}
{\mathbf D}_{\mu~\sigma}^{\beta}(y) \epsilon^{\sigma}(y) \right).
\eqno{(10.28)}
$$
Define
$$
\begin{array}{rcl}
{\mathbf M}^{\alpha}_{~\sigma}(x,y) &=& -g
\frac{\delta}{\delta \epsilon^{\sigma}(y)}
f^{\alpha}(^gC(x)) \\
&&\\
&=&\int {\rm d}^4 z
\frac{\delta f^{\alpha}(C(x))}{\delta C_{\mu}^{\beta}(z)}
{\mathbf D}_{\mu~\sigma}^{\beta}(z) \delta(z-y) .
\end{array}
\eqno{(10.29)}
$$
Then eq.(10.28) is changed into
$$
\begin{array}{rcl}
\Delta_f^{-1} (C) &=&
\int \lbrack {\cal D} \epsilon \rbrack
\prod_{x, \alpha}
\delta \left( - \frac{1}{g} \int {\rm d}^4 y
{\mathbf M}^{\alpha}_{~\sigma}(x,y) \epsilon^{\sigma}(y)
\right)  \\
&&\\
&=& const. \times (det {\mathbf M} )^{-1}.
\end{array}
\eqno{(10.30)}
$$
Therefore,
$$
\Delta_f (C) = const. \times det {\mathbf M}.
\eqno{(10.31)}
$$
Put this constant into normalization constant, then generating
functional eq.(10.21) is changed into
$$
W\lbrack J \rbrack  = N \int \lbrack {\cal D} C\rbrack~~
det {\mathbf M} \cdot exp \left\lbrace i \int {\rm d}^4 x ( {\cal L}
- \frac{1}{2 \alpha} \eta_{\alpha \beta} f^{\alpha} f^{\beta}
+ J^{\mu}_{\alpha} C^{\alpha}_{\mu}) \right\rbrace.
\eqno{(10.32)}
$$
\\

In order to evaluate the contribution from $det {\mathbf M}$,
we introduce ghost fields $\eta^{\alpha}(x)$
and $\bar{\eta}_{\alpha}(x)$. Using the following relation
$$
\int \lbrack {\cal D} \eta \rbrack
\lbrack {\cal D}\bar{\eta} \rbrack
exp \left\lbrace i \int {\rm d}^4 x {\rm d}^4 y ~
\bar{\eta}_{\alpha} (x) {\mathbf M}^{\alpha}_{~\beta}(x,y)
\eta^{\beta} (y) \right\rbrace
 = const. \times det {\mathbf M}
\eqno{(10.33)}
$$
and put the constant into the normalization constant, we can get
$$
W\lbrack J \rbrack  = N \int \lbrack {\cal D} C\rbrack
\lbrack {\cal D} \eta \rbrack
\lbrack {\cal D}\bar{\eta} \rbrack
exp \left\lbrace i \int {\rm d}^4 x ( {\cal L}
- \frac{1}{2 \alpha} \eta_{\alpha \beta} f^{\alpha} f^{\beta}
+ \bar{\eta} {\mathbf M } \eta
+ J^{\mu}_{\alpha} C^{\alpha}_{\mu}) \right\rbrace,
\eqno{(10.34)}
$$
where $\int {\rm d}^4x \bar{\eta} {\mathbf M } \eta$ is a
simplified notation, whose explicit expression is
$$
\int {\rm d}^4x \bar{\eta} {\mathbf M } \eta
= \int {\rm d}^4 x {\rm d}^4 y ~
\bar{\eta}_{\alpha} (x) {\mathbf M}^{\alpha}_{~\beta}(x,y)
\eta^{\beta} (y).
\eqno{(10.35)}
$$
The appearance of the non-trivial ghost fields is a inevitable
result of the non-Able nature of the gravitational gauge group.
\\

Now, let's take Lorentz covariant gauge condition,
$$
f^{\alpha} (C) = \partial^{\mu} C_{\mu}^{\alpha} .
\eqno{(10.36)}
$$
Then
$$
\int {\rm d}^4x \bar{\eta} {\mathbf M } \eta =
- \int {\rm d}^4x \left( \partial^{\mu}
\bar{\eta}_{\alpha} (x) \right)
{\mathbf D}_{\mu~\beta}^{\alpha}(x) \eta^{\beta} (x).
\eqno{(10.37)}
$$
And eq.(10.34) is changed into
$$
\begin{array}{rcl}
W\lbrack J, \beta, \bar{\beta} \rbrack
& = & N \int \lbrack {\cal D} C\rbrack
\lbrack {\cal D} \eta \rbrack
\lbrack {\cal D}\bar{\eta} \rbrack
exp \left\lbrace i \int {\rm d}^4 x ( {\cal L}
- \frac{1}{2 \alpha}
\eta_{\alpha \beta} f^{\alpha} f^{\beta} \right. \\
&&\\
&& \left. - (\partial^{\mu}\bar{\eta}_{\alpha} )
{\mathbf D}_{\mu~\sigma}^{\alpha} \eta^{\sigma}
+ J^{\mu}_{\alpha} C^{\alpha}_{\mu}
+ \bar{\eta}_{\alpha} \beta^{\alpha}
+ \bar{\beta}_{\alpha} \eta^{\alpha}
) \right\rbrace,
\end{array}
\eqno{(10.38)}
$$
where we have introduced external sources $\eta^{\alpha}(x)$
and $\bar{\eta}_{\alpha}(x)$ of ghost fields. \\

The effective Lagrangian ${\cal L}_{eff}$
is defined by
$$
{\cal L}_{eff} \equiv
{\cal L} - \frac{1}{2 \alpha}
\eta_{\alpha \beta} f^{\alpha} f^{\beta}
- (\partial^{\mu}\bar{\eta}_{\alpha} )
{\mathbf D}_{\mu~\sigma}^{\alpha} \eta^{\sigma}.
\eqno{(10.39)}
$$
${\cal L}_{eff}$ can separate into free Lagrangian
${\cal L}_F$  and interaction Lagrangian ${\cal L}_I$,
$$
{\cal L}_{eff} = {\cal L}_F + {\cal L}_I,
\eqno{(10.40)}
$$
where
$$
\begin{array}{rcl}
{\cal L}_F &=& - \frac{1}{2}
\eta^{\mu \rho} \eta^{\nu \sigma} \eta_{2 \alpha \beta}
\left\lbrack (\partial_{\mu} C_{\nu}^{\alpha})
(\partial_{\rho} C_{\sigma}^{\beta})
-(\partial_{\mu} C_{\nu}^{\alpha})
(\partial_{\sigma} C_{\rho}^{\beta})
\right\rbrack   \\
&&\\
&&  -\frac{1}{2 \alpha} \eta_{2 \alpha \beta}
(\partial^{\mu} C_{\mu}^{\alpha})
(\partial^{\nu} C_{\nu}^{\beta})
- (\partial^{\mu} \bar{\eta}_{\alpha})
(\partial_{\mu} \eta^{\alpha}),
\end{array}
\eqno{(10.41)}
$$
$$
\begin{array}{rcl}
{\cal L}_I &=& - \frac{1}{2} (e^{I(C)}-1)
\eta^{\mu \rho} \eta^{\nu \sigma} \eta_{2 \alpha \beta}
\left\lbrack (\partial_{\mu} C_{\nu}^{\alpha})
(\partial_{\rho} C_{\sigma}^{\beta})
-(\partial_{\mu} C_{\nu}^{\alpha})
(\partial_{\sigma} C_{\rho}^{\beta})
\right\rbrack   \\
&&\\
&&   + g e^{I(C)} \eta^{\mu \rho} \eta^{\nu \sigma}
\eta_{2 \alpha \beta} (\partial_{\mu} C_{\nu }^{\alpha}
- \partial_{\nu} C_{\mu}^{\alpha})
C_{\rho}^{\delta} \partial_{\delta} C_{\sigma}^{\beta} \\
&&\\
&&  - \frac{1}{2} g^2 e^{I(C)} \eta^{\mu \rho} \eta^{\nu \sigma}
\eta_{2 \alpha \beta}
(C_{\mu}^{\delta} \partial_{\delta} C_{\nu}^{\alpha}
- C_{\nu}^{\delta} \partial_{\delta} C_{\mu}^{\alpha} )
C_{\rho}^{\epsilon} \partial_{\epsilon} C_{\sigma}^{\beta}\\
&&\\
&& + g(\partial^{\mu} \bar{\eta}_{\alpha})
C_{\mu}^{\beta} (\partial_{\beta} \eta^{\alpha})
- g(\partial^{\mu} \bar{\eta}_{\alpha})
(\partial_{\sigma} C_{\mu}^{\alpha}) \eta^{\sigma}.
\end{array}
\eqno{(10.42)}
$$
From the interaction Lagrangian, we can see that ghost fields
do not couple to $e^{I(C)}$. This is the reflection of the fact
that ghost fields are not physical fields, they are virtual fields.
Besides, the gauge fixing term does not couple to $e^{I(C)}$ either.
Using effective Lagrangian ${\cal L}_{eff}$, the generating
functional $W\lbrack J, \beta, \bar{\beta} \rbrack$ can be
simplified to
$$
W\lbrack J, \beta, \bar{\beta} \rbrack
=  N \int \lbrack {\cal D} C\rbrack
\lbrack {\cal D} \eta \rbrack
\lbrack {\cal D}\bar{\eta} \rbrack
exp \left\lbrace i \int {\rm d}^4 x ( {\cal L}_{eff}
+ J^{\mu}_{\alpha} C^{\alpha}_{\mu}
+ \bar{\eta}_{\alpha} \beta^{\alpha}
+ \bar{\beta}_{\alpha} \eta^{\alpha}
) \right\rbrace,
\eqno{(10.43)}
$$
\\

Use eq.(10.41), we can deduce propagator of gravitational gauge
fields and ghost fields. First, we change its form to
$$
\int {\rm d}^4x {\cal L}_F =
\int {\rm d}^4x \left\lbrace \frac{1}{2}
C_{\nu}^{\alpha} \left\lbrack
\eta_{2 \alpha \beta} \left(
\eta^{\mu \nu} \partial^2 - (1 - \frac{1}{\alpha})
\partial^{\mu} \partial^{\nu} \right)
\right\rbrack C_{\nu}^{\beta}
+ \bar{\eta}_{\alpha}
\partial^2 \eta^{\alpha} \right\rbrace.
\eqno{(10.44)}
$$
Denote the propagator of gravitational gauge field as
$$
-i \Delta_{F \mu \nu}^{\alpha \beta} (x),
\eqno{(10.45)}
$$
and denote the propagator of ghost field as
$$
-i \Delta_{F \beta}^{\alpha } (x).
\eqno{(10.46)}
$$
They satisfy the following equation,
$$
- \left\lbrack
\eta_{2 \alpha \beta} \left(
\eta^{\mu \nu} \partial^2 - (1 - \frac{1}{\alpha})
\partial^{\mu} \partial^{\nu} \right)
\right\rbrack \Delta_{F \nu \rho}^{\beta \gamma} (x)
= \delta(x) \delta^{\gamma}_{\alpha} \delta^{\mu}_{\rho},
\eqno{(10.47)}
$$
$$
- \partial^2 \Delta_{F \beta}^{\alpha } (x)
= \delta_{\beta}^{\alpha } \delta(x).
\eqno{(10.48)}
$$
Make Fourier transformations to momentum space
$$
\Delta_{F \mu \nu}^{\alpha \beta} (x)
= \int \frac{{\rm d}^4k}{(2 \pi)^4}
\swav{\Delta}_{F \mu \nu}^{\alpha \beta}(k) \cdot  e^{ikx},
\eqno{(10.49)}
$$
$$
\Delta_{F \beta}^{\alpha } (x)
= \int \frac{{\rm d}^4k}{(2 \pi)^4}
\swav{\Delta}_{F \beta}^{\alpha }(k) \cdot e^{ikx},
\eqno{(10.50)}
$$
where $\swav{\Delta}_{F \mu \nu}^{\alpha \beta}(k)$
and $\swav{\Delta}_{F \beta}^{\alpha }(k)$ are corresponding
propagators in momentum space. They satisfy the following
equations,
$$
\eta_{2 \alpha \beta}
\left\lbrack
k^2 \eta^{\mu \nu} - (1 - \frac{1}{\alpha}) k^{\mu}k^{\nu}
\right\rbrack
\swav{\Delta}_{F \nu \rho}^{\beta \gamma}(k)
= \delta^{\gamma}_{\alpha} \delta^{\mu}_{\rho},
\eqno{(10.51)}
$$
$$
k^2 \swav{\Delta}_{F \beta}^{\alpha }(k)
= \delta^{\alpha}_{\beta}.
\eqno{(10.52)}
$$
The solutions to these two equations give out the propagators
in momentum space,
$$
-i \swav{\Delta}_{F \mu \nu}^{\alpha \beta}(k)
= \frac{-i}{k^2 - i \epsilon}
\eta_2^{\alpha  \beta}
\left\lbrack
\eta_{\mu \nu} - (1 - \alpha)
\frac{k_{\mu} k_{\nu}}{k^2 - i \epsilon},
\right\rbrack
\eqno{(10.53)}
$$
$$
-i \swav{\Delta}_{F \beta}^{\alpha }(k) =
\frac{-i}{k^2 - i \epsilon} \delta^{\alpha}_{\beta}.
\eqno{(10.54)}
$$
It can be seen that the forms of these propagators are quite
similar to those in traditional non-Able gauge theory. The only
difference is that the metric is different. \\

The interaction Lagrangian ${\cal L}_I$ is a function of
gravitational gauge field $C_{\mu}^{\alpha}$ and ghost
fields $\eta^{\alpha}$ and $\bar \eta_{\alpha}$,
$$
{\cal L}_I =
{\cal L}_I ( C, \eta, \bar\eta ).
\eqno{(10.55)}
$$
Then eq.(10.43) is changed into,
$$
\begin{array}{rcl}
W\lbrack J, \beta, \bar{\beta} \rbrack
& = & N \int \lbrack {\cal D} C\rbrack
\lbrack {\cal D} \eta \rbrack
\lbrack {\cal D}\bar{\eta} \rbrack
~exp \left\lbrace i \int {\rm d}^4 x
 {\cal L}_I ( C, \eta, \bar\eta ) \right\rbrace \\
&&\\
&&\cdot exp \left\lbrace i \int {\rm d}^4 x ( {\cal L}_F
+ J^{\mu}_{\alpha} C^{\alpha}_{\mu}
+ \bar{\eta}_{\alpha} \beta^{\alpha}
+ \bar{\beta}_{\alpha} \eta^{\alpha}
) \right\rbrace  \\
&&\\
&=& exp \left\lbrace i \int {\rm d}^4 x
 {\cal L}_I ( \frac{1}{i}\frac{\delta}{\delta J},
\frac{1}{i}\frac{\delta}{\delta \bar \beta},
\frac{1}{-i}\frac{\delta}{\delta \beta} ) \right\rbrace
\cdot W_0\lbrack J, \beta, \bar{\beta} \rbrack ,
\end{array}
\eqno{(10.56)}
$$
where
$$
\begin{array}{rcl}
W_0\lbrack J, \beta, \bar{\beta} \rbrack
&=&  N \int \lbrack {\cal D} C\rbrack
\lbrack {\cal D} \eta \rbrack
\lbrack {\cal D}\bar{\eta} \rbrack
exp \left\lbrace i \int {\rm d}^4 x ( {\cal L}_F
+ J^{\mu}_{\alpha} C^{\alpha}_{\mu}
+ \bar{\eta}_{\alpha} \beta^{\alpha}
+ \bar{\beta}_{\alpha} \eta^{\alpha}
) \right\rbrace  \\
&&\\
&=&  exp \left\lbrace
\frac{i}{2} \int\int {\rm d}^4 x {\rm d}^4 y
\left\lbrack J^{\mu}_{\alpha} (x)
\Delta_{F \mu \nu}^{\alpha \beta} (x-y)
J^{\nu}_{\beta} (y) \right. \right.  \\
&&\\
&& \left.\left.~~+ \bar\eta_{\alpha}(x)
\Delta_{F \beta}^{\alpha } (x-y) \eta^{\beta}(y).
\right\rbrack  \right\rbrace
\end{array}
\eqno{(10.57)}
$$
\\

Finally, let's discuss Feynman rules. Here, we only give
out the lowest order interactions in gravitational gauge
theory. It is known that, a vertex can involve arbitrary number
of gravitational gauge fields. Therefore, it is impossible
to list all Feynman rules for all kinds of vertex. \\

The interaction
Lagrangian between gravitational gauge field and ghost field
is
$$
+ g(\partial^{\mu} \bar{\eta}_{\alpha})
C_{\mu}^{\beta} (\partial^{\beta} \eta^{\alpha})
- g(\partial^{\mu} \bar{\eta}_{\alpha})
(\partial_{\sigma} C_{\mu}^{\alpha}) \eta^{\sigma}.
\eqno{(10.58)}
$$
This vertex belongs to
$C_{\mu}^{\alpha}(k) \bar{\eta}_{\beta}(q) \eta^{\delta}(p) $
three body interactions, its Feynman rule is
$$
-i g \delta^{\beta}_{\delta} q^{\mu} p_{\alpha}
+ i g \delta^{\beta}_{\alpha} q^{\mu} k_{\delta}.
\eqno{(10.59)}
$$
The lowest order interaction Lagrangian between gravitational gauge
field and Dirac field is
$$
g \bar \psi \gamma^{\mu} \partial_{\alpha} \psi C_{\mu}^{\alpha}
- g \eta_{1 \alpha}^{\mu} \bar \psi
\gamma^{\nu} \partial_{\nu} \psi C_{\mu}^{\alpha}
- g m \eta_{1 \alpha}^{\mu} \bar \psi \psi C_{\mu}^{\alpha}.
\eqno{(10.60)}
$$
This vertex belongs to
$C_{\mu}^{\alpha}(k) \bar \psi(q) \psi(p)$
three body interactions, its Feynman rule is
$$
- g \gamma^{\mu} p_{\alpha}
+ g \eta_{1 \alpha}^{\mu} \gamma^{\nu} p_{\nu}
- i m g \eta_{1 \alpha}^{\mu}.
\eqno{(10.61)}
$$
The lowest order interaction Lagrangian between gravitational
gauge field and real scalar field is
$$
g \eta^{\mu \nu} C_{\mu}^{\alpha}
(\partial_{\nu} \phi) (\partial_{\alpha} \phi)
- \frac{1}{2} g \eta_{1 \alpha}^{\mu}
C_{\mu}^{\alpha} ( (\partial^{\nu} \phi)
(\partial_{\nu} \phi) + m^2 \phi^2 ).
\eqno{(10.62)}
$$
This  vertex  belongs to
$C_{\mu}^{\alpha}(k) \phi(q) \phi(p)$
three body interactions, its Feynman rule is
$$
- i g \eta^{\mu \nu}
( p_{\nu} q_{\alpha} +q_{\nu} p_{\alpha}  )
-i g \eta_{1 \alpha}^{\mu} ( -p^{\nu} q_{\nu} + m^2 ).
\eqno{(10.63)}
$$
The lowest order interaction Lagrangian between gravitational
gauge field and complex scalar field is
$$
g \eta^{\mu \nu} C_{\mu}^{\alpha}
((\partial_{\alpha} \phi) (\partial_{\nu} \phi^*)
+ (\partial_{\alpha} \phi^*) (\partial_{\nu} \phi))
- g \eta_{1 \alpha}^{\mu} C_{\mu}^{\alpha}
( (\partial^{\nu} \phi) (\partial_{\nu} \phi^*)
+ m^2 \phi \phi^*  ).
\eqno{(10.64)}
$$
This vertex belongs to
$C_{\mu}^{\alpha}(k) \phi^*(-q) \phi(p)$
three body interactions, its Feynman rule is
$$
i g \eta^{\mu \nu}
( p_{\nu} q_{\alpha} +q_{\nu} p_{\alpha}  )
-i g \eta_{1 \alpha}^{\mu} ( p^{\nu} q_{\nu} + m^2 ).
\eqno{(10.65)}
$$
The lowest order coupling between vector field and gravitational
gauge field is
$$
\begin{array}{l}
 g \eta^{\mu \rho} \eta^{\nu \sigma} A_{0 \mu \nu}
C_{\rho}^{\alpha} \partial_{\alpha} A_{\sigma} \\
\\
+ (g \eta_{1 \tau}^{\lambda} C_{\lambda}^{\tau} )
( - \frac{1}{4} \eta^{\mu \rho} \eta^{\nu \sigma}
A_{0 \mu \nu} A_{0 \rho \sigma}
- \frac{m^2}{2} \eta^{\mu \nu} A_{\mu} A_{\nu} ),
\end{array}
\eqno{(10.66)}
$$
where
$$
A_{0 \mu \nu} =
\partial_{\mu} A_{\nu} - \partial_{\nu} A_{\mu} .
\eqno{(10.67)}
$$
This vertex belongs to
$C_{\mu}^{\alpha}(k) A_{\rho}(p) A_{\sigma}(q) $
three body interactions. Its Feynman rule is
$$
\begin{array}{l}
- ig \eta^{\mu \beta} \eta^{\rho \sigma}
(p_{\beta} q_{\alpha} +  p_{\alpha} q_{\beta})
+ ig \eta^{\mu \rho} \eta^{\sigma \beta} p_{\beta} q_{\alpha}
+ ig \eta^{\mu \sigma} \eta^{\rho \beta} q_{\beta} p_{\alpha}\\
\\
+ \frac{i}{2} g \eta_{1 \alpha}^{\mu} \eta^{\lambda \beta}
   \eta^{\rho \sigma}
   ( p_{\lambda} q_{\beta} +  q_{\lambda} p_{\beta} ) \\
\\
- \frac{i}{2} g \eta_{1 \alpha}^{\mu} \eta^{\rho \beta}
   \eta^{\nu \sigma}  p_{\nu} q_{\beta}
- \frac{i}{2} g \eta_{1 \alpha}^{\mu} \eta^{\rho \nu}
   \eta^{\beta \sigma}  q_{\nu} p_{\beta}
-i g m^2 \eta_{1 \alpha}^{\mu} \eta^{\rho \sigma}.
\end{array}
\eqno{(10.68)}
$$
The lowest order self coupling of gravitational gauge fields
is
$$
\begin{array}{l}
 g \eta^{\mu \rho} \eta^{\nu \sigma} \eta_{2 \alpha \beta}
\lbrack (\partial_{\mu} C_{\nu}^{\alpha} ) C_{\rho}^{\beta_1}
  (\partial_{\beta_1} C_{\sigma}^{\beta} )
- (\partial_{\mu} C_{\nu}^{\alpha} ) C_{\sigma}^{\beta_1}
  (\partial_{\beta_1} C_{\rho}^{\beta} )
\rbrack  \\
\\
- \frac{1}{4} (g \eta_{1 \tau}^{\lambda} C_{\lambda}^{\tau} )
  \eta^{\mu \rho} \eta^{\nu \sigma} \eta_{2 \alpha \beta}
  F_{0 \mu \nu}^{\alpha} F_{0 \rho \sigma}^{\beta}.
\end{array}
\eqno{(10.69)}
$$
This vertex belongs to
$C_{\nu}^{\alpha}(p) C_{\sigma}^{\beta}(q) C_{\rho}^{\gamma}(r)$
three body interactions. Its Feynman rule is
$$
\begin{array}{l}
-i g \lbrack \eta^{\mu \rho} \eta^{\nu \sigma} \eta_{2 \alpha \beta}
     ( p_{\mu} q_{\gamma} +  q_{\mu} p_{\gamma} )
+     \eta^{\mu \sigma} \eta^{\nu \rho} \eta_{2 \alpha \gamma}
     ( p_{\mu} r_{\beta} +  r_{\mu} p_{\beta}) \\
\\
+     \eta^{\mu \nu} \eta^{\rho \sigma} \eta_{2 \gamma \beta}
     (q_{\mu} r_{\alpha} + r_{\mu} q_{\alpha}) \rbrack \\
\\
+ ig \lbrack \eta^{\mu \sigma} \eta^{\nu \rho} \eta_{2 \alpha \beta}
      p_{\mu} q_{\gamma}
+    \eta^{\mu \nu} \eta^{\rho \sigma} \eta_{2 \alpha \beta}
      q_{\mu} p_{\gamma}
+     \eta^{\mu \rho} \eta^{\nu \sigma} \eta_{2 \alpha \gamma}
      p_{\mu} r_{\beta} \\
\\
+      \eta^{\mu \nu} \eta^{\rho \sigma} \eta_{2 \alpha \gamma}
      r_{\mu} p_{\beta}
+     \eta^{\mu \rho} \eta^{\nu \sigma} \eta_{2 \beta \gamma}
      q_{\mu} r_{\alpha}
+     \eta^{\mu \sigma} \eta^{\nu \rho} \eta_{2 \beta \gamma}
      r_{\mu} q_{\alpha}  \rbrack  \\
\\
+     ig \eta_{1 \gamma}^{\rho}  \eta_{2 \alpha \beta}
      ( p_{\mu} q^{\mu} \eta^{\nu \sigma} - p^{\sigma} q^{\nu} )
+     ig \eta_{1 \alpha}^{\nu}  \eta_{2 \beta \gamma}
      ( q_{\mu} r^{\mu} \eta^{\rho \sigma} - r^{\sigma} q^{\rho} )\\
\\
+   ig \eta_{1 \beta}^{\sigma}  \eta_{2 \alpha \gamma}
      ( r_{\mu} p^{\mu} \eta^{\rho \nu} - p^{\rho} r^{\nu} )
\end{array}
\eqno{(10.70)}
$$
\\

It could be found that all Feynman rules for vertex is
proportional to energy-momenta of one or more particles,
which is one of the most important properties of gravitational
interactions. In fact, this interaction property is expected
for gravitational interactions, for energy-momentum
is the source of gravity. \\

\section{Renormalization }

In gravitational gauge theory, the gravitational coupling
constant has the dimensionality of negative powers of mass.
According to traditional theory of power counting law, it seems
that the gravitational gauge theory is a kind of non-renormalizable
theory. But this result is not correct. The power counting
law does not work here. General speaking, power counting law
does not work when a theory has gauge symmetry. If a theory has
gauge symmetry, the constraints from gauge symmetry will make
some divergence cancel each other. In gravitational gauge theory,
this mechanism works very well. In this chapter, we will give a
strict formal proof on the renormalization of the gravitational
gauge theory. We will find that the effect of renormalization is
just a scale transformation of the original theory. Though
there are infinite number of divergent vertexes in the gravitational
gauge theory, we need not introduce infinite number of interaction
terms that do not exist in the original Lagrangian and infinite
number of parameters. All the divergent vertex can find its
correspondence in the original Lagrangian. Therefor, in
renormalization, what we need to do is not to introduce extra
interaction terms to cancel divergent terms, but to redefine
the fields, coupling constants and some other parameters of
the original theory. Because most of counterterms come
from the factor $e^{I(C)}$, this factor is key important for
renormalization. Without this factor, the theory is
non-renormalizable. In a word, the gravitational gauge theory
is a renormalizable gauge theory. Now, let's start our discussion
on renormalization from the generalized BRST transformations.
Our proof is quite similar to the proof of the renormalizablility
of non-Able gauge field theory.\cite{c01,c02,c03,c04,c05,c06}\\

The generalized BRST transformations are
$$
\delta C_{\mu}^{\alpha}
= -  {\mathbf D}_{\mu~\beta}^{\alpha} \eta^{\beta}
\delta \lambda,
\eqno{(11.1)}
$$
$$
\delta \eta^{\alpha} = g \eta^{\sigma}
(\partial_{\sigma}\eta^{\alpha}) \delta \lambda,
\eqno{(11.2)}
$$
$$
\delta \bar\eta_{\alpha} = \frac{1}{\alpha}
\eta_{\alpha \beta} f^{\beta} \delta \lambda,
\eqno{(11.3)}
$$
$$
\delta \eta^{\mu \nu} = 0,
\eqno{(11.4)}
$$
$$
\delta \eta^{\mu}_{1\alpha } =
- g  \eta^{\mu}_{1\sigma}
(\partial_{\alpha} \eta^{\sigma})  \delta \lambda,
\eqno{(11.5)}
$$
$$
\delta \eta_{2\alpha \beta} = - g \left(
 \eta_{2\alpha \sigma}(\partial_{\beta} \eta^{\sigma})
+\eta_{2\sigma \beta} (\partial_{\alpha} \eta^{\sigma}) \right)
\delta \lambda,
\eqno{(11.6)}
$$
where $\delta \lambda$ is an infinitesimal Grassman constant.
It can be strict proved that the generalized BRST transformations
for fields $C_{\mu}^{\alpha}$ and $\eta^{\alpha}$ are nilpotent:
$$
\delta( {\mathbf D}_{\mu~\beta}^{\alpha} \eta^{\beta} ) =0,
\eqno{(11.7)}
$$
$$
\delta(\eta^{\sigma}(\partial_{\sigma}\eta^{\alpha})) =0.
\eqno{(11.8)}
$$
It means that all second order variations of fields vanish.\\

Using the above transformation rules, it can be strictly proved that
the generalized BRST transformation for gauge field strength tensor
$F^{\alpha}_{\mu \nu}$ is
$$
\delta F^{\alpha}_{\mu \nu} = - g \left( - (\partial_{\sigma}
\eta^{\alpha}) F^{\sigma}_{\mu \nu} + \eta^{\sigma}
(\partial_{\sigma} F^{\alpha}_{\mu \nu}) \right) \delta \lambda,
\eqno{(11.9)}
$$
and the transformation for the factor $e^{I(C)}$ is
$$
\delta e^{I(C)} = - g \left( (\partial_{\alpha} \eta^{\alpha})
e^{I(C)} + \eta^{\alpha} (\partial_{\alpha} e^{I(C)} ) \right)
\delta \lambda.
\eqno{(11.10)}
$$
Therefore, under generalized BRST transformations,
the Lagrangian ${\cal L}$ given by eq.(10.1) transforms as
$$
\delta {\cal L} = - g (\partial_{\alpha} (\eta^{\alpha}  {\cal L}))
\delta \lambda.
\eqno{(11.11)}
$$
It is a total derivative term, its space-time integration
vanish, i.e., the action of eq.(10.2) is invariant under
generalized BRST transformations,
$$
\delta S =\delta \left(\int {\rm d}^4x {\cal L}\right) = 0.
\eqno{(11.12)}
$$
On the other hand, it can be strict proved that
$$
\delta \left( - \frac{1}{2 \alpha} \eta_{\alpha \beta} f^{\alpha}
f^{\beta} + \bar{\eta}_{\alpha} \partial^{\mu} {\mathbf
D}_{\mu~\sigma}^{\alpha} \eta^{\sigma} \right) =0.
\eqno{(11.13)}
$$
\\

The non-renormalized effective Lagrangian  is denoted
as $ {\cal L}_{eff}^{\lbrack 0 \rbrack} $. It is given by
$$
{\cal L}_{eff}^{\lbrack 0 \rbrack} = {\cal L} - \frac{1}{2 \alpha}
\eta_{\alpha \beta} f^{\alpha} f^{\beta} + \bar{\eta}_{\alpha}
\partial^{\mu} {\mathbf D}_{\mu~\sigma}^{\alpha} \eta^{\sigma}.
\eqno{(11.14)}
$$
The effective action is defined by
$$
S_{eff}^{\lbrack 0 \rbrack} = \int {\rm d}^4x {\cal
L}_{eff}^{\lbrack 0 \rbrack}.
\eqno{(11.15)}
$$
Using eqs.(10.12-13), we can prove that this effective action is
invariant under generalized BRST transformations,
$$
\delta S_{eff}^{\lbrack 0 \rbrack} =0.
\eqno{(11.16)}
$$
This is a strict relation without any approximation.
It is known that BRST symmetry plays a key  role
in the renormalization of gauge theory, for it ensures
the validity of the Ward-Takahashi identities. \\

Before we go any further, we have to do another important
work, i.e., to prove that the functional integration measure
$\lbrack {\cal D}C \rbrack \lbrack {\cal D}\eta \rbrack
\lbrack {\cal D}\bar\eta \rbrack $ is also generalized BRST
invariant.  We have said before that the functional integration
measure $\lbrack {\cal D}C \rbrack$ is not a gauge invariant
measure, therefore, it is highly important to prove that
$\lbrack {\cal D}C \rbrack \lbrack {\cal D}\eta \rbrack
\lbrack {\cal D}\bar\eta \rbrack $ is a generalized BRST
invariant measure. BRST transformation is a kind of
transformation which involves both bosonic fields and
fermionic fields. For the sake of simplicity, let's formally
denote all bosonic fields as $B = \lbrace B_i \rbrace$
and denote all fermionic fields as $F = \lbrace F_i \rbrace$.
All fields that involve  in generalized BRST transformation
are simply denoted by $(B,F)$. Then, generalized BRST
transformation is formally expressed as
$$
(B,F) ~~~ \to ~~~(B',F').
\eqno{(11.17)}
$$
The transformation matrix of this transformation is
$$
J =
\left (
\begin{array}{cc}
\frac{\partial B_i}{\partial B'_j} &
\frac{\partial B_i}{\partial F'_l} \\
\frac{\partial F_k}{\partial B'_j} &
\frac{\partial F_k}{\partial F'_l}
\end{array}
\right )
=
\left (
\begin{array}{cc}
a &
\alpha \\
\beta &
b
\end{array}
\right )  ,
\eqno{(11.18)}
$$
where
$$
a = \left( \frac{\partial B_i}{\partial B'_j} \right),
\eqno{(11.19)}
$$
$$
b = \left( \frac{\partial F_k}{\partial F'_l} \right),
\eqno{(11.20)}
$$
$$
\alpha = \left( \frac{\partial B_i}{\partial F'_l} \right),
\eqno{(11.21)}
$$
$$
\beta = \left( \frac{\partial F_k}{\partial B'_j} \right).
\eqno{(11.22)}
$$
Matrixes $a$ and $b$ are bosonic square matrix while $\alpha$ and
$\beta$ generally are not square matrix. In order to calculate the
Jacobian $det(J)$. we realize the transformation (11.17) in two
steps. The first step is a bosonic transformation
$$
(B,F) ~~~ \to ~~~(B',F).
\eqno{(11.23)}
$$
The transformation matrix of this transformation is denoted as $J_1$,
$$
J_1 =
\left (
\begin{array}{cc}
a - \alpha b^{-1} \beta &
\alpha b^{-1} \\
0 &
1
\end{array}
\right )  .
\eqno{(11.24)}
$$
Its Jacobian is
$$
det~ J_1 = det( a - \alpha b^{-1} \beta ).
\eqno{(11.25)}
$$
Therefore,
$$
\int \prod_i {\rm d}B_i \prod_k {\rm d}F_k = \int \prod_i {\rm
d}B'_i \prod_k {\rm d}F_k \cdot det( a - \alpha b^{-1} \beta ).
\eqno{(11.26)}
$$
The second step is a fermionic transformation,
$$
(B',F) ~~~ \to ~~~(B',F').
\eqno{(11.27)}
$$
Its transformation matrix is denoted as $J_2$,
$$
J_2 =
\left (
\begin{array}{cc}
1  &
0  \\
\beta &
b
\end{array}
\right )  .
\eqno{(11.28)}
$$
Its Jacobian is the inverse of the determinant of the transformation
matrix,
$$
(det~ J_2)^{-1} = (det~b)^{-1}.
\eqno{(11.29)}
$$
Using this relation, eq.(11.26) is changed into
$$
\int \prod_i {\rm d}B_i \prod_k {\rm d}F_k = \int \prod_i {\rm
d}B'_i \prod_k {\rm d}F'_k \cdot det( a - \alpha b^{-1} \beta )
(det~b)^{-1} .
\eqno{(11.30)}
$$
For generalized BRST transformation, all non-diagonal matrix elements
are proportional to Grassman constant $\delta \lambda$. Non-diagonal
matrix $\alpha$ and $\beta$ contains only non-diagonal matrix elements,
so,
$$
\alpha b^{-1} \beta  \propto (\delta \lambda)^2 = 0.
\eqno{(11.31)}
$$
It means that
$$
\int \prod_i {\rm d}B_i \prod_k {\rm d}F_k = \int \prod_i {\rm
d}B'_i \prod_k {\rm d}F'_k \cdot det( a  ) \cdot (det~b)^{-1} .
\eqno{(11.32)}
$$
Generally speaking, $C_{\mu}^{\alpha}$ and
$\partial_{\nu} C_{\mu}^{\alpha}$ are independent degrees of freedom,
so are $\eta^{\alpha}$ and $\partial_{\nu} \eta^{\alpha}$. Using
eqs.(11.1-3), we obtain
$$
\begin{array}{rcl}
(det ~ a^{-1}) & = &
det \left\lbrack
( \delta^{\alpha}_{\beta} +
g (\partial_{\beta} \eta^{\alpha} )
\delta \lambda ) \delta^{\mu}_{\nu}
\right\rbrack  \\
&&  \\
& = &
\prod_{\mu,\alpha,x}
\left\lbrack
( \delta^{\alpha}_{\alpha} +
g (\partial_{\alpha} \eta^{\alpha} )
\delta \lambda ) \delta^{\mu}_{\nu}
\right\rbrack  \\
&&\\
&=&
\prod_{x}
( 1 + g (\partial_{\alpha} \eta^{\alpha} )
\delta \lambda).
\end{array}
\eqno{(11.33)}
$$
$$
\begin{array}{rcl}
(det ~ b^{-1}) & = &
det \left(
 \delta^{\alpha}_{\beta} +
g (\partial_{\beta} \eta^{\alpha} )
\delta \lambda \right)  \\
&&  \\
& = &
\prod_{x}
( 1 + g (\partial_{\alpha} \eta^{\alpha} )
\delta \lambda) .
\end{array}
\eqno{(11.34)}
$$
In the second line of eq.(11.33), there is no summation over the
repeated $\alpha$ index. Using these two relations, we have
$$
det( a  ) \cdot (det~b)^{-1} ~=~ \prod_x~{\rm\bf 1} ~=~1.
\eqno{(11.35)}
$$
Therefore, under generalized BRST transformation, functional
integration measure
$\lbrack {\cal D}C \rbrack \lbrack {\cal D}\eta \rbrack
\lbrack {\cal D}\bar\eta \rbrack $
is invariant,
$$
\lbrack {\cal D}C \rbrack \lbrack {\cal D}\eta \rbrack \lbrack
{\cal D}\bar\eta \rbrack  ~=~ \lbrack {\cal D}C' \rbrack \lbrack
{\cal D}\eta' \rbrack \lbrack {\cal D}\bar\eta' \rbrack .
\eqno{(11.36)}
$$
Though both $\lbrack {\cal D}C \rbrack$ and
$\lbrack {\cal D}\eta \rbrack$ are not invariant under generalized
BRST transformation, their product is invariant under generalized
BRST transformation. This result is interesting and important. \\

The generating functional
$W^{\lbrack 0 \rbrack} \lbrack J,\eta_1,\eta_2 \rbrack $ is
$$
W^{\lbrack 0 \rbrack} \lbrack J,\eta_1, \eta_2 \rbrack =  N \int
\lbrack {\cal D} C\rbrack \lbrack {\cal D} \eta \rbrack \lbrack
{\cal D}\bar{\eta} \rbrack exp \left\lbrace i \int {\rm d}^4 x
({\cal L}^{\lbrack 0 \rbrack} _{eff} + J^{\mu}_{\alpha}
C^{\alpha}_{\mu} ) \right\rbrace,
\eqno{(10.37)}
$$
where $\eta_1$ and $\eta_2$ are two tensors which are used to construct
lagrangian density of pure gravitational gauge field. Because
$$
\int {\rm d}\eta^{\beta} {\rm d}\bar\eta^{\sigma} \cdot
\bar\eta^{\alpha} \cdot f(\eta,\bar\eta ) = 0,
\eqno{(11.38)}
$$
where $f(\eta,\bar\eta )$ is a bilinear function of
$\eta$ and $\bar\eta$, we have
$$
\int \lbrack {\cal D} C\rbrack \lbrack {\cal D} \eta \rbrack
\lbrack {\cal D}\bar{\eta} \rbrack \cdot \bar\eta_{\alpha}(x)
\cdot exp \left\lbrace i \int {\rm d}^4 y ( {\cal L}^{\lbrack 0
\rbrack}_{eff}(y) + J^{\mu}_{\alpha}(y) C^{\alpha}_{\mu}(y) )
\right\rbrace = 0.
\eqno{(11.39)}
$$
If all fields are the fields after generalized BRST transformation,
eq.(11.39) still holds, i.e.
$$
\int \lbrack {\cal D} C' \rbrack
\lbrack {\cal D} \eta' \rbrack
\lbrack {\cal D}\bar{\eta}' \rbrack
\cdot \bar\eta^{\prime}_{\alpha}(x) \cdot
exp \left\lbrace i \int {\rm d}^4 y
( {\cal L}^{\prime \lbrack 0 \rbrack}_{eff}(y)
+ J^{\mu}_{\alpha}(y) C^{\prime \alpha}_{\mu}(y)
) \right\rbrace = 0,
\eqno{(11.40)}
$$
where ${\cal L}^{\prime \lbrack 0 \rbrack}_{eff} $ is the
effective Lagrangian after generalized BRST transformation.
Both functional integration measure and effective action
$\int {\rm d}^4 y {\cal L}^{\prime \lbrack 0 \rbrack}_{eff}(y)$
are generalized BRST invariant, so, using eqs.(11.1-3), we get
$$
\begin{array}{ll}
\int \lbrack {\cal D} C \rbrack
\lbrack {\cal D} \eta \rbrack
\lbrack {\cal D}\bar{\eta} \rbrack
& \left\lbrack
\frac{1}{\alpha} \eta_{\alpha\beta}
f^{\beta} (C(x)) \delta\lambda
- i \bar\eta_{\alpha} (x)
\int {\rm d}^4z ( J^{\mu}_{\beta}(z)
{\mathbf D}_{\mu \sigma}^{\beta}(z)
\eta^{\sigma}(z) \delta\lambda )
\right\rbrack  \\
&\\
&\cdot exp \left\lbrace i \int {\rm d}^4 y
( {\cal L}^{ \lbrack 0 \rbrack}_{eff}(y)
+ J^{\mu}_{\alpha}(y) C^{\alpha}_{\mu}(y)
) \right\rbrace = 0.
\end{array}
\eqno{(11.41)}
$$
This equation will lead to
$$
\frac{1}{\alpha} \eta_{\alpha\beta} f^{\beta}
\left(\frac{1}{i} \frac{\delta}{\delta J(x)} \right)
W^{\lbrack 0 \rbrack} \lbrack J,\eta_1,\eta_2 \rbrack
- \int {\rm d}^4y ~ J^{\mu}_{\beta}(y)
{\mathbf D}_{\mu \sigma}^{\beta}
\left( \frac{1}{i} \frac{\delta}{\delta J(x)} \right)
W^{\lbrack 0 \rbrack \sigma}_{~~~ \alpha}
\lbrack y,x, J,\eta_1,\eta_2 \rbrack =0,
\eqno{(11.42)}
$$
where
$$
W^{\lbrack 0 \rbrack\sigma}_{~~~ \alpha}
\lbrack y,x,J,\eta_1,\eta_2 \rbrack
=  N i  \int \lbrack {\cal D} C\rbrack
\lbrack {\cal D} \eta \rbrack
\lbrack {\cal D}\bar{\eta} \rbrack
\bar\eta_{\alpha}(x) \eta^{\sigma}(y)
exp \left\lbrace i \int {\rm d}^4 z
( {\cal L}^{\lbrack 0 \rbrack} _{eff}
+ J^{\mu}_{\alpha} C^{\alpha}_{\mu}
) \right\rbrace.
\eqno{(10.43)}
$$
This is the generalized Ward-Takahashi identity for
generating functional $W^{\lbrack 0 \rbrack} \lbrack J \rbrack$.\\

Now, let's introduce the external sources of ghost fields, then
the generation functional becomes
$$
W^{\lbrack 0 \rbrack}\lbrack J, \beta, \bar{\beta},\eta_1,\eta_2 \rbrack
=  N \int \lbrack {\cal D} C\rbrack
\lbrack {\cal D} \eta \rbrack
\lbrack {\cal D}\bar{\eta} \rbrack
exp \left\lbrace i \int {\rm d}^4 x
( {\cal L}_{eff}^{\lbrack 0 \rbrack}
+ J^{\mu}_{\alpha} C^{\alpha}_{\mu}
+ \bar{\eta}_{\alpha} \beta^{\alpha}
+ \bar{\beta}_{\alpha} \eta^{\alpha}
) \right\rbrace,
\eqno{(11.44)}
$$
In renormalization of the theory, we have to introduce external
sources $K^{\mu}_{\alpha}$ and $L_{\alpha}$
of the following composite operators,
$$
{\mathbf D}_{\mu \beta}^{\alpha} \eta^{\beta}
~~,~~g \eta^{\sigma} (\partial_{\sigma} \eta^{\alpha} ).
\eqno{(11.45)}
$$
Then the effective Lagrangian becomes
$$
\begin{array}{rcl}
{\swav{\cal L}}^{\lbrack 0 \rbrack} (C,\eta,\bar\eta,K,L,\eta_1,\eta_2)
&=& {\cal L} - \frac{1}{2 \alpha}
\eta_{\alpha \beta} f^{\alpha} f^{\beta}
+ \bar{\eta}_{\alpha} \partial^{\mu}
{\mathbf D}_{\mu~\sigma}^{\alpha} \eta^{\sigma}
+K^{\mu}_{\alpha} {\mathbf D}_{\mu \beta}^{\alpha} \eta^{\beta}
+ g L_{\alpha} \eta^{\sigma} (\partial_{\sigma} \eta^{\alpha})  \\
&&\\
&=& {\cal L}_{eff}^{\lbrack 0 \rbrack}
+K^{\mu}_{\alpha} {\mathbf D}_{\mu \beta}^{\alpha} \eta^{\beta}
+ g L_{\alpha} \eta^{\sigma} (\partial_{\sigma} \eta^{\alpha}) .
\end{array}
\eqno{(11.46)}
$$
Then,
$$
\swav{S}^{\lbrack 0 \rbrack} \lbrack C,\eta,\bar\eta,K,L,\eta_1,\eta_2 \rbrack
= \int {\rm d}^4x {\cal \swav{L}}^{\lbrack 0 \rbrack}
(C,\eta,\bar\eta,K,L,\eta_1,\eta_2).
\eqno{(11.47)}
$$
It is easy to deduce that
$$
\frac{\delta \swav{S}^{\lbrack 0 \rbrack}}
{\delta K^{\mu}_{\alpha} } =
{\mathbf D}_{\mu \beta}^{\alpha} \eta^{\beta},
\eqno{(11.48)}
$$
$$
\frac{\delta \swav{S}^{\lbrack 0 \rbrack}}
{\delta L_{\alpha} } =
g \eta^{\sigma} (\partial_{\sigma} \eta^{\alpha}).
\eqno{(11.49)}
$$
The generating functional now becomes,
$$
W^{\lbrack 0 \rbrack}\lbrack J, \beta, \bar{\beta},K,L,\eta_1,\eta_2 \rbrack
=  N \int \lbrack {\cal D} C\rbrack
\lbrack {\cal D} \eta \rbrack
\lbrack {\cal D}\bar{\eta} \rbrack
exp \left\lbrace i \int {\rm d}^4 x
( \swav{{\cal L}}^{\lbrack 0 \rbrack}
+ J^{\mu}_{\alpha} C^{\alpha}_{\mu}
+ \bar{\eta}_{\alpha} \beta^{\alpha}
+ \bar{\beta}_{\alpha} \eta^{\alpha}
) \right\rbrace.
\eqno{(11.50)}
$$
In previous discussion, we have already proved that
$S_{eff}^{\lbrack 0 \rbrack}$ is generalized BRST
invariant. External sources $K^{\mu}_{\alpha}$
and $L_{\alpha}$ keep unchanged under generalized BRST
transformation. Using nilpotent property of generalized
BRST transformation, it is easy to prove that the two new
terms
$K^{\mu}_{\alpha} {\mathbf D}_{\mu \beta}^{\alpha} \eta^{\beta}$
and
$g L_{\alpha} \eta^{\sigma} (\partial_{\sigma} \eta^{\alpha})$
in ${\swav{\cal L}}^{\lbrack 0 \rbrack}$ are also
generalized BRST invariant,
$$
\delta (K^{\mu}_{\alpha}
{\mathbf D}_{\mu \beta}^{\alpha} \eta^{\beta})
=0,
\eqno{(11.51)}
$$
$$
\delta( g L_{\alpha} \eta^{\sigma}
(\partial_{\sigma} \eta^{\alpha}) ) =0.
\eqno{(11.52)}
$$
Therefore, the action given by (11.47) are generalized
BRST invariant,
$$
\delta \swav{S}^{\lbrack 0 \rbrack} = 0.
\eqno{(11.53)}
$$
It gives out
$$
\begin{array}{l}
\int {\rm d}^4x \left\lbrace
 - ({\mathbf D}_{\mu \beta}^{\alpha} \eta^{\beta}(x))
\delta\lambda \frac{\delta}{\delta C_{\mu}^{\alpha}(x)}
+ g \eta^{\sigma}(x) (\partial_{\sigma} \eta^{\alpha}(x))
\delta\lambda  \frac{\delta}{\delta \eta^{\alpha}(x)} \right.  \\
\\
 \left. + \frac{1}{\alpha} \eta_{\alpha\beta}
 f^{\beta}(C(x)) \delta\lambda
\frac{\delta}{\delta \bar\eta_{\alpha}(x)}
-  L_{1 \alpha}^{\mu} \delta \lambda
\frac{\delta}{\delta \eta_{1 \alpha}^{\mu}}
-  L_{2 \alpha \beta} \delta \lambda
\frac{\delta}{\delta \eta_{2 \alpha \beta}}
\right\rbrace \swav{S}^{\lbrack 0 \rbrack} = 0,
\end{array}
\eqno{(11.54)}
$$
where
$$
L^{\mu}_{1 \alpha} = g \eta^{\mu}_{1 \sigma}
( \partial_{\alpha} \eta^{\sigma} ),
\eqno{(11.55)}
$$
$$
L_{2 \alpha\beta} = g \lbrack \eta_{2 \alpha\sigma}
( \partial_{\beta} \eta^{\sigma} ) + \eta_{2 \sigma\beta}
( \partial_{\beta} \eta^{\sigma} ) \rbrack.
\eqno{(11.56)}
$$
Using relations (11.48-49), we can get
$$
\begin{array}{l}
\int {\rm d}^4x \left\lbrace
\frac{\delta \swav{S}^{\lbrack 0 \rbrack}}
{\delta K^{\mu}_{\alpha}(x) }
\frac{\delta \swav{S}^{\lbrack 0 \rbrack}}
{\delta C_{\mu}^{\alpha}(x) }
+ \frac{\delta \swav{S}^{\lbrack 0 \rbrack}}
{\delta L_{\alpha}(x) }
\frac{\delta \swav{S}^{\lbrack 0 \rbrack}}
{\delta \eta^{\alpha}(x) } \right.\\
\\
\left. + \frac{1}{\alpha} \eta_{\alpha\beta}
f^{\beta}(C(x))
\frac{\delta \swav{S}^{\lbrack 0 \rbrack}}
{\delta \bar\eta_{\alpha}(x) }
+ L^{\mu}_{1\alpha}(x) \frac{\delta \swav{S}^{\lbrack 0 \rbrack}}
{\delta \eta^{\mu}_{1\alpha}(x) }
+ L_{2\alpha\beta}(x)  \frac{\delta \swav{S}^{\lbrack 0 \rbrack}}
{\delta \eta_{2\alpha\beta}(x) }
\right\rbrace  = 0.
\end{array}
\eqno{(11.57)}
$$
On the other hand, from (11.46-47), we can obtain that
$$
\frac{\delta \swav{S}^{\lbrack 0 \rbrack}}
{\delta \bar\eta_{\alpha}(x) }
=\partial^{\mu}
\left( {\mathbf D}_{\mu \beta}^{\alpha} \eta^{\beta}(x)\right).
\eqno{(11.58)}
$$
Combine (11.48) with (11.58), we get
$$
\frac{\delta \swav{S}^{\lbrack 0 \rbrack}}
{\delta \bar\eta_{\alpha}(x) }
= \partial^{\mu} \left( \frac{\delta \swav{S}^{\lbrack 0 \rbrack}}
{\delta K^{\mu}_{\alpha}(x) } \right).
\eqno{(11.59)}
$$
\\

In generation functional
$W^{\lbrack 0 \rbrack}\lbrack J, \beta, \bar{\beta},K,L,\eta_1,\eta_2 \rbrack$,
all fields are integrated, so, if we set all fields to the fields
after generalized BRST transformations, the final result should
not be changed, i.e.
$$
\begin{array}{l}
\swav{W}^{\lbrack 0 \rbrack}\lbrack J, \beta, \bar{\beta},K,L,\eta_1,\eta_2 \rbrack
=  N \int \lbrack {\cal D} C' \rbrack
\lbrack {\cal D} \eta' \rbrack
\lbrack {\cal D} \bar{\eta}' \rbrack  \\
\\
~~~~~\cdot exp \left\lbrace i \int {\rm d}^4 x
( {\cal L}^{\lbrack 0 \rbrack} (C',\eta',\bar\eta',K,L,\eta_1,\eta_2)
+ J^{\mu}_{\alpha} C^{\prime \alpha}_{\mu}
+ \bar{\eta}'_{\alpha} \beta^{\alpha}
+ \bar{\beta}_{\alpha} \eta^{\prime \alpha}
) \right\rbrace.
\end{array}
\eqno{(11.60)}
$$
Both action (11.47) and functional integration
measure $\lbrack {\cal D} C \rbrack
\lbrack {\cal D} \eta \rbrack
\lbrack {\cal D} \bar{\eta} \rbrack$
are generalized BRST invariant, so, the above relation
gives out
$$
\begin{array}{l}
\int \lbrack {\cal D} C \rbrack
\lbrack {\cal D} \eta \rbrack
\lbrack {\cal D} \bar{\eta} \rbrack
\left\lbrace
 i \int {\rm d}^4x \left(
J^{\mu}_{\alpha} \frac{\delta \swav{S}^{\lbrack 0 \rbrack}}
{\delta K^{\mu}_{\alpha}(x) }
- \bar\beta_{\alpha} \frac{\delta \swav{S}^{\lbrack 0 \rbrack}}
{\delta L_{\alpha}(x) }\right. \right. \\
\\
\left. \left.
+ \frac{1}{\alpha} \eta_{\alpha \sigma} f^{\alpha} \beta^{\sigma}
- L^{\mu}_{1\alpha}(x) \frac{\delta\swav{S}^{\lbrack 0 \rbrack}}
{\delta \eta^{\mu}_{1\alpha}(x) }
 - L_{2\alpha\beta}(x) \frac{\delta\swav{S}^{\lbrack 0 \rbrack}}
{\delta \eta_{2\alpha\beta}(x) }
\right)\right\rbrace\\
\\
\cdot exp \left\lbrace i \int {\rm d}^4 y
( \swav{{\cal L}}^{\lbrack 0 \rbrack} (C,\eta,\bar\eta,K,L,\eta_1,\eta_2)
+ J^{\mu}_{\alpha} C^{\alpha}_{\mu}
+ \bar{\eta}_{\alpha} \beta^{\alpha}
+ \bar{\beta}_{\alpha} \eta^{\alpha}
) \right\rbrace = 0.
\end{array}
\eqno{(11.61)}
$$
In order to obtain the above relation, the following relation is used,
$$
\begin{array}{l}
\int {\rm d}^4x
\swav{\cal L}^{\lbrack 0 \rbrack} (C',\eta',\bar\eta',K,L,\eta_1,\eta_2)\\
\\
=\int {\rm d}^4x \left\lbrack
\swav{\cal L}^{\lbrack 0 \rbrack} (C,\eta,\bar\eta,K,L,\eta_1,\eta_2)
- \delta \eta^{\mu}_{1 \alpha} \frac{\delta\swav{S}^{\lbrack 0 \rbrack}}
{\delta \eta^{\mu}_{1\alpha}(x) }
- \delta \eta_{2 \alpha\beta} \frac{\delta\swav{S}^{\lbrack 0 \rbrack}}
{\delta \eta_{2\alpha\beta}(x) }.
\right\rbrack
\end{array}
\eqno{(11.62)}
$$
On the other hand, because the ghost field $\bar\eta_{\alpha}$
was integrated in functional\\
$W^{\lbrack 0 \rbrack}\lbrack J, \beta, \bar{\beta},K,L,\eta_1,\eta_2 \rbrack$,
if we use $\bar\eta'_{\alpha}$ in the in functional integration,
it will not change the generating functional. That is
$$
\begin{array}{l}
\swav{W}^{\lbrack 0 \rbrack}\lbrack J, \beta, \bar{\beta},K,L,\eta_1,\eta_2 \rbrack
=  N \int \lbrack {\cal D} C \rbrack
\lbrack {\cal D} \eta \rbrack
\lbrack {\cal D} \bar{\eta}' \rbrack  \\
\\
~~~~~\cdot exp \left\lbrace i \int {\rm d}^4 x
( \swav{\cal L}^{\lbrack 0 \rbrack} (C,\eta,\bar\eta',K,L,\eta_1,\eta_2)
+ J^{\mu}_{\alpha} C^{\alpha}_{\mu}
+ \bar{\eta}'_{\alpha} \beta^{\alpha}
+ \bar{\beta}_{\alpha} \eta^{\alpha}
) \right\rbrace.
\end{array}
\eqno{(11.63)}
$$
Suppose that
$$
\bar{\eta}'_{\alpha} = \bar{\eta}_{\alpha}
+ \delta \bar{\eta}_{\alpha}.
\eqno{(11.64)}
$$
Then (11.63) and (11.50) will gives out
$$
\begin{array}{l}
\int \lbrack {\cal D} C \rbrack
\lbrack {\cal D} \eta \rbrack
\lbrack {\cal D} \bar{\eta} \rbrack
\left\lbrace
\int {\rm d}^4x \delta \bar{\eta}_{\alpha}
( \frac{\delta \swav{S}^{\lbrack 0 \rbrack}}
{\delta \bar\eta_{\alpha}(x) } + \beta^{\alpha} (x)
)\right\rbrace \\
\\
\cdot exp \left\lbrace i \int {\rm d}^4 y
( \swav{\cal L}^{\lbrack 0 \rbrack} (C,\eta,\bar\eta,K,L,\eta_1,\eta_2)
+ J^{\mu}_{\alpha} C^{\alpha}_{\mu}
+ \bar{\eta}_{\alpha} \beta^{\alpha}
+ \bar{\beta}_{\alpha} \eta^{\alpha}
) \right\rbrace = 0.
\end{array}
\eqno{(11.65)}
$$
Because $\delta \bar{\eta}_{\alpha}$ is an arbitrary
variation, from (11.65), we will get
$$
\begin{array}{l}
\int \lbrack {\cal D} C \rbrack
\lbrack {\cal D} \eta \rbrack
\lbrack {\cal D} \bar{\eta} \rbrack
\left( \frac{\delta \swav{S}^{\lbrack 0 \rbrack}}
{\delta \bar\eta_{\alpha}(x) } + \beta^{\alpha} (x)
\right)  \\
\\
\cdot exp \left\lbrace i \int {\rm d}^4 y
\left( \swav{\cal L}^{\lbrack 0 \rbrack}
(C,\eta,\bar\eta,K,L,\eta_1,\eta_2)
+ J^{\mu}_{\alpha} C^{\alpha}_{\mu}
+ \bar{\eta}_{\alpha} \beta^{\alpha}
+ \bar{\beta}_{\alpha} \eta^{\alpha}
\right) \right\rbrace = 0.
\end{array}
\eqno{(11.66)}
$$
\\

The generating functional of connected Green function is given by
$$
\swav{Z}^{\lbrack 0 \rbrack}\lbrack J,
\beta, \bar{\beta},K,L,\eta_1,\eta_2 \rbrack
= - i~ {\rm ln}~
\swav{W}^{\lbrack 0 \rbrack}\lbrack J,
\beta, \bar{\beta},K,L,\eta_1,\eta_2 \rbrack.
\eqno{(11.67)}
$$
After Legendre transformation, we will get the generating
functional of irreducible vertex
$\swav{\Gamma}^{\lbrack 0 \rbrack}\lbrack C ,
\bar\eta,\eta,K,L,\eta_1,\eta_2 \rbrack$,
$$
\begin{array}{rcl}
\swav{\Gamma}^{\lbrack 0 \rbrack}\lbrack C,
\bar\eta,\eta,K,L,\eta_1,\eta_2 \rbrack
&=& \swav{Z}^{\lbrack 0 \rbrack}\lbrack J,
\beta, \bar{\beta},K,L,\eta_1,\eta_2 \rbrack  \\
&&\\
&& - \int {\rm d}^4 x
\left( J^{\mu}_{\alpha} C^{\alpha}_{\mu}
+ \bar{\eta}_{\alpha} \beta^{\alpha}
+ \bar{\beta}_{\alpha} \eta^{\alpha}
\right) .
\end{array}
\eqno{(11.68)}
$$
Functional derivative of the generating functional
$\swav{Z}^{\lbrack 0 \rbrack}$ gives out the classical fields
$C_{\mu}^{\alpha}$, $\eta^{\alpha}$ and $\bar\eta_{\alpha}$,
$$
C_{\mu}^{\alpha} =
\frac{\delta \swav{Z}^{\lbrack 0 \rbrack}}
{\delta J^{\mu}_{\alpha}},
\eqno{(11.69)}
$$
$$
\eta^{\alpha} =
\frac{\delta \swav{Z}^{\lbrack 0 \rbrack}}
{\delta \bar\beta_{\alpha}},
\eqno{(11.70)}
$$
$$
\bar\eta_{\alpha} =
- \frac{\delta \swav{Z}^{\lbrack 0 \rbrack}}
{\delta \beta^{\alpha}}.
\eqno{(11.71)}
$$
Then, functional derivative of the generating functional
$\swav{\Gamma}^{\lbrack 0 \rbrack}$ gives out external
sources $J^{\mu}_{\alpha}$, $\bar\beta_{\alpha}$ and
$\beta^{\alpha}$,
$$
\frac{\delta \swav{\Gamma}^{\lbrack 0 \rbrack}}
{\delta C_{\mu}^{\alpha}}
= - J^{\mu}_{\alpha},
\eqno{(11.72)}
$$
$$
\frac{\delta \swav{\Gamma}^{\lbrack 0 \rbrack}}
{\delta \eta^{\alpha}}
=  \bar\beta_{\alpha},
\eqno{(11.73)}
$$
$$
\frac{\delta \swav{\Gamma}^{\lbrack 0 \rbrack}}
{\delta \bar\eta_{\alpha}}
=  - \beta^{\alpha}.
\eqno{(11.74)}
$$
Besides, there are two other relations which can be strictly
deduced from (11.68),
$$
\frac{\delta\swav{\Gamma}^{\lbrack 0 \rbrack}}
{\delta K^{\mu}_{\alpha}}
= \frac{\delta\swav{Z}^{\lbrack 0 \rbrack}}
{\delta K^{\mu}_{\alpha}}~~,~~
\frac{\delta\swav{\Gamma}^{\lbrack 0 \rbrack}}
{\delta L_{\alpha}}
= \frac{\delta \swav{Z}^{\lbrack 0 \rbrack}}
{\delta L_{\alpha}}.
\eqno{(11.75)}
$$
$$
\frac{\delta\swav{\Gamma}^{\lbrack 0 \rbrack}}
{\delta \eta^{\mu}_{1\alpha}}
= \frac{\delta \swav{Z}^{\lbrack 0 \rbrack}}
{\delta \eta^{\mu}_{1\alpha}} ~~,~~
\frac{\delta\swav{\Gamma}^{\lbrack 0 \rbrack}}
{\delta \eta_{2\alpha\beta}}
= \frac{\delta \swav{Z}^{\lbrack 0 \rbrack}}
{\delta \eta_{2\alpha\beta}}.
\eqno{(11.76)}
$$
\\

It is easy to prove that
$$
\begin{array}{rcl}
&& i \frac{\delta \swav{S}^{\lbrack 0 \rbrack}}
{\delta K^{\mu}_{\alpha}(x) }
exp \left\lbrace i \int {\rm d}^4 y
( \swav{\cal L}^{\lbrack 0 \rbrack} (C,\eta,\bar\eta,K,L,\eta_1,\eta_2)
+ J^{\mu}_{\alpha} C^{\alpha}_{\mu}
+ \bar{\eta}_{\alpha} \beta^{\alpha}
+ \bar{\beta}_{\alpha} \eta^{\alpha}
) \right\rbrace  \\
&&\\
&=& \frac{\delta} {\delta K^{\mu}_{\alpha}(x) }
exp \left\lbrace i \int {\rm d}^4 y
( \swav{\cal L}^{\lbrack 0 \rbrack} (C,\eta,\bar\eta,K,L,\eta_1,\eta_2)
+ J^{\mu}_{\alpha} C^{\alpha}_{\mu}
+ \bar{\eta}_{\alpha} \beta^{\alpha}
+ \bar{\beta}_{\alpha} \eta^{\alpha}
) \right\rbrace,
\end{array}
\eqno{(11.77)}
$$
$$
\begin{array}{rcl}
&& i \frac{\delta \swav{S}^{\lbrack 0 \rbrack}}
{\delta L_{\alpha}(x) }
exp \left\lbrace i \int {\rm d}^4 y
( \swav{\cal L}^{\lbrack 0 \rbrack} (C,\eta,\bar\eta,K,L,\eta_1,\eta_2)
+ J^{\mu}_{\alpha} C^{\alpha}_{\mu}
+ \bar{\eta}_{\alpha} \beta^{\alpha}
+ \bar{\beta}_{\alpha} \eta^{\alpha}
) \right\rbrace  \\
&&\\
&=& \frac{\delta} {\delta L_{\alpha}(x) }
exp \left\lbrace i \int {\rm d}^4 y
( \swav{\cal L}^{\lbrack 0 \rbrack} (C,\eta,\bar\eta,K,L,\eta_1,\eta_2)
+ J^{\mu}_{\alpha} C^{\alpha}_{\mu}
+ \bar{\eta}_{\alpha} \beta^{\alpha}
+ \bar{\beta}_{\alpha} \eta^{\alpha}
) \right\rbrace.
\end{array}
\eqno{(11.78)}
$$
Use these two relations, we can change (11.61) into
$$
\begin{array}{l}
\int \lbrack {\cal D} C \rbrack
\lbrack {\cal D} \eta \rbrack
\lbrack {\cal D} \bar{\eta} \rbrack
\left\lbrace
 \int {\rm d}^4x \left(
J^{\mu}_{\alpha}(x) \frac{\delta } {\delta K^{\mu}_{\alpha}(x) }
- \bar\beta_{\alpha}(x) \frac{\delta }{\delta L_{\alpha}(x) }
\right. \right.\\
\\
\left. \left.
+ \frac{i}{\alpha} \eta_{\alpha \sigma}
f^{\alpha}(\frac{1}{i}\frac{\delta }{\delta J^{\mu}_{\gamma}(x) } )
\beta^{\sigma}(x)
- L_{1 \alpha}^{\mu} \left( \frac{1}{i}
\frac{\delta}{\delta \bar\beta}    \right)
\frac{\delta}{\delta \eta^{\mu}_{1 \alpha}}
- L_{2 \alpha\beta} \left( \frac{1}{i}
\frac{\delta}{\delta \bar\beta}    \right)
\frac{\delta}{\delta \eta_{2 \alpha\beta}}
\right)\right\rbrace \\
\\
\cdot exp \left\lbrace i \int {\rm d}^4 y
( \swav{\cal L}^{\lbrack 0 \rbrack} (C,\eta,\bar\eta,K,L,\eta_1,\eta_2)
+ J^{\mu}_{\alpha} C^{\alpha}_{\mu}
+ \bar{\eta}_{\alpha} \beta^{\alpha}
+ \bar{\beta}_{\alpha} \eta^{\alpha}
) \right\rbrace = 0.
\end{array}
\eqno{(11.79)}
$$
Using the definition of generating functional eq.(11.63), we can obtain
$$
\begin{array}{l}
\int {\rm d}^4x \left\lbrack
J^{\mu}_{\alpha}(x) \frac{\delta \swav{W}^{\lbrack 0 \rbrack}}
{\delta K^{\mu}_{\alpha}(x) }
- \bar\beta_{\alpha}(x) \frac{\delta \swav{W}^{\lbrack 0 \rbrack}}
{\delta L_{\alpha}(x) }
+ \frac{i}{\alpha} \eta_{\alpha \sigma}
f^{\alpha}(\frac{1}{i}\frac{\delta }{\delta J(x) } )
\beta^{\sigma}(x)\swav{W}^{\lbrack 0 \rbrack}\right.  \\
\\
\left.
- L_{1 \alpha}^{\mu} \left( \frac{1}{i}
\frac{\delta}{\delta \bar\beta(x)}    \right)
\frac{\delta \swav{W}^{\lbrack 0 \rbrack}}{\delta \eta^{\mu}_{1 \alpha}(x)}
- L_{2 \alpha\beta} \left( \frac{1}{i}
\frac{\delta }{\delta \bar\beta(x)}    \right)
\frac{\delta \swav{W}^{\lbrack 0 \rbrack}}{\delta \eta_{2 \alpha\beta}(x)}
\right\rbrack =0.
\end{array}
\eqno{(11.80)}
$$
Using the following three relations,
$$
f^{\beta}\left( \frac{1}{i}
\frac{\delta}{\delta J(x)}    \right)
\swav{W}^{\lbrack 0 \rbrack} =
f^{\beta} (C(x))
\swav{W}^{\lbrack 0 \rbrack},
\eqno{(11.81)}
$$
$$
L^{\mu}_{1 \alpha}\left( \frac{1}{i}
\frac{\delta}{\delta \bar\beta(x)}    \right)
\frac{\delta \swav{W}^{\lbrack 0 \rbrack}}{\delta \eta^{\mu}_{1 \alpha}(x)}
= L^{\mu}_{1 \alpha} (\eta(x))
\frac{\delta \swav{W}^{\lbrack 0 \rbrack}}{\delta \eta^{\mu}_{1 \alpha}(x)},
\eqno{(11.82)}
$$
$$
L_{2 \alpha\beta}\left( \frac{1}{i}
\frac{\delta}{\delta \bar\beta(x)}    \right)
\frac{\delta \swav{W}^{\lbrack 0 \rbrack}}{\delta \eta_{2 \alpha\beta}(x)}
= L_{2 \alpha\beta} (\eta(x))
\frac{\delta \swav{W}^{\lbrack 0 \rbrack}}{\delta \eta_{2 \alpha\beta}(x)},
\eqno{(11.83)}
$$
eq.(11.80) can be changed into,
$$
\begin{array}{l}
\int {\rm d}^4x \left\lbrack
J^{\mu}_{\alpha}(x) \frac{\delta \swav{W}^{\lbrack 0 \rbrack}}
{\delta K^{\mu}_{\alpha}(x) }
- \bar\beta_{\alpha}(x) \frac{\delta \swav{W}^{\lbrack 0 \rbrack}}
{\delta L_{\alpha}(x) }
+ \frac{i}{\alpha} \eta_{\alpha \sigma}
f^{\alpha} \beta^{\sigma}(x)\swav{W}^{\lbrack 0 \rbrack}\right.  \\
\\
\left.
- L_{1 \alpha}^{\mu}
\frac{\delta \swav{W}^{\lbrack 0 \rbrack}}{\delta \eta^{\mu}_{1 \alpha}(x)}
- L_{2 \alpha\beta}
\frac{\delta \swav{W}^{\lbrack 0 \rbrack}}{\delta \eta_{2 \alpha\beta}(x)}
\right\rbrack =0.
\end{array}
\eqno{(11.84)}
$$
Using relations (11.72-74), we can rewrite this equation into
$$
\begin{array}{l}
\int {\rm d}^4x \left\lbrace
\frac{\delta \swav{W}^{\lbrack 0 \rbrack} }
{\delta K^{\mu}_{\alpha}(x) }
\frac{\delta \swav{\Gamma}^{\lbrack 0 \rbrack} }
{\delta C_{\mu}^{\alpha}(x) }
+ \frac{\delta \swav{W}^{\lbrack 0 \rbrack} }
{\delta L_{\alpha}(x) }
\frac{\delta \swav{\Gamma}^{\lbrack 0 \rbrack} }
{\delta \eta^{\alpha}(x) }
+ \frac{i}{\alpha} \eta_{\alpha \sigma}
f^{\alpha} \swav{W}^{\lbrack 0 \rbrack}
\frac{\delta \swav{\Gamma}^{\lbrack 0 \rbrack} }
{\delta \bar\eta_{\sigma}(x) } \right. \\
\\
\left.
+ L_{1 \alpha}^{\mu}
\frac{\delta \swav{W}^{\lbrack 0 \rbrack}}{\delta \eta^{\mu}_{1 \alpha}(x)}
+ L_{2 \alpha\beta}
\frac{\delta \swav{W}^{\lbrack 0 \rbrack}}{\delta \eta_{2 \alpha\beta}(x)}
\right\rbrace =0.
\end{array}
\eqno{(11.85)}
$$
Using (11.67), we can obtain that
$$
\frac{\delta \swav{W}^{\lbrack 0 \rbrack} }
{\delta K^{\mu}_{\alpha}(x) }
= i \frac{\delta \swav{\Gamma}^{\lbrack 0 \rbrack} }
{\delta K^{\mu}_{\alpha}(x) } \cdot
\swav{W}^{\lbrack 0 \rbrack},
\eqno{(11.86)}
$$
$$
\frac{\delta \swav{W}^{\lbrack 0 \rbrack} }
{\delta L_{\alpha}(x) }
= i \frac{\delta \swav{\Gamma}^{\lbrack 0 \rbrack} }
{\delta L_{\alpha}(x) } \cdot
\swav{W}^{\lbrack 0 \rbrack},
\eqno{(11.87)}
$$
$$
\frac{\delta \swav{W}^{\lbrack 0 \rbrack} }
{\delta \eta^{\mu}_{1 \alpha}(x) }
= i \frac{\delta \swav{\Gamma}^{\lbrack 0 \rbrack} }
{\delta \eta^{\mu}_{1 \alpha}(x) } \cdot
\swav{W}^{\lbrack 0 \rbrack},
\eqno{(11.88)}
$$
$$
\frac{\delta \swav{W}^{\lbrack 0 \rbrack} }
{\delta \eta_{2 \alpha\beta}(x) }
= i \frac{\delta \swav{\Gamma}^{\lbrack 0 \rbrack} }
{\delta \eta_{2 \alpha\beta}(x) } \cdot
\swav{W}^{\lbrack 0 \rbrack}.
\eqno{(11.89)}
$$
Then (11.76) is changed into
$$
\begin{array}{l}
\int {\rm d}^4x \left\lbrace
\frac{\delta \swav{\Gamma}^{\lbrack 0 \rbrack} }
{\delta K^{\mu}_{\alpha}(x) }
\frac{\delta \swav{\Gamma}^{\lbrack 0 \rbrack} }
{\delta C_{\mu}^{\alpha}(x) }
+ \frac{\delta \swav{\Gamma}^{\lbrack 0 \rbrack} }
{\delta L_{\alpha}(x) }
\frac{\delta \swav{\Gamma}^{\lbrack 0 \rbrack} }
{\delta \eta^{\alpha}(x) }
+ \frac{1}{\alpha} \eta_{\alpha \sigma}
f^{\alpha}
\frac{\delta \swav{\Gamma}^{\lbrack 0 \rbrack} }
{\delta \bar\eta_{\sigma}(x) } \right.
\\
\left.
+ L_{1 \alpha}^{\mu}(x)
\frac{\delta \swav{\Gamma}^{\lbrack 0 \rbrack}}
{\delta \eta^{\mu}_{1 \alpha}(x)}
+ L_{2 \alpha\beta}(x)
\frac{\delta \swav{\Gamma}^{\lbrack 0 \rbrack}}
{\delta \eta_{2 \alpha\beta}(x)}
\right\rbrace =0.
\end{array}
\eqno{(11.90)}
$$
Using (11.59) and (11.77), (11.65) becomes
$$
\begin{array}{l}
\int \lbrack {\cal D} C \rbrack
\lbrack {\cal D} \eta \rbrack
\lbrack {\cal D} \bar{\eta} \rbrack
\left\lbrack  - i \partial^{\mu}
\frac{\delta } {\delta K^{\mu}_{\alpha}(x) }
+ \beta^{\alpha} (x)   \right\rbrack  \\
\\
\cdot exp \left\lbrace i \int {\rm d}^4 y
( {\cal L}^{\lbrack 0 \rbrack} (C,\eta,\bar\eta,K,L,\eta_1,\eta_2)
+ J^{\mu}_{\alpha} C^{\alpha}_{\mu}
+ \bar{\eta}_{\alpha} \beta^{\alpha}
+ \bar{\beta}_{\alpha} \eta^{\alpha}
) \right\rbrace = 0.
\end{array}
\eqno{(11.91)}
$$
In above equation, the factor $- i \partial^{\mu}
\frac{\delta } {\delta K^{\mu}_{\alpha}(x) }
+ \beta^{\alpha} (x)$
can move out of functional integration, then (11.91)
gives out
$$
\partial^{\mu}
\frac{\delta \swav{\Gamma}^{\lbrack 0 \rbrack}  }
{\delta K^{\mu}_{\alpha}(x) }
= \frac{\delta \swav{\Gamma}^{\lbrack 0 \rbrack}  }
{\delta \bar\eta_{\alpha}(x) }.
\eqno{(11.92)}
$$
In order to obtain this relation, (11.63), (11.86) and
(11.74) are used. \\

Define
$$
\bar{\Gamma}^{\lbrack 0 \rbrack}
\lbrack C, \bar\eta, \eta, K, L ,\eta_1,\eta_2 \rbrack
= \swav{\Gamma}^{\lbrack 0 \rbrack}
\lbrack C, \bar\eta, \eta, K, L,\eta_1,\eta_2 \rbrack
+ \frac{1}{2 \alpha}
\int {\rm d}^4x \eta_{\alpha \beta}
f^{\alpha} f^{\beta}.
\eqno{(11.93)}
$$
It is easy to prove that
$$
\frac{\delta \bar{\Gamma}^{\lbrack 0 \rbrack}  }
{\delta K^{\mu}_{\alpha}(x) }
= \frac{\delta \swav{\Gamma}^{\lbrack 0 \rbrack}  }
{\delta K^{\mu}_{\alpha}(x) },
\eqno{(11.94)}
$$
$$
\frac{\delta \bar{\Gamma}^{\lbrack 0 \rbrack}  }
{\delta L_{\alpha}(x) }
= \frac{\delta \swav{\Gamma}^{\lbrack 0 \rbrack}  }
{\delta L_{\alpha}(x) },
\eqno{(11.95)}
$$
$$
\frac{\delta \bar{\Gamma}^{\lbrack 0 \rbrack}  }
{\delta \eta^{\alpha}(x) }
= \frac{\delta \swav{\Gamma}^{\lbrack 0 \rbrack}  }
{\delta \eta^{\alpha}(x) }    ,
\eqno{(11.96)}
$$
$$
\frac{\delta \bar{\Gamma}^{\lbrack 0 \rbrack}  }
{\delta \bar\eta_{\alpha}(x) }
= \frac{\delta \swav{\Gamma}^{\lbrack 0 \rbrack}  }
{\delta \bar\eta_{\alpha}(x) }    ,
\eqno{(11.97)}
$$
$$
\frac{\delta \bar{\Gamma}^{\lbrack 0 \rbrack}  }
{\delta C_{\mu}^{\alpha}(x) }
= \frac{\delta \swav{\Gamma}^{\lbrack 0 \rbrack}  }
{\delta C_{\mu}^{\alpha}(x) }
- \frac{1}{\alpha} \eta_{\alpha \beta}
\partial^{\mu}  f^{\beta},
\eqno{(11.98)}
$$
$$
\frac{\delta \bar{\Gamma}^{\lbrack 0 \rbrack}  }
{\delta \eta^{\mu}_{1\alpha}(x) }
= \frac{\delta \swav{\Gamma}^{\lbrack 0 \rbrack}  }
{\delta \eta^{\mu}_{1\alpha}(x) }    ,
\eqno{(11.99)}
$$
$$
\frac{\delta \bar{\Gamma}^{\lbrack 0 \rbrack}  }
{\delta \eta_{2\alpha\beta}(x) }
= \frac{\delta \swav{\Gamma}^{\lbrack 0 \rbrack}  }
{\delta \eta_{2\alpha\beta}(x) } .
\eqno{(11.100)}
$$
Using these relations, (11.92) and (11.90) are changed into
$$
\partial^{\mu}
\frac{\delta \bar{\Gamma}^{\lbrack 0 \rbrack}  }
{\delta K^{\mu}_{\alpha}(x) }
= \frac{\delta \bar{\Gamma}^{\lbrack 0 \rbrack}  }
{\delta \bar\eta_{\alpha}(x) },
\eqno{(11.101)}
$$
$$
\int {\rm d}^4x \left\lbrace
\frac{\delta \bar{\Gamma}^{\lbrack 0 \rbrack} }
{\delta K^{\mu}_{\alpha}(x) }
\frac{\delta \bar{\Gamma}^{\lbrack 0 \rbrack} }
{\delta C_{\mu}^{\alpha}(x) }
+ \frac{\delta \bar{\Gamma}^{\lbrack 0 \rbrack} }
{\delta L_{\alpha}(x) }
\frac{\delta \bar{\Gamma}^{\lbrack 0 \rbrack} }
{\delta \eta^{\alpha}(x) }
+ L_{1 \alpha}^{\mu}
\frac{\delta \bar{\Gamma}^{\lbrack 0 \rbrack}}{\delta \eta^{\mu}_{1 \alpha}(x)}
+ L_{2 \alpha\beta}
\frac{\delta \bar{\Gamma}^{\lbrack 0 \rbrack}}{\delta \eta_{2 \alpha\beta}(x)}
\right\rbrace = 0.
\eqno{(11.102)}
$$
Eqs.(11.101-102) are generalized Ward-Takahashi identities
of generating functional of regular vertex. It is the foundations
of the renormalization of the gravitational gauge theory.\\

Generating functional $\swav{\Gamma}^{\lbrack 0 \rbrack}$
is the generating functional of regular vertex with external
sources, which is constructed from the Lagrangian
$\swav{\cal L}_{eff}^{\lbrack 0 \rbrack}$. It is a functional
of physical field, therefore, we can make a functional
expansion
$$
\begin{array}{rcl}
\swav{\Gamma}^{\lbrack 0 \rbrack} &=&
\sum_n \int \frac{\delta^n \swav{\Gamma}^{\lbrack 0 \rbrack}}
{\delta C^{\alpha_1}_{\mu_1}(x_1) \cdots
\delta C^{\alpha_n}_{\mu_n}(x_n) } |_{C=\eta=\bar\eta=0}
C^{\alpha_1}_{\mu_1}(x_1) \cdots C^{\alpha_n}_{\mu_n}(x_n)
{\rm d}^4x_1 \cdots {\rm d}^4x_n  \\
&&\\
&& + \sum_n \int \frac{\delta^2}
{\delta \bar\eta_{\beta_1}(y_1) \delta\eta^{\beta_2}(y_2) }
\frac{\delta^n \swav{\Gamma}^{\lbrack 0 \rbrack}}
{\delta C^{\alpha_1}_{\mu_1}(x_1) \cdots
\delta C^{\alpha_n}_{\mu_n}(x_n) } |_{C=\eta=\bar\eta=0}  \\
&&\\
&&~~~ \cdot \bar\eta_{\beta_1}(y_1) \eta^{\beta_2}(y_2)
C^{\alpha_1}_{\mu_1}(x_1) \cdots C^{\alpha_n}_{\mu_n}(x_n)
{\rm d}^4y_1 {\rm d}^4y_2 {\rm d}^4x_1 \cdots {\rm d}^4x_n\\
&&\\
&& + \cdots.
\end{array}
\eqno{(11.103)}
$$
In this functional expansion, the expansion coefficients are
regular vertexes with external sources. Before renormalization,
these coefficients contain divergences. If we calculate these
divergences in the methods of dimensional regularization,
the form of these divergence will not violate gauge symmetry
of the theory\cite{t01,t02}.
In other words, in the method of dimensional
regularization, gravitational gauge symmetry is not
violated and the generating functional of regular vertex
satisfies Ward-Takahashi identities (11.88-89). In order to eliminate
the ultraviolet divergences of the theory, we need to introduce
counterterms into Lagrangian. All these counterterms are
formally denoted by $\delta {\cal L}$. Then the renormalized
Lagrangian is
$$
\swav{\cal L}_{eff} =
\swav{\cal L}_{eff}^{\lbrack 0 \rbrack}
+ \delta {\cal L}.
\eqno{(11.104)}
$$
Because $\delta {\cal L}$ contains all counterterms,
$\swav{\cal L}_{eff}$ is the Lagrangian density after complete
renormalization. The generating functional of regular vertex
which is calculated from $\swav{\cal L}_{eff}$ is denoted
by $\swav{\Gamma}$. The regular vertexes calculated from
this generating functional $\swav{\Gamma}$ contain no
ultraviolet divergence anymore. Then let external sources
$K^{\mu}_{\alpha}$ and $L_{\alpha}$ vanish, we will get
generating functional $\Gamma$ of regular vertex without external
sources,
$$
\Gamma = \swav{\Gamma} |_{K=L=0} .
\eqno{(11.105)}
$$
The regular vertexes which are generated from $\Gamma$ will
contain no ultraviolet divergence either. Therefore, the
S-matrix for all physical process are finite. For a renormalizable
theory, the counterterm $\delta {\cal L}$ only contain finite
unknown parameters which are needed to be determined by
experiments. If conterterm $\delta {\cal L}$ contains infinite
unknown parameters, the theory will lost its predictive power
and it is conventionally regarded as a non-renormalizable
theory. Now, the main task for us is to prove that
the conterterm $\delta {\cal L}$ for the gravitational gauge
theory only contains a few unknown parameters. If we do this,
we will have proved that the gravitational gauge theory
is renormalizable. \\

Now, we use inductive method to prove the renormalizability
of the gravitational gauge theory.
In the previous discussion, we have proved that the generating
functional of regular vertex before renormalization satisfies
Ward-Takahashi identities (11.88-89).
The effective Lagrangian density that contains all counterterms
which cancel all divergences of $l$-loops ($0 \leq l \leq L $)
is denoted by $\swav{\cal  L}^{\lbrack L \rbrack}$.
$\swav{\Gamma}^{\lbrack L \rbrack}$ is the generating functional
of regular vertex which is calculated from
$\swav{\cal  L}^{\lbrack L \rbrack}$. The regular vertex which
is generated by $\swav{\Gamma}^{\lbrack L \rbrack}$ will contain
no divergence if the number of the loops of the diagram is not
greater than $L$. We have proved that the generating functional
$\swav{\Gamma}^{\lbrack L \rbrack}$ satisfies Ward-Takahashi
identities if $L=0$. Hypothesize that Ward-Takahashi identities
are also satisfied when $L=N$, that is
$$
\partial^{\mu}
\frac{\delta \bar{\Gamma}^{\lbrack N \rbrack}  }
{\delta K^{\mu}_{\alpha}(x) }
= \frac{\delta \bar{\Gamma}^{\lbrack N \rbrack}  }
{\delta \bar\eta_{\alpha}(x) },
\eqno{(11.106)}
$$
$$
\int {\rm d}^4x \left\lbrace
\frac{\delta \bar{\Gamma}^{\lbrack N \rbrack} }
{\delta K^{\mu}_{\alpha}(x) }
\frac{\delta \bar{\Gamma}^{\lbrack N \rbrack} }
{\delta C_{\mu}^{\alpha}(x) }
+ \frac{\delta \bar{\Gamma}^{\lbrack N \rbrack} }
{\delta L_{\alpha}(x) }
\frac{\delta \bar{\Gamma}^{\lbrack N \rbrack} }
{\delta \eta^{\alpha}(x) }
+ L_{1 \alpha}^{\mu}
\frac{\delta \bar{\Gamma}^{\lbrack N \rbrack}}{\delta \eta^{\mu}_{1 \alpha}(x)}
+ L_{2 \alpha\beta}
\frac{\delta \bar{\Gamma}^{\lbrack N \rbrack}}{\delta \eta_{2 \alpha\beta}(x)}
\right\rbrace = 0.
\eqno{(11.107)}
$$
Our goal is to prove that  Ward-Takahashi identities
are also satisfied when $L=N+1$.\\

Now, let's introduce a special product which is defined by
$$
A * B \equiv \int {\rm d}^4x \left\lbrace
\frac{\delta A }{\delta K^{\mu}_{\alpha}(x) }
\frac{\delta B }{\delta C_{\mu}^{\alpha}(x) }
+ \frac{\delta A }{\delta L_{\alpha}(x) }
\frac{\delta B }{\delta \eta^{\alpha}(x) }
\right\rbrace .
\eqno{(11.108)}
$$
Then (11.107) will be simplified to
$$
\bar{\Gamma}^{\lbrack N \rbrack} *
\bar{\Gamma}^{\lbrack N \rbrack} +
\int {\rm d}^4x \left(
 L_{1 \alpha}^{\mu}
\frac{\delta \bar{\Gamma}^{\lbrack N \rbrack}}{\delta \eta^{\mu}_{1 \alpha}(x)}
+ L_{2 \alpha\beta}
\frac{\delta \bar{\Gamma}^{\lbrack N \rbrack}}{\delta \eta_{2 \alpha\beta}(x)}
\right)
= 0.
\eqno{(11.109)}
$$
$\bar{\Gamma}^{\lbrack N \rbrack}$ contains all contributions from
all possible diagrams with arbitrary loops. The contribution from
$l$-loop diagram is proportional to $\hbar^l$. We can expand
$\bar{\Gamma}^{\lbrack N \rbrack}$ as a power serials of $\hbar^l$,
$$
\bar{\Gamma}^{\lbrack N \rbrack}
= \sum_M \hbar^M \bar{\Gamma}^{\lbrack N \rbrack}_M,
\eqno{(11.110)}
$$
where $\bar{\Gamma}^{\lbrack N \rbrack}_M$ is the contribution
from all $M$-loop diagrams. According to our inductive hypothesis,
all $\bar{\Gamma}^{\lbrack N \rbrack}_M$ are finite is $M \leq N$.
Therefore, divergence first appear in
$\bar{\Gamma}^{\lbrack N \rbrack}_{N+1}$.
Substitute (11.110) into (11.109), we will get
$$
\sum_{M,L}  \hbar^{M+L}
\bar{\Gamma}_M^{\lbrack N \rbrack} *
\bar{\Gamma}_L^{\lbrack N \rbrack}
+ \sum_M \hbar^M \int {\rm d}^4x\left(
 L_{1 \alpha}^{\mu}
\frac{\delta \bar{\Gamma}^{\lbrack N \rbrack}_M}{\delta \eta^{\mu}_{1 \alpha}(x)}
+ L_{2 \alpha\beta}
\frac{\delta \bar{\Gamma}^{\lbrack N \rbrack}_M}{\delta \eta_{2 \alpha\beta}(x)}
\right)
= 0.
\eqno{(11.111)}
$$
The $(N+1)$-loop contribution of (11.111) is
$$
\sum_{M=0}^{N+1}
\bar{\Gamma}_M^{\lbrack N \rbrack} *
\bar{\Gamma}_{N-M+1}^{\lbrack N \rbrack}
+\int {\rm d}^4x\left(
 L_{1 \alpha}^{\mu}
\frac{\delta \bar{\Gamma}^{\lbrack N \rbrack}_{N+1}}
{\delta \eta^{\mu}_{1 \alpha}(x)}
+ L_{2 \alpha\beta}
\frac{\delta \bar{\Gamma}^{\lbrack N \rbrack}_{N+1}}
{\delta \eta_{2 \alpha\beta}(x)} \right)
= 0.
\eqno{(11.112)}
$$
$\bar{\Gamma}_{N+1}^{\lbrack N \rbrack}$ can be separated into
two parts: finite part
$\bar{\Gamma}_{N+1,F}^{\lbrack N \rbrack}$
and divergent part
$\bar{\Gamma}_{N+1,div}^{\lbrack N \rbrack}$, that is
$$
\bar{\Gamma}_{N+1}^{\lbrack N \rbrack}
= \bar{\Gamma}_{N+1,F}^{\lbrack N \rbrack}
+ \bar{\Gamma}_{N+1,div}^{\lbrack N \rbrack}.
\eqno{(11.113)}
$$
$\bar{\Gamma}_{N+1,div}^{\lbrack N \rbrack}$ is a divergent
function of $(4-D)$ if we calculate loop diagrams in dimensional
regularization. In other words, all terms in
$\bar{\Gamma}_{N+1,div}^{\lbrack N \rbrack}$
are divergent terms when $(4-D)$ approaches zero.
Substitute (11.113) into (11.112), if we only concern divergent
terms, we will get
$$
\bar{\Gamma}_{N+1,div}^{\lbrack N \rbrack} *
\bar{\Gamma}_0^{\lbrack N \rbrack}  +
\bar{\Gamma}_0^{\lbrack N \rbrack} *
\bar{\Gamma}_{N+1,div}^{\lbrack N \rbrack}
+\int {\rm d}^4x\left(
 L_{1 \alpha}^{\mu}
\frac{\delta \bar{\Gamma}^{\lbrack N \rbrack}_{N+1,div}}
{\delta \eta^{\mu}_{1 \alpha}(x)}
+ L_{2 \alpha\beta}
\frac{\delta \bar{\Gamma}^{\lbrack N \rbrack}_{N+1,div}}
{\delta \eta_{2 \alpha\beta}(x)} \right)
= 0.
\eqno{(11.114)}
$$
$\bar{\Gamma}_{N+1,F}^{\lbrack N \rbrack}$ has no contribution
to the divergent part. Because $\bar{\Gamma}_0^{\lbrack N \rbrack}$
represents contribution from tree diagram and counterterm has
no contribution to tree diagram, we have
$$
\bar{\Gamma}_0^{\lbrack N \rbrack}
= \bar{\Gamma}_0^{\lbrack 0 \rbrack}.
\eqno{(11.115)}
$$
Denote
$$
\bar{\Gamma}_0
= \bar{\Gamma}_0^{\lbrack N \rbrack}
= \swav{S}^{\lbrack 0 \rbrack}
+ \frac{1}{2 \alpha}
\int {\rm d}^4x \eta_{\alpha \beta}
f^{\alpha} f^{\beta}.
\eqno{(11.116)}
$$
Then (11.114) is changed into
$$
\bar{\Gamma}_{N+1,div}^{\lbrack N \rbrack} *
\bar{\Gamma}_0  +
\bar{\Gamma}_0 *
\bar{\Gamma}_{N+1,div}^{\lbrack N \rbrack}
+\int {\rm d}^4x\left(
 L_{1 \alpha}^{\mu}
\frac{\delta \bar{\Gamma}^{\lbrack N \rbrack}_{N+1,div}}
{\delta \eta^{\mu}_{1 \alpha}(x)}
+ L_{2 \alpha\beta}
\frac{\delta \bar{\Gamma}^{\lbrack N \rbrack}_{N+1,div}}
{\delta \eta_{2 \alpha\beta}(x)} \right)
= 0.
\eqno{(11.117)}
$$
Substitute (11.110) into (11.106), we get
$$
\partial^{\mu}
\frac{\delta \bar{\Gamma}_{N+1}^{\lbrack N \rbrack}  }
{\delta K^{\mu}_{\alpha}(x) }
= \frac{\delta \bar{\Gamma}_{N+1}^{\lbrack N \rbrack}  }
{\delta \bar\eta_{\alpha}(x) }.
\eqno{(11.118)}
$$
The finite part $\bar{\Gamma}_{N+1,F}^{\lbrack N \rbrack}$
has no contribution to the divergent part, so we have
$$
\partial^{\mu}
\frac{\delta \bar{\Gamma}_{N+1,div}^{\lbrack N \rbrack}  }
{\delta K^{\mu}_{\alpha}(x) }
= \frac{\delta \bar{\Gamma}_{N+1,div}^{\lbrack N \rbrack}  }
{\delta \bar\eta_{\alpha}(x) }.
\eqno{(11.119)}
$$
\\

The operator $\hat{g}$  and $\hat{g}_1$ are defined by
$$
\begin{array}{rcl}
\hat{g} &\define & \int {\rm d}^4x \left\lbrace
\frac{\delta \bar{\Gamma}_0 }{\delta C_{\mu}^{\alpha}(x) }
\frac{\delta  }{\delta K^{\mu}_{\alpha}(x) }
+ \frac{\delta \bar{\Gamma}_0 }{\delta L_{\alpha}(x) }
\frac{\delta  }{\delta \eta^{\alpha}(x) }
+ \frac{\delta \bar{\Gamma}_0 }{\delta K^{\mu}_{\alpha}(x) }
\frac{\delta  }{\delta C_{\mu}^{\alpha}(x) }
\right. \\
&&\\
&& \left.
+ \frac{\delta \bar{\Gamma}_0 }{\delta \eta^{\alpha}(x) }
\frac{\delta  }{\delta  L_{\alpha}(x) }
+ L_{1 \alpha}^{\mu}
\frac{\delta }{\delta \eta^{\mu}_{1 \alpha}(x)}
+ L_{2 \alpha\beta}
\frac{\delta }{\delta \eta_{2 \alpha\beta}(x)}
\right\rbrace .
\end{array}
\eqno{(11.120)}
$$
$$
\begin{array}{rcl}
\hat{g} &\define & \int {\rm d}^4x \left\lbrace
\frac{\delta \bar{\Gamma}_0 }{\delta C_{\mu}^{\alpha}(x) }
\frac{\delta  }{\delta K^{\mu}_{\alpha}(x) }
+ \frac{\delta \bar{\Gamma}_0 }{\delta L_{\alpha}(x) }
\frac{\delta  }{\delta \eta^{\alpha}(x) }
\right. \\
&&\\
&& \left.
+ \frac{\delta \bar{\Gamma}_0 }{\delta K^{\mu}_{\alpha}(x) }
\frac{\delta  }{\delta C_{\mu}^{\alpha}(x) }
+ \frac{\delta \bar{\Gamma}_0 }{\delta \eta^{\alpha}(x) }
\frac{\delta  }{\delta  L_{\alpha}(x) }
\right\rbrace .
\end{array}
\eqno{(11.121)}
$$
Using definition eq.(11.120), (11.117) simplifies to
$$
\hat{g} \bar{\Gamma}_{N+1,div}^{\lbrack N \rbrack} = 0.
\eqno{(11.122)}
$$
\\

Operators $\hat{g}$ and $\hat{g}_1$ are not
nilpotent operators. It can be proved that
$$
\begin{array}{rcl}
\hat{g}_1^2 &=& \int {\rm d}^4x \left\lbrace
\left\lbrack \frac{\delta}{\delta C_{\mu}^{\alpha}(x)}
(\bar\Gamma_0 * \bar\Gamma_0) \right\rbrack
\frac{\delta}{\delta K^{\mu}_{\alpha}(x)}
+\left\lbrack \frac{\delta}{\delta L_{\alpha}(x)}
(\bar\Gamma_0 * \bar\Gamma_0) \right\rbrack
\frac{\delta}{\delta \eta^{\alpha}(x)} \right\rbrace  \\
\\
&&
-\int {\rm d}^4x \left\lbrace
\left\lbrack \frac{\delta}{\delta K^{\mu}_{\alpha}(x)}
(\bar\Gamma_0 * \bar\Gamma_0) \right\rbrack
\frac{\delta}{\delta C_{\mu}^{\alpha}(x)}
+\left\lbrack \frac{\delta}{\delta \eta^{\alpha}(x)}
(\bar\Gamma_0 * \bar\Gamma_0) \right\rbrack
\frac{\delta}{\delta L_{\alpha}(x)} \right\rbrace,
\end{array}
\eqno{(11.123)}
$$
$$
\begin{array}{rcl}
\hat{g} \hat{g}_1 & = & \int {\rm d}^4x {\rm d}^4y
L^{\mu}_{1\alpha}(x) \left(
\frac{\delta \bar\Gamma_0}{\delta C^{\alpha}_{\mu}(y)}
\frac{\delta }{\delta K_{\alpha}^{\mu}(y)}
+\frac{\delta \bar\Gamma_0}{\delta K_{\alpha}^{\mu}(y)}
\frac{\delta }{\delta C^{\alpha}_{\mu}(y)} \right. \\
&&\\
&&\left. +\frac{\delta \bar\Gamma_0}{\delta L_{\alpha}(y)}
\frac{\delta }{\delta \eta^{\alpha}(y)}
+\frac{\delta \bar\Gamma_0}{\delta \eta^{\alpha}(y)}
\frac{\delta }{\delta L_{\alpha}(y)}
\right) \frac{\delta}{\delta \eta^{\mu}_{1 \alpha}(x)}\\
&&\\
&&+ \int {\rm d}^4x {\rm d}^4y
L_{2\alpha\beta}(x) \left(
\frac{\delta \bar\Gamma_0}{\delta C^{\alpha}_{\mu}(y)}
\frac{\delta }{\delta K_{\alpha}^{\mu}(y)}
+\frac{\delta \bar\Gamma_0}{\delta K_{\alpha}^{\mu}(y)}
\frac{\delta }{\delta C^{\alpha}_{\mu}(y)} \right.\\
&&\\
&&\left. +\frac{\delta \bar\Gamma_0}{\delta L_{\alpha}(y)}
\frac{\delta }{\delta \eta^{\alpha}(y)}
+\frac{\delta \bar\Gamma_0}{\delta \eta^{\alpha}(y)}
\frac{\delta }{\delta L_{\alpha}(y)}
\right) \frac{\delta}{\delta \eta_{2 \alpha\beta}(x)}\\
&&\\
&&+ \int {\rm d}^4x {\rm d}^4y \left\lbrack
\frac{\delta L_{1 \alpha}^{\mu}(x)}{\delta \eta^{\sigma}(y)}
\frac{\delta \bar\Gamma_0}{\delta \eta^{\mu}_{1 \alpha}(x)}
\frac{\delta}{\delta L_{\sigma}(y)}
+\frac{\delta L_{2 \alpha\beta}(x)}{\delta \eta^{\sigma}(y)}
\frac{\delta \bar\Gamma_0}{\delta \eta_{2 \alpha\beta}(x)}
\frac{\delta}{\delta L_{\sigma}(y)}
\right\rbrack,
\end{array}
\eqno{(11.124)}
$$
$$
\begin{array}{rcl}
\hat{g}^2 &=&
\int {\rm d}^4x {\rm d}^4y \left\lbrack
\frac{\delta \bar\Gamma_0}{\delta C_{\nu}^{\sigma}(y)}
\frac{\delta L_{1\alpha}^{\mu}(x)}{\delta K^{\nu}_{\sigma}(y)}
\frac{\delta}{\delta \eta^{\mu}_{1\alpha}(x)}
+ \frac{\delta \bar\Gamma_0}{\delta L_{\sigma}(y)}
\frac{\delta L_{1\alpha}^{\mu}(x)}{\delta \eta^{\sigma}(y)}
\frac{\delta}{\delta \eta^{\mu}_{1\alpha}(x)} \right. \\
&&\\
&&\left. +\frac{\delta \bar\Gamma_0}{\delta C_{\mu}^{\sigma}(y)}
\frac{\delta L_{2\alpha\beta}(x)}{\delta K^{\mu}_{\sigma}(y)}
\frac{\delta}{\delta \eta_{2\alpha\beta}(x)}
+ \frac{\delta \bar\Gamma_0}{\delta L_{\sigma}(y)}
\frac{\delta L_{2\alpha\beta}(x)}{\delta \eta^{\sigma}(y)}
\frac{\delta}{\delta \eta_{2\alpha\beta}(x)}
\right\rbrack\\
&&\\
&&+ \int {\rm d}^4x {\rm d}^4y \left\lbrack
\frac{\delta L_{1 \alpha}^{\mu}(x)}{\delta \eta^{\sigma}(y)}
\frac{\delta \bar\Gamma_0}{\delta \eta^{\mu}_{1 \alpha}(x)}
\frac{\delta}{\delta L_{\sigma}(y)}
+\frac{\delta L_{2 \alpha\beta}(x)}{\delta \eta^{\sigma}(y)}
\frac{\delta \bar\Gamma_0}{\delta \eta_{2 \alpha\beta}(x)}
\frac{\delta}{\delta L_{\sigma}(y)}
\right\rbrack\\
&&\\
&& +  \int {\rm d}^4x {\rm d}^4y \left\lbrack
L_{1\alpha}^{\mu}(y) \left(
\frac{\delta L_{1\sigma}^{\nu}(x)}{\delta\eta^{\mu}_{1\alpha}(y)}
+L_{1\sigma}^{\nu}(x)
\frac{\delta }{\delta\eta^{\mu}_{1\alpha}(y)}
\right) \frac{\delta}{\delta \eta_{1 \sigma}^{\nu}(x)}\right.\\
&&\\
&&+ L_{1\alpha}^{\mu}(y) \left(
\frac{\delta L_{2\rho\sigma}(x)}{\delta\eta^{\mu}_{1\alpha}(y)}
+L_{2\rho\sigma}(x)
\frac{\delta }{\delta\eta^{\mu}_{1\alpha}(y)}
\right) \frac{\delta}{\delta \eta_{2 \rho \sigma}(x)} \\
&&\\
&&
+ L_{2\rho\sigma}(y) \left(
\frac{\delta L_{1\alpha}^{\mu}(x)}{\delta\eta_{2\rho\sigma}(y)}
+L_{1\alpha}^{\mu}(x)
\frac{\delta }{\delta\eta_{2\rho\sigma}(y)}
\right) \frac{\delta}{\delta \eta_{1 \alpha}^{\mu}(x)} \\
&&\\
&&\left. + L_{2\rho\sigma}(y) \left(
\frac{\delta L_{2\alpha\beta}(x)}{\delta \eta_{2\rho\sigma}(y)}
+L_{2\alpha\beta}(x)
\frac{\delta }{\delta\eta_{2\rho\sigma}(y)}
\right) \frac{\delta}{\delta \eta_{2 \alpha\beta}(x)}\right\rbrack
\end{array}
\eqno{(11.125)}
$$
\\

Now, we try to determine the general solutions to eq.(11.112)
and eq.(11.118). First, let's see the action
$S\lbrack C,\eta_1,\eta_2 \rbrack$ which is invariant
under local gravitational gauge transformation.
$S\lbrack C,\eta_1,\eta_2 \rbrack$ is also invariant under generalized
BRST transformation. The generalized BRST transformation
of $S\lbrack C,\eta_1,\eta_2 \rbrack$ is
$$
\delta S\lbrack C,\eta_1,\eta_2 \rbrack =
\int {\rm d}^4x \left( \delta C_{\mu}^{\alpha}(x)
\frac{\delta S}{\delta C_{\mu}^{\alpha}(x)}
+\delta \eta^{\mu}_{1 \alpha}(x)
\frac{\delta S}{\delta \eta^{\mu}_{1\alpha}(x)}
+\delta \eta_{2 \alpha\beta}(x)
\frac{\delta S}{\delta \eta_{2\alpha\beta}(x)}
\right).
\eqno{(11.126)}
$$
Using generalized BRST transformation relation eqs.(11.1), (11.5-6), we can get
the following relation,
$$
\delta S\lbrack C,\eta_1,\eta_2 \rbrack =
\int {\rm d}^4x \left(
-{\mathbf D}_{\mu \beta}^{\alpha} \eta^{\beta}
\frac{\delta S}{\delta C_{\mu}^{\alpha}(x)}
-  L^{\mu}_{1 \alpha}(x)
\frac{\delta S}{\delta \eta^{\mu}_{1\alpha}(x)}
-  L_{2 \alpha\beta}(x)
\frac{\delta S}{\delta \eta_{2\alpha\beta}(x)}
\right) \delta\lambda.
\eqno{(11.127)}
$$
Using eq.(11.48) and eq.(11.116), we can change the above relation
into another form
$$
\delta S\lbrack C,\eta_1,\eta_2 \rbrack =
\int {\rm d}^4x \left(
-\frac{\delta \bar\Gamma_0}{\delta K^{\mu}_{\alpha}}
\frac{\delta S}{\delta C_{\mu}^{\alpha}(x)}
-  L^{\mu}_{1 \alpha}(x)
\frac{\delta S}{\delta \eta^{\mu}_{1\alpha}(x)}
-  L_{2 \alpha\beta}(x)
\frac{\delta S}{\delta \eta_{2\alpha\beta}(x)}
\right) \delta\lambda.
\eqno{(11.128)}
$$
Because $C \lbrack C,\eta_1,\eta_2 \rbrack$ is a functional of only
gravitational gauge fields $C_{\mu}^{\alpha}$, $\eta^{\mu}_{1\alpha}$ and
$\eta_{2\alpha\beta}$, its functional derivatives to $K^{\mu}_{\alpha}$,
$L_{\alpha}$ and $\eta^{\alpha}$ vanish
$$
\frac{\delta S \lbrack C,\eta_1,\eta_2 \rbrack }
{\delta K^{\mu}_{\alpha}(x) } =0,
\eqno{(11.129)}
$$
$$
\frac{\delta S \lbrack C,\eta_1,\eta_2 \rbrack }
{\delta L_{\alpha}(x) } =0,
\eqno{(11.130)}
$$
$$
\frac{\delta S \lbrack C,\eta_1,\eta_2 \rbrack }
{\delta \eta^{\alpha}(x) } =0.
\eqno{(11.131)}
$$
Using these relations, we can prove that
$$
\hat{g} S\lbrack C,\eta_1,\eta_2 \rbrack
= \int {\rm d}^4x \left(
\frac{\delta \bar\Gamma_0}{\delta K^{\mu}_{\alpha}}
\frac{\delta S}{\delta C_{\mu}^{\alpha}(x)}
+  L^{\mu}_{1 \alpha}(x)
\frac{\delta S}{\delta \eta^{\mu}_{1\alpha}(x)}
+  L_{2 \alpha\beta}(x)
\frac{\delta S}{\delta \eta_{2\alpha\beta}(x)}
\right)
\eqno{(11.132)}
$$
Compare eq.(11.132) with eq.(11.128), we can get
$$
\delta S\lbrack C,\eta_1,\eta_2 \rbrack =
- \hat{g} S\lbrack C,\eta_1,\eta_2 \rbrack \delta\lambda.
\eqno{(11.133)}
$$
The generalized BRST symmetry of $S\lbrack C,\eta_1,\eta_2 \rbrack$
gives out the following important property of operator $\hat{g}$,
$$
\hat{g} S\lbrack C,\eta_1,\eta_2 \rbrack = 0.
\eqno{(11.134)}
$$
From eq.(11.134), we know that action $S\lbrack C,\eta_1,\eta_2 \rbrack$
is a possible solution to eq.(11.122).
Suppose that there is another functional $f'$ which is functional
of  $C_{\mu}^{\alpha}(x)$, $\eta^{\alpha}(x)$, $\bar\eta_{\alpha}(x)$,
and $K^{\mu}_{\alpha}(x)$ ,
$$
f' =  f'\lbrack C,\eta,\bar\eta,K \rbrack.
\eqno{(11.135)}
$$
Because
$$
\frac{\delta f'}{\delta \eta^{\mu}_{1\alpha}} = 0,
\eqno{(11.136)}
$$
$$
\frac{\delta f'}{\delta \eta_{2\alpha\beta}} = 0,
\eqno{(11.137)}
$$
$$
\frac{\delta f'}{\delta L_{\sigma}} = 0,
\eqno{(11.138)}
$$
from eq.(11.124), we can prove that $f'$ satisfies
$$
\hat{g} \hat{g}_1 f'\lbrack C,\eta,\bar\eta,K \rbrack = 0.
\eqno{(11.139)}
$$
So, $\hat{g} f'$ is also a solution to eq.(11.122).
The most general solution to
eq.(11.122) can be written in the following form
$$
\bar{\Gamma}_{N+1,div}^{\lbrack N \rbrack}
= \alpha_{N+1} (\varepsilon) S\lbrack C,\eta_1,\eta_2 \rbrack
+ \hat{g}_1 f'\lbrack C,\eta,\bar\eta,K \rbrack
+  f\lbrack C,\eta,\bar\eta,K,L,\eta_1,\eta_2 \rbrack,
\eqno{(11.140)}
$$
where $f\lbrack C,\eta,\bar\eta,K,L,\eta_1,\eta_2 \rbrack$ is
an arbitrary functional of fields $C_{\mu}^{\alpha}(x)$, $\eta^{\alpha}(x)$,
$\bar\eta_{\alpha}(x)$ and external sources
$K^{\mu}_{\alpha}(x)$ and $L_{\alpha}(x)$, and $\eta^{\mu}_{1\alpha}(x)$ and
$\eta_{2\alpha\beta}(x)$.
\\

Now, let's consider constrains from eq.(11.119). Using eq.(11.129)
and eq.(11.131), we can see that $S \lbrack C,\eta_1,\eta_2 \rbrack$
satisfies eq.(11.119), so eq.(11.119) has no constrain on
$S\lbrack C,\eta_1,\eta_2 \rbrack$. Define a new variable
$$
B^{\mu}_{\alpha}
= K^{\mu}_{\alpha}
- \partial^{\mu} \bar\eta_{\alpha}.
\eqno{(11.141)}
$$
$f_1 \lbrack B \rbrack$ is an arbitrary functional of $B$.
It can be proved that
$$
\frac{\delta f_1 \lbrack B \rbrack }
{\delta B^{\mu}_{\alpha}(x) }
= \frac{\delta f_1 \lbrack B \rbrack }
{\delta K^{\mu}_{\alpha}(x) } ,
\eqno{(11.142)}
$$
$$
\frac{\delta f_1 \lbrack B \rbrack }
{\delta \bar\eta_{\alpha}(x) }
= \partial^{\mu} \frac{\delta f_1 \lbrack B \rbrack }
{\delta B^{\mu}_{\alpha}(x) }.
\eqno{(11.143)}
$$
Combine these two relations, we will get
$$
\frac{\delta f_1 \lbrack B \rbrack }
{\delta \bar\eta_{\alpha}(x) }
= \partial^{\mu} \frac{\delta f_1 \lbrack B \rbrack }
{\delta K^{\mu}_{\alpha}(x) }.
\eqno{(11.144)}
$$
There $f_1 \lbrack B \rbrack$ is a solution to eq.(11.119).
Suppose that there is another functional $f_2$ that is
given by,
$$
f_2 \lbrack K,C,\eta \rbrack
= \int {\rm d}^4x ~ K^{\mu}_{\alpha}
T_{\mu}^{\alpha}(C,\eta),
\eqno{(11.145)}
$$
where $T_{\mu}^{\alpha}$ is a conserved current
$$
\partial^{\mu} T_{\mu}^{\alpha} =0.
\eqno{(11.146)}
$$
It can be easily proved that $f_2 \lbrack K,C,\eta \rbrack$
is also a solution of eq.(11.119). Because $\bar{\Gamma}_0$ satisfies
eq.(11.119) (please see eq.(11.101)), operator $\hat{g}$ commutes
with $\frac{\delta}{\delta \bar\eta_{\alpha}(x) }
- \partial^{\mu} \frac{\delta }{\delta K^{\mu}_{\alpha}(x) } $.
It means that functional $f'\lbrack C,\eta,\bar\eta,K,L \rbrack$
in eq.(11.140) must satisfy eq.(11.119). According to these
discussion, the solution of $f'\lbrack C,\eta,\bar\eta,K \rbrack$
has the following form,
$$
f'\lbrack C,\eta,\bar\eta,K \rbrack
= f_1 \lbrack C,\eta,
K^{\mu}_{\alpha} - \partial^{\mu} \bar\eta_{\alpha} \rbrack
+ \int {\rm d}^4x ~ K^{\mu}_{\alpha}
T_{\mu}^{\alpha}(C,\eta).
\eqno{(11.147)}
$$
\\

In order to determine $f'\lbrack C,\eta,\bar\eta,K \rbrack$,
we need to study dimensions of various fields and external sources.
Set the dimensionality of mass to $1$, i.e.
$$
D \lbrack \hat{P}_{\mu} \rbrack =1.
\eqno{(11.148)}
$$
 Then we have
$$
D \lbrack C_{\mu}^{\alpha} \rbrack =1,
\eqno{(11.149)}
$$
$$
D \lbrack {\rm d}^4x \rbrack =-4,
\eqno{(11.150)}
$$
$$
D \lbrack D_{\mu} \rbrack =1,
\eqno{(11.151)}
$$
$$
D \lbrack \eta \rbrack  = D \lbrack \bar\eta \rbrack =1,
\eqno{(11.152)}
$$
$$
D \lbrack K \rbrack  = D \lbrack L \rbrack =2,
\eqno{(11.153)}
$$
$$
D \lbrack g \rbrack   = - 1,
\eqno{(11.154)}
$$
$$
D \lbrack \eta_1 \rbrack  = D \lbrack \eta_2 \rbrack =0,
\eqno{(11.155)}
$$
$$
D \lbrack \bar{\Gamma}_{N+1,div}^{\lbrack N \rbrack} \rbrack
= D \lbrack S \rbrack = 0.
\eqno{(11.156)}
$$
Using these relations, we can prove that
$$
D \lbrack \hat{g} \rbrack   =1,
\eqno{(11.157)}
$$
$$
D \lbrack f' \rbrack   =-1.
\eqno{(11.158)}
$$
Define virtual particle number $N_g$ of ghost field
$\eta$ is 1, and that of ghost field $\bar\eta$ is -1, i.e.
$$
N_g \lbrack \eta \rbrack = 1,
\eqno{(11.159)}
$$
$$
N_g \lbrack \bar\eta \rbrack = - 1.
\eqno{(11.160)}
$$
The virtual particle number is a additive conserved quantity, so
Lagrangian and action carry no virtual particle number,
$$
N_g \lbrack S \rbrack =  N_g \lbrack {\cal L} \rbrack    =  0.
\eqno{(11.161)}
$$
The virtual particle number $N_g$ of other fields and external
sources are
$$
N_g \lbrack C \rbrack = N_g \lbrack D_{\mu} \rbrack =  0,
\eqno{(11.162)}
$$
$$
N_g \lbrack g \rbrack =  0,
\eqno{(11.163)}
$$
$$
N_g \lbrack \bar{\Gamma} \rbrack =  0,
\eqno{(11.164)}
$$
$$
N_g \lbrack K \rbrack =  -1,
\eqno{(11.165)}
$$
$$
N_g \lbrack L \rbrack =  -2.
\eqno{(11.166)}
$$
$$
N_g \lbrack \eta_1 \rbrack = N_g \lbrack \eta_2 \rbrack =  0,
\eqno{(11.167)}
$$
Using all these relations, we can determine the virtual particle
number $N_g$ of $\hat g$ and $f'$,
$$
N_g \lbrack \hat g \rbrack =  1,
\eqno{(11.168)}
$$
$$
N_g \lbrack f' \rbrack =  -1.
\eqno{(11.169)}
$$
According to eq.(11.158) and eq.(11.169), we know that the
dimensionality of $f'$ is $-1$ and its virtual particle
number if also $-1$. Besides, $f'$ must be a Lorentz scalar.
Combine all these results, the only two possible solutions of
$f_1 \lbrack C,\eta, K^{\mu}_{\alpha} - \partial^{\mu}
\bar\eta_{\alpha} \rbrack$ in eq.(11.124) are
$$
(K^{\mu}_{\alpha} - \partial^{\mu} \bar\eta_{\alpha}  )
C_{\mu}^{\alpha},
\eqno{(11.170)}
$$
The only possible solution of $T_{\mu}^{\alpha}$ is
$C_{\mu}^{\alpha}$. But in general gauge conditions,
$C_{\mu}^{\alpha}$ does not satisfy the conserved
equation  eq.(11.146). Therefore, the solution to
$f'\lbrack C,\eta,\bar\eta,K \rbrack$ is
$$
f'\lbrack C,\eta,\bar\eta,K,L \rbrack =
\int {\rm d}^4x \left\lbrack
\beta_{N+1}(\varepsilon) (K^{\mu}_{\alpha} - \partial^{\mu}
\bar\eta_{\alpha}  ) C_{\mu}^{\alpha}
\right\rbrack,
\eqno{(11.171)}
$$
where $\varepsilon = (4-D)$, $\beta_{N+1}(\varepsilon)$
is divergent parameter when $\varepsilon$ approaches zero.
Then using the definition of $\hat g$,
we can obtain the following result,
$$
\hat g f'\lbrack C,\eta,\bar\eta,K,L \rbrack
=  - \beta_{N+1}(\varepsilon)
\int {\rm d}^4x \left\lbrack
\frac{\delta \bar{\Gamma}_0 }{\delta C_{\mu}^{\alpha}(x) }
C_{\mu}^{\alpha}(x)
+ \frac{\delta \bar{\Gamma}_0 }{\delta K^{\mu}_{\alpha}(x) }
K^{\mu}_{\alpha}(x)
- \bar\eta_{\alpha} \partial^{\mu}
{\mathbf D}_{\mu \beta}^{\alpha} \eta^{\beta}
\right\rbrack  .
\eqno{(11.172)}
$$
Therefore, the most general solution of
$\bar{\Gamma}_{N+1,div}^{\lbrack N \rbrack}$ is
$$
\begin{array}{rcl}
\bar{\Gamma}_{N+1,div}^{\lbrack N \rbrack}
&=& \alpha_{N+1}(\varepsilon)  S\lbrack C,\eta_1,\eta_2 \rbrack
- \int {\rm d}^4x \left\lbrack
\beta_{N+1}(\varepsilon) \frac{\delta \bar{\Gamma}_0 }
{\delta C_{\mu}^{\alpha}(x) } C_{\mu}^{\alpha}(x)
\right. \\
&&\\
&& \left.
+ \beta_{N+1}(\varepsilon) \frac{\delta \bar{\Gamma}_0 }
{\delta K^{\mu}_{\alpha}(x) } K^{\mu}_{\alpha}(x)
- \beta_{N+1}(\varepsilon) \bar\eta_{\alpha} \partial^{\mu}
{\mathbf D}_{\mu \beta}^{\alpha} \eta^{\beta}
\right\rbrack.
\end{array}
\eqno{(11.173)}
$$
\\

In fact, the action $S\lbrack C,\eta_1,\eta_2 \rbrack$ is a functional
of pure gravitational gauge field. It also contains gravitational
coupling constant $g$. So, we can denote it as
$S\lbrack C,g \rbrack$,
$$
S\lbrack C,g \rbrack = S\lbrack C,\eta_1,\eta_2 \rbrack.
\eqno{(11.174)}
$$
From eq.(4.20), eq.(4.24) and eq.(4.25),
we can prove that the action $S\lbrack C,g \rbrack$ has
the following important properties,
$$
S\lbrack gC,1 \rbrack = g^2 S\lbrack C,g \rbrack.
\eqno{(11.175)}
$$
Differentiate both sides of eq.(11.175) with respect to coupling
constant $g$, we can get
$$
S\lbrack C,g \rbrack
= \frac{1}{2} \int {\rm d}^4x
\frac{\delta S\lbrack C,g \rbrack }{\delta C_{\mu}^{\alpha}(x) }
C_{\mu}^{\alpha}(x)
- \frac{1}{2} g \frac{\partial S\lbrack C,g \rbrack}{\partial g}.
\eqno{(11.176)}
$$
\\

It can be easily proved that
$$
\begin{array}{l}
\int {\rm d}^4x C_{\mu}^{\alpha}(x)
\frac{\delta  }{\delta C_{\mu}^{\alpha}(x) }
\left\lbrack
\int {\rm d}^4y \bar\eta_{\alpha}(y) \partial^{\mu}
{\mathbf D}_{\mu \beta}^{\alpha} \eta^{\beta}(y)
\right\rbrack  \\
\\
= \int {\rm d}^4x \left\lbrack
(\partial^{\mu} \bar\eta_{\beta}(x))
(\partial_{\mu} \eta^{\beta}(x))
+ \bar\eta_{\alpha}(x) \partial^{\mu}
{\mathbf D}_{\mu \beta}^{\alpha} \eta^{\beta}(x)
\right\rbrack,
\end{array}
\eqno{(11.177)}
$$
$$
\begin{array}{l}
\int {\rm d}^4x C_{\mu}^{\alpha}(x)
\frac{\delta  }{\delta C_{\mu}^{\alpha}(x) }
\left\lbrack
\int {\rm d}^4y K^{\mu}_{\alpha}(y)
{\mathbf D}_{\mu \beta}^{\alpha} \eta^{\beta}(y)
\right\rbrack  \\
\\
= - \int {\rm d}^4x \left\lbrack
K^{\mu}_{\alpha}(x) \partial_{\mu} \eta^{\alpha}(x)
- K^{\mu}_{\alpha}(x) {\mathbf D}_{\mu \beta}^{\alpha}
\eta^{\beta}(x)
\right\rbrack,
\end{array}
\eqno{(11.178)}
$$
$$
\begin{array}{l}
g \frac{\partial  }{\partial g }
\left\lbrack
\int {\rm d}^4x \bar\eta_{\alpha}(x) \partial^{\mu}
{\mathbf D}_{\mu \beta}^{\alpha} \eta^{\beta}(x)
\right\rbrack  \\
\\
= \int {\rm d}^4x \left\lbrack
(\partial^{\mu} \bar\eta_{\beta}(x))
(\partial_{\mu} \eta^{\beta}(x))
+ \bar\eta_{\alpha}(x) \partial^{\mu}
{\mathbf D}_{\mu \beta}^{\alpha} \eta^{\beta}(x)
\right\rbrack,
\end{array}
\eqno{(11.179)}
$$
$$
\begin{array}{l}
g \frac{\partial  }{\partial g }
\left\lbrack
\int {\rm d}^4x K^{\mu}_{\alpha}(x)
{\mathbf D}_{\mu \beta}^{\alpha} \eta^{\beta}(x)
\right\rbrack  \\
\\
= - \int {\rm d}^4x \left\lbrack
K^{\mu}_{\alpha}(x) \partial_{\mu} \eta^{\alpha}
- K^{\mu}_{\alpha}(x) {\mathbf D}_{\mu \beta}^{\alpha}
\eta^{\beta}
\right\rbrack,
\end{array}
\eqno{(11.180)}
$$
$$
\begin{array}{l}
g \frac{\partial  }{\partial g }
\left\lbrack
\int {\rm d}^4x ~g L_{\alpha}(x) \eta^{\beta}(x)
(\partial_{\beta} \eta^{\alpha}(x) )
\right\rbrack  \\
\\
=
\int {\rm d}^4x~ g L_{\alpha}(x) \eta^{\beta}(x)
(\partial_{\beta} \eta^{\alpha}(x) ),
\end{array}
\eqno{(11.181)}
$$

Using eqs.(11.177-178), eq.(11.116) and eq.(11.46),
we can prove that
$$
\begin{array}{rcl}
\int {\rm d}^4x
\frac{\delta S\lbrack C,g \rbrack  }{\delta C_{\mu}^{\alpha}(x) }
C_{\mu}^{\alpha}(x)  &=&
\int {\rm d}^4x
\frac{\delta \bar{\Gamma}_0  }{\delta C_{\mu}^{\alpha}(x) }
C_{\mu}^{\alpha}(x)
+ \int {\rm d}^4x \left\lbrack
-(\partial^{\mu} \bar\eta_{\alpha}(x))
(\partial_{\mu} \eta^{\alpha}(x)) \right. \\
&&\\
&&\left. - \bar\eta_{\alpha}(x) \partial^{\mu}
{\mathbf D}_{\mu \beta}^{\alpha} \eta^{\beta}(x)
+K^{\mu}_{\alpha}(x) \partial_{\mu} \eta^{\alpha}(x)
- K^{\mu}_{\alpha}(x) {\mathbf D}_{\mu \beta}^{\alpha}
\eta^{\beta}(x)
\right\rbrack.
\end{array}
\eqno{(11.182)}
$$
Similarly, we can get,
$$
\begin{array}{rcl}
g \frac{\partial S\lbrack C,g \rbrack }{\partial g }
&=&  g \frac{\partial \bar{\Gamma}_0 }{\partial g }
+  \int {\rm d}^4x \left\lbrack
- (\partial^{\mu} \bar\eta_{\alpha}(x))
(\partial_{\mu} \eta^{\alpha}(x))
 - \bar\eta_{\alpha}(x) \partial^{\mu}
{\mathbf D}_{\mu \beta}^{\alpha} \eta^{\beta}(x)\right.  \\
&&\\
&& \left.
+K^{\mu}_{\alpha}(x) \partial_{\mu} \eta^{\alpha}(x)
- K^{\mu}_{\alpha}(x) {\mathbf D}_{\mu \beta}^{\alpha}
\eta^{\beta}(x)
- g L_{\alpha}(x) \eta^{\beta}(x)
(\partial_{\beta} \eta^{\alpha}(x) )
\right\rbrack.
\end{array}
\eqno{(11.183)}
$$
Substitute eqs.(11.182-183) into eq.(11.176), we will get
$$
S\lbrack C,g \rbrack
= \frac{1}{2} \int {\rm d}^4x
\frac{\delta \bar{\Gamma}_0 }{\delta C_{\mu}^{\alpha}(x) }
C_{\mu}^{\alpha}(x)
- \frac{1}{2} g \frac{\partial \bar{\Gamma}_0}{\partial g}
+ \int {\rm d}^4x \left\lbrace
\frac{1}{2} g L_{\alpha}(x) \eta^{\beta}(x)
(\partial_{\beta} \eta^{\alpha}(x) \right\rbrace.
\eqno{(11.184)}
$$
Substitute eq.(11.184) into eq.(11.173). we will get
$$
\begin{array}{rcl}
\bar{\Gamma}_{N+1,div}^{\lbrack N \rbrack}
&=& \int {\rm d}^4x \left\lbrack
\left(\frac{\alpha_{N+1}(\varepsilon)}{2}
- \beta_{N+1}(\varepsilon) \right) C_{\mu}^{\alpha}(x)
\frac{\delta \bar{\Gamma}_0 }{\delta C_{\mu}^{\alpha}(x) }
 \right. \\
&&\\
&&+ \frac{\alpha_{N+1}(\varepsilon)}{2} L_{\alpha}(x)
\frac{\delta \bar{\Gamma}_0 }
{\delta  L_{\alpha}(x) }
+ \beta_{N+1}(\varepsilon) \bar\eta_{\alpha}(x)
\frac{\delta \bar{\Gamma}_0 } {\delta  \bar\eta_{\alpha}(x) } \\
&&\\
&& \left. + \beta_{N+1}(\varepsilon) K^{\mu}_{\alpha}(x)
\frac{\delta \bar{\Gamma}_0 }{\delta K^{\mu}_{\alpha}(x) }
\right\rbrack - \frac{\alpha_{N+1}(\varepsilon)}{2}
g \frac{\partial \bar{\Gamma}_0 }{\partial g }
\end{array}
\eqno{(11.185)}
$$
\\

On the other hand, we can prove the following relations
$$
\begin{array}{l}
\int {\rm d}^4x  \eta^{\alpha}(x)
\frac{\delta  }{\delta \eta^{\alpha}(x) }
\left\lbrack
\int {\rm d}^4y ~ \bar\eta_{\beta}(y) \partial^{\mu}
{\mathbf D}_{\mu \sigma}^{\beta} \eta^{\sigma}(y)
\right\rbrack
= \int {\rm d}^4x~
 \bar\eta_{\beta}(x) \partial^{\mu}
{\mathbf D}_{\mu \sigma}^{\beta} \eta^{\sigma}(x),
\end{array}
\eqno{(11.186)}
$$
$$
\begin{array}{l}
\int {\rm d}^4x  \eta^{\alpha}(x)
\frac{\delta  }{\delta \eta^{\alpha}(x) }
\left\lbrack
\int {\rm d}^4y ~ K^{\mu}_{\beta}(y)
{\mathbf D}_{\mu \sigma}^{\beta} \eta^{\sigma}(y)
\right\rbrack
=  \int {\rm d}^4x ~
 K^{\mu}_{\beta}(x)
{\mathbf D}_{\mu \sigma}^{\beta} \eta^{\sigma}(x),
\end{array}
\eqno{(11.187)}
$$
$$
\begin{array}{l}
\int {\rm d}^4x  \eta^{\alpha}(x)
\frac{\delta  }{\delta \eta^{\alpha}(x) }
\left\lbrack
\int {\rm d}^4y ~  g L_{\beta}(y)
\eta^{\sigma}(y) (\partial_{\sigma} \eta^{\beta}(y) )
\right\rbrack
= 2  \int {\rm d}^4x~
 g L_{\beta}(x)
\eta^{\sigma}(x) (\partial_{\sigma} \eta^{\beta}(x) ),
\end{array}
\eqno{(11.188)}
$$

$$
\begin{array}{l}
\int {\rm d}^4x  \eta^{\alpha}(x)
\frac{\delta \bar\Gamma_0  }{\delta \eta^{\alpha}(x) }
= \int {\rm d}^4x \left\lbrace
 \bar\eta_{\beta}(x) \partial^{\mu}
{\mathbf D}_{\mu \sigma}^{\beta} \eta^{\sigma}(x) \right. \\
\\
  \left.
+  K^{\mu}_{\beta}(x)
{\mathbf D}_{\mu \sigma}^{\beta} \eta^{\sigma}(x)
+ 2  g L_{\beta}(x)
\eta^{\sigma}(x) (\partial_{\sigma} \eta^{\beta}(x))
\right\rbrace.
\end{array}
\eqno{(11.189)}
$$
Substitute eqs.(11.186-188) into eq.(11.189), we will get
$$
\int {\rm d}^4x  \left\lbrace
- \eta^{\alpha}
\frac{\delta \bar\Gamma_0  }{\delta \eta^{\alpha} }
+  \bar\eta_{\alpha}
\frac{\delta \bar\Gamma_0  }{\delta \bar\eta_{\alpha} }
+ K^{\mu}_{\alpha}
\frac{\delta \bar\Gamma_0  }{\delta K^{\mu}_{\alpha} }
+ 2 L_{\alpha}
\frac{\delta \bar\Gamma_0  }{\delta L_{\alpha} }
\right\rbrace = 0.
\eqno{(11.190)}
$$
Eq.(11.190) times $-\frac{\beta_{N+1}}{2}$, then
add up this results and eq.(11.185), we will get
$$
\begin{array}{rcl}
\bar{\Gamma}_{N+1,div}^{\lbrack N \rbrack}
&=& \int {\rm d}^4x \left\lbrack
\left( \frac{\alpha_{N+1}(\varepsilon)}{2}
- \beta_{N+1}(\varepsilon) \right)\left( C_{\mu}^{\alpha}(x)
\frac{\delta \bar{\Gamma}_0 }{\delta C_{\mu}^{\alpha}(x) }
+  L_{\alpha}(x)
\frac{\delta \bar{\Gamma}_0 }
{\delta  L_{\alpha}(x) } \right) \right.  \\
&&\\
&& \left.
+ \frac{\beta_{N+1}(\varepsilon)}{2}
 \left( \eta^{\alpha}(x)
\frac{\delta \bar\Gamma_0  }{\delta \eta^{\alpha}(x) }
+ \bar\eta_{\alpha}(x)
\frac{\delta \bar{\Gamma}_0 } {\delta  \bar\eta_{\alpha}(x) }
 +  K^{\mu}_{\alpha}(x)
\frac{\delta \bar{\Gamma}_0 }{\delta K^{\mu}_{\alpha}(x) }
\right)
\right\rbrack \\
&&\\
&&  - \frac{\alpha_{N+1}(\varepsilon)}{2}
g \frac{\partial \bar{\Gamma}_0 }{\partial g }.
\end{array}
\eqno{(11.191)}
$$
This is the most general form of
$\bar{\Gamma}_{N+1,div}^{\lbrack N \rbrack}$
which satisfies Ward-Takahashi identities. \\

According to minimal subtraction, the counterterm that cancel
the divergent part of $\bar{\Gamma}_{N+1}^{\lbrack N \rbrack}$
is just $-\bar{\Gamma}_{N+1,div}^{\lbrack N \rbrack}$, that is
$$
\swav{S}^{\lbrack N+1 \rbrack}
= \swav{S}^{\lbrack N \rbrack}
- \bar{\Gamma}_{N+1,div}^{\lbrack N \rbrack}
+ o(\hbar^{N+2}),
\eqno{(11.192)}
$$
where the term of $o(\hbar^{N+2})$ has no contribution to
the divergences of $(N+1)$-loop diagrams. Suppose that
$\bar{\Gamma}_{N+1}^{\lbrack N+1 \rbrack}$ is the generating
functional of regular vertex which is calculated from
$\swav{S}^{\lbrack N+1 \rbrack}$. It can be easily proved that
$$
\bar\Gamma_{N+1}^{\lbrack N+1 \rbrack}
= \bar\Gamma_{N+1}^{\lbrack N \rbrack}
- \bar{\Gamma}_{N+1,div}^{\lbrack N \rbrack}.
\eqno{(11.193)}
$$
Using eq.(11.113), we can get
$$
\bar\Gamma_{N+1}^{\lbrack N+1 \rbrack}
= \bar{\Gamma}_{N+1,F}^{\lbrack N \rbrack}.
\eqno{(11.194)}
$$
$\Gamma_{N+1}^{\lbrack N+1 \rbrack}$ contains no divergence
which is just what we expected. \\

Now, let's try to determine the form of
$\swav{S}^{\lbrack N+1 \rbrack}$. Denote the non-renormalized
action of the system as
$$
\swav{S}^{\lbrack 0 \rbrack}
= \swav{S}^{\lbrack 0 \rbrack}
\lbrack
C_{\mu}^{\alpha}, \bar\eta_{\alpha}, \eta^{\alpha},
K^{\mu}_{\alpha}, L_{\alpha}, g, \alpha,\eta_1,\eta_2
\rbrack.
\eqno{(11.195)}
$$
As one of the inductive hypothesis, we suppose that the
action of the system after $\hbar^N$ order renormalization is
$$
\begin{array}{l}
\swav{S}^{\lbrack N \rbrack}\lbrack
C_{\mu}^{\alpha}, \bar\eta_{\alpha}, \eta^{\alpha},
K^{\mu}_{\alpha}, L_{\alpha}, g, \alpha ,\eta_1,\eta_2 \rbrack  \\
\\
= \swav{S}^{\lbrack 0 \rbrack}
\left\lbrack
\sqrt{Z_1^{\lbrack N \rbrack}}C_{\mu}^{\alpha},
\sqrt{Z_2^{\lbrack N \rbrack}}\bar\eta_{\alpha},
\sqrt{Z_3^{\lbrack N \rbrack}}\eta^{\alpha},
\sqrt{Z_4^{\lbrack N \rbrack}}K^{\mu}_{\alpha},
\sqrt{Z_5^{\lbrack N \rbrack}}L_{\alpha},
      Z_g^{\lbrack N \rbrack} g,
      Z_{\alpha}^{\lbrack N \rbrack} \alpha,
      Z_6^{\lbrack N \rbrack} \eta_1,
      Z_7^{\lbrack N \rbrack}\eta_2
\right\rbrack.
\end{array}
\eqno{(11.196)}
$$
Substitute eq(11.191) and eq.(11.196) into eq.(11.192),
we obtain
$$
\begin{array}{rcl}
\swav{S}^{\lbrack N+1 \rbrack} &=&
\swav{S}^{\lbrack 0 \rbrack}
\left\lbrack
\sqrt{Z_1^{\lbrack N \rbrack}}C_{\mu}^{\alpha},
\sqrt{Z_2^{\lbrack N \rbrack}}\bar\eta_{\alpha},
\sqrt{Z_3^{\lbrack N \rbrack}}\eta^{\alpha},
\sqrt{Z_4^{\lbrack N \rbrack}}K^{\mu}_{\alpha}, \right. \\
&&\\
&&\left.~~~~~~~\sqrt{Z_5^{\lbrack N \rbrack}}L_{\alpha},
      Z_g^{\lbrack N \rbrack} g,
      Z_{\alpha}^{\lbrack N \rbrack} \alpha,
      Z_6^{\lbrack N \rbrack} \eta_1,
      Z_7^{\lbrack N \rbrack}\eta_2
\right\rbrack\\
&&\\
&&- \int {\rm d}^4x \left\lbrack
\left(\frac{\alpha_{N+1}(\varepsilon)}{2}
- \beta_{N+1}(\varepsilon) \right)\left( C_{\mu}^{\alpha}(x)
\frac{\delta \bar{\Gamma}_0 }{\delta C_{\mu}^{\alpha}(x) }
+  L_{\alpha}(x)
\frac{\delta \bar{\Gamma}_0 }
{\delta  L_{\alpha}(x) } \right) \right.  \\
&&\\
&& \left.
+ \frac{\beta_{N+1}(\varepsilon)}{2}
 \left( \eta^{\alpha}(x)
\frac{\delta \bar\Gamma_0  }{\delta \eta^{\alpha}(x) }
+ \bar\eta_{\alpha}(x)
\frac{\delta \bar{\Gamma}_0 } {\delta  \bar\eta_{\alpha}(x) }
 +  K^{\mu}_{\alpha}(x)
\frac{\delta \bar{\Gamma}_0 }{\delta K^{\mu}_{\alpha}(x) }
\right)
\right\rbrack \\
&&\\
&&  + \frac{\alpha_{N+1}(\varepsilon)}{2}
g \frac{\partial \bar{\Gamma}_0 }{\partial g }
 + o(\hbar^{N+2}).
\end{array}
\eqno{(11.197)}
$$
Using eq.(11.116), we can prove that
$$
\int {\rm d}^4x C_{\mu}^{\alpha}(x)
\frac{\delta \bar{\Gamma}_0}{\delta C_{\mu}^{\alpha}(x)}
= \int {\rm d}^4x C_{\mu}^{\alpha}(x)
\frac{\delta \swav{S}^{\lbrack 0 \rbrack}}
{\delta C_{\mu}^{\alpha}(x)}
+ 2 \alpha
\frac{\partial  \swav{S}^{\lbrack 0 \rbrack}}
{\partial \alpha},
\eqno{(11.198)}
$$
$$
\int {\rm d}^4x L_{\alpha}(x)
\frac{\delta \bar{\Gamma}_0}{\delta L_{\alpha}(x)}
= \int {\rm d}^4x L_{\alpha}(x)
\frac{\delta \swav{S}^{\lbrack 0 \rbrack}}
{\delta L_{\alpha}(x)},
\eqno{(11.199)}
$$
$$
\int {\rm d}^4x \bar\eta_{\alpha}(x)
\frac{\delta \bar{\Gamma}_0}{\delta \bar\eta_{\alpha}(x)}
= \int {\rm d}^4x   \bar\eta_{\alpha}(x)
\frac{\delta \swav{S}^{\lbrack 0 \rbrack}}
{\delta \bar\eta_{\alpha}(x)},
\eqno{(11.200)}
$$
$$
\int {\rm d}^4x \eta^{\alpha}(x)
\frac{\delta \bar{\Gamma}_0}{\delta \eta^{\alpha}(x)}
= \int {\rm d}^4x   \eta^{\alpha}(x)
\frac{\delta \swav{S}^{\lbrack 0 \rbrack}}
{\delta \eta^{\alpha}(x)},
\eqno{(11.201)}
$$
$$
\int {\rm d}^4x K^{\mu}_{\alpha}(x)
\frac{\delta \bar{\Gamma}_0}{\delta K^{\mu}_{\alpha}(x)}
= \int {\rm d}^4x K^{\mu}_{\alpha}(x)
\frac{\delta \swav{S}^{\lbrack 0 \rbrack}}
{\delta K^{\mu}_{\alpha}(x)},
\eqno{(11.202)}
$$
$$
g \frac{\partial \bar{\Gamma}_0}{\partial g}
= g \frac{\partial \swav{S}^{\lbrack 0 \rbrack}}{\partial g}.
\eqno{(11.203)}
$$
Using these relations, eq.(11.197) is changed into,
$$
\begin{array}{rcl}
\swav{S}^{\lbrack N+1 \rbrack} &=&
\swav{S}^{\lbrack 0 \rbrack}
\lbrack
\sqrt{Z_1^{\lbrack N \rbrack}}C_{\mu}^{\alpha},
\sqrt{Z_2^{\lbrack N \rbrack}}\bar\eta_{\alpha},
\sqrt{Z_3^{\lbrack N \rbrack}}\eta^{\alpha},
\sqrt{Z_4^{\lbrack N \rbrack}}K^{\mu}_{\alpha},  \\
&&\\
&&~~~~~~~
\sqrt{Z_5^{\lbrack N \rbrack}}L_{\alpha},
      Z_g^{\lbrack N \rbrack} g,
      Z_{\alpha}^{\lbrack N \rbrack} \alpha,
      Z_6^{\lbrack N \rbrack} \eta_1,
      Z_7^{\lbrack N \rbrack}\eta_2
\rbrack  \\
&&\\
&& - \int {\rm d}^4x \left\lbrack
(\frac{\alpha_{N+1}(\varepsilon)}{2}
- \beta_{N+1}(\varepsilon) )\left( C_{\mu}^{\alpha}(x)
\frac{\delta  \swav{S}^{\lbrack 0 \rbrack} }
{\delta C_{\mu}^{\alpha}(x) }
+  L_{\alpha}(x)
\frac{\delta  \swav{S}^{\lbrack 0 \rbrack} }
{\delta  L_{\alpha}(x) } \right) \right.  \\
&&\\
&& \left.
+ \frac{\beta_{N+1}(\varepsilon)}{2}
 \left( \eta^{\alpha}(x)
\frac{\delta  \swav{S}^{\lbrack 0 \rbrack} }
{\delta \eta^{\alpha} }
+ \bar\eta_{\alpha}(x)
\frac{\delta  \swav{S}^{\lbrack 0 \rbrack} }
{\delta  \bar\eta_{\alpha}(x) }
 +  K^{\mu}_{\alpha}(x)
\frac{\delta  \swav{S}^{\lbrack 0 \rbrack} }
{\delta K^{\mu}_{\alpha}(x) }
\right)
\right\rbrack \\
&&\\
&&  + \frac{\alpha_{N+1}(\varepsilon)}{2}
g \frac{\partial  \swav{S}^{\lbrack 0 \rbrack} }{\partial g }
- 2 (\frac{\alpha_{N+1}(\varepsilon)}{2}
- \beta_{N+1}(\varepsilon) )
\alpha \frac{\partial \swav{S}^{\lbrack 0 \rbrack} }
{\partial \alpha} + o(\hbar^{N+2})
\end{array}
\eqno{(11.204)}
$$
We can see that this relation has just the form of first order functional
expansion. Using this relation, we can determine the form of
$\swav{S}^{\lbrack N+1 \rbrack}$. It is
$$
\begin{array}{l}
\swav{S}^{\lbrack N+1 \rbrack} \lbrack
C_{\mu}^{\alpha}, \bar\eta_{\alpha}, \eta^{\alpha},
K^{\mu}_{\alpha}, L_{\alpha}, g, \alpha,\eta_1,\eta_2
\rbrack   \\
\\
= \swav{S}^{\lbrack 0 \rbrack}
\lbrack
\sqrt{Z_1^{\lbrack N+1 \rbrack}}C_{\mu}^{\alpha},
\sqrt{Z_2^{\lbrack N+1 \rbrack}}\bar\eta_{\alpha},
\sqrt{Z_3^{\lbrack N+1 \rbrack}}\eta^{\alpha},
\sqrt{Z_4^{\lbrack N+1 \rbrack}}K^{\mu}_{\alpha}, \\
\\
~~~~~~~
\sqrt{Z_5^{\lbrack N+1 \rbrack}}L_{\alpha},
      Z_g^{\lbrack N+1 \rbrack} g,
      Z_{\alpha}^{\lbrack N+1 \rbrack}  \alpha,
      Z_6^{\lbrack N+1 \rbrack} \eta_1,
      Z_7^{\lbrack N+1 \rbrack}\eta_2
\rbrack,
\end{array}
\eqno{(11.205)}
$$
where
$$
\sqrt{Z_1^{\lbrack N+1 \rbrack}}
= \sqrt{Z_5^{\lbrack N+1 \rbrack}}
= \sqrt{Z_{\alpha}^{\lbrack N+1 \rbrack}}
= 1- \sum_{L=1}^{N+1}
\left( \frac{\alpha_L (\varepsilon)}{2}
-\beta_L (\varepsilon) \right),
\eqno{(11.206)}
$$
$$
\sqrt{Z_2^{\lbrack N+1 \rbrack}}
= \sqrt{Z_3^{\lbrack N+1 \rbrack}}
= \sqrt{Z_4^{\lbrack N+1 \rbrack}}
= 1- \sum_{L=1}^{N+1}
\frac{\beta_L (\varepsilon) }{2} ,
\eqno{(11.207)}
$$
$$
 Z_g^{\lbrack N+1 \rbrack}
= 1 + \sum_{L=1}^{N+1}
\frac{\alpha_L (\varepsilon)}{2},
\eqno{(11.208)}
$$
$$
 Z_6^{\lbrack N+1 \rbrack} = 1 ,
\eqno{(11.209)}
$$
$$
Z_7^{\lbrack N+1 \rbrack} = 1 .
\eqno{(11.210)}
$$
Denote
$$
\sqrt{Z^{\lbrack N+1 \rbrack}}
\define \sqrt{Z_1^{\lbrack N+1 \rbrack}}
= \sqrt{Z_5^{\lbrack N+1 \rbrack}}
= \sqrt{Z_{\alpha}^{\lbrack N+1 \rbrack}},
\eqno{(11.211)}
$$
$$
  \sqrt{\swav{Z}^{\lbrack N+1 \rbrack}}
\define \sqrt{Z_2^{\lbrack N+1 \rbrack}}
= \sqrt{Z_3^{\lbrack N+1 \rbrack}}
= \sqrt{Z_4^{\lbrack N+1 \rbrack}}.
\eqno{(11.212)}
$$
The eq.(11.205) is changed into
$$
\begin{array}{l}
\swav{S}^{\lbrack N+1 \rbrack} \lbrack
C_{\mu}^{\alpha}, \bar\eta_{\alpha}, \eta^{\alpha},
K^{\mu}_{\alpha}, L_{\alpha}, g, \alpha,\eta_1,\eta_2
\rbrack   \\
\\
= \swav{S}^{\lbrack 0 \rbrack}
\lbrack
\sqrt{Z^{\lbrack N+1 \rbrack}}C_{\mu}^{\alpha},
\sqrt{\swav{Z}^{\lbrack N+1 \rbrack}}\bar\eta_{\alpha},
\sqrt{\swav{Z}^{\lbrack N+1 \rbrack}}\eta^{\alpha},
\sqrt{\swav{Z}^{\lbrack N+1 \rbrack}}K^{\mu}_{\alpha},\\
\\~~~~~~~
\sqrt{Z^{\lbrack N+1 \rbrack}}L_{\alpha},
      Z_g^{\lbrack N+1 \rbrack} g,
      Z^{\lbrack N+1 \rbrack} \alpha,
      \eta_1,\eta_2
\rbrack.
\end{array}
\eqno{(11.213)}
$$
Using eq.(11.213), we can easily prove that
$$
\begin{array}{l}
\bar{\Gamma}^{\lbrack N+1 \rbrack} \lbrack
C_{\mu}^{\alpha}, \bar\eta_{\alpha}, \eta^{\alpha},
K^{\mu}_{\alpha}, L_{\alpha}, g, \alpha,\eta_1,\eta_2
\rbrack   \\
\\
= \bar{\Gamma}^{\lbrack 0 \rbrack}
\left\lbrack
\sqrt{Z^{\lbrack N+1 \rbrack}}C_{\mu}^{\alpha},
\sqrt{\swav{Z}^{\lbrack N+1 \rbrack}}\bar\eta_{\alpha},
\sqrt{\swav{Z}^{\lbrack N+1 \rbrack}}\eta^{\alpha},\right.\\
\\~~~~~~~\left.
\sqrt{\swav{Z}^{\lbrack N+1 \rbrack}}K^{\mu}_{\alpha},
\sqrt{Z^{\lbrack N+1 \rbrack}}L_{\alpha},
        Z_g^{\lbrack N+1 \rbrack} g,
    Z^{\lbrack N+1 \rbrack} \alpha,
    ,\eta_1,\eta_2
\right\rbrack.
\end{array}
\eqno{(11.214)}
$$
$\eta^{\mu\nu}$ is the Minkowski metric of flat space-time. The functions
of $\eta^{\mu}_{1\alpha}$ and $\eta_{2\alpha\beta}$ are also
similar to those of  metric, which
is used to contract Lorentz indexes and group indexes. From eq.(11.214),
we can see the renormalization of the theory does not affect their value.
It means that normalization of the theory does not affect the space-time
structure, which is consistent with our basic viewpoint that physical
space-time in the gravitational gauge theory is always flat.
\\

Now, we need to prove that all inductive hypotheses hold at $L=N+1$. The main
inductive hypotheses which is used in the above proof are the following three:
when $L=N$, the following three hypotheses hold,
\begin{enumerate}

\item the lowest divergence of $\bar\Gamma^{\lbrack N \rbrack}$ appears
in the $(N+1)$-loop diagram;

\item $\bar\Gamma^{\lbrack N \rbrack}$ satisfies Ward-Takahashi identities
eqs.(11.106-107);

\item after $\hbar^N th$ order renormalization, the action of the system
has the form of eq.(11.196).

\end{enumerate}

First, let's see the first hypothesis. According to eq.(11.194), after
introducing $(N+1) th$ order counterterm, the $(N+1)$-loop diagram
contribution of $\bar\Gamma^{\lbrack N+1 \rbrack}$ is finite. It means
that the lowest order divergence of $\bar\Gamma^{\lbrack N+1 \rbrack}$
appears in the $(N+2)$-loop diagram. So, the first inductive
hypothesis hold when $L=N+1$.\\

It can be proved that the renormalization constants
$Z_g^{\lbrack N \rbrack}$, $Z^{\lbrack N \rbrack}$ and
$\swav{Z}^{\lbrack N \rbrack}$ satisfy the following relation,
$$
Z_g^{\lbrack N \rbrack} \swav{Z}^{\lbrack N \rbrack}
\sqrt{Z^{\lbrack N \rbrack}} = 1,
\eqno{(11.215)}
$$
where $N$ is an arbitrary non-negative number. It is known that the
non-renormalized generating functional of regular vertex
$$
\bar{\Gamma}^{\lbrack 0 \rbrack} =
\bar{\Gamma}^{\lbrack 0 \rbrack}
\lbrack C, \bar\eta, \eta, K, L ,g,\alpha,\eta_1,\eta_2 \rbrack
\eqno{(11.216)}
$$
satisfies Ward-Takahsshi identities eqs.(11.101-102). If we define
$$
\bar{\Gamma} ' =
\bar{\Gamma}^{\lbrack 0 \rbrack}
\lbrack C', \bar\eta', \eta', K', L' ,g',\alpha',
\eta_1,\eta_2 \rbrack,
\eqno{(11.217)}
$$
then, it must satisfy the following Ward-Takahashi identities
$$
\partial^{\mu}
\frac{\delta \bar{\Gamma}'  }
{\delta K^{\prime\mu}_{\alpha}(x) }
= \frac{\delta \bar{\Gamma} '  }
{\delta \bar\eta'_{\alpha}(x) },
\eqno{(11.218)}
$$
$$
\int {\rm d}^4x \left\lbrace
\frac{\delta \bar{\Gamma} ' }
{\delta K^{\prime\mu}_{\alpha}(x) }
\frac{\delta \bar{\Gamma} ' }
{\delta C_{\mu}^{\prime\alpha}(x) }
+ \frac{\delta \bar{\Gamma} ' }
{\delta L'_{\alpha}(x) }
\frac{\delta \bar{\Gamma} ' }
{\delta \eta^{\prime\alpha}(x) }
+ L_{1 \alpha}^{\prime\mu}(x)
\frac{\delta \bar{\Gamma} '}{\delta \eta^{\prime\mu}_{1 \alpha}(x)}
+ L'_{2 \alpha\beta}(x)
\frac{\delta \bar{\Gamma} '}{\delta \eta'_{2 \alpha\beta}(x)}
\right\rbrace = 0,
\eqno{(11.219)}
$$
where
$$
L_{1 \alpha}^{\prime\mu}
= g' \eta^{\mu}_{1\sigma} (\partial_{\alpha} \eta^{\prime\sigma}),
\eqno{(11.220)}
$$
$$
L'_{2 \alpha\beta} = g' \lbrack
\eta_{2\alpha\sigma} (\partial_{\beta} \eta^{\prime\sigma})
+\eta_{2\sigma\beta} (\partial_{\alpha} \eta^{\prime\sigma}) \rbrack.
\eqno{(11.221)}
$$
Set,
$$
C_{\mu}^{\prime\alpha}
= \sqrt{Z^{\lbrack N+1 \rbrack}} C_{\mu}^{\alpha},
\eqno{(11.222)}
$$
$$
K^{\prime\mu}_{\alpha}
= \sqrt{\swav{Z}^{\lbrack N+1 \rbrack}} K^{\mu}_{\alpha},
\eqno{(11.223)}
$$
$$
L'_{\alpha}
= \sqrt{Z^{\lbrack N+1 \rbrack}} L_{\alpha},
\eqno{(11.224)}
$$
$$
\eta^{\prime\alpha}
= \sqrt{\swav{Z}^{\lbrack N+1 \rbrack}} \eta^{\alpha},
\eqno{(11.225)}
$$
$$
\bar\eta'_{\alpha}
= \sqrt{\swav{Z}^{\lbrack N+1 \rbrack}} \bar\eta_{\alpha},
\eqno{(11.226)}
$$
$$
g' = Z_g^{\lbrack N+1 \rbrack} g,
\eqno{(11.227)}
$$
$$
\alpha' = Z^{\lbrack N+1 \rbrack} \alpha.
\eqno{(11.228)}
$$
In this case, we have
$$
L_{1 \alpha}^{\prime\mu} = \sqrt{\swav{Z}^{\lbrack N+1 \rbrack}}
Z_g^{\lbrack N+1 \rbrack} L_{1 \alpha}^{\mu},
\eqno{(11.229)}
$$
$$
L'_{2 \alpha\beta} =\sqrt{\swav{Z}^{\lbrack N+1 \rbrack}}
Z_g^{\lbrack N+1 \rbrack} L_{2 \alpha\beta},
\eqno{(11.230)}
$$
$$
\begin{array}{rcl}
\bar\Gamma' &=& \bar{\Gamma}^{\lbrack 0 \rbrack}
\left\lbrack
\sqrt{Z^{\lbrack N+1 \rbrack}}C_{\mu}^{\alpha},
\sqrt{\swav{Z}^{\lbrack N+1 \rbrack}}\bar\eta_{\alpha},
\sqrt{\swav{Z}^{\lbrack N+1 \rbrack}}\eta^{\alpha},\right.\\
&&\\
&&~~~~~~~\left.
\sqrt{\swav{Z}^{\lbrack N+1 \rbrack}}K^{\mu}_{\alpha},
\sqrt{Z^{\lbrack N+1 \rbrack}}L_{\alpha},
        Z_g^{\lbrack N+1 \rbrack} g,
    Z^{\lbrack N+1 \rbrack} \alpha,
    ,\eta_1,\eta_2
\right\rbrack\\
&&\\
&=& \bar\Gamma^{\lbrack N+1 \rbrack}.
\end{array}
\eqno{(11.231)}
$$
Then eq.(11.218) is changed into
$$
\frac{1}{\sqrt{\swav{Z}^{\lbrack N+1 \rbrack}}}
\partial^{\mu}
\frac{\delta \bar{\Gamma}^{\lbrack N+1 \rbrack}  }
{\delta K^{\mu}_{\alpha}(x) }
= \frac{1}{\sqrt{\swav{Z}^{\lbrack N+1 \rbrack}}}
\frac{\delta \bar{\Gamma}^{\lbrack N+1 \rbrack}  }
{\delta \bar\eta_{\alpha}(x) }.
\eqno{(11.232)}
$$
Because $\frac{1}{\sqrt{\swav{Z}^{\lbrack N+1 \rbrack}}}$ does not
vanish, the above equation gives out
$$
\partial^{\mu}
\frac{\delta \bar{\Gamma}^{\lbrack N+1 \rbrack}  }
{\delta K^{\mu}_{\alpha}(x) }
=\frac{\delta \bar{\Gamma}^{\lbrack N+1 \rbrack}  }
{\delta \bar\eta_{\alpha}(x) }.
\eqno{(11.233)}
$$
Eq.(11.219) gives out
$$
\begin{array}{l}
\int {\rm d}^4x \left\lbrace
\frac{1}{\sqrt{\swav{Z}^{\lbrack N+1 \rbrack}}
\sqrt{Z^{\lbrack N+1 \rbrack}}}
\left\lbrack
\frac{\delta \bar{\Gamma}^{\lbrack N+1 \rbrack} }
{\delta K^{\mu}_{\alpha}(x) }
\frac{\delta \bar{\Gamma}^{\lbrack N+1 \rbrack} }
{\delta C_{\mu}^{\alpha}(x) }
+ \frac{\delta \bar{\Gamma}^{\lbrack N+1 \rbrack} }
{\delta L_{\alpha}(x) }
\frac{\delta \bar{\Gamma}^{\lbrack N+1 \rbrack} }
{\delta \eta^{\alpha}(x) }\right\rbrack\right.  \\
\\
\left.
+ \sqrt{\swav{Z}^{\lbrack N+1 \rbrack}}
Z_g^{\lbrack N+1 \rbrack} \left\lbrack
 L_{1 \alpha}^{\mu}(x)
\frac{\delta \bar{\Gamma}^{\lbrack N+1 \rbrack}}
{\delta \eta^{\mu}_{1 \alpha}(x)}
+ L_{2 \alpha\beta}(x)
\frac{\delta \bar{\Gamma}^{\lbrack N+1 \rbrack}}
{\delta \eta_{2 \alpha\beta}(x)}
\right\rbrack \right\rbrace = 0.
\end{array}
\eqno{(11.234)}
$$
Using eq.(11.215), we will get
$$
\int {\rm d}^4x \left\lbrace
\frac{\delta \bar{\Gamma}^{\lbrack N+1 \rbrack} }
{\delta K^{\mu}_{\alpha}(x) }
\frac{\delta \bar{\Gamma}^{\lbrack N+1 \rbrack} }
{\delta C_{\mu}^{\alpha}(x) }
+ \frac{\delta \bar{\Gamma}^{\lbrack N+1 \rbrack} }
{\delta L_{\alpha}(x) }
\frac{\delta \bar{\Gamma}^{\lbrack N+1 \rbrack} }
{\delta \eta^{\alpha}(x) }
+ L_{1 \alpha}^{\mu}(x)
\frac{\delta \bar{\Gamma}^{\lbrack N+1 \rbrack}}
{\delta \eta^{\mu}_{1 \alpha}(x)}
+ L_{2 \alpha\beta}(x)
\frac{\delta \bar{\Gamma}^{\lbrack N+1 \rbrack}}
{\delta \eta_{2 \alpha\beta}(x)}
 \right\rbrace = 0.
\eqno{(11.235)}
$$
Eq.(11.233) and eq.(11.235) are just the Ward-Takahashi identities
for $L=N+1$. Therefore, the second inductive hypothesis holds when
$L=N+1$. \\

The third inductive hypothesis has already been proved which is shown
in eq.(11.213). Therefore, all three inductive hypothesis hold when
$L=N+1$. According to inductive principle, they will hold when $L$ is
an arbitrary non-negative number, especially they hold when $L$
approaches infinity. \\

In above discussions, we have proved that, if we suppose that
when $L=N$ eq.(11.196) holds, then it also holds when $L=N+1$.
According to inductive principle, we know that eq.(11.213-214)
hold for any positive integer $N$.  When $N$ approaches infinity,
we get
$$
\begin{array}{l}
\swav{S} \lbrack
C_{\mu}^{\alpha}, \bar\eta_{\alpha}, \eta^{\alpha},
K^{\mu}_{\alpha}, L_{\alpha}, g, \alpha,\eta_1,\eta_2
\rbrack   \\
\\
= \swav{S}^{\lbrack 0 \rbrack}
\left\lbrack
\sqrt{Z}C_{\mu}^{\alpha},
\sqrt{\swav{Z}}\bar\eta_{\alpha},
\sqrt{\swav{Z}}\eta^{\alpha},
\sqrt{\swav{Z}}K^{\mu}_{\alpha},
\sqrt{Z}L_{\alpha},
      Z_g g,
      Z \alpha,\eta_1,\eta_2
\right\rbrack,
\end{array}
\eqno{(11.236)}
$$
$$
\begin{array}{l}
\bar{\Gamma} \lbrack
C_{\mu}^{\alpha}, \bar\eta_{\alpha}, \eta^{\alpha},
K^{\mu}_{\alpha}, L_{\alpha}, g, \alpha,\eta_1,\eta_2
\rbrack   \\
\\
= \bar{\Gamma}^{\lbrack 0 \rbrack}
\left\lbrack
\sqrt{Z}C_{\mu}^{\alpha},
\sqrt{\swav{Z}}\bar\eta_{\alpha},
\sqrt{\swav{Z}}\eta^{\alpha},
\sqrt{\swav{Z}}K^{\mu}_{\alpha},
\sqrt{Z}L_{\alpha},
      Z_g  g,
      Z \alpha,\eta_1,\eta_2
\right\rbrack,
\end{array}
\eqno{(11.237)}
$$
where
$$
\sqrt{Z} =  \lim_{N \to \infty}
 \sqrt{Z^{\lbrack N \rbrack}}
= 1- \sum_{L=1}^{\infty}
\left( \frac{\alpha_L (\varepsilon)}{2}
-\beta_L (\varepsilon) \right),
\eqno{(11.238)}
$$
$$
\sqrt{\swav{Z}}
= \lim_{N \to \infty} \sqrt{\swav{Z}^{\lbrack N \rbrack}}
= 1- \sum_{L=1}^{\infty}
\frac{\beta_L (\varepsilon) }{2} ,
\eqno{(11.239)}
$$
$$
Z_g = \lim_{N \to \infty}
Z_g^{\lbrack N \rbrack}
= 1 + \sum_{L=1}^{\infty}
\frac{\alpha_L (\varepsilon)}{2}.
\eqno{(11.240)}
$$
$\swav{S} \lbrack C_{\mu}^{\alpha}, \bar\eta_{\alpha},
\eta^{\alpha}, K^{\mu}_{\alpha}, L_{\alpha}, g,
\alpha ,\eta_1,\eta_2 \rbrack$
and $\bar{\Gamma} \lbrack C_{\mu}^{\alpha}, \bar\eta_{\alpha},
\eta^{\alpha}, K^{\mu}_{\alpha}, L_{\alpha},
g, \alpha ,\eta_1,\eta_2 \rbrack$
are renormalized action and generating functional of regular vertex.
The generating functional of regular vertex $\bar{\Gamma}$ contains
no divergence. All kinds of vertex that generated from $\bar{\Gamma}$
are finite. From eq.(11.236) and eq.(11.237), we can see that we only
introduce three unknown parameters which are $\sqrt{Z}$, $\sqrt{\swav{Z}}$
and $Z_g$. Therefore, gravitational gauge theory is a renormalizable
theory. \\

We have already proved that the renormalized generating functional
of regular vertex satisfies Ward-Takahashi identities,
$$
\partial^{\mu}
\frac{\delta \bar{\Gamma}  }
{\delta K^{\mu}_{\alpha}(x) }
= \frac{\delta \bar{\Gamma}  }
{\delta \bar\eta_{\alpha}(x) },
\eqno{(11.241)}
$$
$$
\int {\rm d}^4x \left\lbrace
\frac{\delta \bar{\Gamma} }
{\delta K^{\mu}_{\alpha}(x) }
\frac{\delta \bar{\Gamma} }
{\delta C_{\mu}^{\alpha}(x) }
+ \frac{\delta \bar{\Gamma} }
{\delta L_{\alpha}(x) }
\frac{\delta \bar{\Gamma} }
{\delta \eta^{\alpha}(x) }
+ L_{1 \alpha}^{\mu}
\frac{\delta \bar{\Gamma}}
{\delta \eta^{\mu}_{1 \alpha}(x)}
+ L_{2 \alpha\beta}
\frac{\delta \bar{\Gamma}}
{\delta \eta_{2 \alpha\beta}(x)}
\right\rbrace = 0.
\eqno{(11.242)}
$$
It means that the renormalized theory has the structure of gauge symmetry.
If we define
$$
C_{0 \mu}^{\alpha} = \sqrt{Z} C_{\mu}^{\alpha},
\eqno{(11.243)}
$$
$$
\eta_0^{\alpha} = \sqrt{\swav{Z}} \eta^{\alpha},
\eqno{(11.244)}
$$
$$
\bar\eta_{0 \alpha} = \sqrt{\swav{Z}} \bar\eta_{\alpha},
\eqno{(11.245)}
$$
$$
K^{\mu}_{0 \alpha} = \sqrt{\swav{Z}} K^{\mu}_{\alpha},
\eqno{(11.246)}
$$
$$
L_{0 \alpha} = \sqrt{Z} L_{\alpha},
\eqno{(11.247)}
$$
$$
g_0 = Z_g g,
\eqno{(11.248)}
$$
$$
\alpha_0 = Z \alpha.
\eqno{(11.249)}
$$
Therefore, eqs.(11.236-237) are changed into
$$
\swav{S} \lbrack
C_{\mu}^{\alpha}, \bar\eta_{\alpha}, \eta^{\alpha},
K^{\mu}_{\alpha}, L_{\alpha}, g, \alpha,\eta_1,\eta_2
\rbrack
= \swav{S}^{\lbrack 0 \rbrack}
\lbrack
C_{0 \mu}^{\alpha}, \bar\eta_{0 \alpha}, \eta_0^{\alpha},
K^{\mu}_{0 \alpha}, L_{0 \alpha}, g_0, \alpha_0,\eta_1,\eta_2
\rbrack,
\eqno{(11.250)}
$$
$$
\bar{\Gamma} \lbrack
C_{\mu}^{\alpha}, \bar\eta_{\alpha}, \eta^{\alpha},
K^{\mu}_{\alpha}, L_{\alpha}, g, \alpha,\eta_1,\eta_2
\rbrack
= \bar{\Gamma}^{\lbrack 0 \rbrack}
\lbrack
C_{0 \mu}^{\alpha}, \bar\eta_{0 \alpha}, \eta_0^{\alpha},
K^{\mu}_{0 \alpha}, L_{0 \alpha}, g_0, \alpha_0,\eta_1,\eta_2
\rbrack.
\eqno{(11.251)}
$$
$C_{0 \mu}^{\alpha}$, $\bar\eta_{0 \alpha}$ and
$\eta_0^{\alpha}$ are renormalized wave function,
$K^{\mu}_{0 \alpha}$ and $L_{0 \alpha}$ are renormalized
external sources, and $g_0$ is the renormalized gravitational
coupling constant. \\

The action $\swav{S}$ which is given by eq.(11.250) is invariant
under the following generalized BRST transformations,
$$
\delta C_{0\mu}^{\alpha}
= -  {\mathbf D}_{0\mu~\beta}^{\alpha} \eta_0^{\beta}
\delta \lambda,
\eqno{(11.252)}
$$
$$
\delta \eta_0^{\alpha} = g_0 \eta_0^{\sigma}
(\partial_{\sigma}\eta_0^{\alpha}) \delta \lambda,
\eqno{(11.253)}
$$
$$
\delta \bar\eta_{0\alpha} = \frac{1}{\alpha_0}
\eta_{\alpha \beta} f_0^{\beta} \delta \lambda,
\eqno{(11.254)}
$$
$$
\delta \eta^{\mu \nu} = 0,
\eqno{(11.255)}
$$
$$
\delta \eta^{\mu}_{1\alpha } = - g_0  \eta^{\mu}_{1\sigma}
(\partial_{\alpha} \eta_0^{\sigma})  \delta \lambda,
\eqno{(11.256)}
$$
$$
\delta \eta_{2\alpha \beta} = - g_0
\left ( \eta_{2\alpha \sigma}
(\partial_{\beta} \eta_0^{\sigma})
+\eta_{2\sigma \beta}
(\partial_{\alpha} \eta_0^{\sigma}) \right ) \delta \lambda,
\eqno{(11.257)}
$$
where,
$$
{\mathbf D}^{\alpha}_{0 \mu \beta} =
\delta_{\beta}^{\alpha} \partial_{\mu}
- g_0 \delta^{\alpha}_{\beta} C^{\sigma}_{0 \mu} \partial_{\sigma}
+ g_0 (\partial_{\beta} C^{\alpha}_{0\mu}),
\eqno{(11.258)}
$$
$$
f_0^{\alpha} = \partial^{\mu} C_{0 \mu}^{\alpha}.
\eqno{(11.259)}
$$
Therefore, the normalized action has generalized BRST symmetry, which
means that the normalized theory has the structure of gauge theory.\\

\section{Theoretical Predictions}

The gravitational gauge field theory which is discussed in this
paper is renormalizable. Its transcendental  foundation is gauge
principle. Gravitational gauge interactions is completely determined
by gauge symmetry. In other words, the Lagrangian of the system
is completely determined by gauge symmetry. Anyone who is familiar
with traditional quantum gravity must have realized that gravitational
gauge theory is quite different from the traditional quantum gravity.
In this chapter, we plan to discuss some predictions of the theory
which is useful for experimental research and is useful for testing
the validity of the theory. Now, let's discuss them
one by one. \\

\begin{enumerate}

\item \lbrack {\bf Gravitational Wave} \rbrack
In gravitational gauge theory, the gravitational gauge field
is represented by $C_{\mu}$.  From the point
of view of quantum field theory, gravitational gauge field
$C_{\mu}$ is a vector field and it obeys dynamics of vector field.
In other words, gravitational wave is vector wave. Suppose that the
gravitational gauge field is very weak in vacuum, then in leading
order approximation, the equation of motion of gravitational wave
is
$$
\partial^{\mu} F_{0 \mu \nu}^{\alpha}= 0 ,
\eqno{(12.1)}
$$
where $F_{0 \mu \nu}^{\alpha}$ is given by eq.(5.9). If we set
$g C_{\mu}^{\alpha}$ equals zero, we can obtain eq.(12.1)
from eq.(4.39). Eq.(12.1) is very similar to the famous
Maxwell equation in vacuum. Define
$$
F^{\alpha}_{ij}= - \varepsilon_{ijk} B^{\alpha}_{k}
~~,~~
F^{\alpha}_{ 0i} = E^{\alpha}_{i},
\eqno{(12.2)}
$$
then eq.(12.1) is changed into
$$
\nabla \cdot \svec{E}^{\alpha}=0,
\eqno{(12.3)}
$$
$$
\frac{\partial}{\partial t} \svec{E}^{\alpha}
- \nabla \times \svec{B}^{\alpha} =0.
\eqno{(12.4)}
$$
From definitions eq.(12.2), we can prove that
$$
\nabla \cdot  \svec{B}^{\alpha} =0,
\eqno{(12.5)}
$$
$$
\frac{\partial}{\partial t} \svec{B}^{\alpha}
+ \nabla \times \svec{E}^{\alpha} =0.
\eqno{(12.6)}
$$
If there were no superscript $\alpha$, eqs.(12.3-6) would be
the ordinary Maxwell equations. In ordinary case, the
strength of gravitational field in vacuum is extremely weak,
so the gravitational wave in vacuum is composed of four independent
vector waves. \\

Though gravitational gauge field is a vector field, its
component fields $C_{\mu}^{\alpha}$ have one Lorentz index
$\mu$ and one group index $\alpha$. Both indexes have
the same behavior under Lorentz transformation. According to
the behavior of Lorentz transformation, gravitational field
likes a tensor field. We call it pseudo-tensor field. The
spin of a field is determined according to its behavior under
Lorentz transformation, so the spin of gravitational field
is 2. In conventional quantum field theory, spin-1 field is a vector
field, and vector field is a spin-1 field. In gravitational gauge field,
this correspondence is violated. The reason is that, in gravitational
gauge field theory, the group index contributes to the spin of a field,
while in ordinary gauge field theory, the group index do not
contribute to the spin of a field. In a word, gravitational field
is a spin-2 vector field.   \\

\item \lbrack {\bf Gravitational Magnetic Field} \rbrack
From eq.(12.3-6), we can see that the equations of motion of
gravitational wave in vacuum are quite similar to those of
electromagnetic wave. The phenomenological behavior of gravitational
wave must also be similar to that of electromagnetic wave. In
gravitational gauge theory, $\svec{B}^{\alpha}$ is called the
gravitational magnetic field. It will transmit gravitational magnetic
interactions between two rotating objects. In first order approximation,
the equation of motion of gravitational gauge field is
$$
\partial^{\mu} F_{\mu \nu}^{\alpha}
= - g \eta_{\nu \tau} \eta_2^{\alpha \beta}
T^{\tau}_{g \beta}.
\eqno{(12.7)}
$$
Using eq.(12.2) and eq.(12.7), we can get the following equations
$$
\nabla \cdot \svec{E}^{\alpha}=
- g \eta_2^{\alpha \beta} T^0_{g \beta},
\eqno{(12.8)}
$$
$$
\frac{\partial}{\partial t} \svec{E}^{\alpha}
- \nabla \times \svec{B}^{\alpha} =
+ g  \eta_2^{\alpha \beta} \svec{T}_{g \beta},
\eqno{(12.9)}
$$
where $ \svec{T}_{g \beta}$ is a simplified notation whose explicit
definition is given by the following relation
$$
( \svec{T}_{g \beta})^i =  \svec{T}^i_{g \beta}.
\eqno{(12.10)}
$$
On the other hand, it is easy to prove that(omit self
interactions of graviton)
$$
\partial_{\mu} F^{\alpha}_{\nu \lambda}
+ \partial_{\nu} F^{\alpha}_{\lambda \mu}
+ \partial_{\lambda} F^{\alpha}_{\mu \nu}
 = 0.
\eqno{(12.11)}
$$
From eq.(12.11), we can get
$$
\nabla \cdot  \svec{B}^{\alpha} =0,
\eqno{(12.12)}
$$
$$
\frac{\partial}{\partial t} \svec{B}^{\alpha}
+ \nabla \times \svec{E}^{\alpha} =0.
\eqno{(12.13)}
$$
Eq.(12.8) means that energy-momentum density of the system is
the source of gravitational electric fields, eq.(12.9) means that
time-varying gravitational electric fields give rise to gravitational
magnetic fields, and eq.(12.13) means that time-varying gravitational
magnetic fields give rise to gravitational electric fields.
Suppose that the angular momentum of an rotating object is $J_i$,
then there will be a coupling between angular momentum and
gravitational magnetic fields. The interaction Hamiltionian of this
coupling is proportional to
$(\frac{P_{\alpha}}{m}\!\svec{J} \cdot \svec{B}^{\alpha})$.
The existence of gravitational magnetic fields is important for
cosmology. It is known that almost all galaxies in the universe
rotate. The global rotation of galaxy will give rise to gravitational
magnetic fields in space-time. The existence of gravitational magnetic
fields will affect the moving of stars in (or near) the galaxy.
This influence contributes to the formation of the galaxy and
can explain why almost all galaxies have global large scale
structures. In other words, the gravitational magnetic fields
contribute great to the large scale structure of galaxy and
universe. \\

\item \lbrack {\bf Lorentz Force} \rbrack
There is a force when a particle is moving in a gravitational
magnetic field. In electromagnetic field theory, this force is usually
called Lorentz force. As an example, we discuss gravitational interactions
between gravitational field and Dirac field. Suppose that the gravitational
field is static.
According to eqs.(6.2-3), the interaction Lagrangian is
$$
{\cal L}_I = g e^{I(C)} \bar\psi
\gamma^{\mu} \partial_{\alpha} \psi C_{\mu}^{\alpha}.
\eqno{(12.14)}
$$
For Dirac field, the gravitational energy-momentum of Dirac field is
$$
T_{g \alpha}^{\mu} = \bar\psi
\gamma^{\mu} \partial_{\alpha} \psi.
\eqno{(12.15)}
$$
Substitute eq.(12.15) into eq.(12.14), we get
$$
{\cal L}_I = g e^{I(C)} T_{g \alpha}^{\mu} C_{\mu}^{\alpha}.
\eqno{(12.16)}
$$
The interaction Hamiltonian density ${\cal H}_I$ is
$$
{\cal H}_I = - {\cal L}_I  =
- g e^{I(C)} T_{g \alpha}^{\mu}(y,\svec{x}) C_{\mu}^{\alpha}(y).
\eqno{(12.17)}
$$
Suppose that the moving particle is a mass point at point $\svec{x} $,
in this case
$$
 T_{g \alpha}^{\mu}(y,\svec{x}) =  T_{g \alpha}^{\mu}
\delta(\svec{y} - \svec{x} ),
\eqno{(12.18)}
$$
where $ T_{g \alpha}^{\mu}$ is independent of space coordinates.
Then, the interaction Hamiltonian $H_I$ is
$$
H_I = \int {\rm d}^3 \svec{y} {\cal H}_I (y)
= - g \int {\rm d}^3 \svec{y}
e^{I(C)}  T_{g \alpha}^{\mu}(y,\svec{x}) C_{\mu}^{\alpha}(y).
\eqno{(12.19)}
$$
The gravitational force that acts on the mass point is
$$
f_i = g \int {\rm d}^3y e^{I(C)} T^{\mu}_{g \alpha}(y,\svec{x})
F^{\alpha}_{i \mu}
+g \int {\rm d}^3y e^{I(C)} T^{\mu}_{g \alpha}(y,\svec{x})
\frac{\partial}{\partial y^{\mu}} C_i^{\alpha}.
\eqno{(12.20)}
$$
For quasi-static system, if we omit higher order contributions, the
second term in the above relation vanish.
For mass point, using the technique of Lorentz covariance analysis,
we can proved that
$$
P_{g \alpha} U^{\mu} = \gamma T_{g \alpha}^{\mu},
\eqno{(12.21)}
$$
where $U^{\mu}$ is velocity, $\gamma$ is the rapidity, and
$P_{g \alpha}$ is the gravitational energy-momentum. According
eq.(12.18), $P_{g \alpha}$ is given by
$$
P_{g \alpha} = \int {\rm d}^3 \svec{y}
 T_{g \alpha}^0(y) =  T_{g \alpha}^0.
\eqno{(12.22)}
$$
Using all these relations and eq.(12.2), we get
$$
\svec{f} = -g e^{I(C)} P_{g \alpha} \svec{E}^{\alpha}
- g e^{I(C)} P_{g \alpha} \svec{v} \times \svec{B}^{\alpha}.
$$
For quasi-static system, the dominant contribution of the above
relation is
$$
\svec{f} = g e^{I(C)} M \svec{E}^0
+ g e^{I(C)} M \svec{v} \times \svec{B}^0,
\eqno{(12.23)}
$$
where $\svec{v}= \svec{U}/\gamma$ is the velocity of the mass point.
The first term of eq.(12.23) is the classical Newton's gravitational
interactions. The second term of eq.(12.23) is the Lorentz force. The
direction of this force is perpendicular to the direction of the motion
of the mass point. When
the mass point is at rest or is moving along the direction
of the gravitational magnetic field, this force vanishes.
Lorentz force is important for cosmology, because the rotation
of galaxy will generate gravitational magnetic field and this gravitational
magnetic field will affect the motion of stars and affect the large scale
structure of galaxy. \\

\item \lbrack {\bf Origin of Terrestrial Magnetism} \rbrack

It the traditional theory, it is hard to explain the origin of terrestrial
magnetism. But now, using the gravitational gauge theory, it is easy to
explain the origin of terrestrial magnetism. According to above discussions,
we know that a rotating celestial object will give rise to gravitational
magnetic field, so rotating earth will generate gravitational magnetic
field around it. According to unified gravitation-electromagnetic
theory\cite{7}, there is direct coupling between spin and gravitational
magnetic field. In other words, electromagnetic magnet will
directly interact with
gravitational magnetic field. Therefore, the origin of terrestrial
magnetism is gravitational magnetic field. In other words, terrestrial
magnetism and solar magnetism are not electromagnetic magnetism, but
gravitational magnetism. \\

\item \lbrack {\bf Negative Energy} \rbrack
First, let's discuss inertial energy of pure gravitational wave.
Suppose that the gravitational wave is not so strong, so the higher
order contribution is very small. We only consider leading order
contribution here. For pure gravitational field, we have
$$
\frac{\partial {\cal L}_0}{\partial \partial_{\mu} C_{\nu}^{\beta}}
= - \eta^{\mu \rho} \eta^{\nu \sigma} \eta_{2 \beta \gamma}
F_{\rho \sigma}^{\gamma}
+ g \eta^{\lambda \rho} \eta^{\nu \sigma} \eta_{2 \beta \gamma}
C^{\mu}_{\lambda} F^{\gamma}_{\rho \sigma}.
\eqno{(12.24)}
$$
From eq.(4.32), we can get the inertial energy-momentum tensor
of gravitational field in the leading order approximation, that is
$$
T^{\mu}_{i \alpha} = e^{I(C)}
\lbrack
+ \eta^{\mu \rho} \eta^{\nu \sigma} \eta_{2 \beta \gamma}
F_{\rho \sigma}^{\gamma} \partial_{\alpha} C_{\nu}^{\beta}
+ \delta^{\mu}_{\alpha}{\cal L}_0
\rbrack.
\eqno{(12.25)}
$$
Using eq.(12.2), Lagrangian given by eq.(4.20) can be changed into
$$
{\cal L}_0 = \frac{1}{2}
\eta_{2 \alpha \beta}
( \svec{E}^{\alpha} \cdot \svec{E}^{\beta}
- \svec{B}^{\alpha} \cdot \svec{B}^{\beta} ).
\eqno{(12.26)}
$$
Space integral of time component of inertial energy-momentum tensor
gives out inertial energy $H_i$ and inertial momentum $\svec{P}_i$.
They are
$$
H_i = \int {\rm d}^3 \svec{x}
e^{I(C)} \left\lbrack  \frac{1}{2}
\eta_{2 \alpha \beta}
( \svec{E}^{\alpha} \cdot \svec{E}^{\beta}
+ \svec{B}^{\alpha} \cdot \svec{B}^{\beta} )
\right\rbrack,
\eqno{(12.27)}
$$
$$
\svec{P}_i =   \int {\rm d}^3 \svec{x} e^{I(C)}
\eta_{2 \alpha \beta}  \svec{E}^{\alpha} \times \svec{B}^{\beta}.
\eqno{(12.28)}
$$
In order to obtain eq.(12.27), eq.(12.3) is used.
Let consider the inertial energy-momentum of gravitaional
field $C_{\mu}^0$. Because,
$$
\eta_{2 0 0} = -1,
\eqno{(12.29)}
$$
eq.(12.27) gives out
$$
H_i (C^0) = - \frac{1}{2} \int {\rm d}^3 \svec{x}
e^{I(C)} ( \svec{E}^0 \cdot \svec{E}^0
+\svec{B}^0 \cdot \svec{B}^0 ).
\eqno{(12.30)}
$$
$H_i(C^0)$ is a negative quantity. It means that the inertial energy
of gravitational field $C_{\mu}^0$ is negative. The gravitational
energy-momentum of pure gravitational gauge field is given by
eq.(4.40). In leading order approximation, it is
$$
\begin{array}{rcl}
T_{g \alpha}^{\mu} & =&
\eta^{\mu \rho} \eta^{\nu \sigma} \eta_{2 \beta \gamma}
F_{\rho \sigma}^{\gamma} ( \partial_{\alpha} C_{\nu}^{\beta} )
+ \eta^{\mu}_{1 \alpha} {\cal L}_0 \\
&&\\
&&- \eta^{\lambda \rho} \eta^{\mu \sigma} \eta_{2 \alpha \beta }
\partial_{\nu} ( C_{\lambda}^{\nu} F_{\rho \sigma}^{\beta} )
+ \eta^{\nu \rho} \eta^{\mu \sigma} \eta_{2 \alpha \beta }
\eta^{\lambda}_{1 \tau}
( \partial_{\nu} C_{\lambda}^{\tau}) F_{\rho \sigma}^{\beta}.
\end{array}
\eqno{(12.31)}
$$
After omitting surface terms,
the gravitational energy of the system is
$$
H_g  =  \int {\rm d}^3 \svec{x}
\left\lbrack  \frac{1}{2}  \eta_{2 \alpha \beta }
( \svec{E}^{\alpha} \cdot \svec{E}^{\beta}
+ \svec{B}^{\alpha} \cdot \svec{B}^{\beta} )
 - \eta^{i j } \partial_{0} (C_j^0 E_i^0 )
\right\rbrack.
\eqno{(12.32)}
$$
The gravitational energy  of gravitational field $C_{\mu}^0$ is,
$$
H_g (C^0) = - \frac{1}{2} \int {\rm d}^3 \svec{x}
( \svec{E}^0 \cdot \svec{E}^0
+ \svec{B}^0 \cdot \svec{B}^0
+ 2 \eta^{i j } \partial_{0}  ( C_j^0 E_i^0)) .
\eqno{(12.33)}
$$
$H_g$ is also negative. It means that the gravitational energy of
gravitational field $C_{\mu}^0$ is negative. In other words, gravitational
gauge field $C_{\mu}^0$ has negative gravitational energy and
negative inertial energy. But,
inertial mass is not equivalent to gravitational mass for pure
gravitational gauge field.
\\

\item \lbrack {\bf Gravitational Radiation} \rbrack
Because gravitational gauge field $C_{\mu}$ is a vector field, its dominant
radiation is gravitational dipole radiation. Now, let's discuss
gravitational dipole radiation. The equation of motion of gravitational
gauge field is
$$
\partial^{\mu} F_{\mu \nu}^{\alpha}
= - g \eta_{\nu \tau} \eta_2^{\alpha \beta}
T_{g \beta}^{\tau}.
\eqno{(12.34)}
$$
For the sake of simplicity, suppose that the strength of gravitational
gauge field is weak, i.e.
$$
g C_{\mu}^{\alpha} \ll 1.
\eqno{(12.35)}
$$
Then in leading order approximation, eq.(12.34) gives out
$$
\partial^{\mu} (\partial_{\mu} C_{\nu}^{\alpha}
- \partial_{\nu} C_{\mu}^{\alpha} )
= - g \eta_{\nu \tau} \eta_2^{\alpha \beta}
T_{g \beta}^{\tau}.
\eqno{(12.36)}
$$
If we adopt Lorentz gauge
$$
\partial^{\mu} C_{\mu}^{\alpha} = 0,
\eqno{(12.37)}
$$
then eq.(12.36) is changed into
$$
\partial^{\mu} \partial_{\mu} C_{\nu}^{\alpha}
= - g \eta_{\nu \tau} \eta_2^{\alpha \beta}
T_{g \beta}^{\tau}.
\eqno{(12.38)}
$$
$T_{g \beta}^{\tau}$ is a function of space-time,
$$
T_{g \beta}^{\tau} = T_{g \beta}^{\tau}(\svec{x},t).
\eqno{(12.39)}
$$
The solution to eq.(12.38) is
$$
C_{\nu}^{\alpha}(\svec{x},t)
= g \eta_{\nu \tau} \eta_2^{\alpha \beta}
\int \frac{T_{g \beta}^{\tau}(\svec{y},t-r)}{4 \pi r}
{\rm d}^3 \svec{y},
\eqno{(12.40)}
$$
where $r$ is the distance between point $\svec{x}$ and point
$\svec{y}$,
$$
r = | \svec{x} - \svec{y} |.
\eqno{(12.41)}
$$
Suppose that the object is a mass point. Then in the center of mass
system of the moving particle, the solution is given by eq.(9.10), i.e.
$$
C_0^0 = \frac{g}{4 \pi r} P_{g 0}.
\eqno{(12.42)}
$$
Make a Lorentz transformation, we can get the corresponding solution
in laboratory system,
$$
C_{\mu}^{\alpha} =
\frac{g}{4 \pi M \gamma} \eta_2^{\alpha \beta}
P_{g \mu} P_{g \beta} \frac{1}{r - \svec{v} \cdot \svec{r}},
\eqno{(12.43)}
$$
where $M$ is the gravitational mass of the mass point, $\svec{v}$
is the velocity of the mass point. In the above function, all variables
are functions of $t' = t - r$. Suppose that the velocity of the mass
point is much less that the speed of light, then we can obtain that
$$
\svec{B}^{\alpha} =
-\frac{g}{4 \pi r^2}
(\frac{d}{dt}( P_g^{\alpha} \svec{v} )) \times \svec{r},
\eqno{(12.44)}
$$
$$
\svec{E}^{\alpha} =
- \frac{g}{4 \pi r^3}P_g^{\alpha} \svec{r}
- \frac{g}{4 \pi r} \tdot{P}_g^{\alpha}
( \frac{\svec{r}}{r} - \svec{v} )
-  \frac{g}{4 \pi r^3}P_g^{\alpha}
\svec{r} \times (\svec{r} \times \tdot{\svec{v}}  ).
\eqno{(12.45)}
$$
The first term in eq.(12.45) is the traditional Newton's gravitational
field, which has no contribution on gravitational wave radiation. We will
omit it in the calculation of radiation power, i.e., we use the following
relation in the calculation of radiation power,
$$
\svec{E}^{\alpha} =
- \frac{g}{4 \pi r} \tdot{P}_g^{\alpha}
( \frac{\svec{r}}{r} - \svec{v} )
-  \frac{g}{4 \pi r^3}P_g^{\alpha}
\svec{r} \times (\svec{r} \times \tdot{\svec{v}}  ).
\eqno{(12.46)}
$$
\\

In order to calculate the radiation power, let's first determine the
radiation energy flux density. Suppose that there is a space region
which is denoted as $\Sigma$ whose surface is denoted as $\Omega$. The
gravitational force density is denoted as $\svec{f}(x)$ and the speed
of the mass point at point $\svec{x}$ is $\svec{v} (x)$. Then in one
unit time, the work that the system obtained from gravitational field
is
$$
\int_{\Sigma} \svec{f} \cdot \svec{v} {\rm d}^3 \svec{x}.
\eqno{(12.47)}
$$
Suppose that $w(x)$ is the energy density of the system. Then is
one unit time, the increased energy of the system is
$$
\frac{\rm d}{{\rm d} t}
\int_{\Sigma} w(x) {\rm d}^3 \svec{x}.
\eqno{(12.48)}
$$
Supposed that the energy flux density is $\svec{S}$. Then in one unit
time, the energy that flow into the system through the surface of the
system is
$$
- \oint_{\Omega} \svec{S} \cdot {\rm d} \svec{\sigma}
= - \int_{\Sigma} (\nabla \cdot \svec{S}) {\rm d}^3 \svec{x}.
\eqno{(12.49)}
$$
Energy conservation law gives out the following equation
$$
- \oint_{\Omega} \svec{S} \cdot {\rm d} \svec{\sigma}
= \int_{\Sigma} \svec{f} \cdot \svec{v} {\rm d}^3 \svec{x}
+ \frac{\rm d}{{\rm d} t}
\int_{\Sigma} w(x) {\rm d}^3 \svec{x}.
\eqno{(12.50)}
$$
Because integration region is an arbitrary region, from eq.(12.50),
we can obtain
$$
\nabla \cdot \svec{S} + \frac{\partial w}{\partial t}
= -  \svec{f} \cdot \svec{v}.
\eqno{(12.51)}
$$
For the sake of simplicity, suppose that the gravitational gauge
field is very weak, then in leading order approximation,
space-time derivative of $e^{I(C)}$ can be neglected.
Using eq.(12.23), we get
$$
\svec{f} \cdot \svec{v} =
-g e^{I(C)} T^i_{g \alpha} E_i^{\alpha}.
\eqno{(12.52)}
$$
Using eq.(12.9) and eq.(12.13), we get
$$
\begin{array}{rcl}
\svec{f} \cdot \svec{v} & = &
- \nabla \cdot (e^{I(C)} \svec{E}^{\alpha} \times \svec{B}_{\alpha} )
- \frac{\partial}{\partial t}
\lbrack \frac{1}{2} e^{I(C)} ( \svec{E}^{\alpha} \cdot \svec{E}_{\alpha}
+  \svec{B}^{\alpha} \cdot \svec{B}_{\alpha}  )
\rbrack  .
\end{array}
\eqno{(12.53)}
$$
Compare eq.(12.53) with eq.(12.51), we will get
$$
\svec{S} = e^{I(C)} \svec{E}^{\alpha} \times \svec{B}_{\alpha},
\eqno{(12.54)}
$$
$$
w = \frac{1}{2} e^{I(C)} ( \svec{E}^{\alpha} \cdot \svec{E}_{\alpha}
+  \svec{B}^{\alpha} \cdot \svec{B}_{\alpha}  ).
\eqno{(12.55)}
$$
$\svec{S}$ is the gravitational energy flux density. We use it to calculate
the gravitational radiation power. Using eq.(12.44) and eq.(12.46),
we can get
$$
\begin{array}{rcl}
\svec{S} &=& \frac{g^2}{16 \pi^2 r^3} e^{I(C)} \tdot{P}_g^{\alpha}
\lbrack
\frac{\rm d}{{\rm d} t}(P_{g \alpha} \svec{v} )
( r - \svec{r} \cdot \svec{v}  ) \\
&&\\
&& - \svec{r} \tdot{P}_{g \alpha}
( \frac{\svec{r} \cdot \svec{v}}{r} - v^2 )
-  \svec{r} P_{g \alpha}
( \frac{\svec{r} \cdot \tdot{\svec{v}}}{r}
- \svec{v} \cdot \tdot{\svec{v}} ) \rbrack \\
&&\\
&&+ \frac{g^2}{16 \pi^2 r^5} e^{I(C)}\lbrack
P_{g \alpha} P_g^{\alpha} \svec{r}
( r^2 \tdot{v}^2 - (\svec{r} \cdot \tdot{\svec{v}})^2 ) \\
&&\\
&& + P_{g \alpha} \tdot{P}_g^{\alpha} \svec{r}
( r^2 \tdot{\svec{v}} \cdot \svec{v}
- (\svec{r} \cdot \tdot{\svec{v}}) (\svec{r} \cdot \svec{v}) )
\rbrack  .
\end{array}
\eqno{(12.56)}
$$
The above relation can be written into another form
$$
\begin{array}{rcl}
\svec{S} &=& \frac{g^2}{16 \pi^2 r^2} e^{I(C)} P_g^{\alpha}
\lbrack \tdot{\svec{v}} \cdot \frac{{\rm d}}{{\rm d}t}
( P_{g \alpha} \svec{v} ) - ( \svec{n} \cdot \tdot{\svec{v}})
(\svec{n} \cdot \frac{{\rm d}}{{\rm d}t} ( P_{g \alpha} \svec{v} ))
\rbrack \svec{n} \\
&&\\
&& -  \frac{g^2}{16 \pi^2 r^2} e^{I(C)} \tdot{P}_g^{\alpha}
\lbrack (\svec{n} \cdot \svec{v} -1 )
\frac{{\rm d}}{{\rm d}t} ( P_{g \alpha} \svec{v} )
- \svec{n} \lbrack (\svec{v} - \svec{n}) \cdot
\frac{{\rm d}}{{\rm d}t} ( P_{g \alpha} \svec{v} )
\rbrack \rbrack,
\end{array}
\eqno{(12.57)}
$$
where
$$
\svec{n} = \frac{\svec{r}}{r}.
\eqno{(12.58)}
$$

\item \lbrack {\bf Repulsive Force} \rbrack
The classical gravitational interactions are attractive interactions.
But in gravitational gauge theory, there are repulsive interactions as
well as attractive interactions. The gravitational force is given by
eq.(12.23). The first term corresponds to classical gravitational
force. It is
$$
f_i =  g e^{I(C)}  T_{g \alpha}^0 (\partial_i C_0^{\alpha}).
\eqno{(12.59)}
$$
For quasi-static gravitational field, it is changed into
$$
\begin{array}{rcl}
f_i &=& -  g e^{I(C)}  P_{g \alpha} E_i^{\alpha}  \\
&&\\
&=& g e^{I(C)} ( M_1 E_i^0 - P_{gj} E_i^j   ),
\end{array}
\eqno{(12.60)}
$$
where $M_1$ is the gravitational mass of the mass point which
is moving in gravitational field. Suppose that the gravitational field
is generated by another mass point whose gravitational
energy-momentum is $Q_g^{\alpha}$ and gravitational mass is $M_2$.
For quasi-static gravitational field, eq.(12.45) gives out
$$
E_i^{\alpha} =
- \frac{g}{4 \pi r^3} Q_g^{\alpha} r_i
\eqno{(12.61)}
$$
Substitute eq.(12.61) into eq.(12.60), we get
$$
\svec{f} = e^{I(C)} \frac{g^2}{4 \pi r^3}
\svec{r} ( - E_{1g} E_{2g} + \svec{P}_g \cdot \svec{Q}_g ),
\eqno{(12.62)}
$$
where $E_{1g}$  and $E_{2g}$ are gravitational energy of two mass point.
From eq.(12.62), we can see that, if $\svec{P}_g \cdot \svec{Q}_g$ is
positive, the corresponding gravitational force between two momentum
is repulsive. This repulsive force is important for the stability of
some celestial object. For relativistic system, all mass point moving
at a high speed which is near the speed of light. Then the term
$\svec{P}_g \cdot \svec{Q}_g$ has approximately the same
order of magnitude as
that of $E_{1g} E_{2g}$, therefore, for relativistic systems, the
gravitational attractive force is not so strong as the force when
all mass points are at rest.
\\

\item \lbrack {\bf Massive Graviton} \rbrack
We have discussed pure gravitational gauge field in chapter 4. From
eq.(4.20), we can see that there is no mass term of graviton in the
Lagrangian. We can see that, if we introduce mass term of graviton
into Lagrangian, the gravitational gauge symmetry of the action
will be violated. Therefore, in that model, the graviton is massless.
However, just from this Lagrangian, we can not say that there is
no massive graviton in Nature. In literature \cite{6}, a new mechanism
for mass generation of gauge field is proposed. The biggest advantage
of this mass generation mechanism is that the mass term of gauge fields
does not violate the local gauge symmetry of the Lagrangian. This
mechanism is also  applicable to gravitational gauge theory.\\

In order to introduce mass term of gravitational gauge fields, we need
two sets of gravitational gauge fields simultaneously. Suppose that
the first set of gauge fields is denoted as $C_{\mu}^{\alpha}$, and
the second set of gauge fields is denoted as $C_{2 \mu}^{\alpha}$.
Under gravitational gauge transformation, they transform as
$$
C_{\mu}(x) \to  C'_{\mu}(x) =
\ehat (x) C_{\mu} (x) \ehat^{-1} (x)
+ \frac{i}{g} \ehat (x) (\partial_{\mu} \ehat^{-1} (x)),
\eqno{(12.63)}
$$
$$
C_{2 \mu}(x) \to  C'_{2 \mu}(x) =
\ehat (x) C_{2 \mu} (x) \ehat^{-1} (x)
- \frac{i}{\alpha g} \ehat (x) (\partial_{\mu} \ehat^{-1} (x)).
\eqno{(12.64)}
$$
Then, there are two gauge covariant derivatives,
$$
D_{\mu} = \partial_{\mu} - i g C_{\mu} (x),
\eqno{(12.65)}
$$
$$
D_{2 \mu} = \partial_{\mu} + i\alpha g C_{2 \mu} (x),
\eqno{(12.66)}
$$
and two different strengths of gauge fields,
$$
F_{\mu \nu} = \frac{1}{-i g}
\lbrack D_{\mu} ~~,~~ D_{\nu} \rbrack,
\eqno{(12.67)}
$$
$$
F_{2 \mu \nu} = \frac{1}{i \alpha g}
\lbrack D_{2 \mu} ~~,~~ D_{2 \nu} \rbrack.
\eqno{(12.68)}
$$
The explicit forms of field strengths are
$$
F_{\mu \nu} = \partial_{\mu} C_{\nu}(x)
- \partial_{\nu} C_{\mu}(x)
- i g C_{\mu}(x) C_{\nu}(x)
+ i g  C_{\nu}(x) C_{\mu}(x),
\eqno{(12.69)}
$$
$$
F_{2 \mu \nu} = \partial_{\mu} C_{2 \nu}(x)
- \partial_{\nu} C_{2 \mu}(x)
+ i \alpha g C_{2 \mu}(x) C_{2 \nu}(x)
- i \alpha g  C_{2 \nu}(x) C_{2 \mu}(x).
\eqno{(12.70)}
$$
The explicit forms of component strengths are
$$
F_{\mu \nu}^{\alpha} = \partial_{\mu} C_{\nu}^{\alpha}
- \partial_{\nu} C_{\mu}^{\alpha}
- g C_{\mu}^{\beta} \partial_{\beta} C_{\nu}^{\alpha}
+ g C_{\nu}^{\beta} \partial_{\beta} C_{\mu}^{\alpha},
\eqno{(12.71)}
$$
$$
F_{2 \mu \nu}^{\alpha} = \partial_{\mu} C_{2 \nu}^{\alpha}
- \partial_{\nu} C_{2 \mu}^{\alpha}
+ \alpha g C_{2 \mu}^{\beta} \partial_{\beta} C_{2 \nu}^{\alpha}
- \alpha g C_{2 \nu}^{\beta} \partial_{\beta} C_{2 \mu}^{\alpha}.
\eqno{(12.72)}
$$
Using eq.(12.63-64), we can obtain the following transformation
properties,
$$
D_{\mu} (x) \to D'_{\mu} (x)
= \ehat D_{\mu} (x) \ehat^{-1},
\eqno{(12.73)}
$$
$$
D_{2 \mu} (x) \to D'_{2 \mu} (x)
= \ehat D_{2 \mu} (x) \ehat^{-1},
\eqno{(12.74)}
$$
$$
F_{\mu \nu} \to F'_{\mu \nu} =
\ehat F_{\mu \nu} \ehat^{-1},
\eqno{(12.75)}
$$
$$
F_{2 \mu \nu} \to F'_{2 \mu \nu} =
\ehat F_{2 \mu \nu} \ehat^{-1},
\eqno{(12.76)}
$$
$$
(C_{\mu} + \alpha C_{2 \mu})
\to (C'_{\mu} + \alpha C'_{2 \mu}) =
\ehat  (C_{\mu} + \alpha C_{2 \mu}) \ehat^{-1}.
\eqno{(12.77)}
$$
The Lagrangian of the system is
$$
\begin{array}{rcl}
{\cal L}_0 &=& - \frac{1}{4} \eta^{\mu \rho} \eta^{\nu \sigma}
\eta_{2  \alpha \beta}
F_{\mu \nu}^{\alpha} F_{\rho \sigma}^{\beta}
- \frac{1}{4} \eta^{\mu \rho} \eta^{\nu \sigma}
\eta_{2 \alpha \beta}
F_{2 \mu \nu}^{\alpha} F_{2 \rho \sigma}^{\beta}.\\
&&\\
&& - \frac{m^2}{2(1+\alpha^2)}
\eta^{\mu \nu} \eta_{2   \beta\gamma}
(C_{\mu}^{\beta}  + \alpha C_{2 \mu}^{\beta})
(C_{\nu}^{\gamma} + \alpha C_{2 \nu}^{\gamma}).
\end{array}
\eqno{(12.78)}
$$
The full Lagrangian is defined by eq.(4.24) and the action is
defined by eq.(4.25). It is easy to prove that the action $S$
has local gravitational gauge symmetry.
From eq.(12.78), we can see that there is mass term of gravitational
gauge fields. Make a rotation,
$$
C_{3 \mu} = {\rm cos}\theta C_{\mu}
+ {\rm sin}\theta C_{2 \mu},
\eqno{(12.79)}
$$
$$
C_{4 \mu} = - {\rm sin}\theta C_{\mu}
+ {\rm cos}\theta C_{2 \mu},
\eqno{(12.80)}
$$
where $\theta$ is given by
$$
{\rm cos}\theta = 1 / \sqrt{1 + \alpha^2 },~~~~
{\rm sin}\theta = \alpha / \sqrt{1 + \alpha^2 }.
\eqno{(12.81)}
$$
Then, the mass term in eq.(12.78)
$$
 - \frac{m^2}{2}
\eta^{\mu \nu} \eta_{2 ~  \beta\gamma}
 C_{3 \mu}^{\beta}   C_{3 \nu}^{\gamma} .
\eqno{(12.82)}
$$
So, the gravitational gauge field $C_{3 \mu}$ is massive whose
mass is $m$ while gravitational gauge field $C_{4 \mu}$ keeps
massless. \\

The existence of massive graviton in Nature is very important for
cosmology. Because the coupling constant for gravitational interactions
and the massive graviton only take part in gravitational interactions,
the massive graviton must be a relative stable particle in Nature. So,
if there is massive graviton in Nature, there must be a huge amount of
massive graviton in Nature and they stay at intergalactic space. \\

\item \lbrack {\bf Gravitational Red Shift} \rbrack
Some celestial objects, such quasars, have great red shift. This new
quantum theory of gravity will help us to understand some kinds of
big red shift. Supposed that a photon which is emitted from an atom
on earth has definite energy $E_0$. It is known that on earth, the
gravitational field is very weak, i.e.
$$
g \eta_{1 \alpha}^{\mu} C_{\mu}^{\alpha} \ll 1,
\eqno{(12.83)}
$$
therefore, the factor $e^{I(C)}$ is almost 1 on earth. If the atom is
not on earth, but on a celestial object which has strong gravitational
field, according to eq.(5.20), eq.(6.18), and eq.(7.17), the inertial
energy of the photon is
$$
e^{I(C)} \cdot E_0.
\eqno{(12.84)}
$$
Suppose that the gravitational mass of celestial object is $M$,
according to eq.(9.10) and eq.(9.17), we have
$$
e^{I(C)} = exp(-\frac{GM}{r}).
\eqno{(12.85)}
$$
According to eq.(12.19),
at that place that the photon is emitted, the gravitational potential
energy of the photon is
$$
- E_0 \cdot exp(-\frac{GM}{r}) \cdot  \frac{GM}{r}.
\eqno{(12.86)}
$$
When this photon arrive earth, its inertial energy is
$$
E_0 \cdot exp(-\frac{GM}{r}) \cdot (1 -  \frac{GM}{r} ).
\eqno{(12.87)}
$$
Therefore, its gravitational red shift is
$$
z =  exp(\frac{GM}{r}) \cdot (1 -  \frac{GM}{r} )^{-1} -1.
\eqno{(12.88)}
$$
If the mass of the celestial object is very big and the radius of the
celestial object is small, its red shift is big. As an example,
suppose that $\frac{GM}{r} \sim 0.6$, its gravitational red shift
is about 3.6. In other words, the gravitational red shift predicted
by gravitational gauge theory is much larger than that predicted
by general relativity.\\

\item \lbrack {\bf Violation of Inverse Square Law
And Black Hole} \rbrack
It is known that classical Newton's theory of gravity predicts that
gravity obeys inverse square law. According gravitational gauge
theory, the inverse square law will be violated by intense
gravitational field. According to eq.(12.65) and eq.(9.10),
the gravitational force between two relative rest objects is
$$
\svec{f} = - exp(- \frac{G M}{r}) \cdot
\frac{G M M_2}{ r^3}  \svec{r} ,
\eqno{(12.89)}
$$
where $M$ and $M_2$ are gravitational masses of two objects.
In eq.(12.89), we have supposed that $M_2$ is much less that $M$.
We can clearly see that the factor $exp(- \frac{GM}{r})$ violates
inverse square law. We can also see that, if the distance $r$ approaches
zero, the gravitational force does not approach infinity. This
result is important for cosmology. If there were no the factor
$exp(- \frac{GM}{r})$ and if the distance $r$ approaches zero,
the gravitational force would approach infinity. For black hole,
there is no force that can resist gravitational force, therefore,
black hole will collapse forever until it becomes a singularity
in space-time. Because of the existence of the factor
$exp(- \frac{GM}{r})$, the gravitational force does not obey
inverse square law, and therefore the black hole will not
collapse forever. In this point of view, black hole is not a
singularity in space-time, and therefore, black hole has its own
structure. \\

According to eq.(12.19), the gravitational potential $\phi(r)$ which
is generated by an celestial object with gravitational mass $M$ is
$$
\phi (r) =  - exp(- \frac{G M}{r})  \frac{G M}{r}.
\eqno{(12.90)}
$$
Suppose that the gravitational mass $M$ is a constant.
From eq.(12.90), we can see that, when distance $r$  approaches zero
or infinity, the gravitational potential approaches zero. But if
distance $r$ is a finite but non-zero, the gravitational potential
is negative. There must be a definite distance $R_0$ where the
gravitational potential reaches its minimum. The  distance $R_0$
is given by
$$
\frac{\rm d}{{\rm d}r} \phi (r) |_{r = R_0} = 0.
\eqno{(12.91)}
$$
Using eq.(12.90), we can easily determine the value of $R_0$,
$$
R_0 = G M.
\eqno{(12.92)}
$$
From eq.(12.89), we know that, when $r = R_0$, the gravitational
red shift is infinity, and therefore outside world can not see
anything happens in this celestial object. In other words, if
the radius of an celestial object is $R_0$, it will be a black
hole, and the gravitational potential reaches its minimum at the
surface of the celestial object. \\

Suppose that there is a mass point with mass $m$ locate at the
surface of the celestial object. According to eq.(12.89),
the gravitational force that acts on the mass point is
$$
f(r) = exp(- \frac{G M}{r}) \cdot \frac{G M m}{ r^2}.
\eqno{(12.93)}
$$
Suppose that the gravitational force $f(r)$ is strongest at distance
$R_1$. The distance $R_1$ is given by
$$
\frac{\rm d}{{\rm d}r}  f(r) |_{r = R_1} = 0.
\eqno{(12.94)}
$$
From eq.(12.93), we get
$$
R_1 = \frac{G M}{2}= \frac{R_0}{2}.
\eqno{(12.95)}
$$
So, when the distance $r$ becomes shorter than $R_1$, the gravitational
force will become weaker. The radius of black hole is determined by
the balance where gravitational force is equivalent to pressure. \\

Besides, the renormalization effects will change the value of
gravitational coupling constant $g$, and therefore affect inverse
square law. This effect is a quantum effect. \\

\item \lbrack {\bf Energy Generating Mechanism} \rbrack
It is known that some celestial objects, such as quasar, pulsar,
$\cdots$, radiate huge amount of energy at one moment. Where is
the energy comes from? We know that the gravitational field of these
celestial objects are very strong, and therefore the gravitational
wave radiation will also be very strong. According to gravitational
gauge theory, the inertial energy of gravitational field $C_{\mu}^0$
is negative. For ordinary celestial objects, their moving speed is
much less than the speed of light. In this case, the dominant
component of gravitational wave is $C_{\mu}^0$, therefore,
gravitational wave carries negative inertial energy. It means that
ordinary celestial objects obtain inertial energy through gravitational
wave radiation. It is the source of part of the thermal radiation
energy of these celestial objects.  It is also the ultimate energy
source of the whole Universe.   \\

Gravitational wave radiation may cause disastrous consequence for
black hole. Because of intense gravitational force of black hole,
any positive energy can not escape from black hole, but 
negative energy can escape from black hole. Because gravitational
gauge field $C_{\mu}^0$ carries negative inertial energy and negative
gravitational energy, it can escape from black hole. In other words,
black hole can radiate gravitational wave, but it can not radiate
electromagnetic wave. Black holes can obtain inertial energy through
gravitational wave radiation, but it can not radiate inertial
energy to outside world. As a result of gravitational wave radiation,
the inertial energy of black hole becomes larger and larger,
its temperature becomes higher and higher, and its pressure becomes
higher and higher. Then at a time, gravitational force can not
resist pressure from inside of the block hole. It will burst and
release huge amount of inertial energy at a relative short time. \\

\item \lbrack {\bf Dark Matter} \rbrack
Dark matter is an important problem in cosmology. In gravitational gauge
field theory, the following effects are helpful to solve this problem:
1) The existence
of massive graviton will contribute some to dark matter.
2) If the gravitational magnetic field is strong inside a celestial
system, the Lorentz force will provide additional centripetal force
for circular motion of a celestial object. 3) The existence of the
factor $e^{I(C)}$ violate inverse square law of classical gravity.
Besides, there are a lot of other possibilities which is widely
discussed in literature. I will not list them here, for they
have nothing to do with the gravitational gauge theory. \\

\item \lbrack {\bf Particle Accelerating Mechanism} \rbrack
It is known that there are some cosmic rays which have extremely
high energy. It is hard to understand why some cosmic particles
can have energy as high as $10^{21}$ eV. Gravitational gauge
theory gives a possible explanation for this phenomenon. According
to gravitational gauge theory, there are strong gravitational
magnetic field inside a galaxy. A cosmic particle will make circular
motion around the center of galaxy under Lorentz force which is
provided by gravitational magnetic field and classical Newton's
gravitational force. These cosmic particles
will radiate gravitational wave when they moving in gravitational
field. Because gravitational wave carries negative inertial
energy, cosmic particles will be accelerated when they radiate
gravitational wave. If this explanation is correct, most cosmic
particles which have extremely high energy will come from the center
of a galaxy.  \\

\item \lbrack {\bf Equivalence Principle} \rbrack
Equivalence principle is one of the most important foundations of
general relativity, but it is not a logic starting point of
gravitational gauge field theory. The logic starting point of
gravitational gauge field theory is gauge principle. However,
one important inevitable result of gauge principle is that
gravitational mass is not equivalent to inertial mass. The origin
of violation of equivalence principle is gravitational field. If there
were no gravitational field, equivalence principle would strictly
hold. But if gravitational field is strong, equivalence principle
will be strongly violated. For some celestial objects which have
strong gravitational field, such as quasar and black hole, their
gravitational mass will be higher than their inertial mass.
But on earth, the gravitational field is very weak, so the equivalence
principle almost exactly holds. We need to test the validity of
equivalence principle in astrophysics experiments. \\

\item \lbrack {\bf Violation of the Second Law of
    Thermodynamics } \rbrack
If we treat the whole universe as an isolated system, according
to the second law of thermodynamics, our Universe will finally
go to the completely statistical equilibrium, which is main point
of the theory of heat death. If we consider the influence of
gravitational interactions, second law of thermodynamics no longer
holds, for an object can obtain erengy through radiating gravitational
wave. For example, black hole can obtain energy from outside world
through gravitational wave radiation, though the temperature of
black hole is much higher than outside world.
In this meaning, black hole
is a perpetual motion machine of the Universe, and because of the
existence of this perpetual motion machine, our Universe will
not go to the state of heat death.
\\

\item \lbrack {\bf New Energy Source} \rbrack
It is known that one of the biggest problem for the development
of civilization of human kind in future
is the energy crisis. However, gravitational
gauge field theory provides an everlasting energy source for both
human kind and the whole universe. Two hundred years ago, human kind
does not know how to utilize electric energy, but now most energy
comes from electric energy. It is believed that, human kind will
eventually know how to utilize gravitational energy in future.
If so, most energy will come from perpetual motion machine
in future.\\

\end{enumerate}

\section{Summary}

In this paper, we have discussed a completely new quantum gauge
theory of gravity. Finally, we give a simple summary to the
whole theory.

\begin{enumerate}

\item In leading order approximation, the gravitational gauge field
theory gives out classical Newton's theory of gravity.

\item In first order approximation and for vacuum, the gravitational
gauge field theory gives out Einstein's general theory of relativity.

\item Gravitational gauge field theory is a renormalizable quantum
theory.

\end{enumerate}

\end{document}